\documentclass[12pt,openany]{book}

\usepackage{amsmath, mathrsfs,amssymb, amsfonts}
\usepackage{subfig}
\usepackage{wrapfig}
\usepackage[]{fancyhdr, eqnarray, mathtools,graphicx, textcomp, relsize,etoolbox}
\usepackage[section]{placeins}
\usepackage[utf8]{inputenc}
\usepackage{lipsum}
\usepackage{tabu,longtable,multirow,multicol,booktabs}
\usepackage{gensymb}
\usepackage{setspace}
\usepackage[inline]{enumitem}
\usepackage[usenames,dvipsnames,svgnames,table]{xcolor}
\usepackage[margin=1in]{geometry}
\usepackage{rotating} 
\usepackage{tabulary}
\usepackage[toc]{appendix}

\usepackage[]{float}

\usepackage{wasysym}
\usepackage{BOONDOX-cal}

\def\normz{\mathcal{z}}
\def\myascnode{\ascnode}

\def\myPhase{\epsilon}
\def\cf{\textit{c.f.}}
\def\myKepler{\textit{Kepler}}
\def\myECIES{ECIE's}
\def\exonest{EXONEST}

\def\1over2{\frac{1}{2}}
\def\piover2{\frac{\pi}{2}}

\def \mycorrection{$ \mathcal{K}_c $}
\def \mychange{$\Delta\mathcal{JLC} $}
\def \mynew{$ \mathcal{JLC} $}
\def \myfillin{$ \mathcal{JLC}_{clarify} $}

\def\myIntensity{I}
\def\myIntensityDis{\mathcal{J}}
\def\myLuminosity{\mathcal{L}}
\def\mySinEta{\chi}
\def\myScatteringAlbedo{\varpi_0}
\def\myPhaseFunc{\Psi(\myPhase)}
\def\myRefCo{\varrho(\gamma, \gamma', \phi)}
\def\mySinLambda{\mathscr{X}}
\def\myexpansionz{\zeta}
\def\myexpansionparameter{\delta}
\def\mypen1limit{pen1}

\def\mytitle{Estimation of Planetary Photometric Emissions for Extremely Close-in Exoplanets} 
\def\myname{Jennifer L. Carter} 
\def\myType{Dissertation}
\def\myDegree{Doctor of Philosophy} 
\def\myCollege{College of Arts and Sciences} 
\def\myDept{Department of Physics} 
\def\myUni{University at Albany} 
\def\myfullUni{\myUni, State University of New York} 
\def\mydate{2018} 

\newcommand*{\Eval}[3]{\left.#1\right\rvert_{#2}^{#3}} 

\newcommand*{\myKintegral}[4]{K \left( #1, \left[#2, #3\right]\right) \left\lbrace #4 \right\rbrace }
\newcommand*{\myQnCoseps}[4]{Q_{C}^{#4} \left( #1, \left[#2, #3\right]\right)}
\newcommand*{\myQnSineps}[4]{Q_{S}^{#4} \left( #1, \left[#2, #3\right]\right)}
\newcommand*{\myRnCoseps}[4]{R_{C}^{#4} \left( #1, \left[#2, #3\right]\right)}
\newcommand*{\myRnSineps}[4]{R_{S}^{#4} \left( #1, \left[#2, #3\right]\right)}

\newcommand*{\myQnCosepsshort}[1]{Q_{C}^{#1}}
\newcommand*{\myQnSinepsshort}[1]{Q_{S}^{#1}}
\newcommand*{\myRnCosepsshort}[1]{R_{C}^{#1}}
\newcommand*{\myRnSinepsshort}[1]{R_{S}^{#1}}

\newcommand*{\myKopalPhi}[3]{\Phi_{#1} (\mySinEta_{#2},#3)}

\newcommand*{\myF}[2]{\mathcal{F}(\mySinEta_{#1},#2)}

\newcommand*{\mycase}[1]{\textbf{Case #1}}

\newcommand*{\invcos}[1]{\cos^{-1}\left(#1\right)}

\newcommand*{\invsin}[1]{\sin^{-1}\left(#1\right)}

\newcommand*{\invtan}[1]{\tan^{-1}\left(#1\right)}

\newcommand*{\invcoth}[1]{\coth^{-1}\left(#1\right)}
\newcommand*{\invcosh}[1]{\cosh^{-1}\left(#1\right)}

\newenvironment{dedication}%
  {%
    \newpage
	\thispagestyle{fancy}
    \mbox{}
    \vfil
    \begin{center}%
  }%
  {%
    \end{center}%
    \vfil
	\newpage
  }

\newenvironment{abstract}
  {
	\thispagestyle{fancy}
	\addcontentsline{toc}{section}{{\bfseries Abstract}}
	\begin{center}
		{\bfseries\Large Abstract}
	\end{center}
  }
  {
	\newpage
  }

\newenvironment{acknowledgments}
  {
	\addcontentsline{toc}{section}{{\bfseries Acknowledgments}}
	\thispagestyle{fancy}
	\begin{center}
		{\bfseries\Large Acknowledgments}
	\end{center}}%
  {
	\newpage
  }
\makeatletter
\renewcommand*\env@matrix[1][\arraystretch]{%
  \edef\arraystretch{#1}%
  \hskip -\arraycolsep
  \let\@ifnextchar\new@ifnextchar
  \array{*\c@MaxMatrixCols c}}
\makeatother

\usepackage[]{hyperref}
\hypersetup{
    unicode=true,          
    pdftoolbar=true,        
    pdfmenubar=true,        
    pdffitwindow=false,     
    pdfstartview={FitH},    
    pdftitle={\mytitle},  
    pdfauthor={\myname},     
    pdfcreator={\myname},   
    pdfnewwindow=true,      
    colorlinks=true,       
    linkcolor=black,          
    citecolor=black,        
    filecolor=black,      
    urlcolor=blue           
}
\usepackage[font = {small}, margin = 10pt, labelfont = bf, labelsep = colon]{caption} 
\usepackage[noabbrev,nameinlink]{cleveref}      

\crefname{equation}{Equation}{Equations}
\crefname{figure}{Figure}{Figures}
\crefname{table}{Table}{Tables}
\crefname{section}{Section}{Sections}
\crefname{chapter}{Chapter}{Chapters}
\crefname{app}{Appendix}{Appendices}

\let\counterwithin\relax 
\usepackage{chngcntr}
\counterwithin{table}{section}
\counterwithin{figure}{section}
\counterwithin{equation}{section}

\usepackage{titlesec}
\titleformat{\chapter}[display]
  {\normalfont\Large\bfseries\centering}{CHAPTER \thechapter:}{0em}{}
\titlespacing*{\chapter}{0pt}{3.5ex plus 1ex minus .2ex}{2.3ex plus .2ex}

\titleformat{\section}{\normalfont\large\bfseries}{\thesection}{1em}{}{}
\titlespacing*{\section} {0pt}{3.25ex plus 1ex minus .2ex}{2.1ex plus .2ex}

\titleformat{\subsection}{\normalfont\bfseries}{\thesubsection}{1em}{}{}
\titlespacing*{\subsection} {0pt}{3ex plus 1ex minus .2ex}{1.5ex plus .2ex}

\titlespacing*{\subsubsection}{0pt}{3ex plus 1ex minus .2ex}{1.5ex plus .2ex}

\makeatletter
\AtBeginEnvironment{appendices}{%
	\clearpage%
		\titleformat{\chapter}[display]
		{\normalfont\Large\bfseries\centering}{APPENDIX \thechapter:}{0em}{}
		\titlespacing*{\chapter}{0pt}{3.5ex plus 1ex minus .2ex}{2.3ex plus .2ex}
}
\makeatother

\usepackage [english]{babel}
\usepackage [autostyle, english = american]{csquotes}
\MakeOuterQuote{"}

\begin{document}

\doublespacing 

\frontmatter
%
%
%
%
%
%


\begin{titlepage}

\vfill

\newcommand{\HRule}{\rule{\linewidth}{0.5mm}} 

\centering 

\MakeUppercase{\bfseries\mytitle}

\vfill
by

\vfill
\myname
\vfill

A \myType\\
Sumbitted to the \myfullUni\\
In Partial Fulfillment of the \\
Requirements for the Degree of\\
\myDegree\\
\vfill
\myCollege\\
\singlespacing
\myDept\\
\doublespacing
\mydate


\vfill 

\end{titlepage}

%

\begin{dedication}
  To my family, friends, colleagues and students. Without you, it just wouldn't be worth it. 
\end{dedication}





\begin{abstract}
The fine precision of photometric data available from missions like \myKepler\ provide researchers with the ability to measure changes in light on the order of tens of parts per million (ppm). This level of precision allows researchers to  measure the loss of light due to exoplanet transits as well as the light emitted by an exoplanet, or planetary photometric emissions. The planetary photometric emissions are due to the thermal emissions of the exoplanet, and reflected stellar light. In many cases it is assumed that the incident stellar light may be modeled as plane parallel rays. 
For extremely close-in exoplanets the finite angular size of the host star must be taken into account and the plane parallel ray model breaks down. One consequence of modeling the incident stellar radiation in this manner is the creation of three distinct zones as opposed to the two zones present in the plane parallel ray model. The three zones are the fully illuminated, penumbral, and un-illuminated zones. 
The existence of the penumbral zone means that more than half of the exoplanet will be at least partially illuminated by the host star. 
In this work we will present a complete derivation of the reflective luminosity of the fully illuminated zone. 
In addition, we will present an outline for the derivation of the reflected intensity distribution of the penumbral zone. Within this work we will also derive a new expression for the thermal luminosity of an exoplanet by treating each of the three zones as a blackbody emitting at a constant temperature. 
Finally, it will be shown that an estimation of the radius of the exoplanet requires proper accounting of light from the penumbral zone during the primary transit. Not doing so risks underestimating the radius of the exoplanet and an overestimation of its geometric albedo and nightside temperature. 

\end{abstract} 

\begin{acknowledgments}
	I will begin by thanking my advisor, Prof. Kevin Knuth, for his guidance and support. His questions about the reflected light and its connection to the day side temperature of Kepler-91b led to the final topic of this thesis and his encouragement kept me going through the rough patches.

	I would also like to thank the members of my thesis committee. Prof. Ariel Caticha served as my first introduction to UAlbany and helped convince me that I was capable of completing a doctorate in physics. I met Prof. Matthew Szydagis midway through my graduate career and found in him someone who actively supported undergraduate research and education, a cause in which I hold a personal interest. His work with undergraduate students continues to encourage me to act as a mentor to undergraduates. 
	Next, I would like to thank Prof. Oleg Lunin not only for his serving on my committee, but also for his teaching me to use Mathematica as a fine precision tool and not a sledge hammer, and for his practical guidance in research and my job search.
	In addition, I would like to thank Dr. Michael Way for sitting on my committee and his kind introduction to Dr. Caleb Scharf. 

	Finally, I would like to thank the members of the KnuthLab and my fellow graduate students. The conversations I shared with Anthony Gai, Bertrand Carado, James Walsh, Kevin Wynne and Steven Young always served to improve my understanding of physics, research, teaching, and exoplanets.
\end{acknowledgments}

\tableofcontents

\mainmatter
%
%
%

\chapter{Introduction}\label{ch:intro}
Since the discovery of the first exoplanets, researchers have sought to characterize their properties and attempted to determine the distribution of exoplanet types in the universe. To that end, evermore precise tools have been created to detect exoplanets. The two most successful methods of detection thus far have been the radial velocity, or RV, and transit method of detection, \cf\ the The Extrasolar Planet Encyclopedia \cite{exoEncyclopedia}. The radial velocity method of exoplanet detection takes advantage of the fact that an exoplanet and its host star will orbit their common center of mass. If the orbit is correctly aligned, the host star will have a velocity along our line of sight allowing researchers to measure the Doppler shift in the spectral lines of the star to infer the presence of the exoplanet, \cf\ Perryman's book \cite{exohandbook}. In contrast, the transit method of detection infers the presence of an exoplanet by observing the small dips in photometric light as the exoplanet passes in front of its host star along our line of sight, \cf\ \cite{exohandbook}.

In 2009, the \myKepler\ space telescope was launched with the goal to discover earth-like exoplanets orbiting sun like stars using the transit detection method. To meet this goal, the telescope observed over 100,000 stars for a period of just over four years at very high precision. The \myKepler\ space telescope was capable of detecting changes in light of 29 parts per million (ppm) on a magnitude 12  star with a 6.5 hour integration time \cite{KeplerPrecision}. As researchers began to characterize the exoplanets detected by \myKepler\ they found it necessary to develop more precise models to describe the wide variety of exoplanets. Transits alone provide information about the radius and orbital parameters of an exoplanet, but other photometric variations may provide information about the mass, reflectivity, and temperature, \cf\  Placek, Knuth and Angerhausen \cite{Placek2014}.

The photometric variations within a light curve may be split into two main categories: stellar photometric variations and planetary photometric variations. The stellar variations, which are induced by the presence of the exoplanet, include boosted light and ellipsoidal variations. The boosted light is due to the relativistic effects of the star moving along our line of sight, similar to the effects used in the RV method of exoplanet detection. The ellipsoidal variations are due to tidal interactions between the exoplanet and the host star. The gravity produced by an orbiting exoplanet will pull on the surface of the host star resulting in the star being non-spherical; therefore, its observed cross-sectional area will change throughout the orbit, \cf\ \cite{exohandbook,Placek2014}. 

The planetary photometric variations arise from light emitted by the planet itself, which comes in two forms: reflected light and thermal light, \cf\ \cite{exohandbook,Placek2014}. 
The main topic of this work will be the determination of the reflected luminosity of extremely close-in exoplanets, or \myECIES, as a function of phase angle. Generally speaking, it is appropriate to approximate the nature of the incident stellar radiation received by an exoplanet as being plane parallel rays such that half of the exoplanet is fully illuminated and half is completely un-illuminated, \cf\ Seager's book \cite{seager}. This approximation holds for large star-planet separations, but as the separation approaches a few stellar radii one must consider the finite angular size of the host star to accurately describe the incident stellar radiation. This results in an exoplanet whose surface experiences three distinct types of illumination as opposed to the two produced by the plane parallel ray approximation, as described by Kopal in \cite{Kopal1953,Kopal1959}. The first is the fully illuminated zone in which the entirety of the apparent disk of the host star is visible. Second, is the un-illuminated zone which receives no radiation from the host star. Finally, between the two zones lies the penumbral zone, which exists in twilight because it receives light from only part of the host star. 

Following is a summary of the contents of this work  and an outline of its contributions to the field of exoplanet research by chapter. To begin, \cref{ch:orbital}\ we will review orbital mechanics and provide a set of parameters required to describe the characteristics of exoplanet orbits and light curves. In addition, a brief discussion of the current state of the RV method of exoplanet detection will be presented. \cref{ch:transitchapter} presents a description of transit light for both a uniformly emitting star and one that exhibits limb-darkening as described by Mandel and Agol in \cite{Mandel2002}. Next, \cref{ch:hoststar} provides a description of the photometric variations of the host star from previous works. 

The bulk of this work is contained within \cref{ch:planetarychapterReflintensity,ch:reflectedlightluminosity}, in which a description of the reflected light for both plane parallel ray incident stellar radiation, \cf\ \cite{seager}, and the incident radiation experienced by extremely close-in exoplanets will be presented. The new approach to modeling the reflected intensity distribution of the penumbral zone described in \cref{ch:planetarychapterReflintensity}\ was developed because it was discovered that the methods described in Kopal's work, \cite{Kopal1953,Kopal1959}, produces negative luminosity, see \cref{ch:reflectedlightluminosity}. It will also be shown that the penumbral zone itself may be split into two penumbral zones. In addition, a problem was found in which the equations presented in \cite{Kopal1953,Kopal1959}\ for the fully illuminated zone also produced negative luminosity. A new analysis of the luminosity of the fully illuminated zone is presented in \cref{ch:reflectedlightluminosity}\ in addition to an overview of the previous work describing the reflected luminosity of exoplanets. As part of the process we provide to the community an analysis of the required integrals in \cref{app:integrations}.

In \cref{ch:thermalradiation}\ we will describe the time dependent thermal variations of an exoplanet with a range of thermal zones, \cf\ \cite{seager}, as applicable to exoplanets in general and for \myECIES\ in particular. Specifically, a new analysis of the thermal radiation will be presented which uses the four zones described in \cref{ch:planetarychapterReflintensity}\ as opposed to the two zones used in previous work.

Finally, \cref{ch:lightcurvechapter}\ describes the method used to add up the photometric variations, \cf\ Placek's description in \cite{PlacekThesis}. A comparison will be made between the photometric variations described with plane parallel incident radiation and with the finite angular size of the host star taken into consideration and the ability to distinguish the two models will be discussed. Finally, a discussion of conclusions and future work will be presented in \cref{ch:conclusion}.

\chapter{Orbital Mechanics}\label{ch:orbital}
To properly describe the relationship between the light curve of a star-planet system and the parameters describing that system one must first have an understanding of orbital mechanics. As such, we will present in this chapter the derivation of the equations that govern the position of an exoplanet as a function of time. First, we will consider the motion about its host star using the Lagrangian for two masses under the influence of gravity, \cf\ Landau's textbook \cite{landauMech}. Next, we will consider Kepler's problem to describe the star-planet separation as a function of the orbital anomalies. Third, we will discuss the Euler angles used to characterize the orientation of an exoplanet's orbit along a line of sight. Finally, we will present the derivation for the equation describing the radial velocity of an exoplanet, \cf\ Perryman's book \cite{exohandbook}.  
 
\section{Equations of Motion}
To begin, consider two masses, $m_1$ and $m_2$, at positions $\vec{r}_1$ and $\vec{r}_2$ respectively. The Lagrangian for the masses is given by 
\begin{equation} 
L = \frac{1}{2}M_{\textrm{total}}\dot{\vec{R}}^{\,2} + \frac{1}{2}\mu \dot{\vec{r}}^{\,2} -V(\vec{r}),
\end{equation}
where the total mass, $M_{\textrm{total}}$, is the sum of the two masses and the reduced mass, $\mu$ is given by 
\begin{equation}
\mu = \frac{m_1m_2}{m_1+m_2}.
\end{equation}
The variable $\dot{\vec{R}}$, is the velocity of the center of mass of the system and $\vec{r} = |\vec{r}_1 -\vec{r}_2|$ is the relative position of the two masses, \cf\ Landau's textbook \cite{landauMech}. Finally, the potential energy of the system is given by the function $V(\vec{r})$. We will consider the case in which the center of mass remains stationary, i.e. $\dot{\vec{R}}$ = 0, and the potential depends only on the distance between the two particles, or $r = \sqrt{|\vec{r}|^{\,2}}$, we may write,
\begin{equation}
\begin{aligned}\label{eq:polarL}
L & = \frac{1}{2}\mu \dot{\vec{r}}^{\,2} -V(r)\\
			& = \frac{1}{2}\mu (\dot{r}^2 + r^2\dot{\theta}^2) - V(r).
\end{aligned}
\end{equation}
Note that the motion of the two particles is such that the two are moving in a plane; therefore, we have chosen to write the Lagrangian in \cref{eq:polarL}\ using the polar coordinates of a single mass, $ \mu $, under the influence of the potential $V(r)$, where $ \theta $ is the angular coordinate of the mass. 

The angular equation of motion is then
\begin{equation}\label{eq:momentum}
\ell = \mu  r^2 \dot{\theta},
\end{equation}
where $\ell = \partial L/\partial\dot{\theta}$ is the angular momentum of $\mu$ and is a constant of the motion. The radial equation of motion is
\begin{equation}\label{eq:radial}
\mu\ddot{r} = \mu r \dot{\theta}^2  + f(r),
\end{equation}
where $f(r) = -\partial V/\partial r$, is the force due to $ V (r) $ acting on the particle. To determine the radial position of the single mass in terms of $\theta$, one must solve~\cref{eq:radial}, which can be re-written by substituting \cref{eq:momentum}\ into  \cref{eq:radial} to find
\begin{equation}\label{eq:radial2}
\mu \ddot{r} = \frac{\ell^2 }{\mu r^3}+f(r).
\end{equation}
\cref{eq:radial2} is a re-statement of Newton's second law and shows that the force directed along the radial coordinate is the sum of the pseudo-force $\ell^2 /(\mu r^3)$, i.e. the centrifugal force, and $f(r)$, the force resulting from the potential $V(r)$.

Our goal is to describe the trajectory of $\mu$ under the influence of a gravitational potential; therefore, we will consider the solution to \cref{eq:radial2}\ for a central potential. Such potentials are of the form of $V = -k/r$.  To determine the position of $m_2$ relative to $m_1$ in a gravitational potential we let $k = Gm_1$. The solution can be derived by making a change of variables of $u = 1/r$. The foregoing substitutions reveal that
\begin{equation}\label{eq:V}
	V=\frac{-G m_{1}}{r}=-G m_{1}u
\end{equation}
and
\begin{equation}\label{eq:ddr}
\ddot{r} = -\left(\frac{\ell u}{\mu}\right)^2\frac{d^2u}{d\theta^2}.
\end{equation}
Which can be substituted into \cref{eq:radial2}\ to reveal
\begin{equation}\label{eq:udoubleprime2}
u'' = -u+\frac{\mu k}{\ell^2}.
\end{equation}
where $u''$ indicates the second derivative of $u$ with respect to $\theta$.

Further substitutions may now be used to finish the derivation. With $w = u-\mu k/\ell^2$ we may re-write \cref{eq:udoubleprime2}\ as simply
\begin{equation}
w''(\theta) = -w(\theta),
\end{equation}
which has the general solution of $ w(\theta) = A\cos(\theta - \theta_0) $, where $A$ and $\theta_0$ are set by some initial conditions. Solving for $u(\theta)$ then gives
\begin{equation}
\begin{aligned}
u(\theta) & = A\cos(\theta-\theta_0)+\frac{\mu k}{\ell^2} \\
	      & = \frac{\mu k}{\ell^2}(1+\epsilon\cos(\theta-\theta_0)),
\end{aligned}
\end{equation}
where $\epsilon = A\ell^2/(k\mu)$. Using the relationship $ u=1/r $ and setting $ C=\ell^2 /(\mu k) $ produces
\begin{equation}\label{eq:radialsolved}
r(\theta) = \frac{C}{1+\epsilon\cos(\theta-\theta_0)}.
\end{equation}

\cref{eq:radialsolved}\ may be re-written in terms of eccentricity, $e$, and the semi-major axis, $a$, by first considering the Cartesian coordinates of the orbit given by
\begin{equation}\label{eq:cart1}
\begin{aligned}
r & = \sqrt{x^2+y^2} \\
x & = r\cos\nu \\
y & = r\sin\nu. 
\end{aligned}
\end{equation}
In~\cref{eq:cart1} we have used the fact that $\theta_0$ is normally taken to be zero because it corresponds to an arbitrary starting point in the orbit, and $\theta$ corresponds to the true anomaly of the orbit and will now be referred to as $\nu$ to coincide with the notation often used in exoplanet research. In celestial mechanics the angles used to describe the location of a point in an elliptical orbit are referred to as orbital anomalies in reference to the anomalous motion of planets in the sky. Next, we shall rearrange \cref{eq:radialsolved} as $r+\epsilon r\cos\nu = C$ and substitute \cref{eq:cart1}. Solving for $ r^2 =x^2 +y^2  $ reveals
\begin{equation}
x^2+y^2 = (C-\epsilon x)^2.
\end{equation}
Rearranging the previous equation produces
\begin{equation} 
\left(x+\left(\frac{\epsilon C}{1-\epsilon^2}\right)\right)^2\left(\frac{1-\epsilon^2}{C}\right)^2+y^2\left(\frac{1-\epsilon^2}{C^2}\right) = 1,
\end{equation}
or 
\begin{equation}\label{eq:ellipse}
\frac{\left(x+d\right)^2}{a^2}+\frac{y^2}{b^2} = 1,
\end{equation}
where $b = C/\sqrt{1-\epsilon^2} = a\sqrt{1-\epsilon^2}$, $a = C/(1-\epsilon^2)$, and $d = a\epsilon$.~\cref{eq:ellipse} is the Cartesian equation of an ellipse and allows us to identify $C$ as the semilatus rectum $C = \ell^2/(k\mu) = A\epsilon$ and $\epsilon$ as the eccentricity, $e$ of an ellipse. We may then write the radial equation for the two body problem as
\begin{equation}\label{eq:radialellipse}
r(\nu) = \frac{a(1-e^2)}{1+e\cos\nu},
\end{equation}
where $ r $ is the radial separation between the two bodies. To determine the separation as a function of time, $ r(t) $, one need only determine $\nu$ as a function of time.  \cref{eq:radialellipse}\ is a statement of Kepler's first law, namely that the orbits of planets follow an ellipse, where $r$ is the separation between the centers of the two bodies.

\section{The Orbital Anomalies}\label{sec:orbitalanomalies}
\begin{figure}[tbh]
\centering
\includegraphics[width=0.5\textwidth]{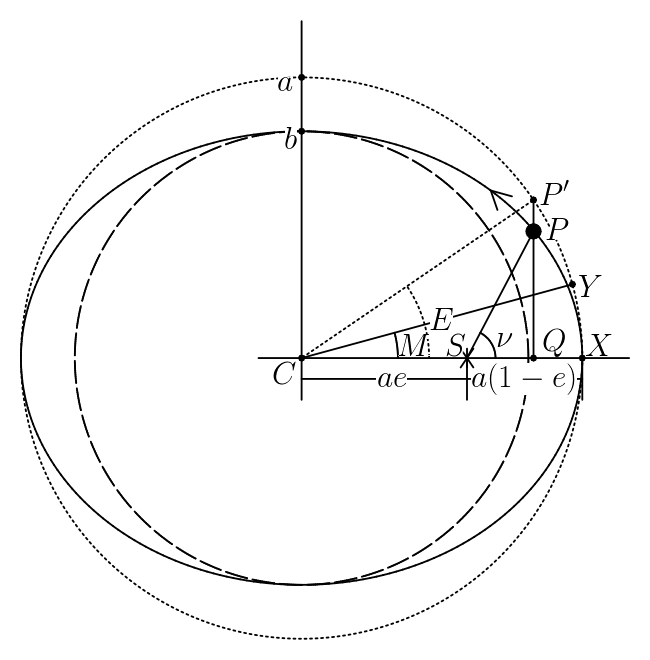}
\caption{Illustration of the relationship between the three orbital anomalies; $M$, the mean anomaly, $E$, the eccentric anomaly and $\nu$, the true anomaly. The geometric center of the orbit is at point $C$ and the star is located at point $S$. The orbit of the exoplanet is represented by the solid line. The dashed line represents the auxiliary circle of radius equal to the semi-minor axis of the exoplanet's orbit, $b$. The dotted line represents a second auxiliary circle of radius equal to the semi-major axis of the exoplanet's orbit, $a$. The exoplanet is located at point $P$ on the ellipse and point $ P' $ marks the projection of $ P $ onto the auxiliary circle of radius $ a $.}
\label{fig:anomalyfigure}
\end{figure}

To determine the true anomaly as a function of time we will make use of two other angles, the mean anomaly, $M$, and the eccentric anomaly, $E$; the relationship between the three angles is illustrated in~\cref{fig:anomalyfigure}. The point of closest approach of the exoplanet to the star is called periastron and is marked by the point $X$. The mean anomaly is named as such because it describes the mean motion of the exoplanet about its host star, i.e. $M$ is the angular distance between the exoplanet's current position and periastron if it was moving in a circular orbit with constant angular velocity. The radius of the circular orbit used to calculate the mean anomaly is the length of the semi-major axis of the elliptical orbit of the planet, $ a $. In addition, the value of $ M $ is such that the area of the triangle $ \bigtriangleup CYX $ is the same as $ \bigtriangleup SPX $. The eccentric anomaly is the angular distance between the exoplanet projected onto an auxiliary circle of radius equal to the semi-major axis and periastron. Both $M$ and $E$ are measured relative to the geometric center of the orbit at point $C$. The host star is located at one of the foci of the ellipse at point $S$ in \cref{fig:anomalyfigure}. The true anomaly, $\nu$, is the angular distance between the exoplanet and periastron measured using the host star as the vertex of the angle. 

To calculate the mean anomaly we make use of its definition given previously. For a complete orbit the exoplanet travels through $2\pi$ radians in a time equal to its period, $T$; therefore, the exoplanet has transversed $2\pi t/T$ radians in time $t$.  The mean anomaly is then given by,
\begin{equation}\label{eq:mean}
\begin{aligned}
M(t) & = M_0+\frac{2\pi t}{T} \\
	 & = \frac{2\pi}{T}(t-t_p),
\end{aligned}
\end{equation}
where $M_0$ is the initial mean anomaly of the orbit and $t_p$ is the the time since periastron as measured at the start of observation. 

The ratio of the areas of the triangles $\bigtriangleup SPX$ and $\bigtriangleup SP'X$ in~\cref{fig:anomalyfigure} can be used to determine the relationship between the mean anomaly and the eccentric anomaly as follows. We begin by considering Kepler's second law, which states that the area swept out by a line joining any planet to the focus of its orbit is equal for equal time intervals, in other words the time derivative of the area is constant. The differential area swept out by a planet is given by
\begin{equation} 
    dA=\1over2 r^2~d\nu
\end{equation}
so that its time derivative is
\begin{equation} \label{eq:difArea}
    \dot{A}=\1over2 r^2 \dot{\nu}.
\end{equation}
From \cref{eq:momentum}\ we may express the time derivative of $ \nu $ in terms of the angular momentum:
\begin{equation} 
    \dot{\nu}=\frac{\ell}{\mu r^2 }.
\end{equation}
Substitution of this expression into the equation for the time derivative of the area reveals that
\begin{equation} \label{eq:keplersecond}
    \dot{A}=\frac{\ell}{2\mu}
\end{equation}
which is a constant.

Next, let us consider the area of the triangle  $ \triangle SPX $ in \cref{fig:anomalyfigure}
\begin{equation}\label{eq:SPX}
\begin{aligned}
|SPX| & = \pi ab\left(\frac{t-t_p}{T}\right)\\
      & = \frac{ab}{2}M
\end{aligned}
\end{equation}
which follows from \cref{eq:keplersecond,eq:mean}. The total area of an ellipse is $A_\textnormal{ellipse} = \pi ab$; therefore, in time $t-t_p$ a fraction of $(t-t_p)/T$ of the ellipse must be swept out. 
The area of the second triangle may be determined via subtraction as follows:
\begin{equation}\label{eq:SPprimeX}
\begin{aligned}
|SP'X| & = |CP'X|-|CP'S| \\
	   & = \frac{1}{2}a^2E-\frac{1}{2}ea^2\sin E\\
       & = \frac{1}{2}a^2\left(E-e\sin E\right).
\end{aligned}
\end{equation}
The ratio of the area of an ellipse to that of a circle with radius equal to the semi-major axis of the ellipse is given by
\begin{equation}
\frac{A_\textnormal{ellipse}}{A_\textnormal{circle}} = \frac{\pi ab}{\pi a^2} = \frac{b}{a}.
\end{equation}
Kepler's second law requires that this also hold for the areas of the triangles $\bigtriangleup SPX$ and $\bigtriangleup SP'X$; therefore, division of \cref{eq:SPX,eq:SPprimeX} reveals the relationship between the eccentric and mean anomalies to be
\begin{equation}\label{eq:meantoeccentric}
E(t) = M(t) + e\sin E(t).
\end{equation}
~\cref{eq:meantoeccentric} is a transcendental equation and can be solved numerically.  

Finally, we may determine the true anomaly from the eccentric anomaly. To do so, consider the two methods to describe the $x$ and $y$ coordinates of the exoplanet as measured from the star: $x = a(\cos E -e) = r\cos\nu$ and $y = b\sin E = r \sin \nu$. The ratio $b:a = \sqrt{1-e^2}$ follows from the equation of an ellipse. These three equations can be used to reveal that,
\begin{equation}\label{eq:EtoNu1}
\cos\nu = \frac{\cos E -2}{1-e\cos E}.
\end{equation} 
However, it is computationally easier to use $\tan\nu$ in determining the true anomaly. To re-write~\cref{eq:EtoNu1} in terms of cosine use
\begin{equation}\label{eq:trigID}
\tan^2\left(\frac{\theta}{2}\right) = \frac{1-\cos\theta}{1+\cos\theta}
\end{equation}
to show that 
\begin{equation}\label{eq:EtoNuFinal}
\tan\left(\frac{E}{2}\right) = \sqrt{\frac{1-e}{1+e}}\tan\left(\frac{\nu}{2}\right).
\end{equation}

To determine $ \nu $, one may first determine the eccentric anomaly by solving~\cref{eq:meantoeccentric} numerically and then using~\cref{eq:EtoNuFinal} to determine the true anomaly. The true anomaly would then be used in~\cref{eq:radialellipse} to determine the star-planet separation. Alternatively, one can determine a relationship between $r$ and $E$ by considering the equation for $\vec{r}$, as measured from the star, in Cartesian coordinates:
\begin{equation}
\label{eq:rCart}
\vec{r} = a(\cos E -e)\hat{i}+b\sin E\hat{j},
\end{equation}
and taking the dot product $r^2 = \vec{r} \cdot \vec{r}$ to reveal that
\begin{equation}
\label{eq:r_trueanomaly}
\begin{aligned}
r(t) &= a(1-e\cos(E(t)))\\
  & = \frac{a(1-e^2)}{1+e\cos(\nu(t))}.
\end{aligned}
\end{equation}
The second line follows from a rearrangement of~\cref{eq:EtoNuFinal}, and shows that~\cref{eq:radialellipse} and the first line of~\cref{eq:r_trueanomaly} are equivalent. 

\section{Euler Angles}
In the foregoing sections we were only concerned with the radial separation between two bodies in planar motion, a one-dimensional problem. To describe the motion of exoplanets within the plane of the sky we require a way to describe the 3-dimensional coordinates of an exoplanet. To do so we will use the Euler angles $\phi,~\theta,~\psi$ as described by Euler, \cf\  \cite{landauMech}. To transform some vector $\vec{r'}$ to $\vec{r}$ one must first rotate about the Z-axis by angle $\phi$, i.e. rotate using the matrix $R(\phi)$. The second rotation, $R(\theta)$, moves through angle $\theta$ about the new X-axis. The final rotation is $R(\psi)$ and rotates about the previous Y-axis by angle $\psi$. The rotations are described by
\begin{equation}
\label{eq:rotations}
\vec{r}(t) = R(\psi)R(\theta)R(\phi)\vec{r'}(t).
\end{equation}

\begin{figure}[tbh]
\centering
\includegraphics[scale = 0.4]{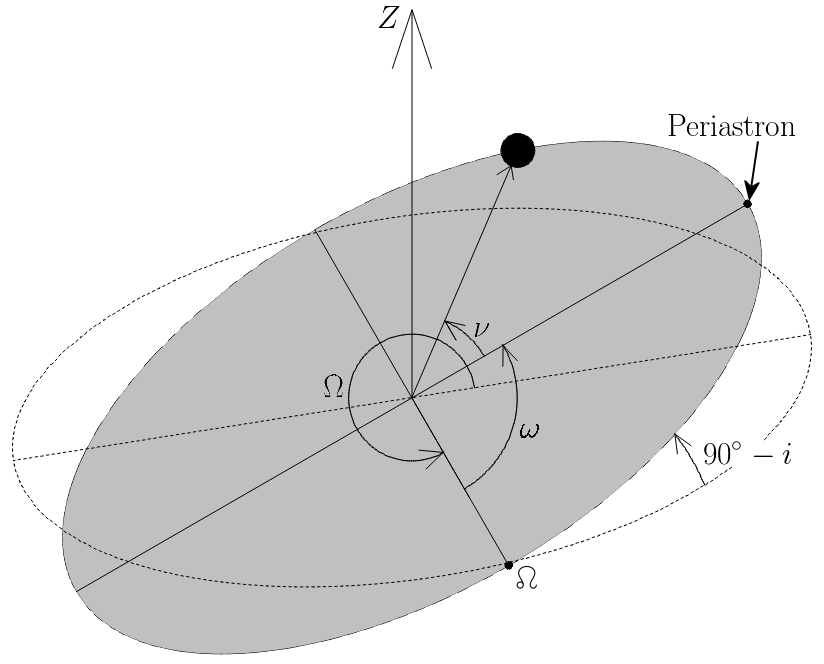}
\caption{A figure illustrating the relationship between a reference plane, shown as the dotted line, and the orbital plane in grey. The orbital plane is projected onto to the reference plane, which is the plane of the sky. The ascending node, $\myascnode$, marks one of the intersections of the orbital plane and the reference plane and is the point of reference for $\omega$, the argument of periastron. The longitude of the ascending node is marked as $ \Omega $, and the true anomaly as $ \nu $. The inclination of the orbit, $ i $, is measured from  $ Z $ and we have labeled the angle between the reference and orbital planes as $ 90\degree - i $.}
\label{fig:eulerfiglabels}
\end{figure}

\begin{figure}[hbt]
\centering
\includegraphics[scale = 0.75]{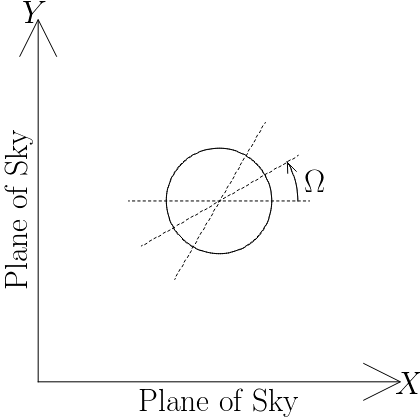}
\caption{A diagram of the plane of the sky with multiple projections of an orbital path shown as dotted lines. Each path corresponds to a change in $\Omega$ only.}
\label{fig:planeofsky}
\end{figure}

The three Euler angles are defined using the orbital elements such that $(\phi, \theta, \psi) = (i, \omega+\nu, \Omega)$, where $i$ is the inclination of the orbit, $\omega$ is the argument of periastron, and $\Omega$ is the longitude of the ascending node, as shown in  \cref{fig:eulerfiglabels}. The inclination, $ i $, is the angle between the reference plane, which we take to be the plane of the sky, and the orbital plane, such that $i$ = 90\degree~indicates an edge-on orbit. The argument of periastron is the angle between the point of closest approach of the exoplanet to its host star and the point at which the orbital plane and reference plane intersect, called the ascending node, $\myascnode$. Finally, the longitude of the ascending node, $\Omega$, is the angle between the projection of periastron onto the reference plane and $\myascnode$. The longitude of the ascending node rotates the orbit about the line of sight, as shown in~\cref{fig:eulerfiglabels,fig:planeofsky}.

Using~\cref{eq:r_trueanomaly,eq:rotations} one may write the Cartesian coordinates of an exoplanet as a function of time as
\begin{equation}\label{eq:cartwithNode}
\begin{pmatrix}
X(t)\\ 
Y(t)\\ 
Z(t)
\end{pmatrix}
= r(t)
\begin{pmatrix}
\cos\Omega\cos(\omega+\nu(t))-\sin\Omega\sin(\omega+\nu(t))\cos i\\ 
\sin\Omega\cos(\omega+\nu(t))+\cos\Omega\sin(\omega+\nu(t))\cos i\\ 
\sin(\omega+\nu(t))\sin i
\end{pmatrix}.
\end{equation}
Typically, $\Omega$ is set to zero because one cannot determine its value without an image of the system and doing so does not affect the description of the total light received from the star-planet system. To see this consider~\cref{fig:planeofsky}, which shows the projection of three orbits on the plane of the sky. The value of $\Omega$ simply rotates the orbit about the line of sight, but has no effect on the total light received as a function of time; therefore, it cannot be measured using photometric measurements. Setting $\Omega$ to zero in~\cref{eq:cartwithNode} gives
\begin{equation}\label{eq:cartfinal}
\begin{pmatrix}
X(t)\\ 
Y(t)\\ 
Z(t)
\end{pmatrix}
= r(t)
\begin{pmatrix}
\cos(\omega+\nu(t))\\ 
\sin(\omega+\nu(t))\cos i\\ 
\sin(\omega+\nu(t))\sin i
\end{pmatrix}
\end{equation}
as our description of the three-dimensional location of the exoplanet projected onto the plane of the sky with the star at the center of the coordinate system.

\begin{figure}[tbh]
\centering
\includegraphics[width=0.7\linewidth]{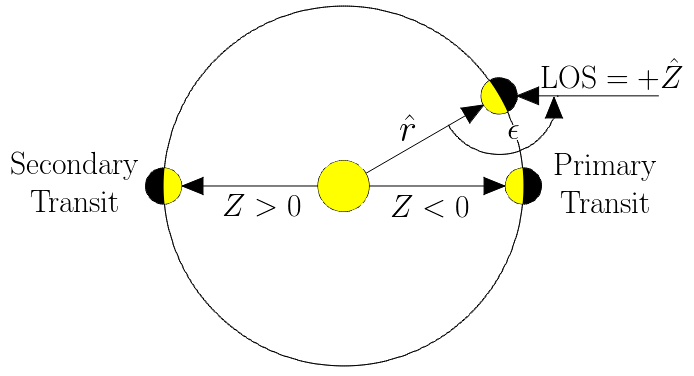}
\caption{An illustration of the geometry of the phase angle, $ \myPhase $ where $ \hat{r} $ points from the center of the host star toward the center of the exoplanet and $ \hat{Z} $ indicates the line of sight, (LOS), and points from the observer to the exoplanet. Note that the exoplanet is behind the host star along our line of sight for $Z\geq 0$, which will also be illustrated in \cref{fig:transitOrientation}.}
\label{fig:losfig3}
\end{figure}

The final angle to consider is the phase angle, $\myPhase$, which is the angle between the vector $\hat{r}$ pointing from the center of the host star to the center of the exoplanet and the line of sight which we will set to $+\hat{Z}$. The phase angle can be calculated using the dot product of these two vectors:
\begin{equation}\label{eq:phaseangle}
\begin{aligned}
\cos\myPhase(t) &= \hat{r}(t)\cdot \hat{Z}\\
		   &= \frac{Z(t)}{r(t)}\\
           &= \sin(\omega +\nu(t))\sin i.
\end{aligned}
\end{equation}
This expression for the phase angle will be useful in determining the flux due to different photometric effects as a function of phase or of time. \cref{fig:losfig3}\ illustrates the relationship between the line of sight and the phase angle. Note that a primary transit, when light is lost due to the exoplanet blocking the host star's light, occurs when $Z(t) <0$ because the line of sight points from the observer to the exoplanet. At full phase, or $\myPhase$ = 0, the exoplanet is behind the host star along our line of sight and a secondary transit occurs due to loss of planetary photometric emissions. Both of these types of transits will be described in greater detail in~\cref{ch:transitchapter}.

\section{Radial Velocity}\label{sec:RV}
\begin{figure}[hbt]
\centering
\includegraphics[scale = .5]{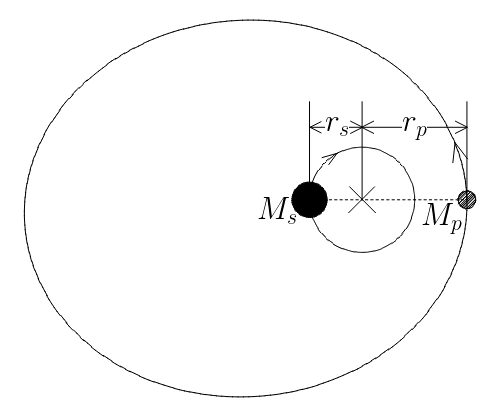}
\caption{The configuration of two masses, $M_s$ and $M_p$, orbiting their common center of mass, labeled with a ``$\times$'' symbol.}
\label{fig:CMfigure}
\end{figure}

An important component of the orbital motion of an exoplanetary system is the radial velocity on the host star induced by the exoplanet. Here, we consider the case of a host star with a single exoplanet,~\cref{fig:CMfigure} shows a possible configuration of a star and exoplanet orbiting their common center of mass. 

For a system in which the mass of the star is given by $M_s$ and that of the exoplanet as $M_p$ we may write the location of their common center of mass as~\cite{landauMech}
\begin{equation}\label{eq:centerOfMass}
\vec{R}_{\textnormal{CM}} = \frac{M_s\vec{r}_s+M_p\vec{r}_p}{M_s+M_p}
\end{equation}
where $\vec{r}_s$ and $\vec{r}_p$ are the position vectors of the star and exoplanet, respectively. If we set the origin of our coordinate system to the center of mass, $\vec{R}_{\textnormal{CM}}$ = 0, then we obtain
\begin{equation}
M_s\vec{r}_s = - M_p\vec{r}_p,
\end{equation}
which indicates that the two bodies are orbiting the center of mass in opposite directions. 

To determine the radial velocity of the star, or its exoplanet, one may use the time derivative of the $Z$-component of position, given in~\cref{eq:cartfinal}:
\begin{equation}\label{eq:radialVelzdot}
V_z = \frac{dZ}{dt} = \left[\dot{r}\sin(\omega + \nu(t))+r\dot{\nu}(t)\cos(\omega+\nu(t))\right]\sin i.
\end{equation}
Using Kepler's second law---orbits sweep out equal areas in equal times---we may determine an expression for $r\dot{\nu}(t)$. To begin, we may 
determine the area swept out in one period, $ T $, via \cref{eq:difArea}\ and setting $\dot{A}$ equal to $A/T$.
The area of an ellipse is given by the equation
\begin{equation}\label{eq:ellipsearea}
A = \pi a^2\sqrt{1-e^2};
\end{equation}
therefore, we find
\begin{equation}\label{eq:rnudot}
r\dot{\nu}(t) = \frac{2\pi a^2\sqrt{1-e^2}}{rT}.
\end{equation}
Differentiation of \cref{eq:r_trueanomaly}\ produces an equation for the change in star-planet separation over time:
\begin{equation}\label{eq:rdot}
\dot{r} = \frac{re\dot{\nu}(t)\sin\nu(t)}{1+e\cos\nu(t)}.
\end{equation}
Substituting \cref{eq:rnudot,eq:rdot}\ into \ref{eq:radialVelzdot}\ gives the radial velocity of an orbiting body in terms of orbital parameters:
\begin{equation}\label{eq:radialVelFinal}
V_z(t) = \frac{2\pi a \sin i}{T\sqrt{1-e^2}}\left[\cos(\omega + \nu(t))+e\cos\omega\right].
\end{equation}
Using Kepler's third law for which $ M_s\gg M_p $ 
\begin{equation}
	T^2=\frac{4\pi^2}{GM_s}a^3,
\end{equation}
one may write the semi-major amplitude of the radial velocity as
\begin{equation}
\label{eq:semiMajorAmp}
\begin{aligned}
K & = \frac{2\pi a \sin i}{T\sqrt{1-e^2}}\\
  & = \left(\frac{2\pi G}{T}\right)^{1/3}\frac{M_p\sin i}{M_s^{2/3}\sqrt{1-e^2}},
\end{aligned}
\end{equation}
thus simplifying \cref{eq:radialVelFinal}\ to
\begin{equation}
	V_z(t) = K\left[\cos(\omega + \nu(t))+e\cos\omega\right].
\end{equation}

\subsection{The Radial Velocity Detection Method}\label{sec:RVmethod}

The radial velocity of the star induced by an orbiting exoplanet may be determined by measuring the redshift, $z$, of the observed spectral absorption lines of the star. This shift occurs due to the Doppler effect, which is the shift in wavelength of emitted star light as the star orbits the star-exoplanet center of mass. The shift in lines is determined by comparing the observed spectra to the expected absorption lines of the host star based on its spectral class.

\begin{figure}[hbt]
\includegraphics[width = \textwidth]{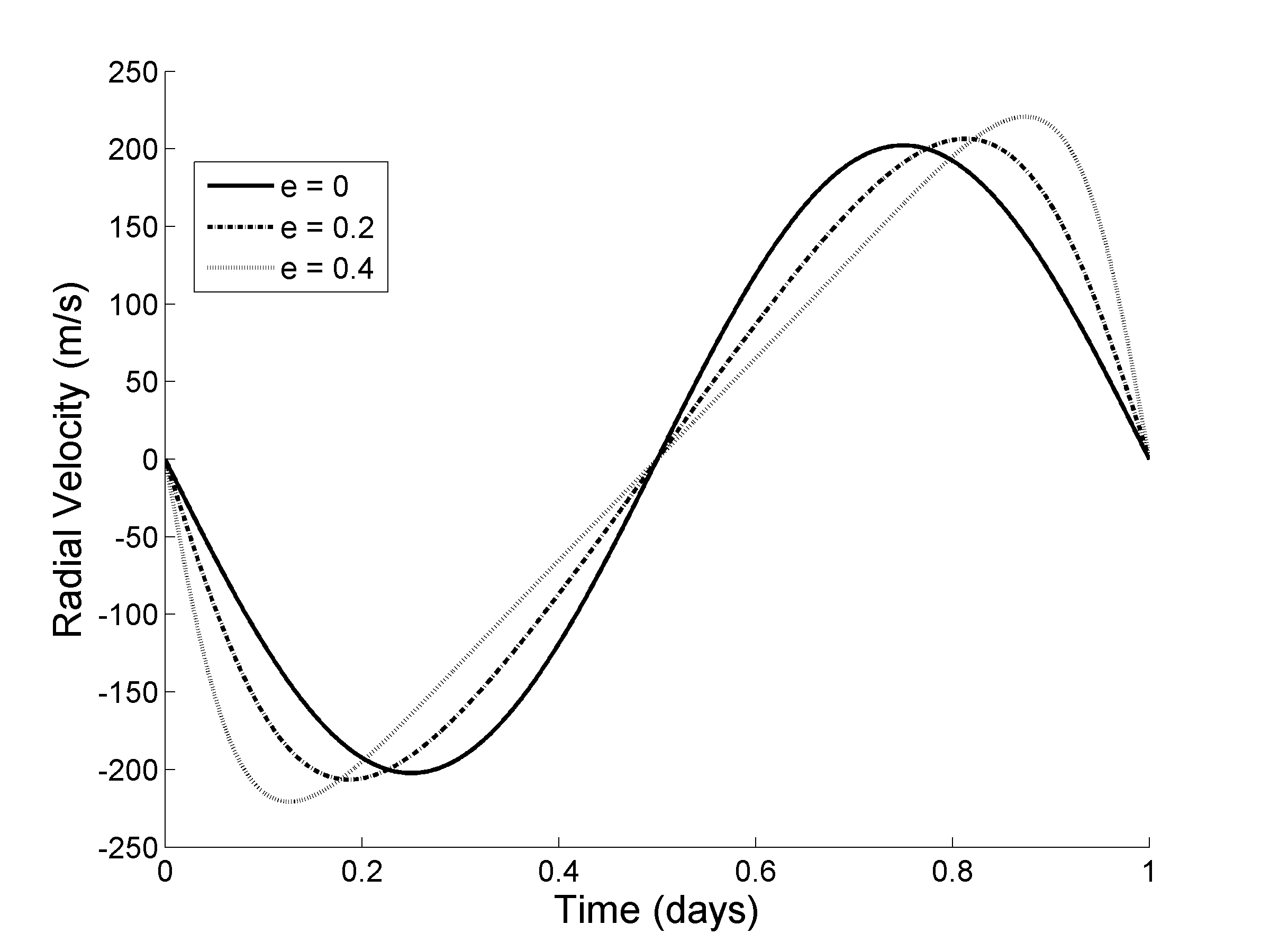}
\caption{The radial velocity curves for varying values of eccentricity, $e$ in the legend. The plot applies to a Jupiter mass exoplanet orbiting a 1 solar mass star with a period of 1 day. For each case the plot begins at $\myPhase$ = 0, or full phase, when the exoplanet is behind the host star. The inclination of each orbit is 90\degree.}
\label{fig:Rvforecc}
\end{figure}

Suppose the emitted wavelength is $\lambda_{emit}$ and the observed wavelength is $\lambda_{obs}$, the two are related to each other by 
\begin{equation}\label{eq:lamb2lamb}
\lambda_{obs} = \lambda_{emit}\sqrt{\frac{1+\beta_r}{1-\beta_r}},
\end{equation}
where $\beta_r$ is the velocity along the line of sight normalized to $c$, the speed of light, \cf\ Perryman's work~\cite{exohandbook}. For our situation, $\beta_r = V_z/c$. The redshift, $z$ is given by 
\begin{equation}\label{eq:redshift}
z = \frac{\lambda_{obs}-\lambda_{emit}}{\lambda_{emit}} = \sqrt{\frac{1+\beta_r}{1-\beta_r}} -1.
\end{equation}
The typical stellar radial velocity is of a order no greater than $10^3$ m s$^{-1}$; therefore, the non-relativistic limit may be applied to reduce~\cref{eq:redshift} to 
\begin{equation}
z \approx \beta_r.
\end{equation}
Thus, a measure of the redshift of stellar absorption lines is a direct measure of the radial velocity of the star. From \cref{eq:radialVelFinal}, we see that a measure of the red shift as a function of time provides information about the period, eccentricity and minimum exoplanet mass, $ M_p \sin i $, in addition to inferring the existence of the exoplanet. \cref{fig:Rvforecc}\ shows example RV curves for three eccentricity values.

\begin{figure}[hbt]
\includegraphics[width = \textwidth]{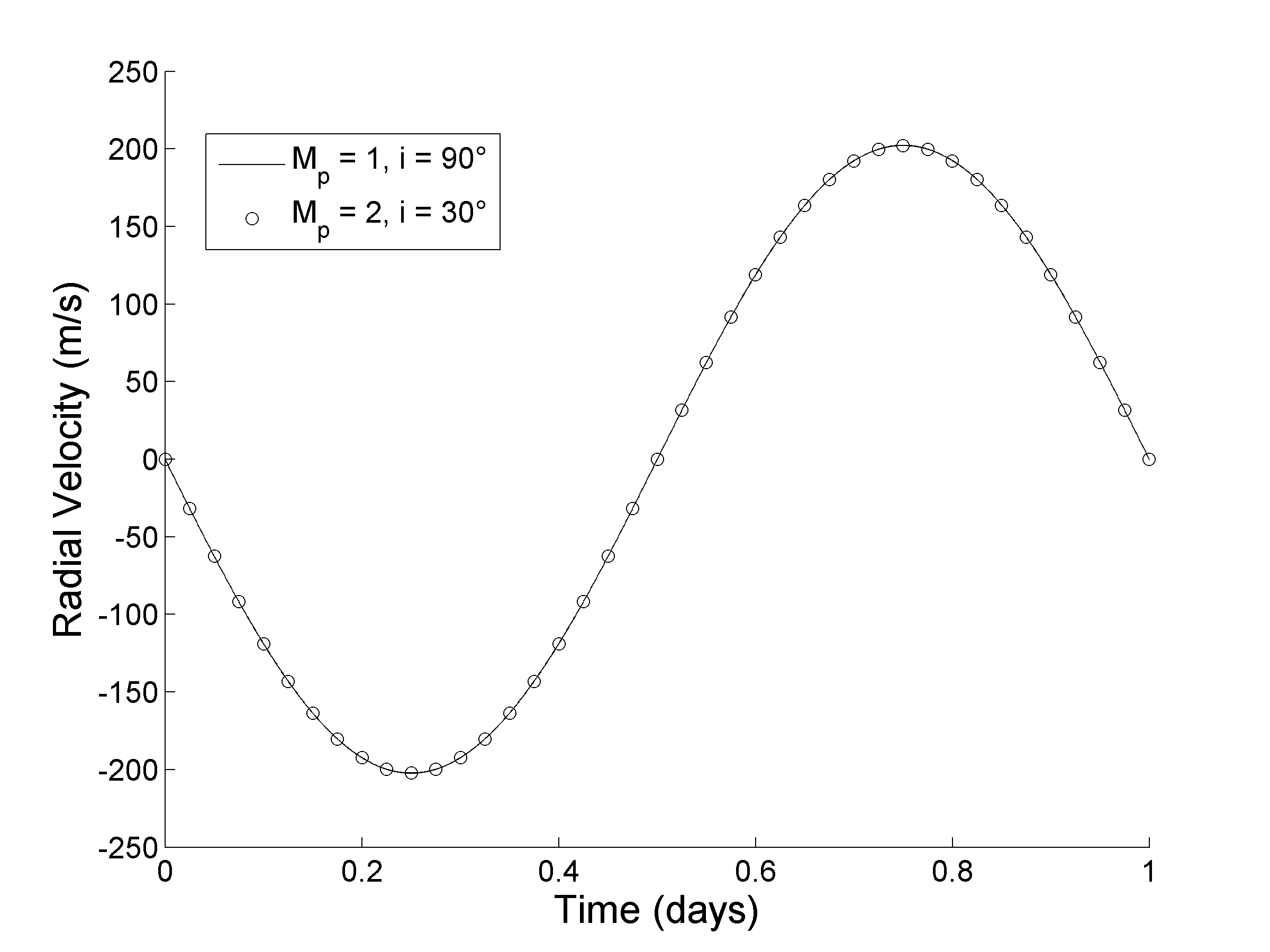}
\caption{An illustration of the degeneracy between the exoplanetary mass, $M_p$, and the inclination, $i$, for the radial velocity detection method. For the cases shown the eccentricity is set to zero and the orbit begins at full phase. For each curve the minimum mass is equal at $ M_p \sin i =1 $ M$_J$ where the solid line shows the curve for the case of a single jupiter mass planet in an edge-on orbit and the circles represent the curve for an exoplanet with mass equal to 2 jupiter masses orbiting at an inclination of 30\degree. Note that the two curves are indistinguishable.}
\label{fig:RVforMpsini}
\end{figure}

The radial velocity, or RV, method has been one of the most successful exoplanetary detection methods to date. As of July 2018, 757 exoplanets in 560 planetary systems with 136 multiple planet systems have been detected using this method~\cite{exoEncyclopedia}. Yet, the RV method does have some limitations. First, multi-planet systems have highly degenerate solutions. Second, the method can only be used to determine a minimum mass of the exoplanet, $M_p\sin i$, see~\cref{eq:semiMajorAmp}and~\cref{fig:RVforMpsini}. It is possible to break the degeneracy between the planetary mass and the inclination by using a second detection method in conjunction with the RV method. For example, the transit method can be used to set limits on the inclination of the system, see~\cref{eq:tranDuration} in~\cref{ch:transitchapter}. 

The previous chapter presented the equations used to describe the orbits of exoplanets. In addition, we reviewed the relationship between radial velocity measurements and the orbital parameters of exoplanets, as well as the mass of the host star and exoplanet. We will now proceed to describe the relationship between light curves and exoplanet parameters, including the period of the orbit and radius of the exoplanet.
\chapter{Transits}\label{ch:transitchapter}
In this chapter we will discuss the most prominent feature of a light curve, a transit, which occurs when an exoplanet passes in front of its host star along an observer's line of sight (LOS). The effect is a temporary decrease in the amount of light observed when the exoplanet blocks out some of its star's light. We will review the relationship between transits and exoplanet characteristics as described previously by Mandel and Agol, \cite{Mandel2002}, such as the relative size of the exoplanet to the host star, and the period of the orbit. A transit is a rare occurrence, as it requires that the observed system be aligned properly along an observer's line of sight. The ideal situation is the case in which the exoplanet's inclination is 90\degree, or edge-on, because such configurations maximize the probability of a transit being observable and we will review the derivation for the probability of a transit. 

\section{Transits of a Uniform Source}\label{sec:uniformtransits}
\begin{figure}[hbt]
\centering
\includegraphics[width = 0.95\textwidth]{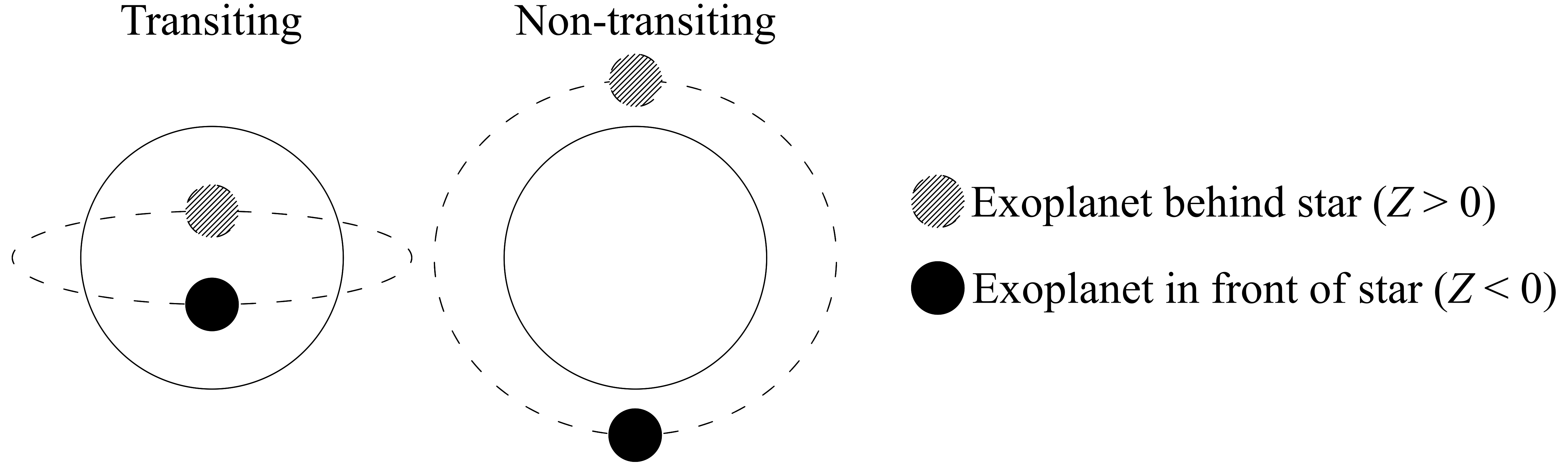}
\caption{Shown are two possible orbital orientations. On the left is an orientation that is nearly edge-on in which both a primary and secondary transit will occur. On the right is a nearly face-on orientation in which neither a primary nor secondary transit will occur. Adapted from Figure 3.2.1 of Ben Placek's thesis, \protect\cite{PlacekThesis}.}\label{fig:transitOrientation}
\end{figure}

A depiction of of two orbital orientations illustrates the two extremes between a near edge-on orbit and a face-on orbit is shown in~\cref{fig:transitOrientation}. 

Also of note is the situation in which the exoplanet passes in back of its host star; such an orientation can result in a secondary transit in which the photometric emissions from the exoplanet are obscured by the star, see~\cref{fig:transitOrientation}. Taken together, the primary and secondary transits provide a great deal of information about an exoplanet. 

\begin{figure}[hbt!]
\centering
\includegraphics[width = 0.625\textwidth]{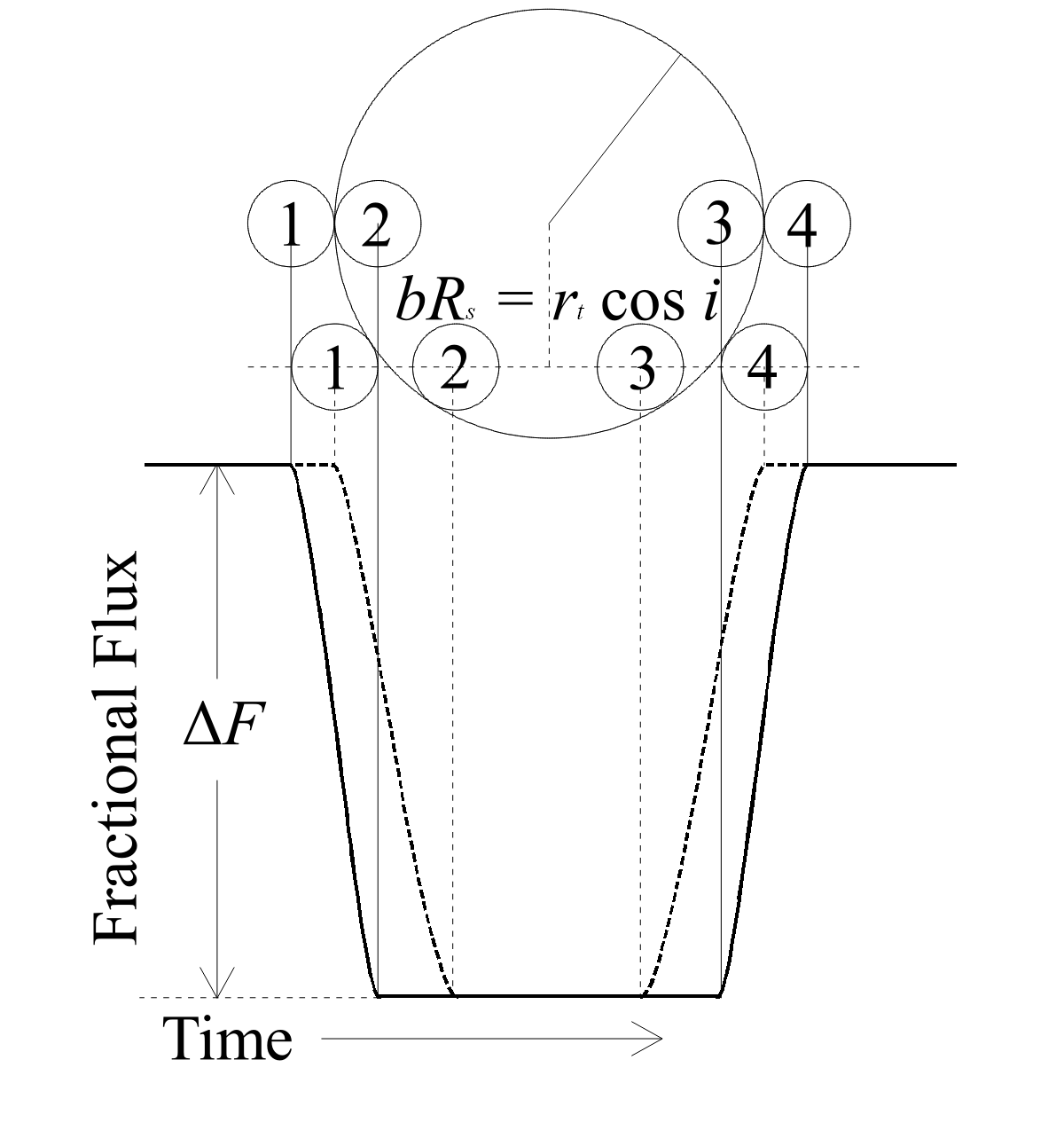}
\caption{Depiction of an exoplanet transit illustrating the relationship between the observed flux and the location of the disk of the exoplanet relative to the disk of the host star. The transit depth, $\Delta F$, is the maximum fractional decrease in flux during the transit. Contact point 1 marks the beginning of the transit and contact point 2 marks the point at which the exoplanetary disk is completely within the star's disk along the LOS. Contact point 3 marks the beginning of the egress from the stellar disk and contact point 4 marks the end of the transit. Limb darkening is not included. The dashed lines correspond to an inclination of 85\degree\ and the solid line depicts an edge-on orbit. The parameter $b$ is the impact parameter of the transit, $ R_s $ is the stellar radius, $ r_t $ is the star-planet separation at the time of mid-transit, and $ i $ is the inclination of the exoplanet's orbit. Adapted from Figure 1 of Seager and Mall{\v e}n-Ornelas, \protect\cite{SeagerMallen2003}.} \label{fig:transitDepth}
\end{figure}

To begin, the primary transit depth, $\Delta F$ in~\cref{fig:transitDepth}, is the maximum dip in fractional flux
\begin{equation}\label{eq:transitdepth}
	\Delta F=\left( \frac{R_p}{R_s}\right)^2;
\end{equation} 
and the secondary transit depth provides information about the total planetary photometric emissions. The period of both transit types is equal to the period of the exoplanet's orbit; whereas, the  primary transit duration is the time between contact points 1 and 2 in \cref{fig:transitDepth}. The primary transit duration is related to the inclination and eccentricity of the orbit through the impact parameter,  $ b $, as illustrated in \cref{fig:transitDepth}. Finally, the eccentricity, the argument of periastron and the inclination of the orbit affect the separation of the primary and secondary transits. 

\cref{eq:transitdepth}\ reveals that the transit depth may be used to determine the radius of the exoplanet if the radius of the host star is known. The stellar properties of \myKepler\ stars are not always well known, for example in Mathur et al.'s work \cite{2017revisedStar}\ it was noted that the typical uncertainty in the radius of a \myKepler\ target star was 27\%. More on this topic will be discussed at the end of \cref{ch:lightcurvechapter}.

In \cite{Mandel2002}, Mandel and Agol  use the geometry of two intersecting discs to describe the transit light curve of a planet eclipsing its host star if the star is approximated as a uniform source of light. The secondary eclipse can be described using similar geometry. In~\cite{Mandel2002}, the authors also applied the transit geometry to a stellar disk exhibiting quadratic limb darkening. The limb darkening further smooths the ingress and egress of the exoplanet across the stellar disk and will be discussed in~\cref{sec:limbdarkening}.

Let $R(t)$ be the center-to-center distance between the stellar and planetary disks along the line of sight, +$\hat{Z}$, 
\begin{equation}
R(t) = \sqrt{X^{2} + Y^{2}} = r(t)\sqrt{\cos^{2}(\omega + \nu(t)) + \cos^{2}(i)\sin^{2}(\omega + \nu(t))}.
\end{equation}
Here, $ r(t) $ is the planet-star separation given by \cref{eq:radialellipse}, $\omega$ is the argument of periastron, $\nu$ is the true anomaly, and $i$ is the inclination of the planet's orbit. 
Following the notation used in \cite{Mandel2002}, let the function $F^{e}(p,z)$ be the ratio of the primary transit flux to that of the unobscured stellar flux, $F_s$:
\begin{equation}
\label{eq:Fe}
F^{e}(p,z) = 1 - \lambda^{e}(p,z),
\end{equation}
where $\lambda^{e}$ is the fractional area of the star obscured by the exoplanet
\begin{equation}
\label{eq:lam_e}
\lambda^{e}(p,z) = 
\begin{cases}
0 & , 1+p<z \\
\frac{1}{\pi} \left[ p^{2}\kappa_{0} + \kappa_{1} - \sqrt{\frac{4z^{2}-(1+z^2-p^2)^2}{4}} \right] & , |1-p|<z\leq 1+p \\
p^2 & , z \leq 1-p \\
1 & , z \leq p-1,
\end{cases}
\end{equation}
and 
\begin{equation}\label{eq:kappas}
	\begin{aligned}
		\kappa_1 &= \invcos{\frac{1 -p^2+z^2}{2z}}\\
		\kappa_0 &= \invcos{\frac{p^2+z^2-1}{2pz}}.
	\end{aligned}
\end{equation}
The unobscured stellar flux may be related to the specific intensity of the host star, $ I_0 $, as follows
\begin{equation} 
    F_s = \pi R_s^2I_0.
\end{equation}
In \cref{eq:lam_e}, the variable $ p $ is the ratio of the exoplanet's radius to that of the host star, $ R_p/R_s $  and that $ z $ is the normalized separation of centers, $ R(t)/R_s $. The superscript $e$ in \cref{eq:Fe,eq:lam_e}\ are used to indicate that the function considers only the geometrical portion of the transit and ignores limb darkening of the stellar disk, which will be discussed in \cref{sec:limbdarkening}.

The function $F^e(p,z)$ describes the output light relative to the stellar light for a uniform source (the star) being eclipsed by an opaque, dark sphere (the exoplanet). Following is a derivation of~\cref{eq:lam_e}. Note that~\cref{eq:Fe} applies when $Z < 0$, as shown in~\cref{fig:transitOrientation}, for which the exoplanet is in front of the host star along the line of sight.

\subsection{Case (1) - No overlap} 
For the first case, $1+p<z$, in which the sum of the radius of the star and the exoplanet is greater than the separation of their centers. In this case there is no overlap of the planetary disk and the star along the line of sight. The fractional area of the star obscured by the planetary disk is then zero.

\subsection{Case (2) - Partial overlap of star by planet} 
\begin{figure}[htb]
\centering
\subfloat[][The intersection of two circles.\label{fig:circles}]{\includegraphics[scale = 0.425]{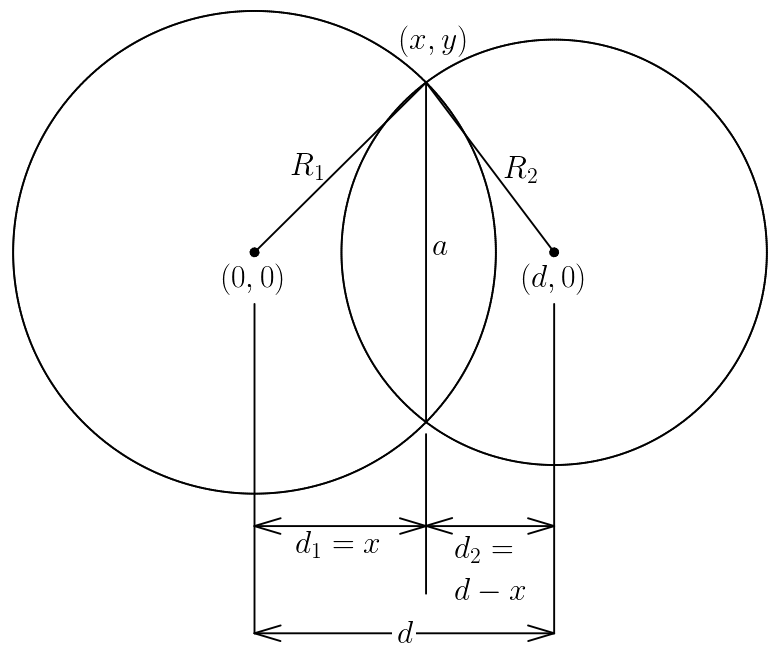}}
\qquad
\subfloat[][The geometry to calculate the area of a circular segment.\label{fig:segment}]{\includegraphics[scale = 0.425]{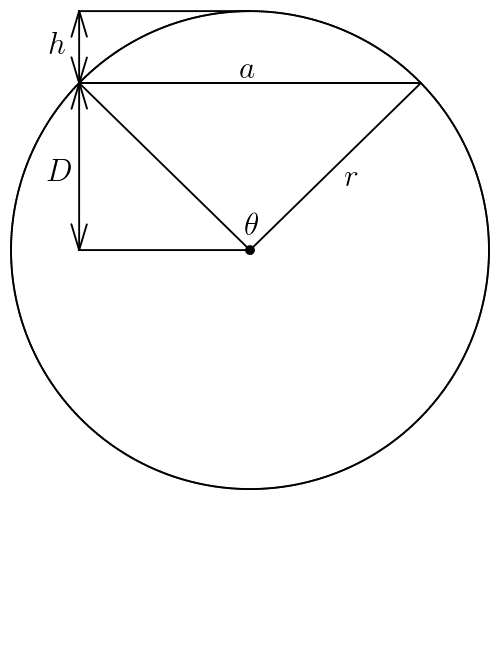}}
\caption{Geometry of two intersecting circles and a circular segment. In~\cref{fig:circles} a circle of radius $R_1$ is centered at (0,0) and the second circle is centered at (0, $d$) with radius $R_2$. The length of the cord is labeled as $a$ and is used to calculate the area of a circular segment as shown in~\cref{fig:segment}.}
\end{figure}

The condition for the second case, $|1-p|<z\leq 1+p$, can also be written as $|R_s - R_p| < R \leq R_s + R_p$. This describes the situation in which the planetary disk is intersecting that of the stellar disk, but is not completely within the disk of the star. The area obscured, $A_{ob}$, is the area of intersection of two circular disks. Neglecting tidal effects of the exoplanet on the star to assume that it is spherical, the cross-sectional area along the line of sight equal to $\pi R_s^{2}$. Under the same assumptions, the cross-sectional area of the exoplanet along the line of sight is $\pi R_{p}^{2}$. The physical situation is depicted in~\cref{fig:circles} and is the situation shown between contact points 1 and 2 (ingress), and 3 and 4 (egress) in~\cref{fig:transitDepth}. 

We may determine the total area obscured by adding the area of each of the circular segments defined by the cord $a$ and the radii. The area of a circular segment is the difference between the area of the circular section defined by $\theta$ and the radius of the circle, and the area of the triangle formed by the cord and the two lines connecting the cord's ends to the center of the circle as shown in~\cref{fig:segment}.

The area of a circular section is $A_{sec} = \frac{1}{2}r^2\theta = r^2 \cos^{-1}\left(\frac{D}{r}\right)$ and the area of a triangle of height $D$ and base $a$ is $A_{tri} = \frac{1}{2}Da$. This gives the area of the circular segment to be $A_{seg} = r^2\cos^{-1}\left(\frac{D}{r}\right) - \frac{1}{2}Da$. Letting $ D=d_1 $ or $ d_2 $ and $ r=R_1 $ or $ R_2 $ as appropriate, we may describe the total area obscured in the situation depicted in ~\cref{fig:circles} as
\begin{equation}
\label{eq:A_ob1}
A_{ob} = R_{1}^{2}\cos^{-1}\left(\frac{d_1}{R_1}\right)+R_{2}^2\cos^{-1}\left(\frac{d_2}{R_2}\right)-\frac{1}{2}a(d_1+d_2).
\end{equation}

To determine $d_1$, $d_2$ and $a$ of~\cref{fig:circles} in terms of $R_1$, $R_2$ and $d$ use the following equations: 
\begin{equation*}
	\begin{aligned}
		x^2+y^2 &= R_1^2\\
	 (x+d)^2+y^2 &= R_2^2\\
		2y &= a
	\end{aligned}
\end{equation*} 
to reveal that
\begin{equation*} 
\begin{aligned}
d_1 &= \frac{d^2 + R_1^2 - R_2^2}{2d} \\
d_2 &= \frac{d^2-R_1^2+R_2^2}{2d} \\
a &= \frac{1}{d}\sqrt{4d^2R_2^2-\left(d^2+R_1^2-R_2^2\right)^2}
\end{aligned}
\end{equation*}
Substituting the foregoing equations into~\cref{eq:A_ob1} reveals that the total area obscured may be expressed as
\begin{equation}
\begin{aligned}
A_{ob} =& R_1^2\cos^{-1}\left(\frac{d^2+R_1^2-R_2^2}{2dR_1}\right)+R_2^2\cos^{-1}\left(\frac{d^2-R_1^2+R_2^2}{2dR_2}\right)\\
& -\sqrt{\frac{4d^2R_1^2-\left(d^2+R_1^2-R_2^2\right)^2}{4}}.
\end{aligned}
\label{eq:Aob0}
\end{equation}
finally, we may rewrite \cref{eq:Aob0}\ in terms of the exoplanet and stellar parameters. Let $R_1 = R_s$, $R_2=R_p$ and $d = R$, then substitute $z=R/R_s$ and $p = R_p/R_s$ to find 
\begin{equation}
\begin{aligned}
A_{ob} =R_s^2 & \left[\cos^{-1}\left(\frac{z^2+1-p^2}{2z}\right) + p^2 \cos^{-1}\left(\frac{z^2-1+p^2}{2zp}\right)\right. \\
& \left.- \sqrt{\frac{4z^2-\left(z^2+1-p^2\right)^2}{4}}\ \right].
\end{aligned}
\label{eq:Aob}
\end{equation}
Finally, we may simplify \cref{eq:Aob}\ by writing it in terms of $\kappa_0$ and $\kappa_1$ from \cref{eq:kappas}:
\begin{equation}
\label{A_ob}
A_{ob} = R_s^{2}\left[ p^2\kappa_0 + \kappa_1 - \sqrt{\frac{4z^2 - (1+z^2-p^2)}{4}} \right].
\end{equation}
Revealing that the fractional area of the star obscured is $\lambda^e  = A_{ob}/(\pi R_s^{2})$; therefore, we recover the equation presented in the second case of~\cref{eq:lam_e}.

\subsection{Case (3) - Partial eclipse of star}
For the third case of~\cref{eq:lam_e}, we have $z\leq 1 - p$ or $R + R_p\leq R_s$; therefore, the disk of the exoplanet is fully within the disk of the star, and its radius is less than that of the star. The situation corresponds to contact points 2 and 3 in \cref{fig:transitDepth}\ where we are assuming that the star-planet separation does not change much during the transit. The fractional cross sectional area obscured is then $\lambda^e = \pi R_{p}^{2}/\pi R_s^{2} = p^2$.

\subsection{Case (4) - Full eclipse of star}
The fourth case occurs when $z \leq p-1$ or $R + R_s \leq R_p$, i.e. the stellar disk is the same size or smaller than the  planetary disk. If the star is fully obscured then the fractional area obscured by the exoplanet is $\lambda^e = 1$ and the transit light curve depends on the flux of the exoplanet. If this is the case, we are likely not looking at a planet, but rather a binary star system. 

\subsection{The Secondary Transit}\label{sec:secondarytransit}
The secondary transit takes place at points in the orbit for which $Z > 0$, as shown in~\cref{fig:transitOrientation}. Let the function, $F^p(p,z)$ describe the fractional flux of the star plus that of the exoplanet obscured by the star;
\begin{equation}
\label{eq:Fp}
F^{p}(p,z) = 1 - \Phi_p\lambda^{ep}(p,z),
\end{equation}
where, $\lambda^{ep}$ is the fractional area of the exoplanet obscured by the star and $\Phi_p$ is the total fractional planetary photometric flux, to be discussed in \cref{ch:lightcurvechapter}. Here the superscript $p$ is used to indicate loss of light due to the eclipse of the planet. The equation for $\lambda^{ep}$ is similar to that of $\lambda^{e}$ and may be determined via $ A_{obs}/(\pi R_p^2 ) $ for each case described below:
\begin{equation}
\label{eq:lam_ep}
\lambda^{ep}(p,z) = 
\begin{cases}
0 & ,  1+p<z \\ 
\frac{1}{\pi p^2} \left[ p^{2}\kappa_{0} + \kappa_{1} - \sqrt{\frac{4z^{2}-(1+z^2-p^2)^2}{4}} \right] & ,  |1-p|<z\leq 1+p \\ 
1 & ,  p-1 < z \leq 1-p \\ 
\frac{1}{p^2} & ,  z \leq p-1. 
\end{cases}
\end{equation} 

The depth of the secondary eclipse is the maximum total fractional flux of the planetary emissions, or $\Phi_{p,\textnormal{max}}$, during the secondary eclipse. For circular orbits, this maximum occurs at $ z = 0 $, \cf\  Sudarsky et al. \cite{sudarsky}, which coincides with the third case in \cref{eq:lam_ep}, i.e. $  p-1 < z \leq 1-p  $. The third case corresponds to a situation in which the exoplanet is completely behind the host star and its radius is less than that of the host star so that it is completely obscured. In the fourth case, we again have an object that is larger than the host star---it is likely a binary star system and not an exoplanetary system.
 
\begin{figure}[hbt]
\centering
\includegraphics[width = 0.98\textwidth]{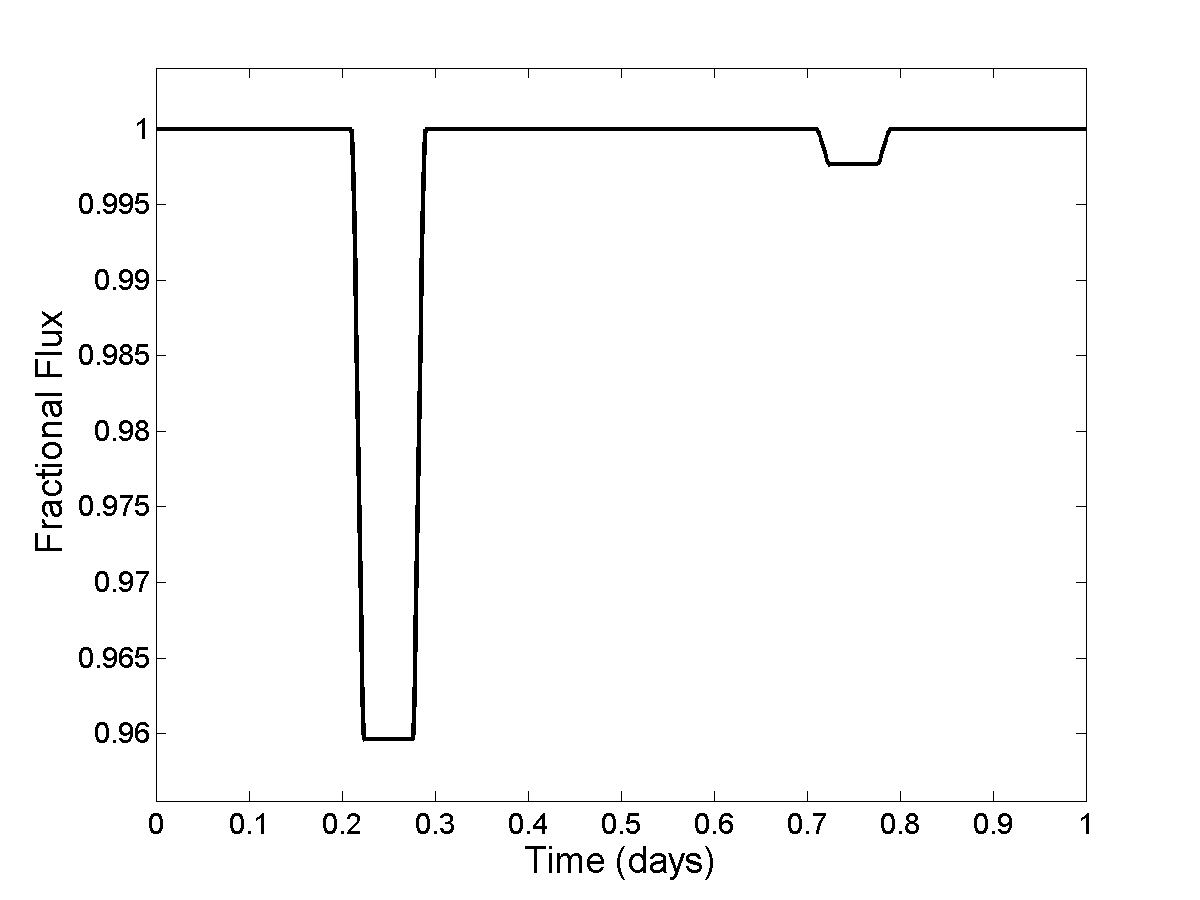}
\caption{Plotted is a simulated, noiseless light curve including the primary and secondary transits of an exoplanet. The parameters used to produce the light curve are as follows: $p$ =  0.2010, $T$ = 1 day, $\Phi_{p,\textnormal{max}}$ = 0.0023, $\cos i$ = 0 and $e$ = 0. The orbit begins at quarter phase, waning; therefore, the primary transit is centered at 0.25 days, for which $ \myPhase=\pi $, and the secondary transit one half period later at 0.75 days. Adapted from Figure 3.2.2 in \protect\cite{PlacekThesis}.} \label{fig:secondaryandprimary}
\end{figure}

If we were only to consider the transit light curve described in \cref{sec:uniformtransits}\ and the loss of planetary photometric emissions during the secondary eclipse, then the total fractional flux of the exoplanetary system can be described by the addition of~\cref{eq:Fe,eq:Fp}, as shown in~\cref{fig:secondaryandprimary}. The method to determine the net fractional flux given by a light curve will be discussed in~\cref{ch:lightcurvechapter}.

\section{Transits with Limb-darkening}\label{sec:limbdarkening}
Both \cref{fig:transitDepth,fig:secondaryandprimary}\ show a primary transit with sharp changes in flux as the exoplanet disk moves through ingress and egress in front of the host star. Neither of these models includes the limb-darkening exhibited by stars which are not uniformly bright as assumed in \cref{sec:uniformtransits}. Limb-darkening is the result of an observer detecting photons from different depths of the stellar photosphere. Near the center of the star the photons detected by the observer originate from a hotter plasma deeper in the star than photons emitted from the cooler surface plasma at the limb. The result is a dimming in the brightness of the stellar disk as one moves away from its center.  

In~\cite{Claret2000}, Claret proposes a limb-darkening law that applies to a wide range of stellar models,
\begin{equation}\label{eq:Ilimb}
I(r) = I_0\left( 1-\sum^4_{n=1}c_n(1-\mu^{n/2})\right), 
\end{equation}
where $\mu = \cos\theta = \sqrt{1-r^2}$, $r$ is the normalized radial coordinate on the disk of the star, and $ I_{0} $ is the specific intensity. The coordinate $\theta$ is the angle between the observer and the normal to the stellar surface. The equation to determine the light curve is then
\begin{equation}\label{eq:Flimb}
F^{s}(p,z) = \left[\int^1_0 dr 2r I(r)\right]^{-1}\int_0^1 dr I(r) \frac{d\left[F^e(p/r, z/r)r^2\right]}{dr},
\end{equation}
as described in~\cite{Mandel2002}. 

Here we shall follow the work of \cite{Mandel2002}, in which the authors describe the full solution to the transit light curve for quadratic limb darking, for which \cref{eq:Ilimb}\ becomes 
\begin{equation}\label{eq:Iquad}
I(r) = I_0( 1 - \gamma_1(1-\mu)-\gamma_2(1-\mu)^2),
\end{equation}
and ~\cref{eq:Flimb} simplifies to
\begin{equation}\label{eq:Fquadratic}
F^{s}(p, z) = 1 - \frac{1}{4\varOmega}\left[ (1-c_2)\lambda^e + c_2\left(\lambda^d + \frac{2}{3}\Theta(p-z)\right) -c_4\eta^d \right],
\end{equation}
where $ \lambda^e $ is given by \cref{eq:lam_e}, $\lambda^d$ and $\eta^d$ are described in \cref{tab:lambdaandeta,eq:lambdaandeta,eq:andeta}, and the $s$ superscript indicates the transit of the host star. The variable $\varOmega$ is defined as $\varOmega = \sum\limits_{n=0}^{4} c_n (n+4)^{-1}$ for convenience and $c_n$ are the limb-darkening coefficients for $n\neq 0$ where $c_0 \equiv 1 - c_1 -c_2 -c_3-c_4$. Finally, $\Theta(p-z)$ is the heaviside function of $p-z$,  \cite{Mandel2002}. The coefficients in~\cref{eq:Iquad} are the result of reducing the law given in~\cref{eq:Ilimb} to $c_1=c_3 = 0$, $c_2 = \gamma_1+2\gamma_2$, and $c_4 = -\gamma_2$. 

\begin{table}[hbt]\centering
	\tabulinesep = 1mm
	\caption{\label{tab:lambdaandeta}Table to determine $ \lambda^d $ and $ \eta^d $ in \cref{eq:Fquadratic}, see Table 1 of \protect\cite{Mandel2002}. Each case describes a unique combination of expressions for $\lambda^d$ and $ \eta^d $.}
		\begin{tabu} to 0.7\textwidth {
			X[1.5,l]
			X[3,l]
			X[10,l]
			X[3,l]
			X[1,c]
		}
		\toprule
		\textbf{Case}&\multicolumn{1}{c}{ $ p $ }&\multicolumn{1}{c}{$ z $ }  & \multicolumn{1}{c}{$ \lambda^d(z) $}&\multicolumn{1}{c}{ $ \eta^d(z) $ } \\
		\midrule
		I& $ (0,\infty) $ & $ [1+p,\infty) $ &0&0\\
		 & 0& $ [0,\infty) $&0&0\\
		II& $ (0,\infty) $ & $  (\1over2+|p -\1over2|,1 +p) $ & $ \lambda_1 $ & $ \eta_1 $\\
		III& $ (0,\1over2)  $ & $  (p,1 -p) $ & $ \lambda_2 $ & $ \eta_2 $ \\
		IV & $ (0,\1over2)  $ & $ 1-p $ & $ \lambda_5 $ & $ \eta_2 $ \\
		V & $  (0,\1over2)  $ & $ p $ & $\lambda_4 $ & $ \eta_2 $ \\
		VI & $ \1over2 $ & $\1over2$& $ \frac{1}{3} -\frac{4}{9\pi} $ & $ \frac{3}{32} $ \\
		VII & $  (\1over2,\infty) $& $ p $ & $ \lambda_3 $ & $ \eta_1 $ \\
		VIII & $ (\1over2,\infty)  $  & $ [|1 -p|,p) $ & $ \lambda_1 $ & $ \eta_1 $ \\
		IX & $  (0, 1) $& $  (0,\1over2 -|p -\1over2|) $&  $ \lambda_2 $ & $ \eta_2 $ \\
		X & $  (0, 1) $& $ 0 $ & $ \lambda_6 $ & $ \eta_2 $ \\
		XI & $  (1,\infty) $& $ [0,p -1) $ &1 & 1 \\
		\bottomrule
	\end{tabu}
\end{table}

In \cref{tab:lambdaandeta}\ the functions $ \lambda_i $ and $ \eta_i $ are defined as
\begin{equation}\label{eq:lambdaandeta}
	\begin{aligned}
		\lambda_1=&\frac{1}{9\pi\sqrt{pz}}\bigg[ ( (1-b)(2b+a-3)-3q (b-2))K(k)+\\
			&\hspace{0.55in}4pz (z^2 +7p^2 -4)E (k) -3\frac{q}{a}\Pi\left( \frac{a-1}{a},k\right)\bigg],\\
		\lambda_2= &\frac{2}{9\pi\sqrt{1 -a}}\bigg[ (1 -5z^2 +p^2 +q^2)K(k^{-1}) +\\
		&\hspace{0.7in}(1 -a)(z^2 +7p^2 -4)E(k^{-1}) -3\frac{q}{a}\Pi\left( \frac{a - b}{a}, k^{-1}\right)\bigg],\\
		\lambda_3= &\frac{1}{3} +\frac{16p}{9\pi}(2p^2 -1)E\left( \frac{1}{2 k}\right) -\frac{(1-4 p^2) (3 - 8p^2)}{9\pi p} K\left( \frac{1}{2k}\right), \\
		\lambda_4 = &\frac{1}{3} +\frac{2}{9\pi}\left[ 4\left( 2 p^2 -1\right) E(2 k) +\left( 1 - 4p^2\right) K(2k)\right],\\
		\lambda_5 = &\frac{2}{3\pi}\invcos{1 - 2p} -\frac{4}{9\pi}\left( 3+ 2p -8 p^2\right),\\
		\lambda_6 = & -\frac{2}{3}\left( 1 - p^2\right)^{3/2},
	\end{aligned}
\end{equation}
and
\begin{equation}\label{eq:andeta}
\begin{aligned}
		\eta_1= &(2\pi)^{-1}\bigg[  \kappa_1 +2\eta_2\kappa_0 -\frac{1}{4}\left( 1+ 5p^2 + z^2 \right)\sqrt{(1 - a)(b -1)} \bigg],\\
		\eta_2 = &\frac{p^2 }{2}\left( p^2 + 2z^2 \right),
\end{aligned}
\end{equation}
where $ a=(z-p)^2  $, $ b=(z+p)^2  $,  $ k=\sqrt{(1 - a)/(4zp)} $ and $ q=p^2-z^2 $. In addition the functions $ K(k) $, $ E(k) $ and $ \Pi(n,k) $ are the complete elliptic integrals of the first, second, and third kind respectively, Equation 7 in \cite{Mandel2002}. 

\begin{figure}[bth]
\centering
\includegraphics[width=0.7\textwidth]{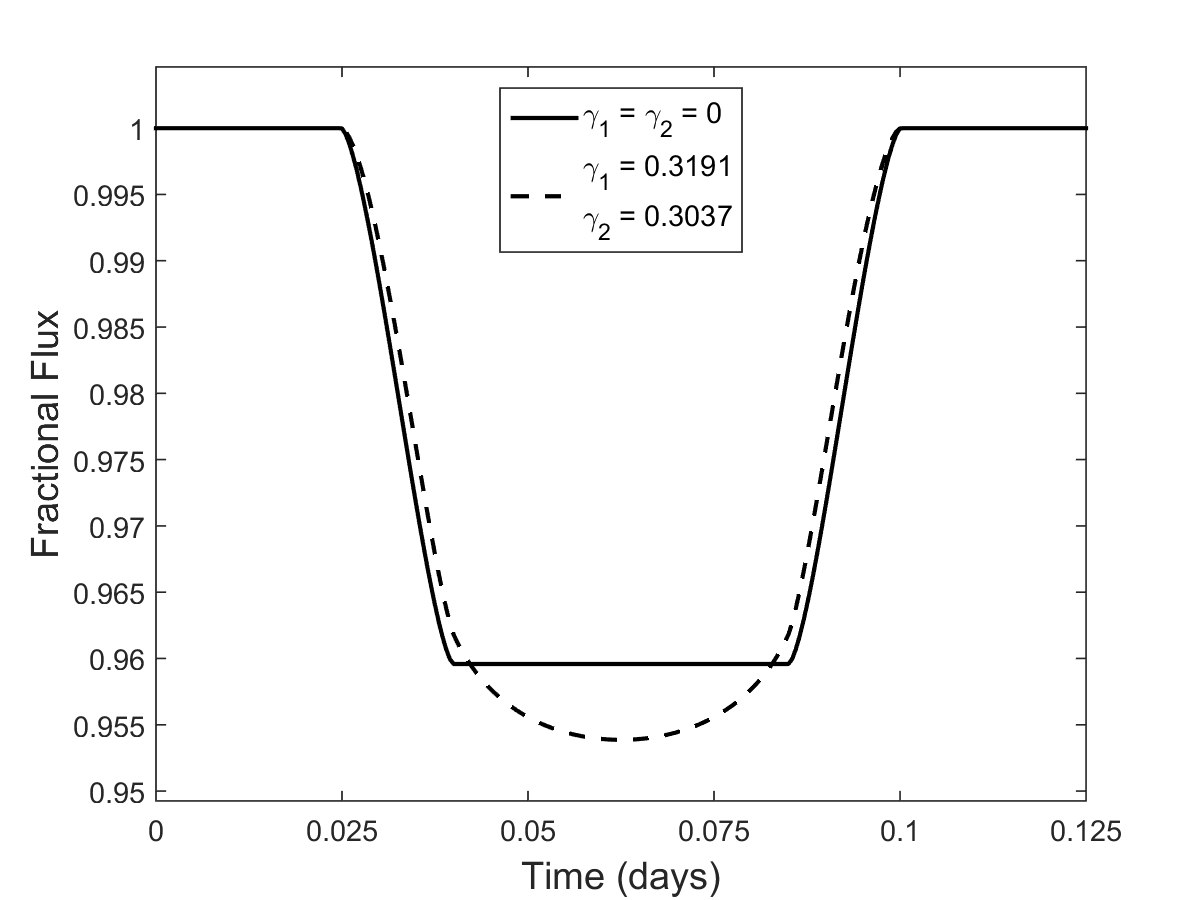}
\caption{A depiction comparing a primary transit for zero limb-darkening (solid line) and quadratic limb-darkening (dashed line). The parameters used to produce the transit are as follows: $p$ = 0.2010, $T$ = 1 day, $i$ = 85\degree\ and $e$ = 0. The limb-darkening coefficients are given in the figure.}
\label{fig:limbdarkening}
\end{figure}

The inclusion of limb darkening for the primary transit results in a smoother transition of ingress and egress of the exoplanet. \cref{fig:limbdarkening}\ is a comparison between a transit with zero limb-darkening and one for the quadratic limb-darkening model described in the foregoing sections following the methods described in \cite{Mandel2002}. 

\begin{figure}[tbh]
\centering
\includegraphics[width=0.7\linewidth]{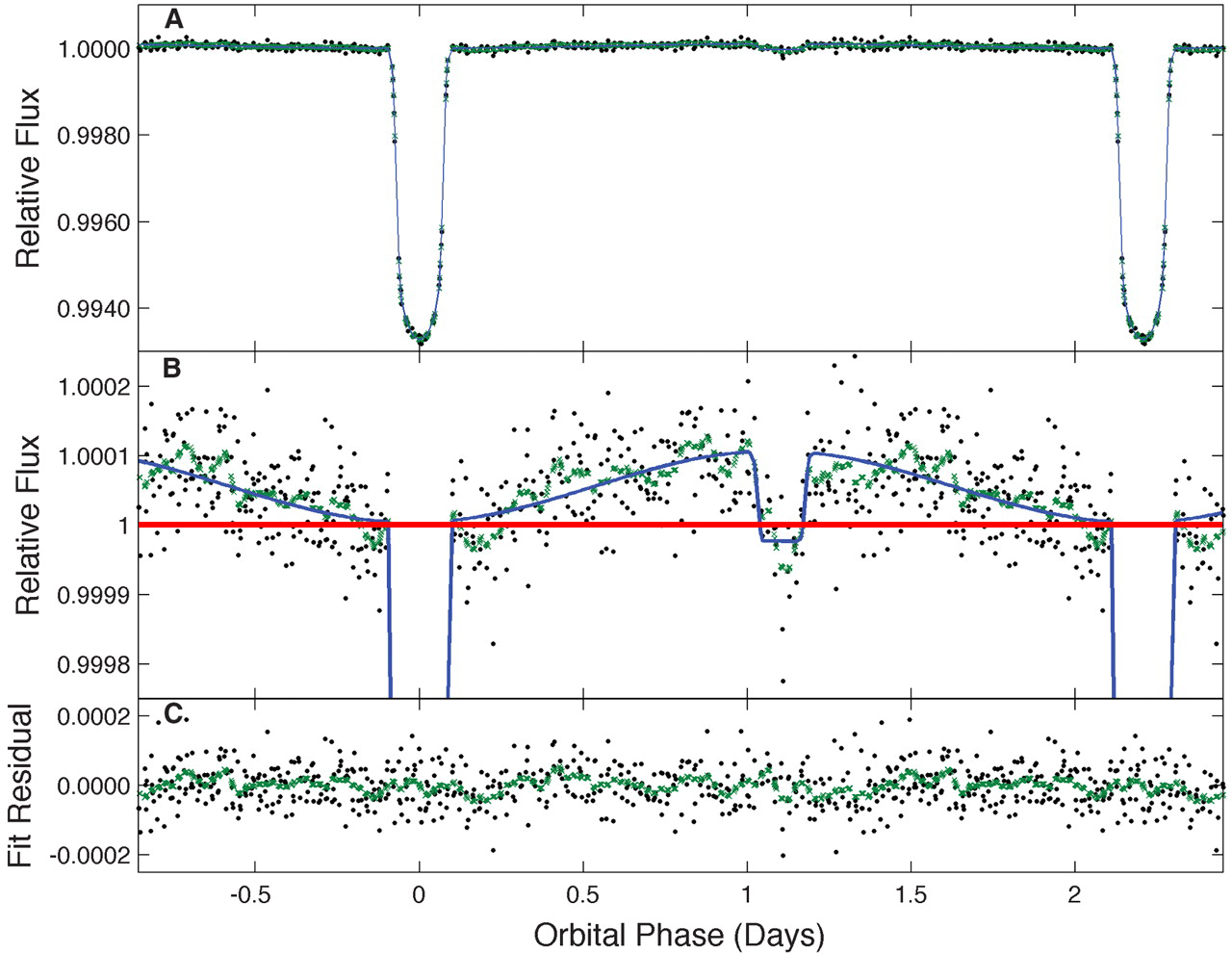}
\caption{(\textbf{A}) The light curve of HAT-P-7b recorded by \myKepler. The above was processed and published by \protect\cite{Borucki2009}\ (reprinted here with permission). (\textbf{B}) Shows the same light curve at a greater magnification where the red line marks the baseline flux of the host star just before the primary transit. (\textbf{C}) is a plot of the residual, or the difference between the data in the model. We see here a lack of obvious structure in the residual, thus indicating that most of the effects have been modeled. The effects include the primary transit, and the photometric variations we will discuss in this work.}
\label{fig:hat-p-7b---borucki---science---baseline}
\end{figure}

The plots within this work will be noiseless and are produced to explore the effects of the photometric variations on light curves. We present \cref{fig:hat-p-7b---borucki---science---baseline}\ to illustrate real light curve data from \myKepler\ of the exoplanet HAT-P-7b, \cite{Borucki2009}. Also within the plots is the fitted light curve which includes the effects of the primary transit, limb darkening, and the photometric variations we will be discussing in later chapters. The photometric variations include Doppler boosting, the ellipsoidal variations, reflective light, and thermal light.

\section{Transit Separation and Duration}\label{sec:tranSepDuration}
\begin{figure}[tbh]
\includegraphics[width=0.98\textwidth]{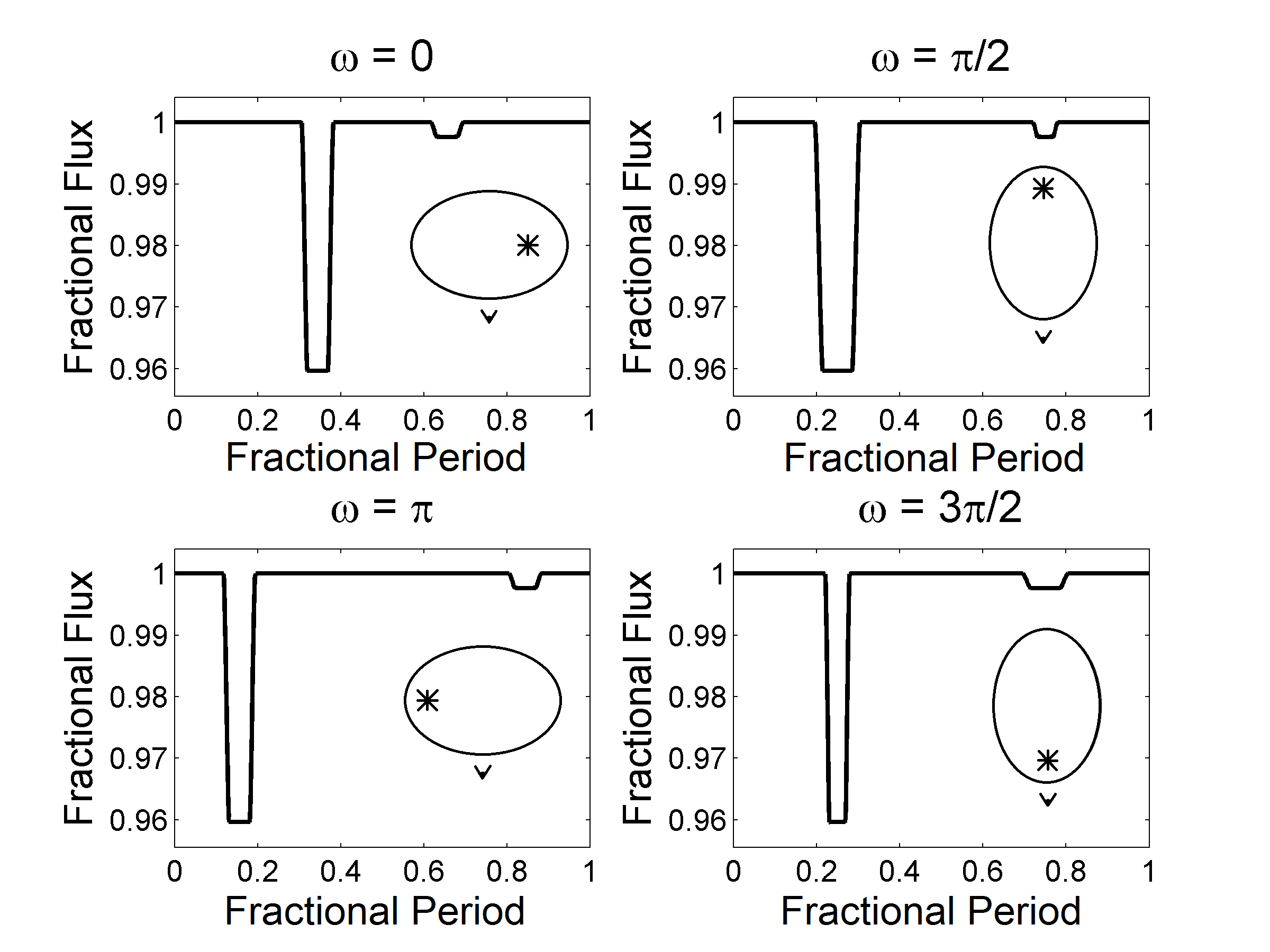}
\caption{Shown is an illustration of the effect of the argument of periastron, $ \omega $, on the transit separation. In each case the eccentricity is set to 0.3. Inset in each sub-figure is a picture of the orientation of the orbit where the observer is labeled with a small eye at the bottom of the sub-figure.}
\label{fig:separationfig}
\end{figure}

\cref{fig:secondaryandprimary}\ shows the primary and secondary transit for a circular orbit for which the separation in time of the two events is one half period. It is only the case for circular orbits that the time between the point of primary mid-transit ($t_I$) and secondary mid-transit ($t_{II}$) is one half period; however, eccentric orbits exhibit a time offset that depends on the eccentricity, the argument of periastron, $\omega$, and the inclination of the orbit as described by Kallrath and Milone in \cite{KallrathMilone99}. The relationship between the mid-transit times is given by 
\begin{equation}\label{eq:timesep}
\begin{aligned}
	\frac{\pi}{2T}\left(t_I-t_{II}-\frac{T}{2}\right)&\approx e\cos\omega\\
	\frac{\pi}{2T}\Delta t_{II}&\approx e\cos\omega,
\end{aligned}
\end{equation}
as derived in Charbonneau et. al.'s work~\cite{charbonneau05}. The time offset, $\Delta t_{II}$, of eccentric orbits is zero for $\omega$ equal to $\pi/2$ and $3\pi/2$, as shown in \cref{fig:separationfig}, for which the transit separation is a half period.

\begin{figure}[hbt]
\centering
\includegraphics[width = 0.75\textwidth]{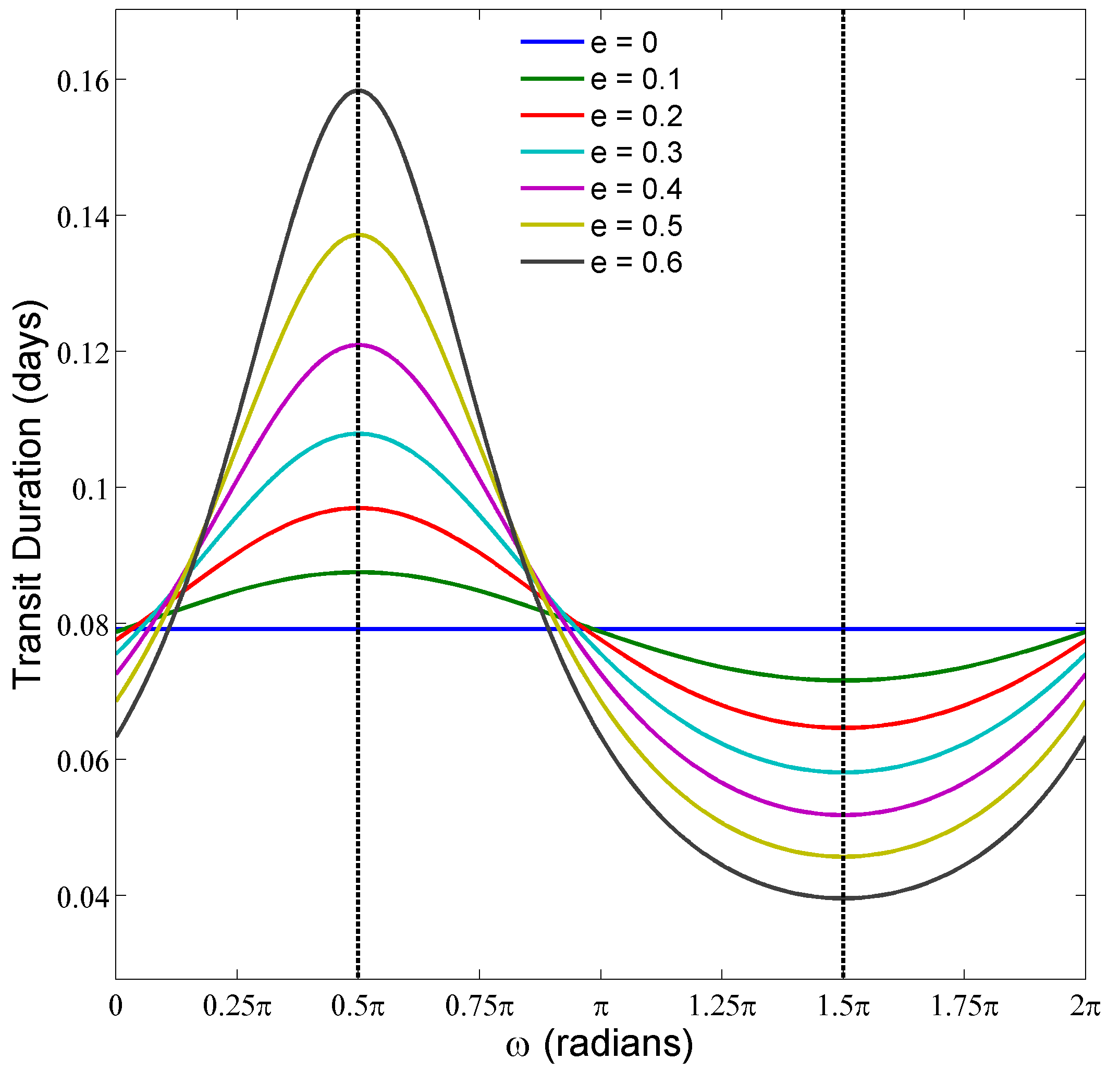}
\caption{\label{fig:durationFig} An illustration of the influence of both eccentricity, e, and the argument of periastron, $\omega$, on the transit duration for a 1.431$R_J$ radius exoplanet in a 2.205 day period orbit about a 2$R_{\astrosun}$ star at an inclination of $i$ = 84.1\degree. The vertical, black dashed lines show where $\omega$ is equal to a half unit of $\pi$. For non-circular orbits, the transit duration can be less than or greater than that expected from a circular orbit of radius equal to the semi-major axis. The duration is a minimum for $\omega = 3\pi/2$ (the transit occurs at periastron) and a maximum for $\omega$ = $\pi/2$ (the transit occurs at apoastron) for a given eccentricity.}
\end{figure}

Like the time offset, the primary transit duration depends on the orbital parameters of the exoplanet. The duration, $t_D$ is given by \cite{ford08,TingleySackett}, 
\begin{equation}\label{eq:tranDuration}
\frac{t_D}{T} = \frac{R_s}{\pi a \sqrt{1-e^2}}\sqrt{(1-p)^2-b^2}\left(\frac{r_t}{a}\right),
\end{equation}
where $r_t$ is the star-planet separation at the time of mid-transit and is assumed to be approximately constant throughout the transit. The equation for the mid-transit star-planet separation is,
\begin{equation}
	r_t =\frac{ a\left(1-e^2\right)}{1+e\cos(\pi/2 +\omega)},
\end{equation} 
where $ (\pi/2 +\omega) $ is the true anomaly at mid-transit in our coordinate system, see Figure 1 in \cite{TingleySackett}; therefore,
\begin{equation}
r_t = \frac{a\left(1-e^2\right)}{1-e\sin\omega}.
\label{eq:rt}
\end{equation}
Finally, $b$ in \cref{eq:tranDuration}\ is the impact parameter, $b = (r_t/R_s)\cos i$. \cref{fig:durationFig}\ illustrates the relationship between the transit duration, the argument of periastron and eccentricity of the orbit described in~\cref{eq:tranDuration}. 

Together the time offset and transit duration can be used to determine the eccentricity and $\omega$ for a given inclination, but this is no easy task. As can be seen in~\cref{fig:durationFig}, there are multiple combinations of  eccentricity and $\omega$ that are able to produce similar transit durations as that of a circular orbit with radius equal to the semi-major axis of the eccentric orbit for cases in which $\omega$ is near zero or $\pi$. The degeneracy between eccentricity and $\omega$ can be difficult to break without another detection method to determine eccentricity or the presence of a secondary transit to constrain $\omega$ via the time offset.  Alternative parameterizations of the transits use joint parameters of the eccentricity and argument of periastron, such as $e\sin\omega$ and $e\cos\omega$, see Ford's work \cite{Ford06MCMC}.

\section{Transit Probability} 
\begin{figure}[hbt]
\centering
\includegraphics[width = 0.8\textwidth]{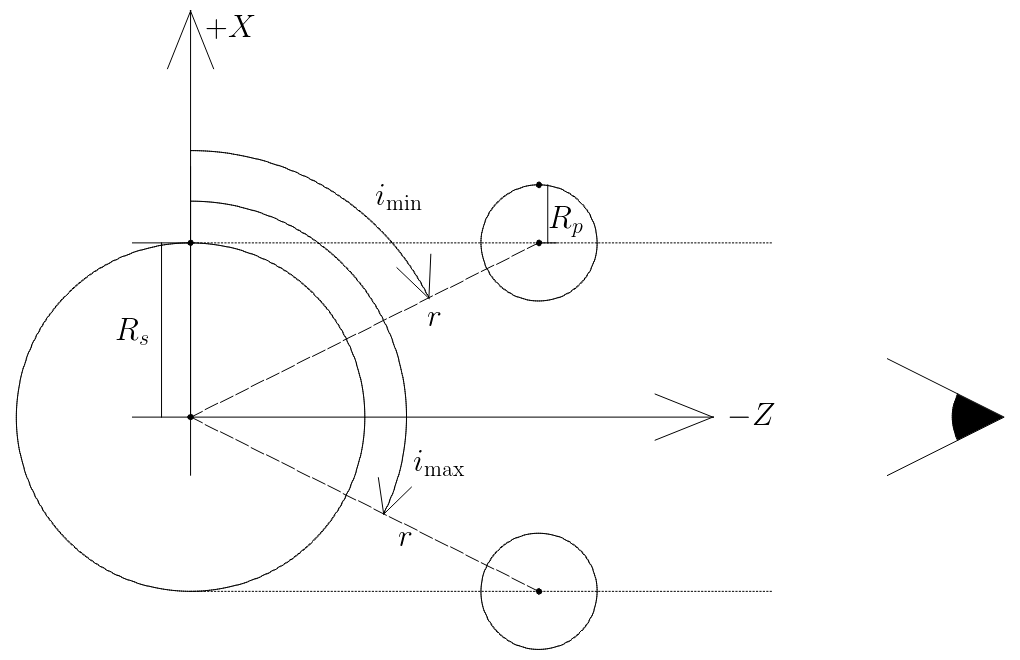}
\caption{A figure illustrating the relationship between the orbital orientation and the minimum and maximum inclinations for a transit along our line of sight. The line of sight points along the $-Z$ axis and the inclinations are measured from $+X$. The minimum and maximum inclinations for a transit, $i_{\textnormal{min}}$ and $i_{\textnormal{max}}$ respectively, shown here include grazing transits in which the entirety of the disk of the exoplanet is not within the disk of the host star.}
\label{fig:inclinationProb}
\end{figure}

A transit will only occur if an exoplanet passes in front of its host star along our line of sight. The probability of such an event, the extremes of which are depicted in \cref{fig:inclinationProb}, was first derived by Borucki and Summer in their work~\cite{BoruckiSummers1984} for circular orbits and was found to be given by
\begin{equation}
\label{eq:circProb}
P(transit|circular, I) = \frac{R_s}{a},
\end{equation}
where $a$ is the radius of the orbit, and $ I $ symbolizes all prior information, such as the model being used to describe the orientation of the exoplanet along the line of sight. \cref{eq:circProb}\ shows that the probability of a transit increases for larger stars and decreases as the exoplanet's orbital radius increases. This is because the transit probability can be related to the shadow swept out by the exoplanet which will be less for greater orbital radius, as will now be shown.

Following, is a derivation for the transit probability for eccentric orbits using the geometry shown in \cref{fig:inclinationProb}\ and following the work of Barnes and of Knuth et al., \cite{Barnes2007,2017Entropy}, with modifications for the coordinate system used in this work. To begin, we assume that there is no preferred orbital orientation for exoplanets. The three parameters of interest are the star-planet separation, $r$, the orbital inclination, $i$, and the true anomaly, $\nu$, where $r$ is related to $\nu$ via~\cref{eq:r_trueanomaly}. Their joint probability is given by $P(\nu, i| I) = 1/C$, where $C$ is a constant that can be calculated by normalizing $P(\nu, i| I)$:
\begin{equation}
\begin{aligned}
1 &= \int_{0}^{2\pi}\int_0^{\pi} P(\nu, i| I)\sin i~di~d\nu \\
  & = \int_{0}^{2\pi}\int_0^{\pi}\frac{1}{C}\sin i~di~d\nu \\
\end{aligned}
\end{equation}
Solving for $C$ gives the normalization constant
\begin{equation}\label{eq:normalC}
C = \int_{0}^{2\pi}\int_0^{\pi}\sin i~di~d\nu = 4\pi,
\end{equation}
which shows that the joint probability of $\nu$ and $i$ is $P(\nu, i| I) = 1/(4\pi)$, i.e. the joint prior of $ \nu $ and $ i $ is the inverse of the solid angle of a sphere. The result in \cref{eq:normalC}\ holds true for both eccentric and circular orbits as shown in \cite{2017Entropy}. 

~\cref{fig:inclinationProb} shows that a transit's occurrence depends on its inclination, where $\cos i_{\textnormal{min}} = R_s/r$ and $\cos i_{\textnormal{max}} = \cos(\pi-i_{\textnormal{min}}) = -\cos i_{\textnormal{min}}$. The probability of a transit is then 
\begin{equation}
\label{eq:transitProb1}
\begin{aligned}
P(transit|I) &= P\left(i_{\textnormal{min}} \leq i \leq \piover2 \textnormal{ OR } i_{\textnormal{max}} \geq i \geq \piover2 |I\right) \\
             & = P\left(i_{\textnormal{min}} \leq i \leq \piover2|I\right) + P\left(i_{\textnormal{max}} \geq i \geq \piover2 |I\right) \\
             & = \frac{1}{4\pi}\left(\int_0^{2\pi}\int_{i_{\textnormal{min}}}^{\pi/2}\sin i ~di~d\nu + \int_0^{2\pi}\int_{\pi/2}^{i_{\textnormal{max}}}\sin i ~di~d\nu\right) \\
             & = \frac{1}{4\pi}\int_0^{2\pi}(\cos i_\textnormal{min} - \cos i_\textnormal{max})~d\nu \\
             & = \frac{1}{2\pi}\int_0^{2\pi}\frac{R_s}{r (\nu)}~d\nu,
\end{aligned}
\end{equation}
where we have normalized the transit probability to the probability of any given orbital inclination and true anomaly using~\cref{eq:normalC}. In the second line of \cref{eq:transitProb1}\ we have applied the exclusive OR rule for probability, i.e. $ P(A\textnormal{ OR }B|I)=P(A| I) + P(B| I) $. Substituting \cref{eq:r_trueanomaly}\ for $r (\nu)$ in \cref{eq:transitProb1}\ produces
\begin{equation}
\label{eq:transitProbFinal}
\begin{aligned}
	P(transit|I) &= \frac{R_s}{2\pi}\int_0^{2\pi}\frac{1+e\cos\nu}{a(1-e^2)}~d\nu \\
				 &= \frac{R_s}{a(1-e^2)},
\end{aligned}
\end{equation}
which reduces to $P(transit|I) = R_s/a$ for zero eccentricity as derived in \cite{BoruckiSummers1984}\ and expected from \cref{eq:circProb}. 

~\cref{eq:transitProbFinal} was first derived in~\cite{Barnes2007} and includes grazing orbits for which the impact parameter, $b$, is less than or equal to one. The exclusion of grazing orbits requires the replacement of the numerator in~\cref{eq:transitProbFinal} with $R_s - R_p$. To include all transits, even those in which less than half of the disk of the exoplanet occults the stellar disk, replace the numerator with $R_s + R_p$. 

We may assume that $e\leq 1$ for orbits of interest in exoplanet research; therefore, \cref{eq:transitProbFinal}\ shows that an eccentric orbit is more likely than circular orbit of radius equal to the semi-major axis, $a$, to result in a transit. Yet, this fact may not mean an increase in detections. The star-planet separation, $r$, and hence $\omega$, controls the transit time and shorter transits are more difficult to detect than longer transits. As a result, transits that occur near apoastron will be easier to detect than those near periastron. This may also be thought of as a consequence of Kepler's second law; an exoplanet sweeping out a smaller area in a given time period, i.e. near periastron, will be harder to detect than one sweeping out a larger area, i.e. near apoastron.

Also of interest is the possibility of detecting exoplanets with only a secondary transit due being in a highly inclined and eccentric orbit. Such an occurrence requires that the exoplanet have strong planetary photometric emissions and will likely orbit closely to its host star. Finally,~\cref{eq:transitProbFinal} implies that a bias to detect eccentric exoplanets will exist within transit data. Such a bias could be taken into account by using an informative prior for eccentricity, as described in Kipping's work ~\cite{kipping2014}.

In the foregoing chapter,  we reviewed the relationship between the transit depth, its shape, and important exoplanet characteristics, such as the exoplanet's radius. In addition, we explored the difference between a transit of a uniform source versus a transit of a source that exhibits quadratic limb darkening,  \cite{Mandel2002}. We then explored the relationship between the separation of the primary and secondary transits and exoplanet orbital parameters such as the eccentricity and argument of periastron, \cf\ \cite{KallrathMilone99,charbonneau05}. We also explored the relationship between said parameters and the transit duration, \cf \cite{ford08,TingleySackett}. Finally, we concluded with a discussion of the probability that an exoplanet will transit along and observer's line of sight, \cf\ \cite{Barnes2007,BoruckiSummers1984}.
\chapter{Photometric Variations of the Host Star}\label{ch:hoststar}

In addition to transits, photometric measurements can also reveal other variations. In this chapter we will look at two variations that affect the stellar emissions and are induced by the presence of an exoplanet. The first is the result of relativistic effects on the emitted light of the host star and is referred to as either Doppler Boosting (Boosted light) or Beaming. The second effect is the ellipsoidal variation which is the result of tidal forces acting on the surface of the the host star. Four models for the ellipsoidal variation will be outlined, three of which are trigonometric and one is implemented by modeling the deviation of the host star's shape from spherical.

\section{Doppler Boosting or Beaming}
The Boosted light is the result of the radial velocity of the host star along our line of sight as described in~\cref{sec:RV}. As the star moves toward an observer the aberration of light creates a beaming effect in which the light rays exhibit a decrease in angle and an increase in emission rate in the frame of a stationary observer.  As the star moves away there will be a decrease in the observed emission rate. Thus, the Boosted light is characterized by an increase and decrease in flux over the course of one period. Note that the effect is only sinusoidal for circular orbits. The following will derive the equation for this effect for the non-relativistic limit as laid out by Rybicki and Lightman~\cite{rybickiandlightman}.

First, we must transform the four-vector, $k^\mu = (\omega, \vec{k})$, describing the emitted electromagnetic wave from rest frame $S'$ to frame $S$ moving relative to $S'$. Here $\omega$ is angular frequency of the wave  and $\vec{k}$ is the wave vector where its magnitude describes how many oscillations it completes per unit of space, i.e. $ |\vec{k}|=2\pi/\lambda $ where $ \lambda $ is the wavelength. 

It is convenient to use the fact that $k^\mu k_\mu = 0$ to write the magnitude of the wave vector as 
\begin{equation*}
	|\vec{k}| = \frac{\omega}{c}.
\end{equation*}
Furthermore, we may write the magnitude of the wave vector in the $x$-direction as
\begin{equation}\label{eq:kxvec} 
    k_x = \frac{\omega}{c}\cos\theta, 
\end{equation}
as described in \cite[Ch. 4]{rybickiandlightman}. We will consider the $ x $-direction to be aligned along our line of sight. The transformation of $k^\mu = (\omega, \vec{k})$ from $S'$ to $S$ is
\begin{equation}\label{eq:ktransform}
k^{\mu'} = \Lambda_\mu^{\mu'}k^\mu,
\end{equation}
where $\Lambda_\mu^{\mu'}$ is the Lorentz transformation matrix in the $x$-direction and is given by 
\begin{equation}\label{eq:lorentzmatrix}
\Lambda_\mu^{\mu'} = 
\begin{pmatrix}
\gamma & -\beta\gamma  & 0 &0 \\ 
 -\beta\gamma& \gamma & 0 & 0\\ 
0 & 0 & 1 & 0\\ 
0 & 0 & 0 & 1
\end{pmatrix}.
\end{equation}
In \cref{eq:lorentzmatrix}, $\gamma = 1/\sqrt{1-\beta^2}$ is the Lorentz factor, $\beta = v/c$ is the ratio of the speed along the $x$-axis to that of light~\cite{rybickiandlightman}. 

\cref{eq:ktransform}\ produces two equations of interest. First, the shift in angular frequency is given by
\begin{equation}\label{eq:omegatransform} 
\omega ' = \gamma(1-\beta\cos\theta)\omega
\end{equation}
and the transformation of $k_x$ to $k_x'$ can be re-written as
\begin{equation}\label{eq:thetatransform} 
\omega ' \cos\theta ' = \gamma(\cos\theta -\beta)\omega.
\end{equation}
Substitution of \cref{eq:omegatransform}\ into \cref{eq:thetatransform}\ and solving for $ \cos{\theta'}$ produces the equation for light aberration: 
\begin{equation}\label{eq:abberation}
\cos\theta' = \frac{\cos\theta-\beta}{1-\beta\cos\theta}
\end{equation}
which describes the how angles transform from frame $S$ to frame $S'$ with respect to the $x$-axis. 

We will assume that the emitted light of the host star is that of a sphere radiating isotropically in its rest frame $S'$. The star's motion will be along our line of sight, or along the $x$-axis. The observed flux, $dF$, of an observer in frame $S$ is given by the amount of energy, $dE$, at emission rate, $d\Gamma$, per unit solid angle, $ d\Omega $:
\begin{equation}
dF = \frac{dEd\Gamma}{d\Omega}.
\end{equation}
We are interested in the ratio of the observed to emitted flux:
\begin{equation}\label{eq:fluxratio}
\frac{dF}{dF'} = \frac{dE}{dE'}\frac{d\Gamma}{d\Gamma '}\frac{d\Omega '}{d \Omega}.
\end{equation}
The aberration of light will affect the solid angle, which is given by
\begin{equation}
d\Omega ' =  d(-\cos\theta') d\theta'.
\end{equation}
and is shifted to the new frame such that,
\begin{equation}
d\Omega =d(-\cos\theta) d\theta
\end{equation}
The relationship between $d\Omega$ and $d\Omega'$ can be written as
\begin{equation}
\frac{d\Omega}{d\Omega'} = \gamma^2(1-\beta\cos\theta)^2
\end{equation}
by differentiation of \cref{eq:abberation}.

The ratio of the energy observed to the emitted energy can be determined from \cref{eq:omegatransform} to find that
\begin{equation}
	\frac{dE}{dE'} = \frac{1}{\gamma(1-\beta\cos{\theta})}.
\end{equation}
Time-dilation may be used to determine the ratio of the observed rate of emission to that of the star with the result being
\begin{equation}
\frac{d\Gamma}{d\Gamma '} = \frac{1}{\gamma(1-\beta\cos\theta)}.
\end{equation}
Substitution of the above equations into \cref{eq:fluxratio}\ produces the observed flux, $F_{\textnormal{boost}}$ in terms of the emitted flux $F_s$:
\begin{equation}\label{eq:Fboostexact}
F_{\textnormal{boost}}(t) = F_s\left(\frac{1}{\gamma(1-\beta \cos\theta(t))}\right)^4.
\end{equation}
The typical radial velocity of a star is less than 10$^3$ m/s, so it is appropriate to take the non-relativistic limit:
\begin{equation}
	F_{\textnormal{boost}}(t) \approx F_s(1+4\beta\cos{\theta})
\end{equation}
where  $ \beta\cos\theta $ is the component of $ \beta $ along the line of sight. Letting $ \beta_r=\beta\cos\theta $ we may write the fractional flux of the boosted light as
\begin{equation}\label{eq:phiBoost}
\Phi_{\mathrm{boost}} = \frac{F_{\textnormal{boost}}(t)}{F_s} \approx 4\beta_r(t)
\end{equation}
where $\beta_r = V_z/c$ and $V_z$ is given in \cref{eq:radialVelFinal}, and we have dropped the $+1$ term to reflect the fact that the final addition of transit and photometric variations already includes this term. 

\begin{figure}[hbt]
\centering
\includegraphics[scale = 0.7]{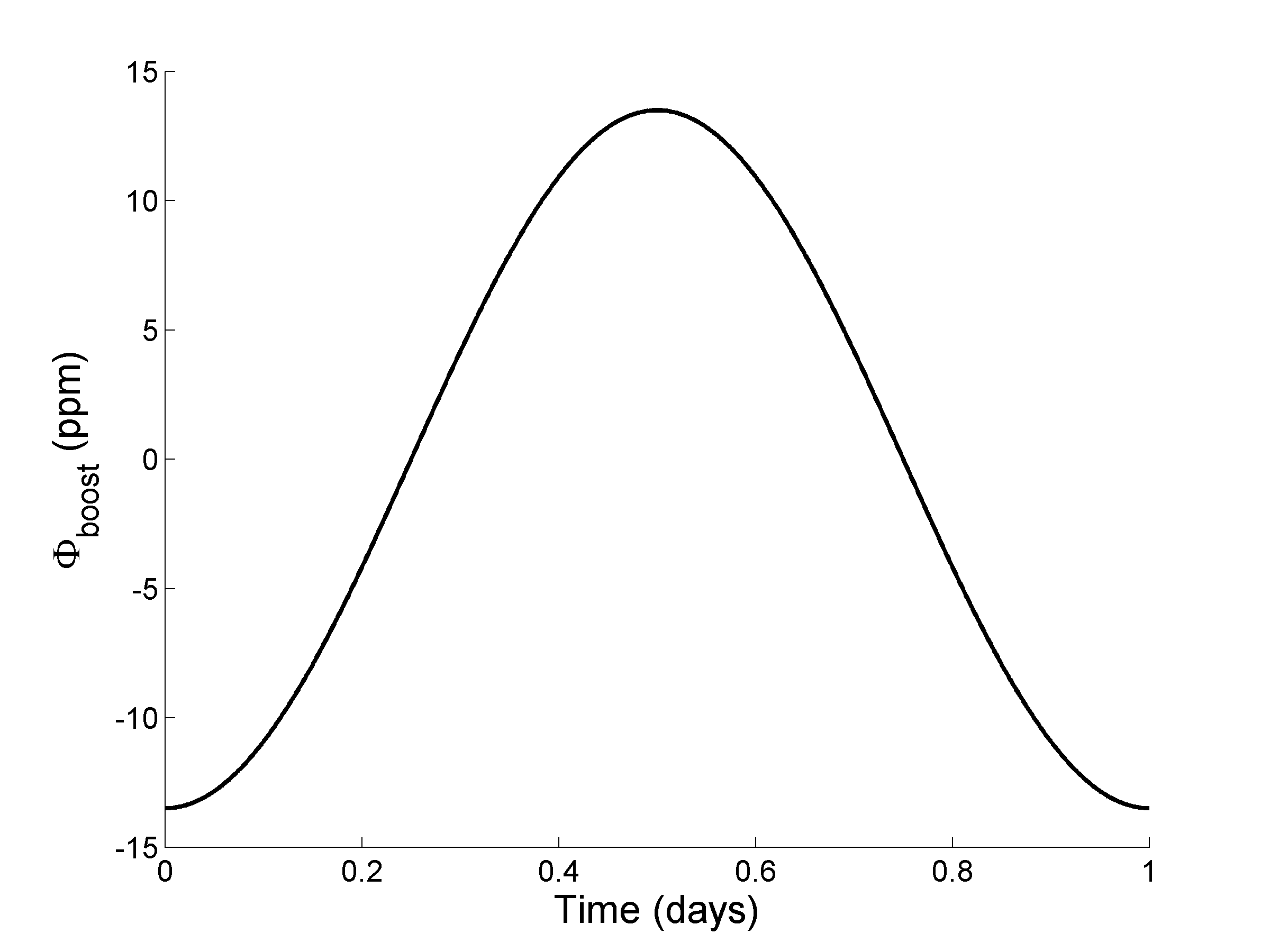}
\caption{A plot of the fractional flux, $\Phi_{\textnormal{boost}}$, of the boosted variation induced on a solar mass star by an exoplanet with for which $M_p\sin{i}= 5 M_J$ orbiting with a period of 1 day. The orbit is oriented edge on and is circular. The vertical axis is in parts per million (ppm). See text for full description.}
\label{fig:boostedFig}
\end{figure}

In \cref{fig:boostedFig}, the exoplanet is beginning its orbit at mid-primary transit and is moving toward the observer; therefore, the host star is moving away from the observer and a negative change in flux occurs. The exoplanet reaches full phase at approximately 0.25 days and we observe zero boosted light, then the star begins to move toward the observer as the exoplanet moves away toward the secondary transit. The maximum variation occurs during the secondary transit at 0.50 days. 

The boosted light may be used to determine the minimum mass of the exoplanet because \cref{eq:phiBoost}\ is proportional to $V_r$, which in turn in proportional to $M_p\sin i$, the minimum mass, see \cref{eq:radialVelFinal}. A detection of the boosted variation allows exoplanet researchers to obtain a mass measurement using only photometric data; however, the variation is only detectable for high mass exoplanets. In \cite{Loeb2003}, Loeb and Gaudi suggest that Kepler would be capable of detecting the boosted light of exoplanets with minimum mass $\gtrsim 5 M_J$ and periods $\lesssim 7$ days for main-sequence stars. If the boosted light variation is accompanied by a transit, then it is possible that the transit could provide a measure of the inclination via transit-timing, see~\cref{eq:tranDuration}. With the inclination set, the true mass of the exoplanet could then be determined.  

\section{Ellipsoidal Variations} 
The ellipsoidal variations of the host star are induced by the gravitational pull of an exoplanet and depend on the extent to which the star deviates from spherical. First, we will consider the amplitude of the variation as a fraction of the flux of a spherical star of equal mass, where $\Phi_{\textnormal{ellip}} = F_{\textnormal{ellip}}/F_s$ is the fractional change in flux of the star induced by the ellipsoidal variations.  Then, we will discuss four approaches to modeling the time dependence of the ellipsoidal variation. 

To estimate the amplitude of the ellipsoidal variations, consider the tidal acceleration acting on a small section of stellar surface at a distance from the center of exoplanet of $r-R_s$
\begin{equation}\label{eq:aT}
\begin{aligned}
a_T &= GM_p\left(\frac{1}{(r-R_s)^2}-\frac{1}{r^2}\right) \\
    & = \frac{GM_p}{r^2}\left(\frac{1}{(1-R_s/r)^2}-1\right)\\
    & \approx \frac{2GM_pR_s}{r^3}
\end{aligned}
\end{equation}
The last line of~\cref{eq:aT} follows from keeping the first order terms of $R_s/r$. The ratio of the tidal acceleration to that of the surface gravity on the host star, $a_g = GM_s/R_s^2$, gives us an approximation of the amplitude of the ellipsoidal variations:
\begin{equation}
\frac{a_T}{a_g} \propto \frac{M_p}{M_s}\left(\frac{R_s}{r}\right)^3.
\end{equation}
The amplitude of the ellipsoidal variations will also depend on the spectral characteristics of the host star and is given by Loeb and Gaudi in \cite{Loeb2003},
\begin{equation}\label{eq:amplitude}
A_{\textnormal{ellip}} = \beta\frac{M_p}{M_s}\left(\frac{R_s}{a}\right)^3, 
\end{equation}
where $\beta$ is the gravity-darkening exponent. The value of $\beta$ can be determined in multiple ways. In this work $\beta$ depends on the linear gravity-darkening coefficient, $g$, and the linear limb-darkening coefficient, $u$, as described in \cite{Morris1985}\ by Morris:
\begin{equation}
\beta = 0.15\frac{(15+u)(1+g)}{3-u}.
\end{equation}
The amplitude of the effect in~\cite{Loeb2003}, shown in \cref{eq:amplitude}, was determined for circular orbits with radius $a$. Here we will make the substitution of $a = r$ to account for eccentric orbits, where $r$ is given by \cref{eq:r_trueanomaly}. The values of $g$ and $u$ can be determined from the spectral characteristics of the star, \cf\ in \cite{ClaretBloemen2011}\ by Claret and Bloemen. For solar-type stars $\beta \approx$ 0.45. 

The ellipsoidal variations arise due to the deviation of stellar shape from spherical. The variation will be at a maximum when the largest face of the star is facing the observer, at $ \myPhase=\pi/2 $, and will be at a minimum during full and new phase when $\myPhase=0$ or $\pi$. As such, the variation should occur out of phase from the planetary photometric emissions, which peak during the full phase which will be discussed in \cref{ch:reflectedlightluminosity,ch:thermalradiation}, and will go through two cycles per orbit. Three trigonometric methods have been used to describe this behavior, \cite{Faigler2010,KaneGelino2012,Placek2014}, and a fourth method seeks to determine the deviation at each point of a projected stellar grid, \cite{Jackson2012}. To begin, we will consider the three trigonometric methods. 

The first trigonometric method, proposed in \cite{Faigler2010} by Faigler and Mazeh, is used in the BEaming, Ellipsoidal and Reflection (BEER) method of modeling photometric light curves of large, short period exoplanets. In the BEER method, the time dependent nature of the variation is taken into account through physical arguments, e.g. it should be out of phase from the reflected light and peak twice per orbit. To account for inclination, the authors include a $\sin^2 i$ term, giving:
\begin{equation}\label{eq:BEER}
\Phi_{\textnormal{ellip,BEER}} \approx -A_\textnormal{ellip}\sin^2 i \cos\left(\frac{2\pi}{T/2}t\right).
\end{equation}
The minus sign reflects the fact that the ellipsoidal variations are out of phase from the planetary photometric emissions from \cref{ch:reflectedlightluminosity,ch:thermalradiation} and the $T/2$ term reflects the fact that the effect peaks twice per orbit. For circular orbits the phase angle, $\myPhase$, is equal to $2\pi t/T$, but this does not hold true for highly eccentric orbits, see \cref{sec:orbitalanomalies}. As an approximation, we may re-write \cref{eq:BEER}\ as 
\begin{equation}\label{eq:BEERphase}
\Phi_{\textnormal{ellip,BEER}} \approx -\beta \frac{M_p}{M_s}\left(\frac{R_s}{r}\right)^3\sin^2 i \cos(2\myPhase).
\end{equation}

\begin{figure}[hbt]
\centering
\includegraphics[scale = 0.7]{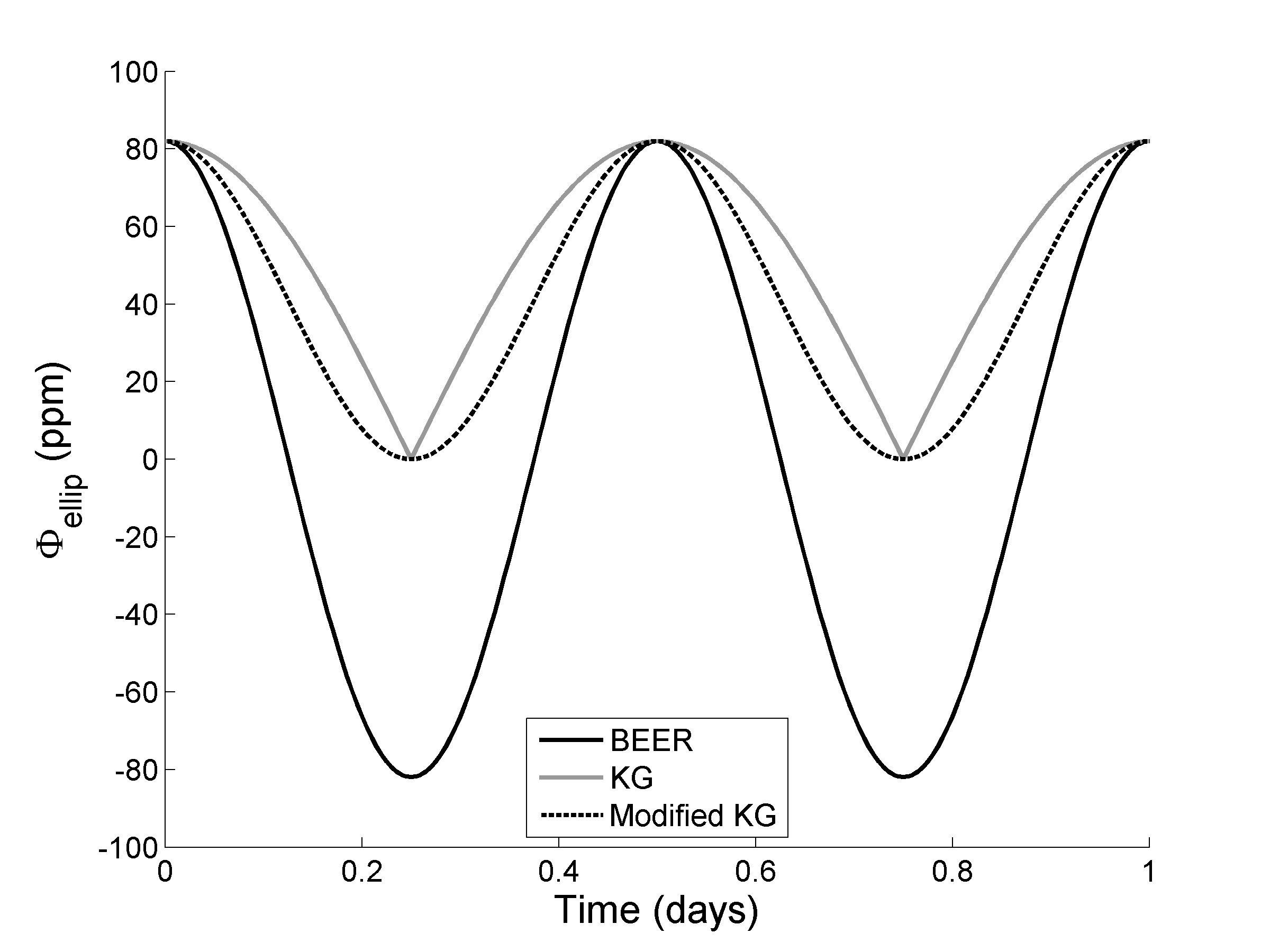}
\caption{Depiction of the three trigonometric versions of the ellipsoidal variation for an exoplanet with $M_p\sin i$ = 5 $M_J$ orbiting a solar mass star with a one day period. The solid black line shows the BEER version described by~\cref{eq:BEERphase}, the solid grey line that of Kane and Gelino from~\cref{eq:KGfinal} and the dashed-black line the modified Kane and Gelino variation of~\cref{eq:KGmodified}. Note that the minimum value for the BEER version is negative rather than zero. This negative minimum may result in an underestimation for the exoplanetary  mass because it takes half the mass to produce the same amplitude effect.}
\label{fig:ellipFigAll}
\end{figure}

The next trigonometric method of estimating the fractional change in flux due to the ellipsoidal variations was proposed by Kane and Gelino in \cite{KaneGelino2012}\ and considers the unit separation of centers along our line of sight:
\begin{equation}\label{eq:normSEP}
\begin{aligned}
\frac{R(t)}{r(t)} &= \frac{\sqrt{X(t)^2+Y(t)^2}}{r(t)} \\
  &= \sqrt{\cos^2(\omega+\nu(t))+\sin^2(\omega+\nu(t))\cos^2 i},
\end{aligned}
\end{equation}
where $X$ and $Y$ are given in \cref{eq:cartfinal}. One may then write the time dependent variations as
\begin{equation}\label{eq:KG1}
\frac{F_{ellip,KG}}{F_s} = \Phi_\textnormal{ellip,KG} \approx A_\textnormal{ellip}\sqrt{\cos^2(\omega+\nu)+\sin^2(\omega+\nu)\cos^2 i}.
\end{equation}
Substituting \cref{eq:phaseangle,eq:amplitude}\ into \cref{eq:KG1}\ produces the a simpler form of the equation:
\begin{equation}\label{eq:KGfinal}
\Phi_\textnormal{ellip,KG} \approx \beta \frac{M_p}{M_s}\left(\frac{R_s}{r}\right)^3\sin\myPhase.
\end{equation}
As seen in \cref{fig:ellipFigAll}, \cref{eq:KGfinal}\ is discontinuous in the first derivative at its minima when $ \myPhase=0 $ or $ \pi $. 

A slight modification can be made to~\cref{eq:KGfinal} to be rid of the discontinuity as described by Placek et. al. in~\cite{Placek2014}. Rather than considering the separation of the stellar disk center from the exoplanet disk center along our line of sight, we instead consider the square separation, producing
\begin{equation}\label{eq:KGmodified}
\frac{F_{ellip,mod}}{F_s} =\Phi_\textnormal{ellip,mod} \approx \beta \frac{M_p}{M_s}\left(\frac{R_s}{r}\right)^3\sin^2\myPhase.
\end{equation}

The three trigonometric models are shown in~\cref{fig:ellipFigAll}. In each case the maxima occur at the same phase and have the same value; however, the minima, though occurring at the same phase, differ. Both of the models that use the unit separation of centers along our line of sight share a minimum value of zero, but the BEER method has a negative value at this point. In effect, the BEER method has twice the amplitude of the other two models; therefore, an exoplanet with half the mass could produce an equal amplitude as documented by Gai and Knuth in \cite{gaiandknuth2018}. The behavior of the BEER method implies that it is not an accurate description of the ellipsoidal variations of stars induced by the presence of an exoplanet. 

Finally, a brief look at the fourth method of modeling the ellipsoidal variation, known as Ellipsoidal Variations Induced by a Low-Mass Companion (EVIL-MC), developed by Jackson et. al. and described in~\cite{Jackson2012} in IDL for circular orbits and adapted to Matlab and eccentric orbits by Gai, \cite{gaimasters}. EVIL-MC determines the shape of the star and determines its deviation from spherical for each point on a projected stellar grid. This deviation is calculated as
\begin{equation}\label{eq:deviation}
\delta R = q\left(\left[\normz^2-2\normz\cos\psi + 1\right]^{-1/2}-\left[\normz^2+1\right]^{-1/2}-\frac{\cos\psi}{\normz^2}\right)-\frac{\omega^2}{2\normz^3}\cos^2\lambda,
\end{equation}
where $q = M_p/M_s$ is the mass ratio of the exoplanet and the star, $\normz$ is the star-planet separation normalized to the stellar radius ($\normz = r(t)/R_s$), $\cos\psi = \hat{R}_s \cdot \hat{r}$ (where $\hat{R}_s$ is the direction of a point on the stellar surface), $\omega$ is the stellar rotation rate and $\cos\lambda = \hat{R}_s \cdot \hat{\omega}$. The observed ellipsoidal variation is determined by summing the blackbody flux over the observed area of the star and is given by,
\begin{equation}\label{eq:EVIL}
\frac{F_{ellip}}{F_s} =\Phi_\textnormal{ellip} = 1- \frac{\phi_\textnormal{sphere}}{\phi_\textnormal{star}},
\end{equation}
where $\phi_\textnormal{sphere}$ is the flux from a spherical star of equal mass and $\phi_\textnormal{star}$ is the flux from the ellipsoidal star. 

Because the star is more of a tear drop than an ellipsoid, the fractional change in flux is greater (i.e. a lower local minimum value) when the exoplanet is at new phase than at full phase because the exoplanet pulls the stellar surface more strongly toward the observer during the primary transit, or new phase at $ \myPhase=\pi $. As the stellar surface is pulled away from its center, the effective temperature observed decreases, thus lowering the observed flux, see Figure 2 in \cite{Jackson2012}. 

Due to its computational complexity, the EVIL-MC model will not be considered in this work. In~\cite{gaimasters}, it was shown that the modified Kane and Gelino model described in~\cref{eq:KGmodified} most closely represents the EVIL-MC model, which was determined to be the preferred model for the Kepler-13 system. For this reason, it will be the one considered when modeling the light curve of an exoplanetary system in this work. 

The ellipsoidal variation provides researchers with another tool to determine the mass of an exoplanet, so long as the stellar parameters are known. Its unique dependence on inclination allows one to calculate the true mass of an exoplanetary companion should the minimum mass ($M_p\sin i$) or the inclination be determined from another variation, e.g. the boosted light or the radial velocity method. Should the radius of the exoplanet be determined, e.g. from the transit depth, researchers may then determine the density of the exoplanet and gain valuable insight into its possible composition. The density sets limits on the rocky or gaseous nature of an exoplanet. 

To conclude, in this chapter we explored the relationship between exoplanet characteristics and photometric variations of the host star induced by the presence of an exoplanet. First, we reviewed the physics behind Doppler Boosting and saw that a detection of this effects can provide information about the mass of an exoplanet in the absence of spectroscopic data, \cite{rybickiandlightman}. In addition, the ellipsoidal variations due to the non-spheroidal shape of the host star created by tidal effects between the host star in exoplanet also provides information about the mass of the exoplanet, \cf \cite{gaiandknuth2018}.

\chapter{Photometric Planetary Emissions: Reflected Light and its Intensity Distribution}\label{ch:planetarychapterReflintensity} 
The next two chapters, \cref{ch:planetarychapterReflintensity,ch:reflectedlightluminosity}, will present a variety of material that serves to: review previous literature; add details to the reasoning behind previous results; present a different approach to achieve the same results as previous literature; make corrections to previous literature; and add unique findings to the field of exoplanet research. To clarify the situation we introduce the following notation as a shorthand to distinguish the above situations. To begin, standard citation shall be used when reviewing previous literature. Secondly, the symbol \myfillin, shall be used to note situations in which we are clarifying the reasoning behind previous literature. Much of the work in \cref{sec:finitesize,sec:luminosityFiniteSizeog,sec:luminosityFiniteSizenew}\ follows the general approach presented by Kopal in \cite{Kopal1953,Kopal1959}. To denote a change in approach to the work presented in \cite{Kopal1953,Kopal1959}\ or others we shall use the symbol \mychange. In addition, corrections to Kopal's work shall be signified by the symbol \mycorrection. Finally, material new to this work will be indicated by the symbol \mynew.

Here we present a summary of the above contribution types within this chapter. A derivation of the reflected intensity distribution of an exoplanet that is exposed to plane parallel ray incident light is presented to both review the literature and to allow for comparisons to the case in which the finite angular size of the host star is considered in determining the reflected intensity distribution. 

\myfillin, A detailed description of the general problem of determining the reflected intensity distribution is presented in \cref{sec:generalApproach}. In \cref{sec:incidentFluxFull,sec:incidentFluxPenKopalMethod} we review the geometry involved in determining the said distribution as presented in \cite{Kopal1953,Kopal1959}\ and provide further details than given in his earlier work. 

\mychange, In each of the following sections a slightly modified coordinate system will be used than that which is typically used in the analysis of reflected light for exoplanets and will be described in \cref{sec:generalApproach}. In addition, the work presented in \cite{Kopal1953,Kopal1959}\ was concerned with the reflection effect between two stars for which the heat albedo was considered to be unity. In this work we make a slight modification by instead considering an exoplanet with single scattering albedo, $ \myScatteringAlbedo $, which is a parameter that describes how much light is reflected by a single surface element of the exoplanet. 

\mycorrection, Corrections to Kopal's previous work include showing that there is a typo in the equations defining $ J_{1,2}' $ in Equations (46) and (47) of \cite{Kopal1953}\ in \cref{sec:incidentFluxPenKopalMethod}. In addition, there will be accompanying corrections to the equations for the reflected intensity distribution, Equations (59) and (60) and (81)-(85) \cite{Kopal1953}.

\mynew, New contributions to the field of reflected light from exoplanets includes the discovery that some of the equations presented in \cite{Kopal1953,Kopal1959}\ produce negative luminosity and other non-physical effects for the fully illuminated zone, as will be discussed in \cref{ch:reflectedlightluminosity}. In addition, it will be shown that the analysis used to describe the reflected intensity distribution in the penumbral zone produces negative luminosity and an outline of how to correct this issue will be presented in \cref{sec:incidentFluxPennew}. Further contributions will be described in \cref{ch:reflectedlightluminosity}.

\section{The General Problem}\label{sec:generalApproach}
The photometric planetary emissions originate from the exoplanet itself and are the result of conservation of energy, \cf\ the work of Sobolev or Seager \cite{sobolev,seager}. The total luminosity of an exoplanet is given by 
\begin{equation}
	\label{eq:totallum}
	\myLuminosity_p = \myLuminosity_{p, refl}+\myLuminosity_{p, therm}
\end{equation}
where $\myLuminosity_{p, refl}$ is the amount of light reflected by the exoplanet due to the incoming luminosity of its host star. The second term, $\myLuminosity_{p, therm}$, is the sum of luminosity produced by the exoplanet itself from either internally generated heat or heating due to the host star and will be discussed in~\cref{ch:thermalradiation}.
The nature of the reflected light will be explored for two methods of characterizing the incident flux of the host star. We will begin by considering the intensity distribution of the reflected light,  $ \myIntensityDis_{refl} $, first for the plane parallel case, in which the incident flux of the host star reaches the exoplanet as plane parallel rays, and then for Extremely Close-in Exoplanets (\myECIES), for which the normalized star-planet separation, $ a/R_s $, is on the order of one and the finite angular size of the host star must be considered. We will see that for \myECIES, the intensity distribution of the reflected light must be split into two illuminated zones, one of which is fully illuminated and the other is partially illuminated. Finally, the reflected luminosity as a function of phase, $ \myLuminosity_{refl}(\myPhase) $, will be determined for the plane parallel ray case as well as for \myECIES.

\begin{figure}
	\centering
	\includegraphics[width=0.5\linewidth]{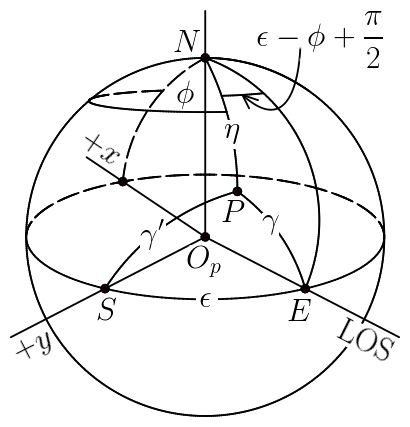}
	\caption{The geometry of a point, $P$, on the exoplanet surface is described by the polar angle, $\eta$, and azimuthal angle, $\phi$. The phase angle, $\myPhase$, is the angle between the substellar point $ S $ and the sub-planetary point, $ E $. The line of sight, or LOS, passes through the center of the exoplanet, $O_p$, and the sub-planetary point. The portions of the exoplanet that are visible to an observer is determined by the phase angle. The angle $\gamma'$ describes the angle between the substellar point and the point $P$ and corresponds to the angle of incidence for plane parallel ray illumination. The angle of foreshortening, $\gamma$, is also called the angle of reflection and is the angle between a point $P$ and the sub-planetary point.  Adapted from \protect\cite{Kopal1953,Kopal1959}.}
	\label{fig:planetcoordinateslabels}
\end{figure}

To calculate the reflected luminosity of an exoplanet, $\myLuminosity_{p, refl}$, as a function of phase, $\myPhase$, one must integrate the reflected intensity distribution, $\myIntensityDis_{refl}$, over the area of the exoplanet visible along an observer's line of sight that is illuminated, $\mathbf{C}$, \cf\ \cite{sobolev,seager}. \cref{fig:planetcoordinateslabels}\ shows the planetocentric coordinate system used to describe the location of a point, $P$, on an exoplanet, adapted from Kopal's work \cite{Kopal1953,Kopal1959}. The center of the coordinate system is marked as $ O_p $ in \cref{fig:planetcoordinateslabels}\ and we will refer to the center of the star as $ O_s $. The angle $\eta$ denotes the polar angle and $\phi$ is the azimuthal angle measured from the $ +x $-axis, where the $ +y $ direction points from the exoplanet center to the sub-stellar point, i.e. the center of the planetary disk as seen from the star marked as $S$, and the $xy$-plane corresponds to the intensity equator. The angle $\gamma'$ denotes the angle $\angle O_sO_pP$ and is related to the angle of incidence of stellar radiation. The angle $\gamma$ is the angle of foreshortening, also called the angle of reflection. 

The point  $ P $ has planetocentric Cartesian coordinates
\begin{equation}\label{eq:planetOriginCoordinates}
    \begin{pmatrix}
		x\\
		y\\
		z
	\end{pmatrix} = R_{p}\begin{pmatrix}
								\sin\eta\cos\phi\\
								\sin\eta\sin\phi\\
								\cos\eta
						 \end{pmatrix}.
\end{equation}
%
In this coordinate system the polar angle is defined such that $ 0\le \eta\le \pi $ and the azimuthal angle such that $ 0\le \phi\le 2\pi $. It is worth noting that in other works either the sub-stellar point, marked as $S$ \cite{Kopal1953,Kopal1959}, or the sub-planetary point, marked as $E$ \cite{sobolev,seager}, is typically used to define the $ x $-axis. The azimuthal angle used here will therefore differ by $ \pi/2 $ from those works that set the $ x $-axis as passing through the sub-stellar point. The coordinate system shown in \cref{fig:planetcoordinateslabels}\ is preferable for \myECIES\ because it will provide additional symmetries to simplify the integration of the reflected intensity distribution over the surface area of the exoplanet. Moreover, we will see in \cref{sec:Lforplaneparallel}\ that the foregoing coordinate system produces the identical reflected luminosity function as those found by \cite{sobolev,seager}, namely the Lambertian phase function.

We will now use \cref{fig:planetcoordinateslabels}\ to determine the relationships between the various angles and the coordinate system described in \cref{eq:planetOriginCoordinates}. Using the law of cosines for a spherical triangle, one finds that the angle of foreshortening, $ \gamma $, can be written in terms of $\eta$ and $\phi$ using
\begin{equation}
	\label{eq:gamma}
	\cos\gamma = \sin\eta\sin(\phi-\myPhase).
\end{equation}
The angle $ \gamma $ is also called the angle of reflection, \cite{sobolev,seager}. The angle $\gamma'$, which is the angle of incidence for the plane parallel ray model of incident stellar radiation \cite{sobolev,seager}, may also be determined using the law of cosines and is given by
\begin{equation}
	\label{eq:gammaprime}
	\cos\gamma' = \sin\eta\sin\phi.
\end{equation}
Finally, the differential surface area of the exoplanet may be expressed as
\begin{equation}
	\label{eq:dAp}
	dA_p = R_p^2\sin\eta~d\eta~d\phi.
\end{equation}
We may then write $\myLuminosity_{p, refl}$ as an integration of the reflected intensity distribution, $\myIntensityDis_{refl}(\eta,\phi)$, over the visible portion of the illuminated surface area of the exoplanet, \cf\ \cite{sobolev,seager}:
\begin{equation}\label{eq:lumplanet}
	\begin{aligned}
		\myLuminosity_{p, refl}(\myPhase) &=  \int_{\mathbf{C}}\myIntensityDis_{refl}(\eta,\phi)\cos\gamma~dA_p\\	
								& = R_p^2\int_{\mathbf{C}}\myIntensityDis_{refl}(\eta,\phi)\sin(\phi-\myPhase)\sin^2\eta~d\eta~d\phi.
	\end{aligned}	
\end{equation}

\begin{figure}[!hbt]
	\centering
	\subfloat[][Incident radiation for plane parallel ray model creates only two zones on the exoplanet. Adapted from \cite{seager}.\label{fig:planeParallel}]{\includegraphics[width=0.45\linewidth]{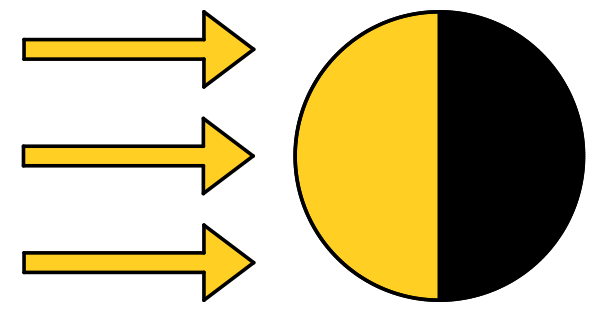}}
	\subfloat[][For \myECIES, the finite angular size of the star (at left) determines the location of three zones of the exoplanet (at right). \label{fig:threeZones}]{\includegraphics[width=0.45\linewidth]{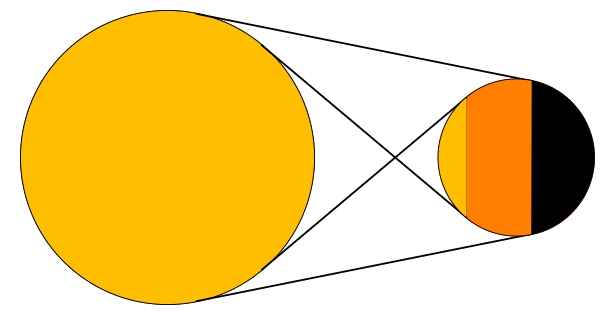}}
	\caption{A comparison of the two models describing the incident stellar radiation. The plane parallel ray approximation shown in \cref{fig:planeParallel}\ applies only for large star-planet separations; whereas, the model depicted in \cref{fig:threeZones}\ is best suited for exoplanets that orbit closely to their host star and may be used for all star-planet separations.}
	\label{fig:zoneCompare}
\end{figure}

\cref{eq:lumplanet}\ will depend on the nature of $\myIntensityDis_{refl}$ and $\mathbf{C}$, which is the portion of the illuminated surface of the exoplanet visible to an observer along the LOS. \cref{fig:zoneCompare}\ illustrates the difference between $\mathbf{C}$ for the two models of incident radiation at $ \myPhase=\pi/2 $. For the plane parallel case, the domain of $\mathbf{C}$ is the portion of the half sphere illuminated by the host star that is visible for a given phase angle, $\myPhase$, as shown in  \cref{fig:planeParallel}. The other situation, shown in \cref{fig:threeZones}, considers the finite angular size of the host star. In the latter case one must consider the inner and outer tangents between the stellar and exoplanetary disks in the determination of $\mathbf{C}$ as described in \cite{Kopal1953,Kopal1959}. 

In each of the foregoing situations, we will assume that the reflected intensity is isotropic, i.e. it follows Lambert's cosine law, such that
\begin{equation}\label{eq:lambertLaw}
	\myIntensityDis_{refl}(\phi, \eta) = \myRefCo S\cos\alpha,
\end{equation}
where  $\myRefCo$ is the reflection coefficient, $\pi S$ is the flux of the host star per unit normal  and $\alpha$ is the angle of incidence, \cf\ \cite{Kopal1953,Kopal1959,sobolev}. The quantity $\pi S$ is equivalent to the radiant flux density for plane parallel rays, see Fairbairn Chapter 1 \cite{fairbairn}. The determination of $\alpha$ in terms of $\phi$ and $\eta$ will depend on the model of illumination of the exoplanet by the host star. Finally, the determination of $\myRefCo$ will depend on the nature of the reflection of the exoplanet. For example, a Lambertian sphere has $ \myRefCo=\myScatteringAlbedo $, where  $ \myScatteringAlbedo $ is the single scattering albedo that describes the reflectivity of a small surface element of the exoplanet. A lossless Lambertian sphere is one for which $ \myScatteringAlbedo =1 $, \cite{sobolev}.

\begin{figure}[hbt]
\centering
\includegraphics[width=0.98\linewidth]{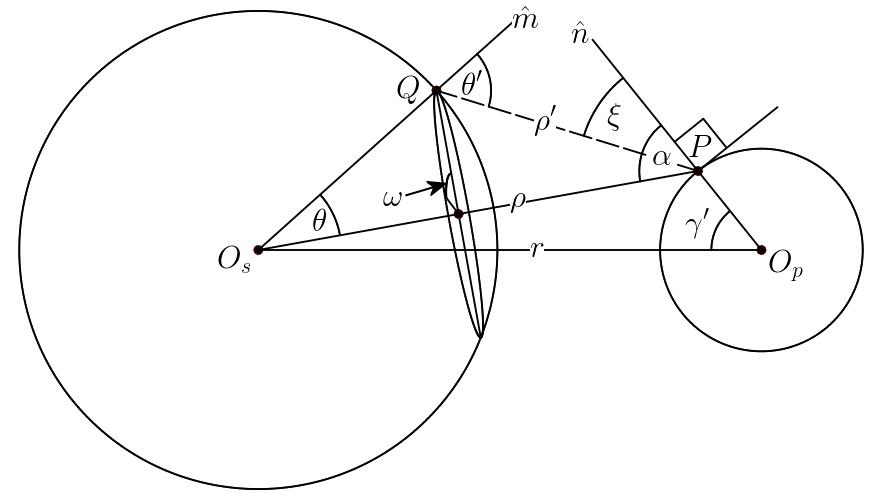}
\caption{Shown is an illustration of the geometry required to determine the intensity of light received at some point $ P $ on the exoplanet from a point $ Q $ of the host star, adapted from \protect\cite{Kopal1953,Kopal1959}. See text for details.}
\label{fig:intensitydis}
\end{figure}

In general, to determine the intensity distribution of reflected light, $ \myIntensityDis(\eta, \phi)$, we must consider the intensity of stellar light absorbed at a point $ P $ on the exoplanet, \cf\ \cite{seager}. At this point, we must consider the solid angle subtended by the host star and visible from $ P $; therefore, in the following the will use a coordinate system centered at point $ P $, as shown in \cref{fig:intensitydis}, adapted from \cite{Kopal1953,Kopal1959}. 

\myfillin, A more general form of \cref{eq:lambertLaw}\ may be written as
\begin{equation} 
    \pi\myIntensityDis_{refl} (\phi,\eta) = \myRefCo\iint_{\mathbf{C}_{s,@P}} \myIntensity(\theta') (\hat{m}\cdot\hat{\rho}')~d\Omega_{s,@P}
\end{equation}
where  $ \theta' $ is defined in \cref{fig:intensitydis}\ and is the angle between the line connecting the point $ P $ on the exoplanet and the point $ Q $ of the host star from which light emerges. The intensity of the light emerging from point $ Q $ is given by the function  $ \myIntensity  (\theta')$. The amount of this intensity absorbed at the point $ P $ is $ \myIntensity  (\theta') (\hat{m}\cdot\hat{\rho})$, which must be integrated over the visible portion of the stellar disk at point $ P $, $ \mathbf{C}_{s,@P} $. Here we will let the subscript $s$ indicate the hemispherical surface of the host star pointed toward the exoplanet. It is convenient to rewrite the solid angle subtended by the host star at point $ P $, $d\Omega_{s,@P}$, in terms of the surface area of the host star visible from that point. With the assistance of \cref{fig:intensitydis}\ we find that
\begin{equation} 
	d\Omega_{s,@P}=\frac{\hat{n}\cdot d\vec{A}_s}{\rho'^2 }=\frac{\hat{n}\cdot\hat{\rho}'}{\rho'^2 } (R_s^2 \sin{\theta}~d\theta~d\omega).
\end{equation}
Therefore, the general expression for the reflected intensity distribution of light from an exoplanet illuminated by its host star is
\begin{equation} \label{eq:generalIntDis}
    \pi\myIntensityDis_{refl} (\phi,\eta)= \myRefCo R_s^2 \iint_{\mathbf{C}_{s,@P}} \myIntensity(\theta') \left( \frac{\cos{\theta'}\cos{\xi}}{\rho'^2 }\right)\sin{\theta}~d\theta~d\omega.
\end{equation}
\cref{eq:generalIntDis}\ is equivalent to the expressions given by Kopal in Equations (6-23) to (6-26) \cite{Kopal1953}.

\section{Plane Parallel Rays}\label{sec:planeparallel}
\begin{figure}[!htb]
	\centering
	\includegraphics[width=\textwidth]{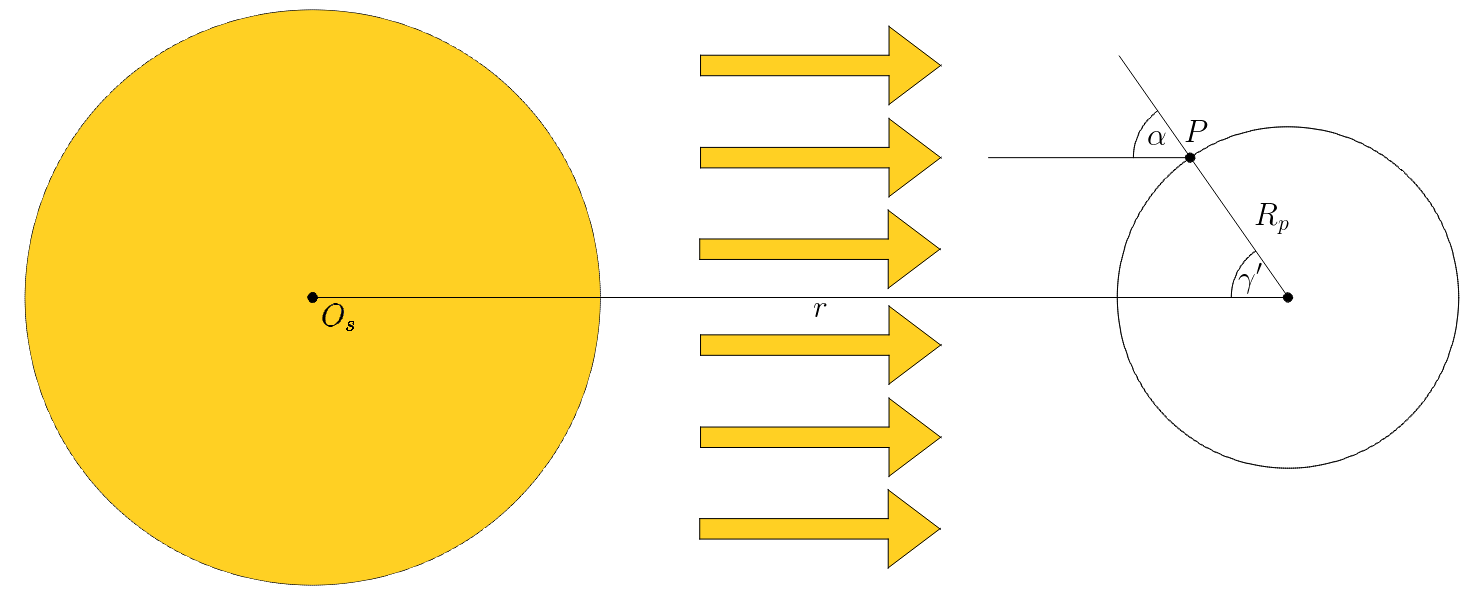}
	\caption{An illustration of an incident ray hitting the exoplanet at point $P$ with angle of incidence $\alpha$ for the plane parallel ray case of incident light.}
	\label{fig:planeparallelincidentrayslabels}
\end{figure}

\myfillin, Let us now explore \cref{eq:generalIntDis}\ for the plane parallel ray case, in which the incident stellar radiation reaches the exoplanet as plane parallel rays as shown in \cref{fig:planeParallel}. The following derivation follows that given in Chapter 9 of \cite{sobolev}\ with adaptations for our coordinate system, \mychange. \cref{fig:planeparallelincidentrayslabels}\ is an illustration of an incident ray of light originating from the host star and hitting the exoplanet at a point $P$. From the definition of $\gamma'$ given in~\cref{eq:gammaprime} and a comparison to \cref{fig:planetcoordinateslabels}\ we see that the angle of incidence, $\alpha$ is given by
\begin{equation}\label{eq:angleIncPlaneParallel}
	\cos \alpha  = \cos\gamma' = \sin\eta\sin\phi.
\end{equation}
for the plane parallel ray model of illumination. Furthermore, comparison of \cref{fig:intensitydis,fig:planeparallelincidentrayslabels}\ reveals that we may make the following approximations
\begin{equation}\label{eq:planeapprox1}
	\begin{aligned}
		\rho' \approx \rho &\approx r\\
        \theta &\approx \theta'\\
        \xi& \approx \alpha.
	\end{aligned}
\end{equation}
With the aid of \cref{eq:angleIncPlaneParallel,eq:planeapprox1}\ we may rewrite \cref{eq:generalIntDis} for plane parallel illumination of an exoplanet as
\begin{equation}
	\begin{aligned}
		\myIntensityDis_{refl,plane} &= \frac{R_s^2 }{\pi}\myRefCo\int_{0}^{\pi/2}\int_{0}^{2\pi}\myIntensity(\theta)\left( \frac{\cos{\theta}\cos{\gamma'}}{r^2 }\right)\sin\theta ~d\theta~d\omega\\
				& =2\myRefCo\left( \frac{R_s}{r}\right)^2 \cos{\gamma'}\int_{0}^{\pi/2}\myIntensity(\theta)\cos\theta\sin\theta ~d\theta\\
				& =2\myRefCo\left( \frac{R_s}{r}\right)^2 \cos{\alpha}\int_{0}^{1}\myIntensity(\psi)\psi~d\psi,
	\end{aligned}
\end{equation}
where we have made the substitutions $ \psi=\cos{\theta} $ and $ d\psi = -\sin\theta~d\theta $.

First, consider the simple case in which the stellar disk is of uniform intensity, i.e. $ \myIntensity(\psi)=\myIntensity_0 $. The resulting integration is then
\begin{equation}\label{eq:intDisPlaneUniform}
  \begin{aligned}
        \myIntensityDis_{refl,plane} &= 2\myRefCo\myIntensity_0\left( \frac{R_s}{r}\right)^2 \cos{\alpha}\int_{0}^{1}\psi~d\psi\\
						& =\myIntensity_0\myRefCo\left( \frac{R_s}{r}\right)^2 \cos{\alpha}\\
						& =\left( \frac{L_s}{\pi R_s^2 } \right)\myRefCo\left( \frac{R_s}{r}\right)^2\cos{\alpha}\\
						& =S\myRefCo\cos{\alpha},
  \end{aligned}
\end{equation}
where we have used the luminosity of the host star $L_s=\pi R_s^2 \myIntensity_0 $ and $ S = L_s/(\pi r^2) $ is the flux of the host star to recover \cref{eq:lambertLaw}. 

Here we have presented the derivation in Chapter 9 of \cite{sobolev}\ using the coordinate system described in \cref{sec:generalApproach}. As discussed in \cref{sec:uniformtransits}, stars are not uniform sources of light, but instead exhibit limb darkening. For linear limb darkening we have, \cf\ \cite{Kopal1953,Kopal1959,Claret2000},
\begin{equation} \label{eq:lineardarkeningdis}
    \myIntensity(\theta')=\myIntensity_0 (1-u+u\cos{\theta'})\approx \myIntensity_0(1-u+u\psi)
\end{equation}
where  $ u $ is the linear limb darkening coefficient and  $ \myIntensity_0 $ is the intensity of light at the center of the stellar disk. The evaluation of \cref{eq:generalIntDis}\ where $ I(\theta') $ is given by \cref{eq:lineardarkeningdis}\ produces
\begin{equation}\label{eq:intDisPlaneLinearDark}
	\begin{aligned}
		\myIntensityDis_{refl,plane} &= \myIntensity_0\myRefCo\left( \frac{R_s}{r}\right)^2\left( 1-\frac{u}{3}\right)\cos{\alpha}\\
						& =\left( \frac{L_s}{\pi r^2 }\right)\myRefCo\cos{\alpha},
	\end{aligned}
\end{equation}
where the luminosity of the host star is now written as $ L_s=\pi R_s^2 \myIntensity_0 (1-u/3) $ and the flux of the host star remains $ S = L_s/(\pi r^2) $.

\section{Finite Angular Size}\label{sec:finitesize}


In~\cite{lilloBox2013}, Lillo-Box et. al. confirm the planetary nature of Kepler-91b and determine its mass, radius and eccentricity. At the time of writing, Kepler-91b is one of the closest orbiting exoplanets at a star-planet separation of $a/R_s = 2.45_{-0.30}^{+0.15}$, see Table 8 of \cite{lilloBox2013}. The star subtends an angle of 48\degree\ and would cover around 10\% of the exoplanet's sky; furthermore, the authors calculate that approximately 70\% of the exoplanet's surface would be illuminated by its host star. In \cite{guillot2010}, Guillot points out that the use of the plane-parallel approximation, which ignores the spherical geometry of the host star, will generally overestimate the magnitude of the difference between the temperatures of the day and night sides of an exoplanet whose orbit is at a low inclination. Kepler-91b has an estimated inclination of  ${ 68.5^{+1.0}_{-1.63}}\degree $ and may exhibit this overestimation of temperature differences. In addition, it was noted by Robinson in \cite{2017robinson}\ that the analysis of the light scattered by close-in exoplanets would be influenced by the finite angular size of the host star and would allow one to probe information at different optical depth of the atmosphere.

Another exploration of the Kepler-91 system by Placek, Knuth, Angerhausen and Jenkins in \cite{K91Trojan}, analyzes the Kepler-91 system with the inclusion of a Trojan companion to Kepler-91b. Although the Bayesian evidence weighed in favor of a Trojan partner to Kepler-91b, the authors were led to believe that the Trojan partner was a false positive because the estimated planetary temperatures were too great to be physically realistic. The dayside temperature of the Trojan candidate was estimated to be $5184.6\pm 531.5 $ K, see Table 3 of \cite{K91Trojan}; whereas, the host star has an estimated effective temperature of 4550  $ \pm $ 75 K~\cite{lilloBox2013}. It is possible that the low inclination and extremely close-in orbit of Kepler-91b and its hypothetical Trojan partner are contributing to this overestimation of the dayside temperature as suggested in Knuth et al., 2017 \cite{2017Entropy}. In this work, we will use Kepler-91b as a model to illustrate the effects of extremely close-in orbits on the photometric emissions of an exoplanet; \cref{tab:K91params}\ lists the relevant parameters of the Kepler-91 system.

Finally, the estimation of the radius of the exoplanet requires proper accounting of light from the penumbral zone during the primary transit. Recall from \cref{sec:uniformtransits}, \cref{eq:Fe,eq:lam_e}\ indicate that the radius of the exoplanet maybe determined by measuring the transit depth given by \cref{eq:transitdepth}\ if the radius of the host star is known. If an exoplanet exhibits a large penumbral zone, i.e. a small un-illuminated zone, the apparent loss of light during the primary transit will be reduced because the exoplanet will be reflecting light along the line of sight. In effect, the estimation of the radius of the exoplanet will be smaller than its true radius. As will be described in \cref{sec:Lforplaneparallel}, an underestimation of the radius of the exoplanet will result in an overestimation of the reflectivity of the exoplanet and its night side temperature.

\begin{table}[!htb]\tabulinesep = 3mm
	\centering
	\caption{\label{tab:K91params}Stellar parameters for Kepler-91 and estimated planetary parameters for Kepler-91b, \protect\cite{lilloBox2013,lilloBox2014,K91Trojan}. The normalized semi-major axis, $ a/R_s $, was calculated using the parameters given in this table. The single scattering albedo, $ \myScatteringAlbedo $, was calculated from the geometric albedo as will be described in \cref{sec:addingVariations}. Note that the gravity darkening exponent, limb darkening coefficients, eccentricity, and albedos are unitless quantities.}
	\begin{tabular}{lr}	
		\toprule
		\multicolumn{2}{l}{Stellar Parameters}\\
		\toprule
		Mass,  ($M_{\astrosun}$)& 1.31 $ \pm $ 0.10\\
		Radius, ($R_{\astrosun}$)&6.30 $\pm$ 0.16\\
		Effective Temperature, (K)& 4550 $ \pm $ 75\\
		Surface Gravity, $ \log g_* $, (c.g.s.)& 2.953 $ \pm$ 0.007\\
		Gravity Darkening Exponent ($g$) & 0.733\\
		Linear Limb Darkening Coefficient ($u$)& 0.549\\
		Quadratic Limb Darkening Coefficients ($ \gamma_1 $, $ \gamma_2 $) & (0.69, 0.05)\\
		\midrule
		Estimated Planetary Parameters\\
		\midrule
		Mass,  ($M_J$)& 1.09 $ \pm $ 0.20\\
		Radius, ($R_J$)& $1.384_{-0.054}^{+0.011}$\\
		Period, (days) & 6.24650 $ \pm $ 0.000082\\
		Inclination,  (\degree) & $ 68.5_{-1.6}^{+1.0} $ \\
		Eccentricity& $ 0.066_{-0.017}^{+0.013} $ \\
		 $ a/R_s$ & $ 2.48\pm 0.12 $ \\
		Semi-major Axis, $a$, (AU)& $0.0726\pm 0.0019 $ \\
		Equilibrium Temperature, (K) & $ 2460_{-40}^{+120} $ \\
		Geometric Albedo ($A_g$) &$ 0.39\pm 0.15 $\\
		Single Scattering Albedo ($\myScatteringAlbedo$)&$ 0.58 \pm 0.09$\\
		\bottomrule
	\end{tabular}
\end{table}


Following is a description of the geometry of \myECIES\ and the approach taken to derive the intensity distribution of reflected light for such exoplanets. 

\subsection{The Geometry}\label{sec:thegeometry}
\begin{figure}[hbt]
	\centering
	\subfloat[][The inner tangents.\label{fig:innertangents}]{\includegraphics[width=0.5\textwidth]{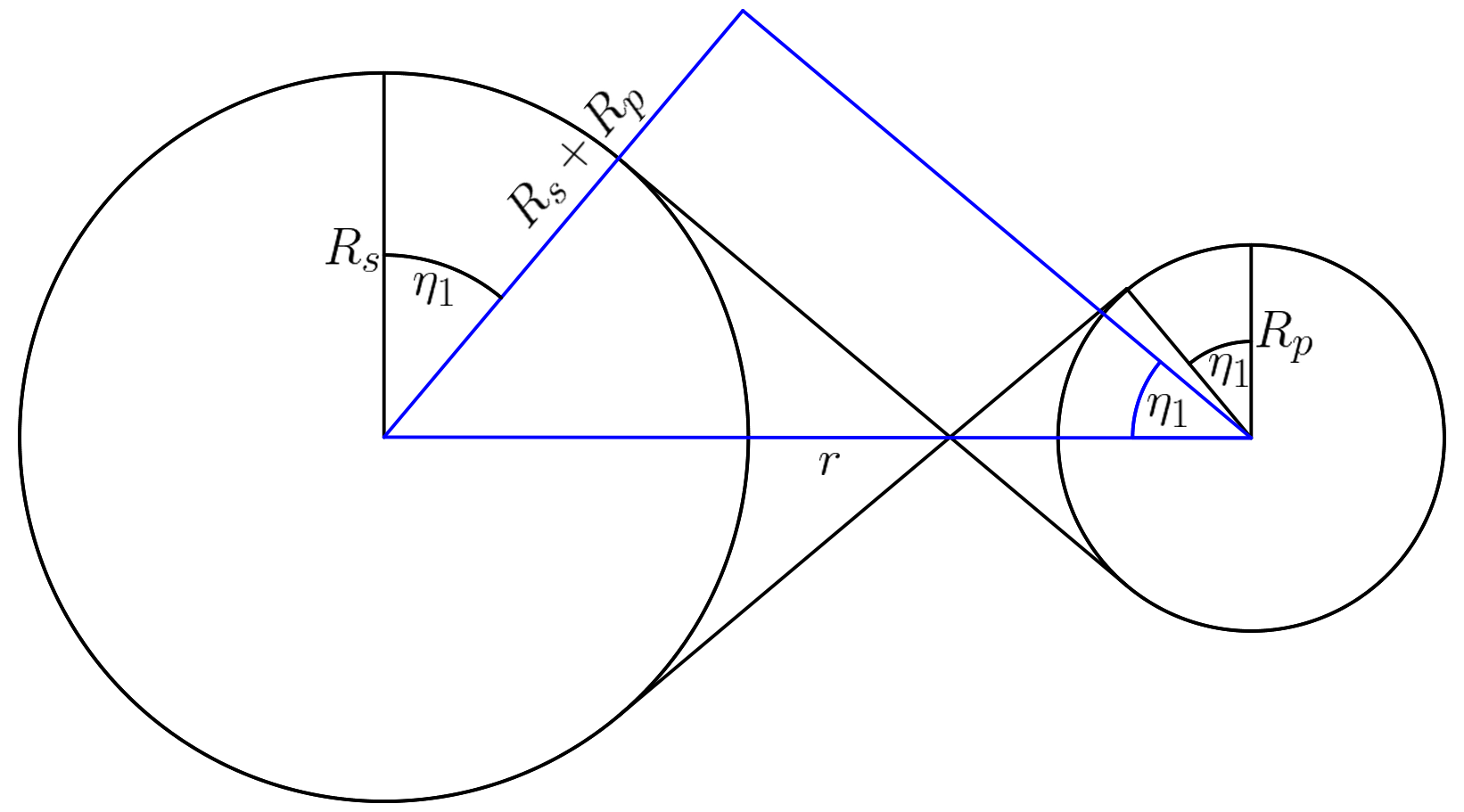}}
	\hfill
	\subfloat[][The outer tangents\label{fig:outertangents}]{\includegraphics[width=0.5\textwidth]{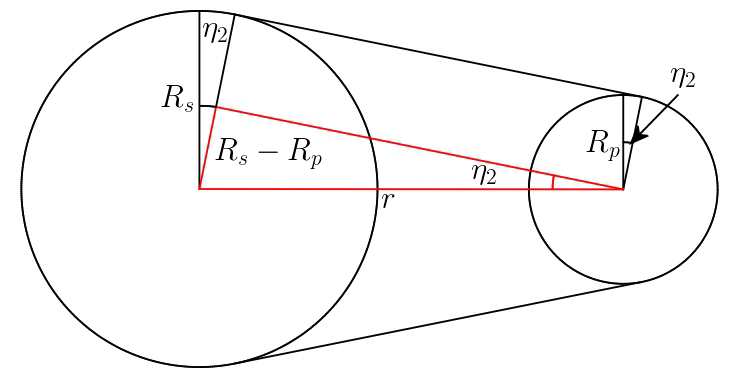}}
	\caption{The inner tangents set the boundaries on the un-illuminated zone and the outer set those of the fully illuminated zone.}
	\label{fig:innerandoutertangents}
\end{figure}

\myfillin, Unlike the plane parallel case shown in~\cref{fig:planeParallel}, the exoplanet can be split into three distinct zones as shown in~\cref{fig:threeZones}. The limits of the zones are determined by the inner and outer tangents between the host star and exoplanet, \cite{Kopal1953,Kopal1959}, as shown in~\cref{fig:innerandoutertangents}. 

The inner tangents, shown in~\cref{fig:innertangents}, delimit the fully illuminated zone. Any point within the fully illuminated zone receives flux from the entire apparent disk of the host star. The limit on the polar angle, $\eta$, is given by
\begin{equation}
\begin{aligned}\label{eq:sineta1}
\sin\eta_1 &= \frac{R_p+R_s}{r} \\
&= \frac{p+1}{\normz}
\end{aligned}
\end{equation}
where $p = R_p/R_s$, is the ratio of  the planet-star radii, and $\normz = r/R_s$, is the normalized star-planet separation, see Equation (6-8) \cite{Kopal1959}. 

The limits on the un-illuminated zone are determined by the outer tangents, as shown in \cref{fig:outertangents}\ and are delimited as follows, Equation (6-9) \cite{Kopal1959}:
\begin{equation}
\begin{aligned}\label{eq:sineta2}
\sin\eta_2 & = \left\vert\frac{R_p-R_s}{r}\right\vert\\
& = \frac{R_s-R_p}{r}\\
& = \frac{1-p}{\normz}.
\end{aligned}
\end{equation}
The absolute value in \cref{eq:sineta2}\ is used because our coordinate system requires that the polar angle be a positive quantity and recognizes that $R_p-R_s$ is always negative for exoplanets because $R_p < R_s$; therefore, the un-illuminated zone begins in what would be considered the night side of an exoplanet if the exoplanet were illuminated by plane parallel rays, which is clear in~\cref{fig:outertangents}. In other words, the un-illuminated zone does not begin at $ y= 0 $ as it does for the plane parallel ray case, but rather at some point for which $ y <0 $ in our planetocentric coordinate system given in \cref{fig:planetcoordinateslabels}. The penumbral zone lies between $\eta_1$ and $\eta_2$. Within the penumbral zone the apparent disk the host star will gradually sink below the horizon.

\begin{figure}[hbt]
	\centering
	\includegraphics[width = 0.5\textwidth]{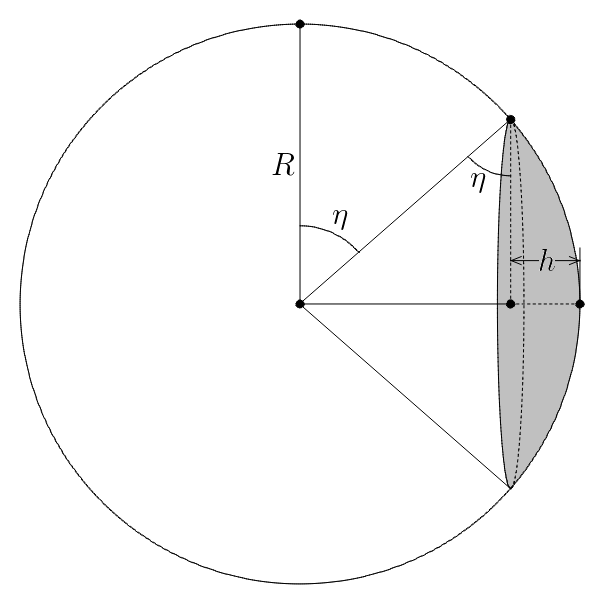}
	\caption{Illustration of the spherical cap, in gray, defined by radius $R$ and angle $\pi/2 - \eta$. The height, $h$, is given by~\cref{eq:height} and its area by~\cref{eq:Scap}.}
	\label{fig:cap}
\end{figure}

\mynew, To compare the plane parallel case to the model required for \myECIES, let us consider the fractional area of each zone
\begin{equation}\label{eq:sigma}
	\sigma_{zone} = \frac{A_{zone}}{4\pi R_p^2}.
\end{equation}
The area, $ A_{zone} $, is the surface area of each zone. For the fully illuminated and un-illuminated zones  $ A_{zone} = S_{cap} $ where the value of $S_{cap}$ is determined from the surface area of a spherical cap with apex angle $ \pi/2-\eta $ . Consider a sphere of radius $R$ and a cap of height $h$, in which the polar angle determines the height, as shown in~\cref{fig:cap}. The height is given by 
\begin{equation}\label{eq:height}
	h = R(1-\sin\eta)
\end{equation}
and the area of the cap is given by 
\begin{equation}\label{eq:Scap}
	\begin{aligned}
		S_{cap} & = 2\pi Rh\\
		& = 2\pi R^2(1-\sin\eta).
	\end{aligned}
\end{equation}

Using \cref{eq:sigma,eq:Scap}, it can be shown that the fractional area of the fully illuminated zone is given by
\begin{equation}\label{eq:fullarea}
\begin{aligned}
\sigma_{full} & = \frac{1}{2}(1-\sin\eta_1)\\
& = \frac{1}{2}\left(1 - \frac{p+1}{\normz}\right);
\end{aligned}
\end{equation}
and the fractional area of the un-illuminated zone, i.e. the night side, is given by 
\begin{equation}\label{eq:nightarea}
\begin{aligned}
\sigma_{un} & = \frac{1}{2}(1-\sin\eta_2)\\
& = \frac{1}{2}\left(1-\frac{1-p}{\normz}\right).
\end{aligned}
\end{equation}
The fractional area of the penumbral zone is the remaining fractional area of the exoplanet, i.e. $\sigma_{pen} = 1-\sigma_{full}-\sigma_{night}$, and is given by 
\begin{equation}\label{eq:penarea}
	\begin{aligned}
		\sigma_{pen} & = \frac{4\pi R_p^2 - A_{full}-A_{un}}{4\pi R_p^2 }\\
			& = \frac{1}{2}(\sin\eta_1+\sin\eta_2)\\
			& = \frac{R_s}{r}= \frac{1}{\normz}.
	\end{aligned}
\end{equation}
\cref{eq:penarea}\ shows that the extent of the penumbral zone does not depend on the radius of the exoplanet, but instead only  on the normalized star-planet separation. It is also possible for 
the fractional surface area of the penumbral zone to be greater than that of the fully illuminated zone if 
\begin{equation}
	\normz <p+3,
\end{equation}
or
\begin{equation}
	r<R_p+3R_s.
\end{equation} 

Let us also consider the fractional area that receives at least partial illumination, or the fractional area that is lit:
\begin{equation}\label{eq:litarea}
	\begin{aligned}
		\sigma_{lit} & = \sigma_{full}+\sigma_{pen}\\
		& = \frac{1}{2}(1+\sin\eta_2)\\
		& = \frac{1}{2}\left(1+\frac{1-p}{\normz}\right).
	\end{aligned}
\end{equation}
\cref{eq:fullarea,eq:penarea,eq:litarea}\ appear in Knuth et al., see Equations (15)-(17)\cite{2017Entropy}.

In the plane parallel model of illumination, the exoplanet always experiences 50\% illumination. For large $\normz$, the $\sigma_{pen}$ approaches zero and the penumbral zone disappears; furthermore, \cref{eq:nightarea,eq:fullarea}\ each approach one half. Therefore, for large $\normz$ the finite angular size of the host star may be neglected and we return to the plane parallel ray model. 

\cref{tab:targets}\ shows a list of confirmed \myKepler\ exoplanets for which $\sigma_{lit}>0.600$ and $\sigma_{pen}>\sigma_{full}$. Of the confirmed \myKepler\ exoplanets with periods less than 10 days, a total of 87 have more than 60.0\% of their surface illuminated, note that Kepler-91b ranks fourth in terms of  $ \sigma_{lit} $.  \cref{tab:targets}\ also contains values for Mercury and Earth for comparative purposes of which neither planet has a significant penumbral zone. The table was prepared in collaboration with Ben Placek and Anthony D. Gai. 

\begin{table}[!htb]
	\centering
	\caption{Partial list of confirmed \myKepler\ exoplanets for which $ \sigma_{pen}>\sigma_{full} $ and $\sigma_{lit}>0.600$. Also shown for comparative purposes are the values for Mercury and Earth. The table includes the polar angles delimiting the fully illuminated zone, $ \eta_{1} $, and the un-illuminated zone, $ \eta_{2} $. We have also listed the fractional surface area of the fully illuminated zone, $ \sigma_{full} $ given by \cref{eq:fullarea}, the penumbral zone, $ \sigma_{pen} $ given by \cref{eq:penarea}, and the total area that is at least partially lit, $ \sigma_{lit} $. The fractional area at least partially lit is the sum of $ \sigma_{full} $ and $ \sigma_{pen} $.\label{tab:targets}}
	\begin{tabular}{rccccc}
		Name & $\eta_1$ (\degree) & $\eta_2$ (\degree) & $\sigma_\textnormal{full}$ &$\sigma_\textnormal{pen}$ &$\sigma_\textnormal{lit}$\\ \hline
		Kepler-1613 b &	41.0	& 39.4	& 0.172 &	0.645	& 0.817\\
		Kepler-1520 b &	29.2	& 24.9&	 0.256	&0.455	&0.711\\
		Kepler-32 f &	25.9	& 24.0	& 0.282&	0.422	&0.704 \\
		Kepler-91 b &	22.9	& 21.8	 &0.306&	0.380&	0.686\\
		Kepler-1368 b	&22.6	 &21.9&	 0.308&	0.379	&0.687\\
		Kepler-1270 b	&22.3	& 21.9	& 0.310	&0.376	&0.686\\
		Kepler-1189 b	&21.0	 &20.4	& 0.321	&0.353	&0.674\\
		Kepler-990 c	&20.7	& 20.2	& 0.323	&0.349	&0.672\\
		Kepler-929 b	&20.7	& 20.2	& 0.324&	0.349&	0.673\\
		Kepler-1078 b   &20.7   & 19.9	& 0.323&	0.347&	0.670 \\
		Kepler-912 b	&20.7	& 19.7	& 0.324	&0.345	&0.668 \\
		Kepler-780 b	&20.3	& 19.9	& 0.327	&0.344	&0.671\\
		Kepler-845 b	&20.4	& 19.4	& 0.326	&0.340	&0.666\\
		Kepler-1340 b	&19.6	& 19.2	 &0.332	&0.332	&0.665\\	
		\hline
		Mercury & 0.691&0.686 & 0.493 & 0.012 & 0.506\\
		Earth & 0.269 & 0.264 & 0.498 & 0.005 & 0.503
	\end{tabular}
\end{table}

We will now proceed in the determination of the reflected intensity distribution within the fully illuminated zone and then continue to the penumbral zone.

\subsection{$\myIntensityDis_{p,refl} (\mu)  $  within the Fully Illuminated Zone}\label{sec:incidentFluxFull}
\begin{sidewaysfigure}
	\centering
	\includegraphics[width=0.98\linewidth]{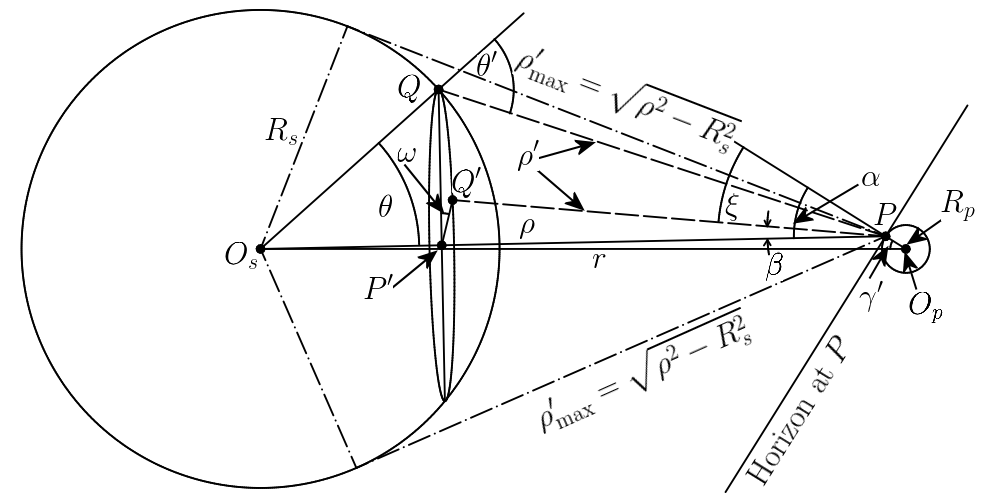}
	\caption{Shown here is an illustration of the geometry of a point $ P $ within the fully illuminated zone, adapted from Figure 4-4 \protect\cite{Kopal1959}. The angle $ \alpha = \angle NPO_s$ is the angle of incidence. See text for explanation.}
	\label{fig:Kopal4p4Full}
\end{sidewaysfigure}

\myfillin, As in the plane parallel case described in~\cref{sec:planeparallel}, we must determine $ \myIntensityDis_{p,refl}(\phi, \eta) $ for each point $ P $ within the fully illuminated zone. We again adopt a coordinate system centered at  $ P $, but now must consider the distance between  $ P $ and the point $ Q $ on the stellar surface. We will assume that the intensity varies according to Lambert's cosine law, i.e.~\cref{eq:lambertLaw}, but we may no longer make the approximations given in \cref{eq:planeapprox1}. As a result, it is now the case that $ \alpha $ is a function of $ \phi $ and $ \eta $, as shown in \cref{fig:Kopal4p4Full}. Finally, we must take into account the fact that the flux per unit area, $ \pi S $, is now delimited by polar angles $ \eta_1 $ and $ \pi-\eta_1 $ such that we must integrate the brightness distribution of the star, $ \myIntensity(\theta') $, over the visible surface area of the star as viewed from a point $ P $ on the exoplanet surface. Here we seek to recover Equations (6-29) and (6-30) \cite{Kopal1959}\ with the additional factor of the reflection coefficient, $ \myRefCo $, given \cite{sobolev,seager}.


It is useful to define a coordinate system centered at the point $ P $ on the surface of the exoplanet, as shown in \cref{fig:Kopal4p4Full}.
The $ z $-axis is along the line $ \overline{O_sP} $, which has length $ \rho $. The plane of the page, i.e. of the triangle $ \bigtriangleup O_sPQ $, is the $ xz $-plane. The $ xy $-plane is such that the $ x $-axis originates from $ P $ and is tangent to $ \overline{O_sP}$ and the $ y $-axis is directed out of the page. The angle $ \theta $ is the colatitude measured from the $ z $-axis and the angle $ \omega $ is the azimuthal angle measured about the  $ z $-axis from the positive $ x $-axis. Thus, the direction cosines for an arbitrary line as measured from point $ P $ is given by
\begin{equation} 
    \begin{pmatrix}
		\sin\theta\cos\omega\\
		\sin\theta\sin{\omega}\\
		\cos\theta
	\end{pmatrix}.
\end{equation}

To determine $\myIntensityDis_{refl,full}(\eta,\phi)$ we must take into account the fact that the flux per unit area, $ \pi S $, is now delimited by the lines labeled as $\rho'_\textnormal{max}$. These lines are tangent to the stellar surface and connect to the point $ P $. The surface area of the star visible at point $ P $ forms a spherical cap with apex angle $ \pi/2 -\eta_{1} $. The area of integration is symmetric about the line $ \rho $ in \cref{fig:Kopal4p4Full}\ such that we may integrate $ \myIntensity(\theta') $ about the angle $ \omega $ from zero to $ 2\pi $ and along the line $ \rho' $ from its minimum $ \rho'_\textnormal{min}=\rho-R_s $ to its maximum $ \rho'_\textnormal{max} = \sqrt{\rho^2-R_s^2 } $. In the following we shall assume that $ \myIntensity(\theta') $ follows a linear limb darkening law given  \cref{eq:lineardarkeningdis}, but in this case we cannot assume that $ \theta'\approx\theta $. The reflected intensity distribution at point $ P $ is then given by 
\begin{equation}\label{eq:Jfull}
\begin{aligned} 
    \myIntensityDis(\phi, \eta) &= \myRefCo \int_{\rho'_\textnormal{min}}^{\rho'_\textnormal{max}}\int_{0}^{2\pi} \myIntensity(\theta')\left( \frac{\cos{\xi}\cos{\theta'}}{\pi\rho'^{2}}\right)~dA_{s}\\
								&= I_0\myRefCo((1-u)J_1+uJ_2)
\end{aligned}
\end{equation}
where
\begin{equation}\label{eq:Jn}
\pi J_n = \int_{\rho'_\textnormal{min}}^{\rho'_\textnormal{max}}\int_{0}^{2\pi} \frac{\cos\xi\cos^n\theta'}{\rho'^2}dA_s,
\end{equation}
which is equivalent to Equation 6-25 \cite{Kopal1959} for the fully illuminated zone.
%
We must now determine a relationship between $ \xi $, $\theta'  $, $ \rho' $ and $ \alpha $ with the aid of \cref{fig:Kopal4p4Full}.

The direction cosines for the lines $ \overline{O_pP} $ and $ \overline{PQ'} $ are given by 
\begin{equation}
\overline{O_pP} = \begin{pmatrix}
\sin\alpha\\ 
0\\ 
\cos\alpha 
\end{pmatrix} 
\hspace{4em}
\overline{PQ'} = \begin{pmatrix}
\sin\beta\cos\omega\\
\sin\beta\sin\omega\\
\cos\beta
\end{pmatrix}
\end{equation}
which are used to determine the angle $ \xi = \angle NPQ' $ via the equation
\begin{equation}\label{eq:cosXi}
\cos\xi = \overline{O_pP}\cdot \overline{PQ} = \cos\alpha\cos\beta+\sin\alpha\sin\beta\cos\omega.
\end{equation}
We may determine relationships between $\beta $ and the other geometrical components using the law of cosines and the law of sines for the triangle $\triangle O_sPQ$:
\begin{equation}\label{eq:beta}
\cos\beta = \frac{\rho-R_s\cos\theta}{\rho'} \hspace{4em} \sin\beta = \frac{R_s}{\rho}\sin\theta'.
\end{equation} 
In addition, the same triangle permits a relationship between $\theta' $,  $\theta $, $\rho $ and $\rho' $:
\begin{equation}
	\rho^2=\rho'^2-R_{s}^2 -2\rho R_{s}\cos{\theta'}
\end{equation}
or
\begin{equation}\label{eq:cosThetaPrime}
\cos\theta' = \frac{\rho^ 2 -R_s^ 2 -\rho'^ 2}{2\rho'R_s} =\frac{\rho\cos\theta -R_s}{\rho'}
\end{equation}
and
\begin{equation}\label{eq:rhoPrimeSqr}
\rho'^2 =\rho^2+R_s^2-2\rho R_s\cos\theta.
\end{equation}
We may now find a relationship between $\alpha $ and the variables shown in~\cref{fig:Kopal4p4Full} by using the law of cosines for the triangle $ \triangle O_pO_sP $:
\begin{equation}\label{eq:rhoSqr}
\rho^2 = r^2+R_p^2-2rR_p\cos\gamma'
\end{equation}
and 
\begin{equation}\label{eq:rSqr}
r^2 = \rho^2+R_p^2+2\rho R_p\cos\alpha
\end{equation}
which may solved for $ \cos\alpha $ to determine that.  
\begin{equation}\label{eq:cosAlpha_rho}
\cos\alpha = \frac{r\cos\gamma'-R_p}{\rho}.
\end{equation}
\cref{eq:cosAlpha_rho}\ may also be written in terms of $ \phi $ and $ \eta $ with the application of \cref{eq:gammaprime}\ and \cref{eq:rhoSqr}:
\begin{equation}\label{eq:cosAlpha_phiEta}
\begin{aligned} 
    \cos\alpha &= \frac{r\sin\eta\sin\phi-R_p}{\sqrt{r^2+R_p^2-2rR_p\sin\eta\sin\phi}}\\
			   &=\frac{r\mu - R_{p}}{\sqrt{r^2+ R_{p}^2 -2 rR_{p}\mu}},
\end{aligned}
\end{equation}
where
\begin{equation}\label{eq:mudef}
	\mu =\cos{\gamma'} =\sin \eta  \sin\phi.
\end{equation}
Note that for large star-planet separations we may use  $ \rho \approx r $ and $ r\gg R_p $ to find that  $ \cos{\alpha}=\cos{\gamma'} $ as given in \cref{eq:angleIncPlaneParallel}.

The previous equations may be used to evaluate \cref{eq:Jn}\ within the fully illuminated zone. First, it is convenient to write the differential area of the star in terms of $ \rho' $ and $ \omega $ due to the symmetries described previously. The differential spherical area of the host star is given by 
\begin{equation}
	dA_s = R_s^2\sin\theta~d\theta~d\omega,
\end{equation}
but~\cref{eq:rhoPrimeSqr} may be used to show that 
\begin{equation}
	\sin\theta~d\theta =\frac{\rho'}{R_s\rho}~d\rho'
\end{equation}
because $ \rho $ is constant over the integration of the host star from point $ P $. Thus,
\begin{equation}
	dA_s = R_s\left( \frac{\rho'}{\rho} \right)~d\rho'~d\omega.
\end{equation}
Next, we may simplify the integration by eliminating the second term in \cref{eq:cosXi} because 
\begin{equation}
	\int_{0}^{2\pi} \sin\alpha \sin\beta \cos\omega~d\omega =0.
\end{equation}
\cref{eq:Jn}\ may then be re-written as follows:
\begin{equation}\label{eq:JnReWrite}
	\begin{aligned}
		J_n & = \frac{1}{\pi}\int_{\rho'_\textnormal{min}}^{\rho'_\textnormal{max}}\int_{0}^{2\pi}\frac{\cos\alpha\cos\beta}{\rho'^2}\cos^n\theta'\left( R_s\left( \frac{\rho'}{\rho}~d\rho'~d\omega \right) \right)\\
		& = \frac{2R_s\cos\alpha}{\rho}\int_{\rho'_\textnormal{min}}^{\rho'_\textnormal{max}}\frac{1}{\rho'}\left( \frac{\rho'^2+\rho^2-R_s^2}{2\rho'\rho} \right)\left( \frac{\rho^2-R_s^2-\rho'^2}{2\rho'R_s} \right)^n~d\rho'\\
		& = \frac{\cos\alpha}{2^nR_s^{n-1}\rho^2}\int_{\rho'_\textnormal{min}}^{\rho'_\textnormal{max}}\frac{(\rho'^2+\rho^2-R_s^2)(\rho^2-R_s^2-\rho'^2)^n}{\rho'^{2+n}}~d\rho',
	\end{aligned}
\end{equation}
\mychange, It is worth noting that \cref{eq:JnReWrite}\ may be use to determine the reflected intensity distribution for any limb darkening model for which the stellar intensity distribution is described by powers of $ \cos\theta' $. In addition, it may be used for non-spherical geometries if one is able to eliminate $ \omega $ from the integration and is able to obtain an expression for $ \rho'_\textnormal{min} $ and $ \rho'_\textnormal{max} $ in terms of $ \rho' $ for said geometry. If we consider spherical bodies and let $ n=1 $, which corresponds to zero limb darkening, and $ n=2 $, which is required for linear limb darkening, we recover Equations (6-29) and (6-30) \cite{Kopal1959}.

\myfillin, With the assistance of~\cref{eq:JnReWrite} we may now determine the expressions for $ J_1 $ and $ J_2 $ in~\cref{eq:Jfull} as follows. First, we have
\begin{equation}\label{eq:J1full}
	J_1 = \frac{\cos\alpha}{2\rho^2}\int_{\rho'_\textnormal{min}}^{\rho'_\textnormal{max}}\frac{(\rho^2-R_s^2)^2-\rho'^4}{\rho'^3}~d\rho' = \left( \frac{R_s}{\rho} \right)^2\cos\alpha
\end{equation}
and
\begin{equation}\label{eq:J2full}
	J_2 = \frac{\cos\alpha}{4R_s\rho^2}\int_{\rho'_\textnormal{min}}^{\rho'_\textnormal{max}}\frac{(\rho'^2+\rho^2-R_s^2)(\rho^2-\rho'^2-R_s^2)^2}{\rho'^4}~d\rho'=\frac{2}{3}\left( \frac{R_s}{\rho} \right)^2\cos\alpha.
\end{equation}
Substitution of~\cref{eq:J1full,eq:J2full} into~\cref{eq:Jfull} reveals that 
\begin{equation}
	\myIntensityDis(\phi,\eta) = I_0\myRefCo\left( \frac{R_s}{\rho} \right)^2\cos\alpha\left( 1-\frac{u}{3} \right).
\end{equation}
If we let  $ L_s =\pi R_s^2I_0(1-u/3)$, as was done for the plane parallel ray case, be the apparent luminosity of the host star and substitute \cref{eq:cosAlpha_phiEta} for $ \cos\alpha $, we find that 
\begin{equation}\label{eq:intensityDisFull}
	\myIntensityDis_{refl, full}(\mu) = \frac{\myRefCo L_s(r\mu-R_p)}{\pi\rho^3} = \frac{\myRefCo L_s(r\mu-R_p)}{\pi\left( r^2+R_p^2-2rR_p\mu \right)^{3/2}}
\end{equation}  
is the intensity of light reflected through an angle of $ \gamma' $ within the fully illuminated zone, see Equation (6-33) \cite{Kopal1959}. Furthermore, should we neglect linear limb darkening the expression for the intensity distribution within the fully illuminated zone is still given by \cref{eq:intensityDisFull}\ with the exception that the luminosity of the star is now expressed as $ L_s=\pi R_s^2 I_0 $. 

\myfillin, It will be convenient when we evaluate the reflected luminosity of the exoplanet in \cref{ch:reflectedlightluminosity}\ to expand \cref{eq:intensityDisFull}\ in ascending powers of $ R_p/r $ using the Legendre polynomials. First, consider 
\begin{equation}
 G(\mu)=\frac{r\mu-R_p}{(r^2+R_p^2-2rR_p\mu)^{3/2}}, 
\end{equation}
where $ \myIntensityDis_{refl, full} = L_s\myRefCo G(\mu)/\pi $. Let $ h = R_p/r $ to write
\begin{equation}
 G(\mu) = \frac{(\mu-h)r}{r^3(1+h^2-2h\mu)^{3/2}} = \left(\frac{\mu-h}{r^2(1+h^2-2h\mu)}\right)\frac{1}{\sqrt{1+ h^2 - 2h\mu}}.
\end{equation}
The second term of $ G(\mu) $ is the generating function for Legendre polynomials, \cf\ Chapter 12, section 5 of \cite{boas},
\begin{equation}
\Phi (\mu, h) =\frac{1}{\sqrt{1+ h^2 - 2h\mu}} =\sum_{\ell =0}^{\infty}h^\ell P_\ell (\mu);
\end{equation}
therefore,
\begin{equation}
G(\mu) = \left(\frac{\mu-h}{r^2(1+h^2-2h\mu)}\right)\Phi(\mu, h).
\end{equation}
Next, consider two forms of the partial derivative of $ \Phi(\mu, h) $
\begin{equation}
\begin{aligned}
	\frac{\partial \Phi}{\partial h} & = \left( \frac{\mu - h}{1+ h^2 - 2h\mu} \right)\Phi(\mu, h)\\
	& = \sum_{\ell=1}^\infty \ell h^{\ell-1}P_\ell(\mu),
\end{aligned}
\end{equation}
to reveal that 
\begin{equation}\label{eq:G}
G(\mu) = \frac{1}{r^2}\sum_{\ell=1}^\infty \ell h^{\ell-1}P_\ell(\mu).
\end{equation}
Substitution of \cref{eq:G}\ into \cref{eq:intensityDisFull} and recalling that $ h = R_p/r $ permits us to write 
\begin{equation}\label{eq:fullLegendre}
\begin{aligned}
	 \myIntensityDis_{refl, full}(\mu) &= \frac{L_s\myRefCo}{\pi r^2}\sum_{\ell=1}^\infty \ell \left( \frac{R_p}{r} \right)^{\ell-1}P_\ell(\mu)\\
	 &=\frac{L_s\myRefCo}{\pi r^2}\left(P_1 (\mu) +2\left(\frac{R_p}{r}\right) P_2 (\mu) +3\left(\frac{R_p}{r}\right)^2 P_3 (\mu) +\ldots\right),
\end{aligned}
\end{equation}
as given by Equation (6-60) \cite{Kopal1959}, with the addition of $ \myRefCo $.

We will now present the method to determining the reflected intensity distribution of the penumbral zone as presented in \cite{Kopal1953,Kopal1959}.

\subsection{Original Derivation of $\myIntensityDis_{p,refl} (\mu)  $  within the Penumbral Zone}\label{sec:incidentFluxPenKopalMethod}
\myfillin, The following is a detailed derivation of the reflected intensity distribution of the penumbral zone as described in \cite{Kopal1953,Kopal1959}. It should be noted that in \cref{sec:luminosityFiniteSizeog}\ it will be shown that this analysis will produce negative luminosity within the penumbral zone and is therefore not an accurate description of the reflected intensity distribution for the penumbral zone. Finally, possible reasons for this phenomenon will also be presented.

\begin{sidewaysfigure}
	\includegraphics[width = \textwidth]{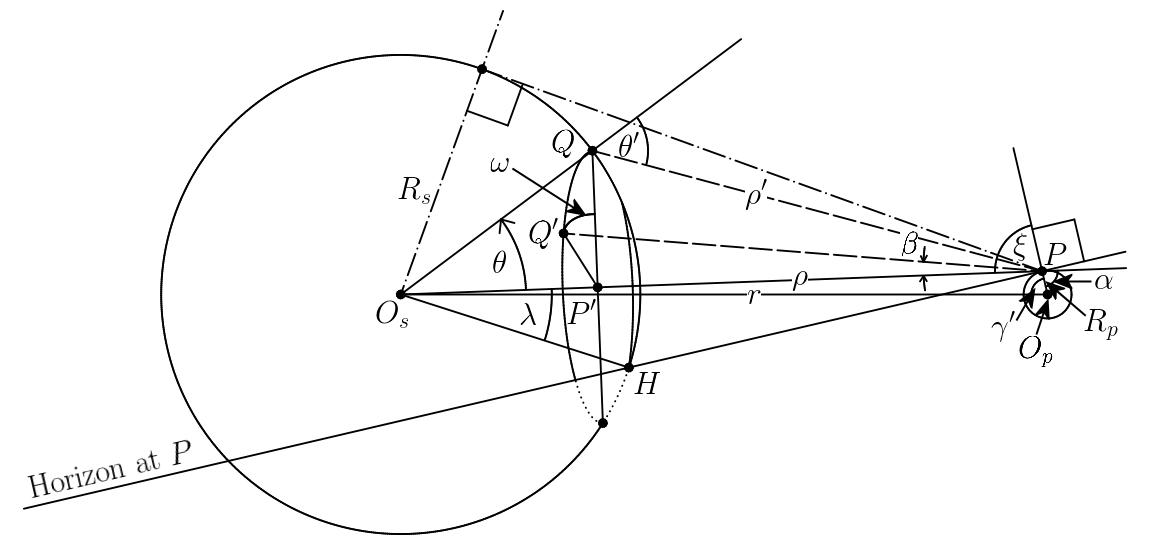}
	\caption{Adaption of Figure 4.4 from \protect\cite{Kopal1953}. In the above, $P$ is a point on the planet within the penumbral zone and on the ``day side'' of the exoplanet, i.e. its  $ y- $coordinate in the planetocentric coordinate system is greater than zero. See text for explanation.}
	\label{fig:PenAll}
\end{sidewaysfigure}

\begin{figure}[hbt!]
	\centering
	\includegraphics[width = 0.8\textwidth]{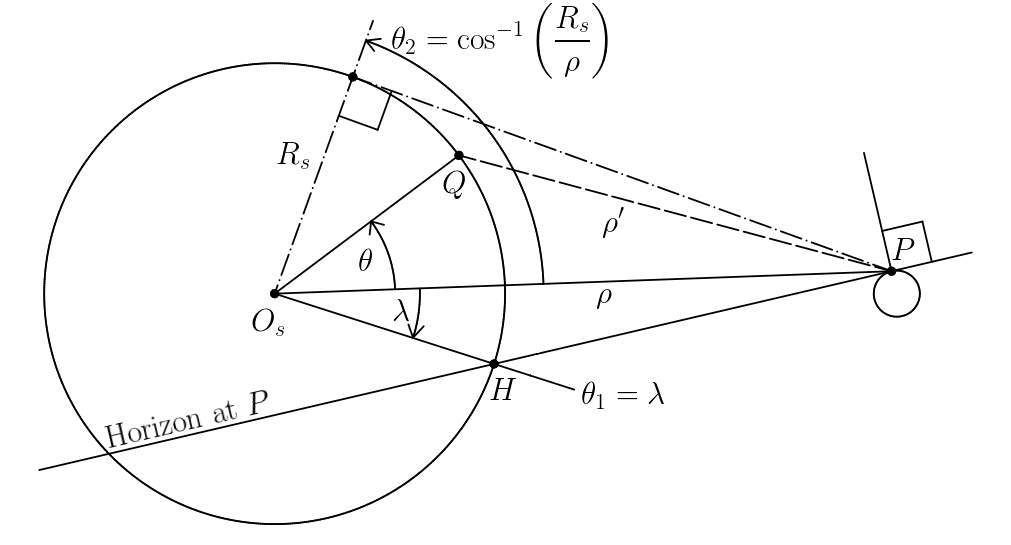}
	\caption{Adaption of Figure 4.4 from \protect\cite{Kopal1953} to illustrate the relationship between  $ \theta $ and $\lambda $ for use in~\cref{eq:penJs}. In the above, $P$ is a point on the planet within the penumbral zone.}
	\label{fig:strippedPen}
\end{figure}

\cref{fig:PenAll}\ shows the geometrical situation for points within in the penumbral zone. For such points the visible surface area of the host star is reduced because the horizon line at $ P $ intersects the stellar disk. As a result the evaluation of \cref{eq:Jn}\ is complicated by the lack of symmetry about $ \rho $. Using the same strategy to determine \cref{eq:Jfull}, we find that within the penumbral zone the intensity of the light reflected at a point $P$ is given by 
\begin{equation}\label{eq:penIntensityReceived}
	\myIntensityDis'(\phi, \eta) = I_0\myRefCo((1-u)J_1'+uJ_2'),
\end{equation}
where
\begin{equation}\label{eq:penJs}
	\pi J_n' = 2R_s^2 \int_{\theta_1}^{\theta_2} \int_{\omega_1}^{\omega_2} \frac{\cos\xi\cos^n\theta'\sin\theta}{\rho'^2}~d\theta~d\omega,
\end{equation}
where we have used the symmetry of $ \omega $ about point $ P' $ to give a factor of two, see \cref{fig:PenAll,fig:strippedPen,fig:omegalimits}.  \cref{eq:penJs}\ may be used for any limb darkening model which depends on powers of $ \cos{\theta'} $. In addition, it may be used for non-spherical geometries if one is able to obtain expressions of $ \theta_1, \theta_2, \omega_1 $ and $ \omega_2 $ for said geometry. Here we concern ourselves only with spherical stars and exoplanets and will recover Equations (47) and (48) \cite{Kopal1953}.

%


The limits of integration in \cref{eq:penJs}\ extend over the visible range of the stellar surface area as viewed from point $P$ on the exoplanet, for which the horizon line delimits the lower boundary on $\theta$ at point $H$ in \cref{fig:PenAll,fig:strippedPen}\ and the dot-dash line tangent to the stellar surface and connecting $P$ sets its upper boundary. We may then write $\theta_1 = \lambda$ and $\theta_2 = \invcos{R_s/\rho}$ as in \cref{fig:strippedPen}. The triangle $\bigtriangleup O_sHP$ in \cref{fig:PenAll}\ may be used to describe the angle $ \lambda $ using the law of sines,
\begin{equation}\label{eq:lambda}
	\lambda = \alpha -\invcos{\frac{\rho}{R_s}\cos\alpha} = \alpha - \invcos{\normz\cos\gamma'-p}, 
\end{equation}
where $\normz = r/R_s$ is the normalized star-planet separation, $\gamma'$ is the angle of foreshortening and $p = R_p/R_s$, Equation (6-37) \cite{Kopal1959}.

The angle $\lambda$ can be thought of as the ``geometric depth of the eclipse,'' \cite{Kopal1953,Kopal1959}. It lies within the range of $\pm\invcos{R_s/\rho}$, where its lower limit is located at the end of the eclipse, i.e. at $P$ located at the penumbral/un-illuminated zone boundary for which $\eta = \eta_2 = \invsin{(p-1)/\normz}$ and the azimuthal angle is between $ \pi $ and $ 2\pi $. The limits of $ \theta $ are illustrated in \cref{fig:strippedPen} and are set using the integration procedure presented in \cite{Kopal1953}\ rather than that of \cite{Kopal1959}. 
%


\begin{figure}
\centering
\includegraphics[width=0.5\linewidth]{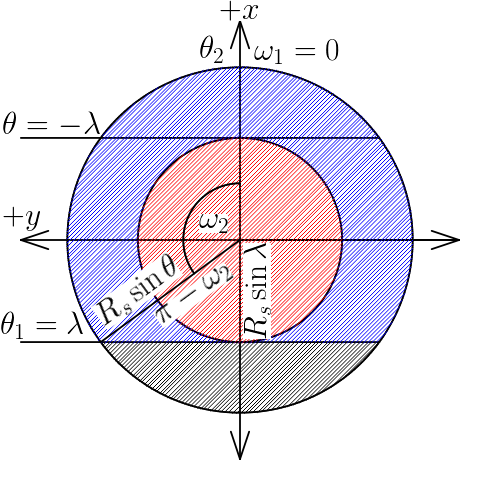}
\caption{Illustrated is a projection of the stellar disk as viewed from a point $ P $ within the penumbral zone for which $ \lambda<0 $. The black area is the portion of the disk that has sunk below the horizon and is no longer visible and the visible portion is made up of the blue and red zones. The above illustration may be used to determine the limits on $ \omega $ in \cref{eq:penJs}. Within the red portion of the figure \cref{eq:Mn,eq:Nn}\ are imaginary because $ \sin{\theta} >\sin{\lambda} $ for $ -\lambda<\theta<\lambda $ .}
\label{fig:projectionkopallabels}
\end{figure}

\myfillin, Finally, the limits of $ \omega $ may be determined in terms of $ \theta $ with the assistance of \cref{fig:projectionkopallabels,fig:omegalimits} where \cref{fig:projectionkopallabels}\ illustrates a projection of the stellar disk as it would be viewed by a person within the penumbral zone of the exoplanet. The radius of the stellar disk viewed by this person is  $ R_s\sin{\theta} $ in the radius of the red circle is given by $ R_s\sin{\lambda} $. Solving for $ \omega_2 $ reveals that the upper limit on $ \omega $ is given by
\begin{equation} \label{eq:omegalimitscorrect}
    \omega_2 = \pi-\invcos{\frac{\sin\lambda}{\sin\theta}},
\end{equation} 
at point $ H $. The resulting limits of $\omega$ are then $\omega_1 = 0$ and $\omega_2 = \pi-\invcos{\sin\lambda\csc\theta}$. It should be noted that for $ -\lambda<\theta<\lambda $ or the red area in \cref{fig:projectionkopallabels}, \cref{eq:omegalimits}\ is imaginary. This is likely one of the reasons behind the negative penumbral zone luminosity produced using this method of analysis for exoplanets. In \cref{sec:incidentFluxPennew}\ we will seek to avoid this problem by instead integrating over the whole spherical cap and then subtracting the integration over the black region in \cref{fig:projectionkopallabels}. Finally, in both \cite{Kopal1953,Kopal1959}, 
\begin{equation} \label{eq:omegalimits}
    \omega_2 =\invcos{\frac{\sin{\lambda}}{\sin{\theta}}} 
\end{equation}
rather than the limit implied by \cref{eq:omegalimitscorrect}.
In the analysis presented in this section we will use the limit given in  \cref{eq:omegalimits}\ to remain consistent with the methods presented in \cite{Kopal1953,Kopal1959}.

\begin{figure}[hbt!]
	\centering 
	\includegraphics[width = 0.95\textwidth]{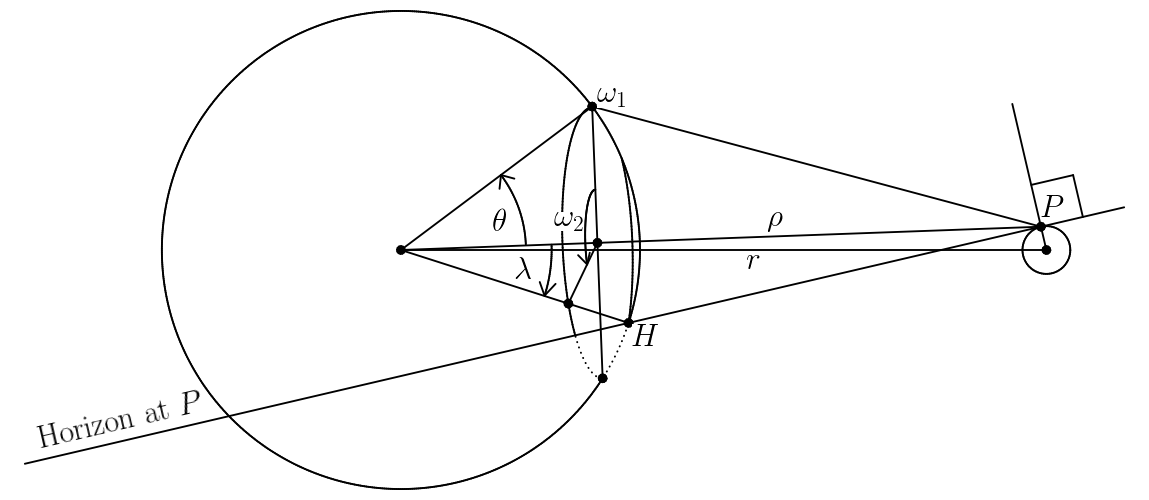}
	\caption{A simplified depiction of~\cref{fig:PenAll} to illustrate the limits on $\omega$ in~\cref{eq:penJs}.}
	\label{fig:omegalimits}
\end{figure}
%


By integrating over $ \omega $ we may simplify~\cref{eq:penJs} to
\begin{equation}\label{eq:shortJs}
	J_n' = M_n\cos\alpha+N_n\sin\alpha,
\end{equation}
where 
\begin{equation}\label{eq:Mn}
	\pi M_n = 2R_s^2\int_{\theta_1}^{\theta_2}\frac{\cos\beta}{\rho'^2}\invcos{\frac{\sin\lambda}{\sin\theta}}\cos^n\theta'~d\theta
\end{equation}
and
\begin{equation}\label{eq:Nn}
	\pi N_n = 2R_s^2\int_{\theta_1}^{\theta_2} \frac{\sin\beta}{\rho'^2}\sqrt{\sin^2\theta-\sin^2\lambda}\cos^n\theta' ~d\theta,
\end{equation}
see Equations (6-38) to (6-40) \cite{Kopal1959}.

\myfillin, \cref{eq:shortJs,eq:Mn,eq:Nn} may be derived from \cref{eq:penJs} as follows
\begin{equation}
	\begin{aligned}
		J_n' & = \frac{2R_s^2}{\pi}\int_{\theta_1}^{\theta_2}\int_{\omega_1}^{\omega_2}\left(\frac{\cos\alpha\cos\beta+\sin\alpha\sin\beta\cos\omega}{\rho'^2}\right)\cos^n\theta'\sin\theta~d\theta~d\omega \\
		& = \frac{2R_s^2}{\pi}\int_{\theta_1}^{\theta_2}\sin\theta\cos^n\theta'\left(\frac{\cos\alpha\cos\beta\invcos{\sin\lambda\csc\theta}}{\rho'^2}\right.\\
		& \hspace{11em}\left. +\frac{\sin\alpha\sin\beta\sin\left[ \invcos{\sin\lambda\csc\theta}\right]}{\rho'^2}\right)~d\theta \\
		& = \left[\frac{2R_s^2}{\pi}\int_{\theta_1}^{\theta_2}\frac{\cos\beta}{\rho'^2}\invcos{\frac{\sin\lambda}{\sin\theta}}\cos^n\theta'\sin\theta~d\theta\right]\cos\alpha\\
		&	\hspace{1em}+\left[\frac{2R_s^2}{\pi}\int_{\theta_1}^{\theta_2}\frac{\sin\beta}{\rho'^2}\sqrt{\sin^2\theta-\sin^2\lambda}\cos^n\theta'~d\theta\right]\sin\alpha\\
		& = M_n\cos\alpha+N_n\sin\alpha,
	\end{aligned}
\end{equation}
where the identity $\sin\left(\invcos{x}\right) =\sqrt{1 -x^2} $ was used to show that 
\begin{equation}
	\sin\left[\invcos{\sin\lambda\csc\theta}\right]\sin\theta =\sqrt{\sin^2\theta-\sin^2\lambda},
\end{equation}
if $\sin\theta \ge 0$, which is only true for $\alpha\ge 0$ as will be shown in \cref{sec:incidentFluxPennew}. This too may contribute to the un-physical luminosity within the penumbral zone found in \cref{sec:luminosityFiniteSizeog}.


Unfortunately,~\cref{eq:Mn,eq:Nn} have no closed form solutions except in certain limits, \cite{Kopal1953,Kopal1959}; however, they could be evaluated numerically to a desired precision or for other geometries in which the limits on $ \omega $ are described by \cref{eq:omegalimits}. If we wish to maintain an accuracy of $p^4$ in our final calculations, i.e. we may neglect tidal distortion and treat the star and exoplanet as spheres, we may approximate our variables to first order in $p$ and  $ \normz $ as follows: 
\begin{equation}\label{eq:approximation}
	\begin{aligned}
		\rho' &\approx \rho\\
		\cos\theta' &\approx \cos\theta \\
		\sin\beta  &\approx \frac{R_s}{r}\sin\theta \\
		\cos\beta  &\approx 1.
	\end{aligned}
\end{equation}
Note that in this approximation scheme, the limits on $\lambda$ become approximately, $-\pi/2 \leq \lambda \leq \pi/2$ and we can set $\theta_2$ to approximately $\pi/2$. 

\mynew, The approximations given in \cref{eq:approximation}\ are inexact and are applicable if the penumbral zone is relatively small such that its square and higher-order terms are negligible \cite{Kopal1953,Kopal1959}. In the case of \myECIES\ it may be that the above approximations are too extreme because the star-planet separation is of the same order as the stellar radius. For example, for Kepler-91b the normalized semi-major axis is  $ a/R_s= 2.48\pm0.12 $, \cref{tab:K91params}, so that terms like $\rho/R_s $ are approximately equal to 0.4. For now we will continue as described in \cite{Kopal1953,Kopal1959}.


\myfillin, Let us begin by solving for $M_1$ and $M_2$, in the limit described in \cref{eq:approximation}, \cref{eq:Mn}\ becomes
\begin{equation}\label{eq:Mnapprox}
	M_n \approx \frac{2R_s^2}{\pi\rho^2}\int_{\theta_1}^{\theta_2}\cos^{-1}\left(\frac{\sin\lambda}{\sin\theta}\right)\sin\theta \cos^n\theta ~d\theta = \left(\frac{2R_s^2}{\pi\rho^2}\right)I_{M_n},
\end{equation}
where $ \theta_1 = \lambda $ and $ \theta_2 = \invcos{R_s/\rho} \approx \pi/2 $ in this approximation. Furthermore, we may assume that $ -\pi/2 \le \lambda \le \pi/2 $ because $ \lambda $ has a range of $ \pm\invcos{R_s/\rho} $ as shown in \cref{fig:PenAll}. Using the following substitutions: $a = \sin\lambda$, $x = a/\sin\theta$, we may write $I_{M_n}$ as the integral in~\cref{eq:Mnapprox};
\begin{equation}\label{eq:IMn}
	I_{M_n} = -a^2\int_1^a\invcos{x}\frac{(x^2-a^2)^{(n-1)/2}}{x^{(n+2)}}~dx.
\end{equation}

We begin by considering the case for $n = 1$,
\begin{equation}\label{eq:IM1}
	I_{M_1}  = -a^2 \int_{1}^{a}\frac{\cos^{-1}(x)}{x^3}~dx.
\end{equation}
\cref{eq:IM1}\ may be solved with the assistance of Gradshteyn and Ryzhik, see Equation 2.832 \cite{mathTable}: 
\begin{equation}
	\int_{x_1}^{x_2}x^n\cos^{-1}(x)dx = \Eval{\frac{x^{n+1}}{n+1}\cos^{-1}(x)}{x_1}{x_2}+\frac{1}{1+n}\int_{x_1}^{x_2}\frac{x^{1+n}}{\sqrt{1+x^2}}~dx.
\end{equation}
In the case of \cref{eq:IM1}, we have $n = -3$, so that
\begin{equation}\label{eq:IM1solved}
	\begin{aligned}
		I_{M_1} &= \frac{a^2}{2}\left[\Eval{\frac{\invcos{x}}{x^2}}{1}{a}+\int_1^a\frac{dx}{x^2\sqrt{1-x^2}}\right] \\
			&= \frac{\sin^2\lambda}{2}\left[\Eval{\frac{\invcos{x}}{x^2}-\frac{\sqrt{1-x^2}}{x}\right]}{1}{\sin\lambda} \\
			&= \frac{\sin^2\lambda}{2}\left[\frac{\invcos{\sin\lambda}}{\sin^2\lambda}-\frac{\sqrt{1-\sin^2\lambda}}{\sin\lambda}\right]\\
			& = \frac{1}{2}\left[\frac{\pi}{2}-\lambda -\sin\lambda\cos\lambda\right]\\
			& = \frac{\pi}{2}\left[\frac{1}{2}-\frac{\lambda+\sin\lambda\cos\lambda}{\pi}\right]
	\end{aligned}
\end{equation}
where we have used the fact that $\cos^{-1}(\sin\lambda) = \pi/2-\lambda$ within the domain of $ \lambda $. Substitution into~\cref{eq:Mnapprox} reveals that 
\begin{equation}\label{eq:M1solved}
	M_1 = \left(\frac{R_s}{\rho}\right)^2\left[\frac{1}{2}-\frac{\lambda+\sin\lambda\cos\lambda}{\pi}\right],
\end{equation}
as expected from Equation (43) \cite{Kopal1953}. 

The $M_n$ integral for the linear darkening case is obtained by setting $n=2$ in~\cref{eq:IMn} to find that 
\begin{equation}\label{eq:Im2}
	I_{M_2} = -a^2\int_{1}^a \invcos{x}\frac{\sqrt{x^2-a^2}}{x^4}~dx,
\end{equation}
To solve the integral we will use integration by parts where
\begin{equation}
	\begin{aligned}
		u & = \invcos{x} & dv & = \frac{\sqrt{x^2-a^2}}{x^4}\\
		du & = \frac{-dx}{\sqrt{1-x^2}} & v & = \frac{(x^2-a^2)^{3/2}}{3a^2x^3},
	\end{aligned}
\end{equation}
to find that 
\begin{equation}
	\begin{aligned}
		I_{M_2} & = -a^2\left[\Eval{\frac{\invcos{x}}{3a^2x^3}(x^2-a^2)^{3/2}}{1}{a}+\int_1^a\frac{(x^2-a^2)^{3/2}}{3a^2x^3\sqrt{1-x^2}}~dx\right]\\
		& = -\frac{1}{3}\int_1^a\sqrt{\frac{(x^2-a^2)^3}{x^6-x^8}}~dx,
	\end{aligned}
\end{equation}
which can be solved using contour integration. The result is 
\begin{equation}\label{eq:Im2eval}
	\begin{aligned}
		I_{M_2} & = \frac{\pi}{12}\left(2-3a+a^3\right)\\
		& = \frac{\pi}{12}\left(2-3\sin\lambda+\sin^3\lambda\right) \\
		& = \frac{\pi}{12}\left(2-3\sin\lambda+\sin\lambda(1-\cos^2\lambda)\right)\\
		& = \frac{\pi}{6}\left(1-\sin\lambda\left(1+\frac{\cos^2\lambda}{2}\right)\right).
	\end{aligned}
\end{equation}
\cref{eq:Im2eval}\ holds for $-1<a<0$ and $0<a<1$, which is equivalent to $-\pi/2<\lambda<0$ and $0<\lambda<\pi/2$. In the realm of our approximation, $-\pi/2\leq \lambda \leq \pi/2$, so our limits on $\lambda$ must be shifted to the ranges of $-\pi/2<\lambda<0$ and $0<\lambda<\pi/2$. To complete our analysis for the $M_n$ integrals, substitute~\cref{eq:Im2eval} into~\cref{eq:Mnapprox}:
\begin{equation}\label{eq:M2solved}
	M_2 = \frac{1}{3}\left(\frac{R_s}{\rho}\right)^2\left(1-\sin\lambda\left(1+\frac{\cos^2\lambda}{2}\right)\right),
\end{equation}
see Equation (45) \cite{Kopal1953}. 

Let us now proceed to the evaluation of the $ N_n $ integrals within the same approximations; where,
\begin{equation}
	N_n \approx \frac{2}{\pi}\left(\frac{R_s}{\rho}\right)^3 \int_\lambda^{\pi/2}\sqrt{\sin^2\theta-\sin^2\lambda}\sin\theta\cos^n\theta~d\theta = \left( \frac{2R_s^3}{\pi \rho^3} \right)I_{N_n}.
\end{equation}
We will re-write the integral, $I_{N_n}$, by substituting $x =\cos\theta$ and $b = \cos\lambda$:
\begin{equation}
	I_{N_n} = \int_0^{b}x^n\sqrt{b^2 -x^2}~dx.
\end{equation}
Setting $n = 1$, we find
\begin{equation}\label{eq:In1solved}
	\begin{aligned}
		I_{N_1} & = \int_0^b x\sqrt{b^2-x^2}~dx \\
			& = \Eval{-\frac{1}{3}\left(b^2-x^2\right)^{3/2}}{0}{b}\\
			& = \frac{1}{3}\cos^3\lambda;
	\end{aligned}
\end{equation}
therefore,
\begin{equation}\label{eq:N1solved}
	N_1 = \frac{2}{3\pi}\left(\frac{R_s}{\rho}\right)^3\cos^3\lambda,
\end{equation}
see Equation 44 \cite{Kopal1953}\ or Equation (6-43) \cite{Kopal1959}.

Next, for $n = 2$, we have
\begin{equation}
	\begin{aligned}
		I_{N_2} & = \int_0^b x^2\sqrt{b^2-x^2}\\
		& = \Eval{-\frac{x\left(b^2-x^2\right)^{3/2}}{4}+\frac{xb^2\sqrt{b^2-x^2}}{8}+\frac{b^4}{8}\invsin{\frac{x}{b}}}{0}{b}\\
		& = \frac{\pi}{16}\cos^4\lambda; 
	\end{aligned}
\end{equation}
therefore,
\begin{equation}\label{eq:N2solved}
	N_2 = \frac{1}{8}\left(\frac{R_s}{\rho}\right)^3\cos^4\lambda.
\end{equation}
see Equation 46 in \cite{Kopal1953}\ or Equation  (6-45) in \cite{Kopal1959}.

We may now insert~\cref{eq:M1solved,eq:N1solved,eq:M2solved,eq:N2solved} into~\cref{eq:shortJs} to obtain
\begin{equation}\label{eq:J1prime}
	J_1' = \left(\frac{R_s}{\rho}\right)^2\left(\frac{1}{2}-\frac{\lambda+\sin\lambda\cos\lambda}{\pi}\right)\cos\alpha +\frac{2}{3\pi}\left(\frac{R_s}{\rho}\right)^3\cos^3\lambda \sin\alpha
\end{equation}
and
\begin{equation}\label{eq:J2prime}
	J_2' = \frac{1}{3}\left(\frac{R_s}{\rho}\right)^2\left(1-\sin\lambda\left(1+\frac{\cos^2\lambda}{2}\right)\right)\cos\alpha+\frac{1}{8}\left(\frac{R_s}{\rho}\right)^3\cos^4\lambda\sin\alpha,
\end{equation}
Equation 47 and 48 in \cite{Kopal1953}. 
To check for consistency, we set $\lambda = -\pi/2$ to obtain
\begin{equation}
	\begin{aligned}
		J_1'\left(-\frac{\pi}{2}\right) & = \left(\frac{R_s}{\rho}\right)^2\cos\alpha,\\
		J_2'\left(-\frac{\pi}{2}\right) & = \frac{2}{3}\left(\frac{R_s}{\rho}\right)^2\cos\alpha,
	\end{aligned}
\end{equation}
which is consistent with the result for the fully illuminated zone, as expected. In addition, $J_1'(\pi/2) = J_2'(\pi/2) = 0$, as expected for the boundary of the un-illuminated zone, see Equations 50-52 in \cite{Kopal1953}. 

Continuing to follow the methods described in \cite{Kopal1953,Kopal1959}, we will simplify the equations for $ J_{1, 2}' $ by using the fact that within the penumbral zone $ \alpha $ is approximately $ \pi/2 $ because $ \rho\approx r $ and $ r\gg R_p $; therefore, we may approximate $ \sin{\lambda} $ as
\begin{equation}\label{eq:sinlambdaapprox}
  \begin{aligned}
        \mySinLambda &=\sin{\lambda}\\
& =\sin\left( \alpha-\left[\piover2-\invcos{\frac{\rho}{R_s}\cos{\alpha}}\right]\right)\\
&\approx \left( \frac{r}{R_s}\cos{\alpha}\right).
  \end{aligned}
\end{equation} 
We may then write $ \cos\alpha\approx (R_s/r)\mySinLambda $ and approximate $ \sin\alpha\approx1 $. Substitution of these approximations into \cref{eq:J1prime,eq:J2prime} reveals 
\begin{equation} \label{eq:J1primefinal}
\begin{aligned}
	J_1' &= \left( \frac{R_s}{r} \right)^2\left( \1over2 -\frac{\invsin{\mySinLambda}+\mySinLambda\sqrt{1-\mySinLambda^2}}{\pi} \right)\left(\frac{R_s}{r}\mySinLambda\right)+\frac{2}{3\pi}\left( \frac{R_s}{r} \right)^3\left( 1-\mySinLambda^2\right)^{3/2}\\
	& = \left( \frac{R_s}{r}  \right)^3\left[\mySinLambda\left( \1over2 - \frac{\invsin{\mySinLambda}+\mySinLambda\sqrt{1-\mySinLambda^2}}{\pi}\right) + \frac{2}{3\pi}\left( 1-\mySinLambda^2\right)^{3/2}\right]
\end{aligned}
\end{equation}
and 
\begin{equation} \label{eq:J2primefinal}
\begin{aligned}
J_2' &= \frac{1}{3}\left( \frac{R_s}{r} \right)^2\left[ 1-\mySinLambda\left( 1+\1over2 (1-\mySinLambda^2) \right) \right]\left(\frac{R_s}{r}\mySinLambda\right)+\frac{1}{8}\left( \frac{R_s}{r} \right)^3\left( 1-\mySinLambda^2 \right)^2\\
& = \left( \frac{R_s}{r} \right)^3\left[ \frac{\mySinLambda}{3}-\frac{\mySinLambda^2}{3}\left( 1+\1over2 -\frac{\mySinLambda}{2} \right) +\frac{1}{8}\left( 1-\mySinLambda^2 \right)^2 \right] \\
& = \left( \frac{R_s}{r} \right)^3\left[ \frac{\mySinLambda}{6}(1-\mySinLambda)(-\mySinLambda^2-\mySinLambda+2)+\frac{1}{8}\left( 1-\mySinLambda^2 \right)^2\right].
\end{aligned}
\end{equation}
\mycorrection, It is at this point that we find that our analysis differs slightly than that given in \cite{Kopal1953,Kopal1959}\ where \cref{eq:J1primefinal,eq:J2primefinal}\ differ from previous work by a minus sign. It appears that in \cite{Kopal1953,Kopal1959}\ a substitution of $ \cos\alpha = -(R_s/r)\mySinLambda $ may have been used by Kopal rather than $ \cos\alpha\approx (R_s/r)\mySinLambda $ as suggested. Within this work, we will continue applying the methods described in \cite{Kopal1953,Kopal1959}, but will use \cref{eq:J1primefinal,eq:J2primefinal}\ as opposed to those used by Kopal.


\myfillin, Recalling further that $ L_s = \pi R_s^2 I_0\left(1 -u/3\right) $ we may then write~\cref{eq:penIntensityReceived} as
\begin{equation}\label{eq:intensityDisPen}
	\begin{aligned}
		\myIntensityDis_{refl,pen}(\mu) &= \frac{L_s\myRefCo}{\pi R_s^2}\left( \frac{3}{3-u} \right)\left[(1-u)J_1'+uJ_2'\right]\\
			&= \frac{3(1-u)\myRefCo}{3-u}\myIntensityDis^\mathbf{U}(\phi, \eta)+\frac{2u\myRefCo}{3-u}\myIntensityDis^\mathbf{D}(\phi, \eta),
	\end{aligned}
\end{equation}
where 
\begin{equation}\label{eq:Ju} 
	\myIntensityDis^\mathbf{U}(\mu) = \frac{L_sR_s}{6\pi^2r^3}\left[\mySinLambda\left(3\pi-6\invsin{\mySinLambda}-6\mySinLambda\sqrt{1-\mySinLambda^2}\right)+4\left(1-\mySinLambda^2\right)^{3/2}\right]
\end{equation}
and
\begin{equation}\label{eq:Jd} 
	\myIntensityDis^\mathbf{D}(\mu) = \frac{L_sR_s}{16\pi r^3}\left( \mySinLambda-1 \right)^2\left( 7\mySinLambda^2+14\mySinLambda+3 \right).
\end{equation}

As was true of \cref{eq:intensityDisFull}, integration of \cref{eq:Ju,eq:Jd}\ will be aided by expansion of the foregoing equations in ascending powers of $ \mySinLambda $. In order to maintain the level of precision of this work we must expand to fourth order, thus transforming \cref{eq:Ju,eq:Jd}\ to
\begin{equation}\label{eq:Juapprox}
		\myIntensityDis^\mathbf{U}(\mySinLambda)\approx \frac{L_sR_s}{\pi r^3} \left(\frac{2}{3 \pi} +\frac{\mySinLambda}{2}-\frac{3 \mySinLambda^2}{\pi}+  \frac{7 \mySinLambda^4}{12 \pi}\right)
\end{equation}
and
\begin{equation}\label{eq:Jdapprox}
		\myIntensityDis^\mathbf{D}(\mySinLambda) \approx \frac{L_sR_s}{\pi r^3}\left(\frac{3}{16}+\frac{\mySinLambda}{2} -\frac{9 \mySinLambda^2}{8}+\frac{7 \mySinLambda^4}{16}\right).
\end{equation}
Next, we must rewrite $ \myIntensityDis^\mathbf{U,D}(\mu) $ in terms of  $ \mu $ through use of \cref{eq:sinlambdaapprox,eq:cosAlpha_phiEta}, producing
\begin{equation}\label{eq:JUDSum}
	\myIntensityDis^\mathbf{U,D}(\mu) \approx \frac{L_sR_s}{\pi r^3}\sum_{n=0}^{4} C_n^\mathbf{U,D}\mu^n
\end{equation}
where
\begin{equation}\label{eq:CUs}
	\begin{aligned}
		C_0^\mathbf{U} &=\frac{1}{12\pi}\left( \frac{7 R_p^4}{R_s^4}-\frac{36 R_p^2}{R_s^2}-\frac{6 \pi  R_p}{R_s}+8\right) \\
		C_1^\mathbf{U} &= \frac{r\left( 36 R_p R_s^2-14 R_p^3+3 \pi  R_s^3\right)}{6 \pi R_s^4}\\
		C_2^\mathbf{U} &= \frac{r^2 \left(7 R_p^2-6 R_s^2\right)}{2 \pi  R_s^4}\\
		C_3^\mathbf{U} &= -\frac{7 r^3 R_p}{3 \pi  R_s^4}\\
		C_4^\mathbf{U} &= \frac{7 r^4}{12 \pi  R_s^4}\\
	\end{aligned}
\end{equation}
and 
\begin{equation}\label{eq:CDs}
	\begin{aligned}
		C_0^\mathbf{D} &= \frac{1}{16} \left(\frac{7 R_p^4}{R_s^4}-\frac{18 R_p^2}{R_s^2}-\frac{8 R_p}{R_s}+3\right)\\
		C_1^\mathbf{D} &= \frac{r \left(9 R_p R_s^2-7 R_p^3+2 R_s^3\right)}{4 R_s^4}\\
		C_2^\mathbf{D} &= \frac{3 r^2 \left(7 R_p^2-3 R_s^2\right)}{8 R_s^4}\\
		C_3^\mathbf{D} &= -\frac{7 r^3 R_p}{4 R_s^4}\\
		C_4^\mathbf{D} &= \frac{7 r^4}{16 R_s^4}.
	\end{aligned}
\end{equation}
\mycorrection, As mentioned previously, the equations presented here differ from Kopal's work because this analysis corrects for a lost minus sign in \cite{Kopal1959}\ and because we have not included the $ (r/R_s)^n $ term within the summation as was done in Equation (6-77) of \cite{Kopal1959}. 

\mynew, It should be reiterated that the foregoing analysis will be shown to be inappropriate for exoplanets in \cref{sec:luminosityFiniteSizeog}. Possible reasons include the fact that the integration over $ \theta $ includes values of $ \theta $ for which the integrand is imaginary and the fact that the upper limit of $\omega$ in \cref{eq:omegalimits}\ neglects the red portion of the stellar disk shown in \cref{fig:projectionkopallabels}. In addition, the approximations made for \cref{eq:approximation,eq:sinlambdaapprox}\ are inexact and may not apply well to \myECIES\ because $r$ and $R_s$ are of the same magnitude such that it cannot be said that the penumbral zone is small. Finally, the above analysis does not take into account some important sign changes that will be described in the following section.

\subsection{New Derivation of $\myIntensityDis_{p,refl} (\mu)  $  within the Penumbral Zone}\label{sec:incidentFluxPennew}
\mynew, Within the penumbral zone there are two distinct situations one must consider to determine the reflected intensity distribution,  $ \myIntensityDis_{pen} $. The two situations are distinguished by the sign of $ \alpha $. In the first situation, $ \alpha $ is negative and an observer at point $ P $ would observe half or more of the apparent disk of the host star. Second, for positive values of $ \alpha $ less than half of the apparent disk of host star would be visible. \cref{fig:PenAll}\ depicts a situation in which more than half of the apparent stellar disc is visible. First, we shall consider the determination of the reflected intensity for negative values of $ \alpha $.

\subsubsection{Reflected Intensity Distribution of Penumbral Zone 1}\label{sec:pen1intensitydis}
\begin{figure}[hbt!]
	\centering
	\subfloat[][\label{fig:startEclipse} Depiction of the start of the penumbral zone, for which $ \lambda $ is $ \invcos{R_s/\rho} $ .]{\includegraphics[width = 0.48\linewidth]{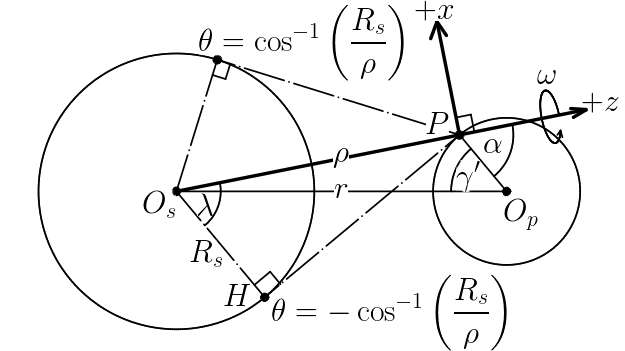}}
	\hfill
	\subfloat[][\label{fig:lambdazero} Depiction of the end of Penumbral Zone One, for which $ \lambda $ is zero and $ \alpha $ is  $ \pm \pi/2 $.]{\includegraphics[width = 0.48\linewidth]{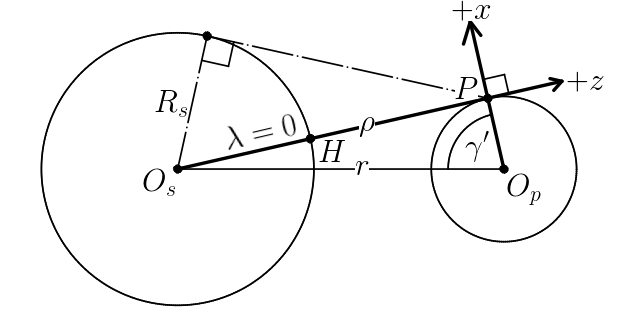}}
	\caption{\label{fig:PenzoneNegAlpha}Limiting cases for Penumbral Zone One for which \cref{fig:PenAll}\ is a general case.}
\end{figure}

\mynew, The first portion of the penumbral zone is such that an observer would see half or more of the apparent stellar disk above the horizon. The values of $ \alpha $ for which this is true are $ [-\piover2,-\invcos{R_s/\rho}] $ where the eclipse of the host star begins at $ \alpha =  -\invcos{R_s/\rho}$ and the point at which less than half of the star will be visible occurs at $ \alpha = \pm \pi/2 $. \cref{fig:PenzoneNegAlpha}\ depicts these two situations for which \cref{fig:PenAll}\ is a general case.

Let us begin by determining a relationship between the geometric depth of the eclipse, $ \lambda $, and $ \alpha $, where we will measure $ \lambda $ in the same direction as $ \alpha $ and $ \theta $; therefore, $ \lambda\le 0 $ in Penumbral Zone One. All three of these angles are measures of colatitude. From the triangle  $  \bigtriangleup PHO_s $ and by properly accounting for the fact that  $ \alpha\le 0 $ we may use the law of sines as follows
\begin{equation}\label{eq:lambdapen1}
  \begin{aligned}
        \frac{\rho}{\sin{(\pi/2+(\lambda-\alpha))}}&=\frac{R_s}{\sin{(\pi/2 +\alpha)}}\\
        \frac{\rho}{\cos{(\lambda -\alpha)}}& =\frac{R_s}{\cos{\alpha}}\\
        \lambda =\alpha&+\invcos{\frac{\rho}{R_s}\cos{\alpha}}.
  \end{aligned}
\end{equation}
Setting $ \alpha = \pm \pi/2$ we find that
\begin{equation}
  \begin{aligned}
        \lambda& =\pm\piover2+\invcos{0}\\
                & =\pm\piover2 +\piover2\\
                & =0 \textnormal{ or }\pi,
  \end{aligned}
\end{equation}
where we choose the value of 0 by recalling that  $ \lambda $ is a measure of colatitude. Furthermore, at the start of the penumbral zone the angle $ \angle O_s H P $ is  $ \pi/2 $. Trigonometry may be used to show that
\begin{equation}
	\begin{aligned}
		\cos{\alpha_0}&=\frac{R_s}{\rho}\\
		\alpha_0&=\pm \invcos{\frac{R_s}{\rho}}
	\end{aligned}
\end{equation}
where we must choose the negative value by virtue our selected coordinate system. Substitution into \cref{eq:lambdapen1}\ reveals that
\begin{equation}
	\lambda_0=-\invcos{\frac{R_s}{\rho}}
\end{equation}
as expected from  \cref{fig:startEclipse}. It should be noted that with this coordinate system the lower limit of the geometric depth of the eclipse, $ \lambda_0 $, is at the start of the eclipse of the host star as opposed to the end of the eclipse as was the case in \cref{sec:incidentFluxPenKopalMethod}\ and \cite{Kopal1953}.

The zone begins at the fully illuminated/penumbral zone boundary at $\eta=\eta_{1}$ whose azimuthal coordinates will be discussed in \cref{sec:luminosityFiniteSizeog}. \cref{eq:cosAlpha_rho}\ may be used to solve for the value of $ \gamma' $ for which $ \alpha =-\pi/2 $, which marks the end of Penumbral Zone One:
\begin{equation}
	\begin{aligned}
		0&= \frac{r\cos\gamma'-R_p}{\rho}\\
		\cos\gamma' &= \frac{R_p}{r}\\
		\sin\eta\sin\phi&= \frac{R_p}{r}.
	\end{aligned}
\end{equation}
Let us first consider situations like those depicted in \cref{fig:innerandoutertangents}\ for which $ \phi=\pi/2 $ to find that the polar coordinates of Penumbral Zone One lie between $\eta=\eta_{1}$ and 
\begin{equation}\label{eq:pen12boundary}
	\eta_{\mypen1limit}=\invsin{\frac{R_p}{r}}.
\end{equation}
Note that for large $r$ this zone disappears. One may also derive the relationship given in \cref{eq:pen12boundary}\ from the right triangle shown in \cref{fig:lambdazero}. The azimuthal coordinates of this zone will be discussed in \cref{sec:luminosityFiniteSizenew}.

\begin{figure}[tbh]
\centering
\includegraphics[width=0.5\linewidth]{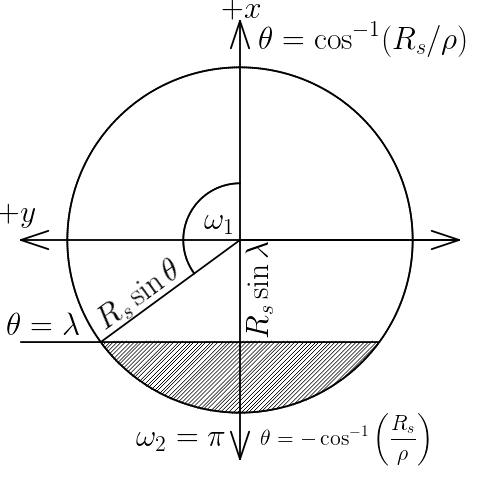}
\caption{Shown is a projection of the apparent stellar disk in the plane of the sky of an observer within Penumbral Zone One. The radius of the apparent stellar disk is $ R_s\sin{\theta} $ and the apparent distance between the center of the disk and the portion of it no longer visible is $R_s\sin{\lambda} $.}
\label{fig:projection1}
\end{figure}

To determine the reflected intensity distribution within this zone two distinct integrations must be performed. First, we must integrate over the spherical cap of the host star delimited by the colatitude angles $ \theta = [-\invcos{R_s/\rho}, \invcos{R_s/\rho}] $. In addition, we must integrate over the spherical segment for which $ \omega=[\pi-\invcos{\sin{\lambda}/\sin{\theta}},\pi]$ and  $  \theta = [\lambda, -\invcos{R_s/\rho}] $ and subtract this portion from the integration over the spherical cap. Here we take advantage of the fact that the integration is even about $ \omega = \pi $ to multiply the integral by two. A projection of the apparent stellar disk as viewed from a point $P$ within Penumbral Zone One is shown in \cref{fig:projection1}\ illustrates the geometry required to determine the limits on $ \omega $. Note that both $ \theta $ and $ \lambda $ are negative, so we have dropped the minus sign on the expression of the radius of the disk and depth of the eclipse, or the apparent distance between the center of the disk and the portion of it no longer visible.

As was the case in the fully illuminated zone, we may write the integration over the spherical cap as 
\begin{equation}\label{eq:Jpen1capsubtraction}
	\begin{aligned}
		J_{pen1,cap,n} &=  {\pi}\int_{0}^{\invcos{R_s/\rho}}\int_{0}^{2\pi}\left( \frac{\cos{\xi}\cos^n{\theta'}}{\pi\rho'^{2}}\right)~dA_{s}\\
				  &= \frac{\cos\alpha}{2^nR_s^{n-1}\rho^2}\int_{\rho-R_s}^{\sqrt{\rho^2 -R_s}}\frac{(\rho'^2+\rho^2-R_s^2)(\rho^2-R_s^2-\rho'^2)^n}{\rho'^{2+n}}~d\rho'\\
	\end{aligned}
\end{equation}
Then, we must subtract the black section in \cref{fig:projection1}, which is described by
\begin{equation}\label{eq:Jpen1segsubtraction}
	\begin{aligned}
		J_{pen1,seg,n} &=  \frac{2R_s^2}{\pi}\int_{\lambda}^{-\invcos{R_s/\rho}}\int_{\pi-\invcos{\sin\lambda/\sin\theta}}^{\pi}\left( \frac{\cos{\xi}\cos^n{\theta'}}{\pi\rho'^{2}}\right)\sin{
\theta}~d\theta~d\omega\\
					&= \frac{2 R_s^2}{\pi}\int_{\lambda}^{-\invcos{R_s/\rho}} \left( \frac{\sin\theta (\rho\cos\theta-R_s)^n}{(\rho^2 +R_s^2 -2\rho R_s\cos\theta)^{(3+n)/2}}\right)\bigg[\rho  \cos\alpha \invcos{\frac{\sin\lambda}{\sin\theta}}-\\
					&\hspace{1in}R_s \left(\cos\alpha \cos (\theta )\invcos{\frac{\sin\lambda}{\sin\theta}}+\sin\alpha \sqrt{\sin ^2(\theta )-\sin ^2\lambda}\right)\bigg]~d\theta.
	\end{aligned}
\end{equation}
Here we shall restrict ourselves to the analysis for zero limb darkening and leave linear limb darkening to a future work. By inspection we see that  \cref{eq:Jpen1capsubtraction}\ is the same as \cref{eq:JnReWrite}; therefore, we may use the results from \cref{eq:J1full}:
\begin{equation}\label{eq:Jpen1capnis1}
	J_{pen1,cap,1}=\left( \frac{R_s}{\rho}\right)^2 \cos{\alpha}.
\end{equation}

Unfortunately, the integrand of \cref{eq:Jpen1segsubtraction}\ is not easily evaluated; therefore, we shall use a Taylor series to approximate \cref{eq:Jpen1segsubtraction}, \cf\ Chapter 1 in \cite{boas}. Within the first penumbral zone we know that $ \cos{\alpha} $ is between zero and  $ R_s/\rho $, in addition we may safely assume that for physical systems  $ R_s/\rho<1 $. We will then expand about both variables $ y=\cos{\alpha}  $ and  $ x=R_s/\rho $ simultaneously. 

In general, for a function $ f(x,y,z)  $ that depends on three variables for which $ x $ and $ y $ are small we may approximate $ f(x,y,z) $ as follows
\begin{equation} 
    f(x,y,z)\approx C_{0,0}(z) +C_{1, 0}(z) x+C_{0,1}(z) y+C_{1,1}(z)xy+C_{2, 0}(z) x^2 +C_{0,2}(z) y^2 +\ldots 
\end{equation}
where  $ C_{n,m} (z)$ are coefficients that depend on $ z $.  For cases in which $ x\sim y $ it is convenience to rewrite the above equation by letting $ x=x'\myexpansionparameter $ and $ y= y'\myexpansionparameter $ as follows
\begin{equation} 
    f(x,y,z)\approx C_{0,0}(z) +\left[C_{1, 0}(z) x'+C_{0,1}(z) y'\right]\myexpansionparameter+\left[C_{1,1}(z)x'y' +C_{2, 0}(z) x'^2+C_{0,2}(z) y'^2\right]\myexpansionparameter^2  +\ldots 
\end{equation}

We shall apply this method to \cref{eq:Jpen1segsubtraction}\ using the expansion parameter $ \myexpansionparameter $ where
\begin{equation}\label{eq:expansionparameters}
	\begin{aligned}
		x&=\frac{R_s}{\rho}=x'\myexpansionparameter\\
		y& =\cos{\alpha}=y' \myexpansionparameter.
	\end{aligned}
\end{equation} 
To maintain the order of accuracy of $( R_{s}/r)^4  $ we will expand about  $ \myexpansionparameter=0 $ to fourth order so that we may approximate $ J_{pen1,seg,1} $ as
\begin{equation}\label{eq:Jpen1seg1expanded}
	\begin{aligned}
		J_{pen1,seg,1}=\int_{\lambda}^{-\invcos{x'\myexpansionparameter}} &\frac{\myexpansionparameter^4 x'^3}{2 \pi
   }\bigg[y' (3 \sin (3 \theta )-\sin \theta) \invcos{\frac{\sin{\lambda}}{\sin{\theta}}}-\\
	&\hspace{0.45in}4 x' \sin (3 \theta ) \sqrt{\sin ^2\theta-\sin ^2\lambda}\bigg]+\\
	&\frac{\myexpansionparameter^3 x'^2}{\pi } \bigg[\sin (2 \theta ) y' \invcos{\frac{\sin{\lambda}}{\sin{\theta}}}-x' \sqrt{\sin ^2\theta-\sin ^2\lambda}\bigg]~d\theta
	\end{aligned}
\end{equation}

After evaluating the indefinite integral we may substitute the exact upper and lower limits of $ \theta $. For the lower limit we will first substitute $ \theta=\lambda $ and for the upper limit we will substitute $ \theta=-\invcos{x' \myexpansionparameter} $. Next, the resulting definite solution to \cref{eq:Jpen1seg1expanded}\ is again expanded about $ \myexpansionparameter $ up to fourth order. Finally, we may substitute the definition of $ \lambda $ in terms of our expansion parameter:
\begin{equation} 
    \lambda =\invcos{\myexpansionz_1}+\invcos{y' \myexpansionparameter}
\end{equation}
where
\begin{equation} \label{eq:myexpansionz1def}
    \myexpansionz_1= \frac{y}{x}=\frac{r\mu-R_p}{R_s},
\end{equation}
and we may rewrite  $ J_{pen1,cap,1} $ 
\begin{equation}\label{eq:Jpen1capnis1expansion}
	J_{pen1,cap,1}(\myexpansionz_1) =x'^3\myexpansionz_1\myexpansionparameter^3.
\end{equation}

In a future work we will determine the intensity distribution within Penumbral Zone One from
\begin{equation}\label{eq:Jpen11}
	J_{pen1,1}(\myexpansionz_1)=J_{pen1,cap,1}(\myexpansionz_1)-J_{pen1,seg,1}(\myexpansionz_1).
\end{equation}
 To determine  $ \myIntensityDis_{pen1} $ from \cref{eq:penIntensityReceived}\ for zero limit darkening we must substitute \cref{eq:Jpen11}\ into \cref{eq:penIntensityReceived}\ for $ J_1' $ and set  $ u=0 $, or
\begin{equation}\label{eq:pen1intensitydistribution}
	\myIntensityDis_{pen1}=\frac{L_s\myRefCo}{\pi R_s^2} J_{pen1,1}(\myexpansionz_1).
\end{equation}
We must also substitute the expressions of $ x', \rho$ and $ \mu $ in terms of  $ \myexpansionz_1 $ from  \cref{eq:expansionparameters,eq:myexpansionz1def}\ which will give us an equation in terms of $ \myexpansionz_{1}, R_s,R_p,r $ and $ \myexpansionparameter $. For the purposes of determining the reflected luminosity of this zone it may be preferable to integrate over the parameter $ \myexpansionz_1 $, in which case it is beneficial to rescale \cref{eq:Jpen11}\ in terms of $ r'=r/\myexpansionparameter $ to account for the fact that terms like $R_p/r$ and $R_s/r$ are both small and on the order of $ \myexpansionparameter $.

\subsubsection{Reflected Intensity Distribution of Penumbral Zone 2}\label{sec:penzone2intensitydis}
\begin{figure}
\centering
\includegraphics[width=0.7\linewidth]{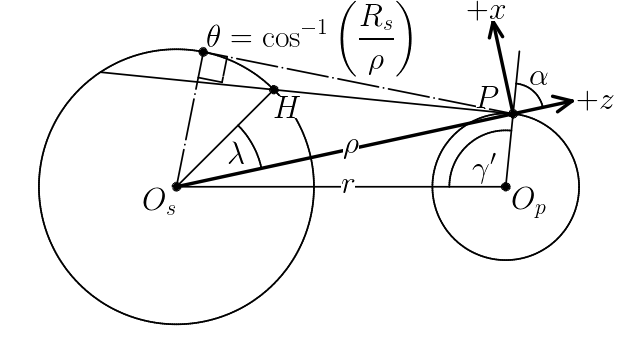}
\caption{Shown is a figure illustrating the geometry for points within Penumbral Zone Two of the exoplanet for which $ \alpha, \theta $ and $ \lambda $ are all positive.}
\label{fig:penzonenegalpha}
\end{figure}

Within this second penumbral zone $ \alpha $ and $ \lambda $ are positive and less than half of the apparent stellar disk will be above the horizon as viewed by an observer, as shown in \cref{fig:penzonenegalpha}. For such situations $ \lambda $ is related to $ \alpha $ by the equation
\begin{equation} \label{eq:lambdapen2}
    \lambda = \alpha -\invcos{\frac{\rho}{R_s}\cos{\alpha}}.
\end{equation}
The region begins at $ \alpha=\pi /2 $ and  $ \alpha $ decreases to a minimum value of $ \invcos{R_s/\rho} $. At the start of this region we find that $ \lambda =  0$ and at the end of the region $ \lambda = \invcos{R_s/\rho} $. Furthermore, the relationship between $\alpha$ and $\mu$ is now
\begin{equation}\label{eq:cosalphaPos}
	\cos{\alpha}=\frac{R_p-r\mu}{\rho}
\end{equation}
which may be determined via the triangle $\bigtriangleup PO_sO_p$ in \cref{fig:penzonenegalpha}, where it is now the case that
\begin{equation}
	r^2 =\rho^2 +R_p^2 -2rR_p\cos{\alpha},
\end{equation}
but the equation for $ \rho $ remains that given by \cref{eq:rhoSqr}.

\begin{figure}[tbh]
\centering
\includegraphics[width=0.5\linewidth]{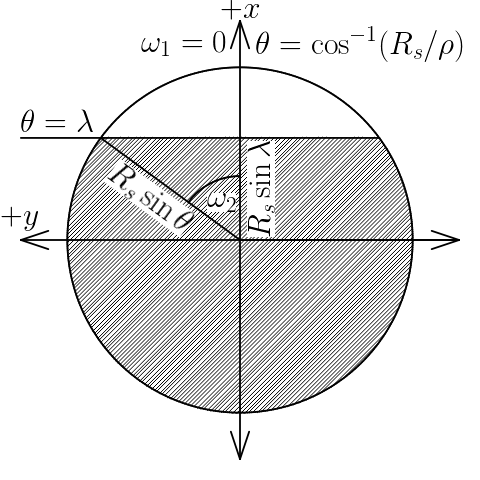}
\caption{Shown is a projection of the apparent stellar disk in the plane of the sky of an observer within Penumbral Zone Two. The radius of the apparent stellar disk is $ R_s\sin{\theta} $ and the projected distance between the center of the apparent disk and portion of the disk still visible is $ R_s\sin{\lambda} $.}
\label{fig:projection2}
\end{figure}

To determine the reflected intensity distribution for points in which $ \alpha=[\invcos{R_s/\rho},\pi/2] $ we need only integrate over a spherical segment, shown in \cref{fig:projection2} in white.
The new integration to be performed is
\begin{equation}\label{eq:Jpen2seg}
	\begin{aligned}
		J_{pen2,n} &=  \frac{2R_s^2}{\pi}\int_{\lambda}^{\invcos{R_s/\rho}}\int_{0}^{\invcos{\sin\lambda/\sin\theta}}\left( \frac{\cos{\xi}\cos^n{\theta'}}{\pi\rho'^{2}}\right)\sin{
\theta}~d\theta~d\omega,
	\end{aligned}
\end{equation}
which is the same expression given by \cite{Kopal1953}\ and we see here that Kopal had the correct expression only for points for which less than half of the apparent disk of the host star was visible. For now, we concern ourselves only with the case for which there is zero when darkening; therefore, we need only evaluate \cref{eq:Jpen2seg}\ for $ n =1 $. We will follow the same procedure as that described for the evaluation of \cref{eq:Jpen2seg}\ in \cref{sec:pen1intensitydis}\ to determine that
\begin{equation}\label{eq:Jpen2segnis1theta}
  \begin{aligned}
        J_{pen2,1}=\int_{\lambda}^{\invcos{x'\myexpansionparameter}}&\frac{\myexpansionparameter ^4 x'^3}{\pi } \sin \theta \Bigg[2 x' \left(2 \left(2 \cos ^2\theta-1\right)+1\right) \sqrt{\cos ^2\lambda-\cos ^2\theta}+\\
& \hspace{0.75in}y' \left(3 \left(2 \cos ^2\theta-1\right)+1\right)\invcos{\frac{\sin{\lambda}}{\sin{\theta}}}\Bigg]+\\
&\frac{2 \myexpansionparameter ^3 x'^2}{\pi } \sin \theta \cos \theta \left[x' \sqrt{\cos ^2\lambda-\cos ^2\theta}+y' \invcos{\frac{\sin{\lambda}}{\sin{\theta}}}\right].
  \end{aligned}
\end{equation}

The evaluation of \cref{eq:Jpen2segnis1theta}\ will be left for future work. We note here that within Penumbral Zone Two the expressions for $ \lambda $ and $ \myexpansionz $ will change because $ \alpha >0 $ and are given by
\begin{equation} 
    \lambda =\invcos{y' \myexpansionparameter}-\invcos{\myexpansionz_2}
\end{equation}
and
\begin{equation} \label{eq:myexpansionz2def}
    \myexpansionz_2= \frac{y'}{x'}=\frac{R_p-r\mu}{R_s}=-\myexpansionz_1.
\end{equation}
The reflected intensity distribution of the second penumbral zone for zero limb darkening would then be given by
\begin{equation}\label{eq:pen2intensitydistribution}
	\myIntensityDis_{pen2}=\frac{L_s\myRefCo}{\pi R_s^2} J_{pen2,1}(\myexpansionz_2).
\end{equation}

In the previous chapter we have outlined the general problem of describing the reflected luminosity of an exoplanet, and we have described the reflected intensity distribution using a variety of methods. First, we reviewed the standard solution assuming plane parallel ray illumination of an exoplanet, \cf\  \cite{sobolev,seager}. Then, we explored some of the consequences of including the finite angular size of the host star for \myECIES, including the fractional surface area of each of three zones, which are the fully illuminated, penumbral and un-illuminated zones. Next, we reviewed the derivation for the fully illuminated zone as described by Kopal in \cite{Kopal1953,Kopal1959}. Finally, we explored the derivation for the penumbral zone and found that previous work was in error and have outlined a new approach to determining the reflected luminosity of this zone.

\chapter{Photometric Planetary Emissions: Reflected Light Luminosity}\label{ch:reflectedlightluminosity} 
With the distribution of reflected luminosity in hand, we are now prepared to determine the reflected luminosity of an exoplanet as a function of phase angle with aid of  \cref{eq:lumplanet}
. To determine $\myLuminosity_{p, refl}$ one must integrate \cref{eq:lumplanet}\ over the visible area of the exoplanet, $\mathbf{C}$, which will depend on the model used to describe the illumination of the exoplanet, and on the phase angle. 

Before continuing on to the derivation of the reflected luminosity of an exoplanet we will summarize the contributions made to the field of exoplanet science described within this chapter. We will begin by reviewing previous work concerned with the plane parallel ray case in \cref{sec:Lforplaneparallel}, \cf\ \cite{sobolev,seager}.

\myfillin, Next, \cref{sec:luminosityFiniteSizeog}\ describes the approach used in \cite{Kopal1953,Kopal1959}\ to determine the reflected luminosity for a spherical body illuminated by another spherical body as a function of phase angle. Additional details are provided than those given in previous work.

\mychange, In \cref{sec:luminosityFiniteSizenew}\ we will present a description of the luminosity of an exoplanet for seven unique cases as opposed to the five cases originally presented in \cite{Kopal1953,Kopal1959}\ and described in \cref{sec:luminosityFiniteSizeog}. 

\mynew, Unique contributions include the discovery that the use of Equations  (62)-(69) in \cite{Kopal1953}\ produces negative luminosity near new phase within the fully illuminated zone, as described in \cref{sec:luminosityFiniteSizeog}. In addition, complete analysis of the reflected luminosity of the penumbral zone, as opposed to approximations used in Equation (86) of \cite{Kopal1953} for all five cases is presented in \cref{sec:luminosityFiniteSizeog}. A corrected analysis of the reflect the luminosity of the fully illuminated zone is presented in \cref{sec:luminosityFiniteSizenew}.This is an extension to the single case given by Equation (90) of \cite{Kopal1953}. As part of the analysis to eliminate the appearance of negative luminosity we will present the results of performing the integrations described in \cref{sec:luminosityFiniteSizeog} for use by future researchers in \cref{app:integrations}. Finally, in \cref{sec:geometricalbedo}\ we will fully discuss the relevance of the geometric albedo in the analysis of reflected light of extremely close-in exoplanets.

We will now proceed to discuss the reflected luminosity for plane parallel ray illumination, and then proceed to a discussion applicable to \myECIES.

\section{Plane Parallel Rays}\label{sec:Lforplaneparallel}
From  \cref{eq:lumplanet,eq:intDisPlaneUniform,eq:intDisPlaneLinearDark}, we find that the reflected luminosity for exoplanets that are illuminated by plane parallel rays is 
\begin{equation}\label{eq:luminosityphasefunctionplaneparallel}
	\begin{aligned}
		\myLuminosity_{p, refl}(\gamma, \gamma', \phi) & = \int_{\mathbf{C}}\myIntensityDis_{p,refl}(\gamma, \gamma', \phi)\cos\gamma ~dA_p\\
														& = SR_p^2\int_{\mathbf{C}}\myRefCo\sin(\phi-\myPhase)\sin\phi\sin^3\eta ~d\eta ~d\phi,
	\end{aligned}
\end{equation}
where the stellar flux is  $ S = I_0(R_s/r)^2  $ for a uniform star and $ S = I_0 (1-u/3) (R_s/r)^2  $ for a star described by linear limb darkening, \cf\ \cite{sobolev}. One of the distinguishing characteristics of the plane parallel ray model of illumination of an exoplanet is the fact that the stellar flux incident on the illuminated side of the exoplanet is constant; therefore, it may be pulled out from the integration to determine the luminosity of the exoplanet as a function of phase angle.

\begin{figure}[!htb]
	\centering
	\includegraphics[width=0.7\textwidth]{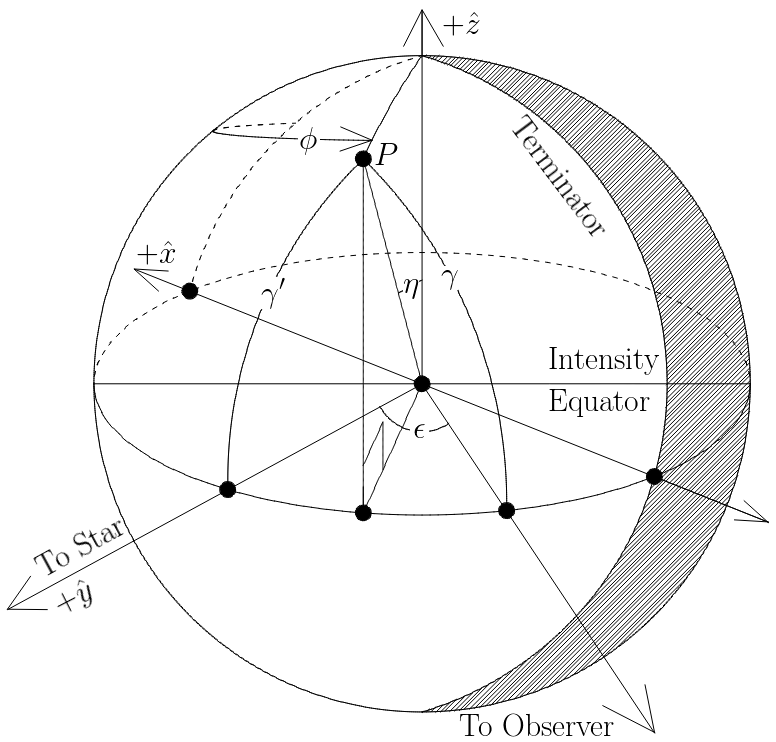}
	\caption{The geometry for the plane parallel ray model for which the terminator is located at $ \phi =\pi $ is shown. Points for which $ \eta $ is between zero and $\pi$ radians and $ \phi $ is between $ \myPhase $ and $\pi$ radians reflect light toward the observer. }
	\label{fig:planeparallelterminatorlabels}
\end{figure}

\myfillin, In the following, we will return to the planetocentric coordinate system described by \cref{eq:planetOriginCoordinates}. The limits of integration of \cref{eq:luminosityphasefunctionplaneparallel}\ for the plane parallel case are illustrated in~\cref{fig:planeparallelterminatorlabels} in which the terminator is located at the azimuthal angle of $\pi$ along the intensity equator of the exoplanet and the limb is located at $\myPhase$. The limits on $\eta$ are 0 to $\pi$ because the whole half hemisphere of the exoplanet facing its host star is always fully illuminated for plane parallel ray illumination. Adapting from \cite{sobolev}, we may now write the reflected luminosity of the exoplanet as
\begin{equation}\label{eq:luminosityH}
	\myLuminosity_{p, refl}(\myPhase) = SR_p^2H(\myPhase),
\end{equation}
where
\begin{equation}\label{eq:Hphase}
	H(\myPhase) = \int_{\myPhase}^{\pi}\int_{0}^{\pi} \myRefCo\sin(\phi-\myPhase)\sin\phi\sin^3\eta~d\eta~d\phi.
\end{equation}
Taking $\myLuminosity_{p, refl}(0)$ to be the luminosity of the exoplanet at full phase---i.e. when $\myPhase = 0$ one may also write $\myLuminosity_{p, refl}(\myPhase)$ as
\begin{equation}\label{eq:luminosityPhaseFunc}
		\myLuminosity_{p, refl}(\myPhase) = \myLuminosity_{p, refl}(0)\myPhaseFunc,
\end{equation}
where 
\begin{equation}\label{eq:phaseEqDef}
	\myPhaseFunc = \left(\frac{H(\myPhase)}{H(0)}\right)
\end{equation} 
and is called the phase function. The phase function allows one to calculate a variety of valuable quantities, such as a the geometric albedo and the fractional flux received by an observer. 

The bond albedo, or spherical albedo, is the ratio of the power reflected  by the whole exoplanet to that intercepted by it, \cf\ Chapter 1 of Sobolev's textbook \cite{sobolev}. The power reflected is given by 
\begin{equation*}
	\begin{aligned}
		P_{refl}&=2\pi \int_{0}^{\pi} \myLuminosity_{p,refl}(\myPhase)\sin\myPhase~d\myPhase \\
		   &= 2\pi \myLuminosity_{p,refl}(0)\int_{0}^{\pi}\myPhaseFunc \sin\myPhase~d\myPhase
	\end{aligned}
\end{equation*} 
and the power intercepted in the plane parallel case is $P_{inc} = \pi S R_p^2$. The spherical albedo is then
\begin{equation}\label{eq:sphericalAlbedo}
	A_s  = \left(\frac{\myLuminosity_{p, refl}(0)}{S R_p^2}\right)q,
\end{equation}
where $q$ is the phase integral
\begin{equation}\label{eq:phaseIntegral}
	q = 2\int_{0}^{\pi}\myPhaseFunc\sin\myPhase~d\myPhase.
\end{equation}
The phase integral, thus, depends on the directional reflecting properties via the $ \myRefCo $ in \cref{eq:Hphase,eq:phaseEqDef}. 

The geometric albedo, $A_g$, is the ratio of the luminosity of a planet at zero phase to that produced by a plane, lossless Lambert disk of the same radius placed at the same position. From~\cref{eq:luminosityH} and the fact that the luminoisty of such a Lambert disk is $\myLuminosity_{Lambert} = \pi S R_p^2$ we find that 
\begin{equation}\label{eq:geometricAlbedo}
	\begin{aligned}
		A_g &= \frac{\myLuminosity_{p, refl}(0)}{\myLuminosity_{Lambert}}\\
			& = \frac{H(0)}{\pi}.
	\end{aligned}
\end{equation}
\cref{eq:geometricAlbedo} provides the following relationship between the geometric albedo and the spherical albedo:
\begin{equation}\label{eq:sphericalAndGeometricAlbedo}
	A_s = qA_g.
\end{equation}
The definition of $A_g$ permits us to rewrite the luminosity reflected by an exoplanet. Solving~\cref{eq:geometricAlbedo} for $\myLuminosity_{p, refl}(0)$ and substituting this into~\cref{eq:luminosityPhaseFunc} we find that 
\begin{equation}\label{eq:luminosityAg}
	\myLuminosity_{p, refl}(\myPhase) = \pi S R_p^2 A_g \myPhaseFunc.
\end{equation}
Finally, the fractional flux reflected by the exoplanet and received by an observer at distance $d$ can be determined by dividing the flux received from the exoplanet by that received from the host star. The flux received from the exoplanet is given by
\begin{equation}
	\begin{aligned}
		F_{p, refl}(\myPhase) &= \frac{\myLuminosity_{p, refl}(\myPhase)}{\pi d^2}\\
							  & =  SA_g\myPhaseFunc\left(\frac{R_p}{d}\right)^2
	\end{aligned}
\end{equation}
and that of the star by
\begin{equation}\label{eq:fluxAtEarth}
	\begin{aligned}
		F_s &= \frac{L_s}{\pi d^2}\\
			& =S \left(\frac{r}{d}\right)^2 
\end{aligned}
\end{equation}
so that the fractional flux reflected by an exoplanet is given by
\begin{equation}\label{eq:phirPhaseFunc}
	\begin{aligned}
		\Phi_\textnormal{refl} (\myPhase)&= \frac{F_{p,refl}(\myPhase)}{F_s}\\
										 & = A_g \left(\frac{R_p}{r}\right)^2\myPhaseFunc.
	\end{aligned}
\end{equation}
~\cref{eq:phirPhaseFunc} reveals that the determination of the fractional flux received by an observer due to light reflected by the exoplanet requires only that one solve the integral given in~\cref{eq:Hphase} to determine the phase function, $ \myPhaseFunc $. 

~\cref{eq:phirPhaseFunc} probes the scattering properties of the exoplanet via $A_g$ and $\myPhaseFunc$ and may be used to determine $\sqrt{A_g}(R_p)$ if the orbital parameters of $r(\myPhase)$ are known. The value of $ R_p $ is the true radius of the exoplanet and the value of the geometric albedo, $ A_g$, is generally less than or equal to unity; therefore, a determination of $ \Phi_\textnormal{refl}(\myPhase) $ is only sufficient to determine the joint parameter $\sqrt{A_g}(R_p)$, which is called the minimum radius of the exoplanet. If the radius of the exoplanet may be determined, e.g. by the transit method, then the geometric albedo may also be calculated. It should also be noted that the geometric albedo is often wavelength dependent, as described in Leigh et. al. \cite{leigh2003a}; therefore, the minimum radius measured will also be wavelength dependent. In order to properly characterize the radius and reflective properties of an exoplanet it is best to take measurements at multiple wavelengths, producing a light curve for each wavelength. Finally, it is worth noting that if the transit method is used to determine the true radius of an extremely close-in exoplanet one must take into account the light originating from the penumbral zone or risk underestimating the true radius of the exoplanet. Such an underestimation in the true radius will result in an over estimation of the geometric albedo and the night side temperature.

\myfillin, For a Lambertian sphere we may assume that the reflection coefficient is constant and equal to the single scattering albedo, i.e. $\myRefCo = \myScatteringAlbedo$. The single scattering albedo represents the probability that a photon will be scattered rather than absorbed upon interacting with a single volume element, Equation (1.2) in \cite{sobolev}. For example, a lossless sphere is one in which $\myScatteringAlbedo = 1$. The Lambertian intensity distribution from \cref{eq:intDisPlaneUniform,eq:intDisPlaneLinearDark}\ may be substituted into \cref{eq:Hphase}\ to show that 
\begin{equation}\label{eq:lambertPhaseFunc}
	\myPhaseFunc_{Lambert} = \frac{(\pi-\myPhase)\cos\myPhase+\sin\myPhase}{\pi}
\end{equation}
as follows. To begin, substitute $\myRefCo = \myScatteringAlbedo$ into \cref{eq:Hphase}
\begin{equation}\label{eq:HphaseLambert}
	\begin{aligned}
		H_{Lambert}(\myPhase) &= \myScatteringAlbedo\int_{\myPhase}^{\pi}\int_{0}^{\pi}\sin(\phi-\myPhase)\sin\phi\sin^3\eta~d\eta~d\phi\\
							  &= \frac{4}{3}\myScatteringAlbedo\int_{\myPhase}^{\pi}\sin(\phi-\myPhase)\sin\phi~d\phi\\
							  &= \frac{2}{3}\myScatteringAlbedo\left((\pi-\myPhase)\cos\myPhase+\sin\myPhase\right).\\
	\end{aligned}
\end{equation}
The definition of the phase function given in  \cref{eq:phaseEqDef}\ reveals that 
\begin{equation}
\begin{aligned}
\myPhaseFunc_{Lambert} & = \frac{H_{Lambert}(\myPhase)}{H_{Lambert}(0)}\\
& = \frac{(2/3)\myScatteringAlbedo\left((\pi-\myPhase)\cos\myPhase+\sin\myPhase\right)}{(2/3) \myScatteringAlbedo\pi}\\
&=\frac{(\pi-\myPhase)\cos\myPhase+\sin\myPhase}{\pi}.
\end{aligned}
\end{equation}
Substitution of \cref{eq:lambertPhaseFunc}\ into \cref{eq:luminosityPhaseFunc}\ gives the luminosity of a Lambertian sphere with single scattering of albedo $ \myScatteringAlbedo $ for plane parallel ray incident stellar radiation:
\begin{equation}\label{eq:lumPlane}
	\myLuminosity_{refl,plane}(\myPhase)=\frac{2 }{3}\left( \frac{R_p}{r}\right)^2 L_s\myScatteringAlbedo\left[\frac{(\pi-\myPhase)\cos\myPhase+\sin\myPhase}{\pi}\right],
\end{equation}
\cf\ \cite{sobolev,seager}.

Furthermore, the geometric albedo, given in \cref{eq:geometricAlbedo}, of a Lambertian sphere is then $A_{g,Lambert} = (2/3)\myScatteringAlbedo$ and the phase integral given in \cref{eq:phaseIntegral}\ is then $q_{Lambert}= (3/2)$. Finally, the spherical albedo  may be calculated via \cref{eq:sphericalAlbedo}\ to reveal that $A_{s,Lambert} = \myScatteringAlbedo$ for a Lambertian sphere. From the definition of $ \myScatteringAlbedo $, we see that the geometric albedo of a lossless Lambertian sphere is $2/3$. In addition, the spherical albedo is unity as expected from its definition as the ratio of the power reflected to that intercepted by an exoplanet. We see here that our coordinate system does produce the expected results for plane parallel rays as described in previous works, \cf\ \cite{sobolev,seager}.

\begin{figure}[hbt]
	\centering
	\includegraphics[width=\textwidth]{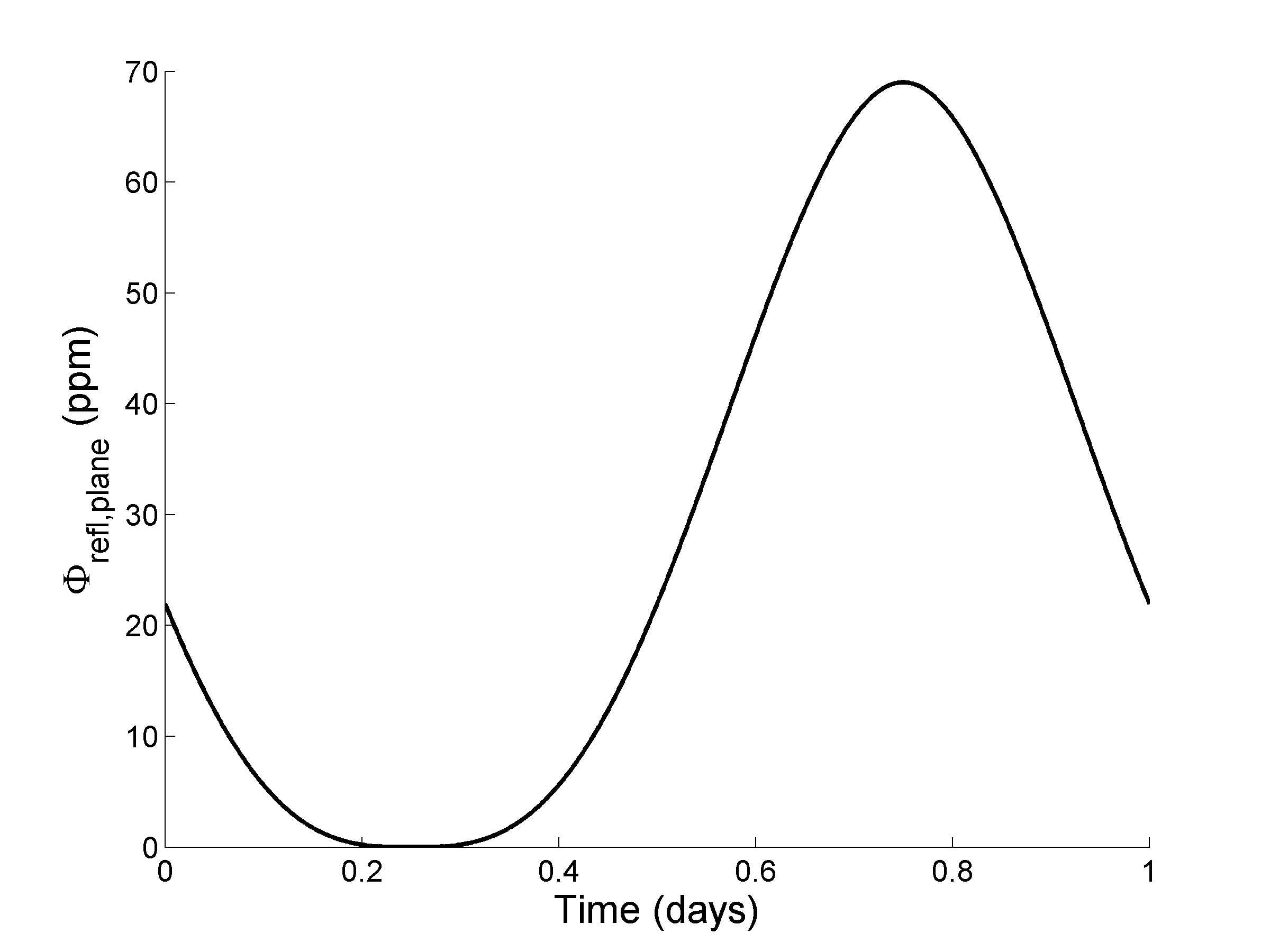}
	\caption{A plot of the fractional light reflected by an exoplanet for $A_g = 0.1$, $R_p = 1.1 R_J$, and a star-planet separation of $a = 4.2R_s$, or a period of 1 day for a solar mass star, using the plane parallel ray assumption for illumination. The exoplanet starts at half phase and moves to new phase at 0.25 days. Full phase occurs at 0.75 days.}
	\label{fig:phiReflPlaneParallel}
\end{figure}

For the purposes of comparison to the model of exoplanet illumination used for \myECIES\ let us determine the fractional reflected flux of a Lambertian sphere. This may be accomplished by substitution of \cref{eq:lambertPhaseFunc}\ into \cref{eq:phirPhaseFunc}:
\begin{equation}\label{eq:phirLambertPlaneParallel}
	\Phi_\textnormal{refl,plane} \equiv \frac{F_{p,refl}(\myPhase)}{F_s} = A_g\left(\frac{R_p}{r(\myPhase)}\right)^2\left(\frac{(\pi-\myPhase)\cos\myPhase+\sin\myPhase}{\pi}\right).
\end{equation}
A plot of $\Phi_\textnormal{refl,plane}$ for a hot Jupiter with a low geometric albedo as a function of time is given in \cref{fig:phiReflPlaneParallel}.

\section{Previous Derivation for Finite Angular Size}\label{sec:luminosityFiniteSizeog}
%

\myfillin, To determine the amount of light reflected by \myECIES\ we will make use of~\cref{eq:lumplanet}, but the limits on $ \mathbf{C} $ must now be altered to account for the inner and outer tangents delimiting the three illumination zones as illustrated in~\cref{fig:threeZones}. The following will present the arguments given in~\cite{Kopal1953,Kopal1959}, in which Kopal describes the reflection effect for binary stars, with corrections, additional details and clarifications as applicable to exoplanets, and with adaptations to the coordinate system described in \cref{sec:generalApproach}, \mychange.

Within the following, it is assumed that both the exoplanet and host star are spherical. This remains the case so long as tidal distortions between the two are small. As described in \cite{Kopal1959}, tidal distortions are of the order of $ (R_{p,s}/r)^5 $; therefore, in the following we wish for our final expressions for the reflected flux of the exoplanet to be accurate to the order of $ (R_{p,s}/r)^4 $.

\myfillin\ and \mychange\ (coordinate system), We begin the determination of $ \mathbf{C} $ for \myECIES\ by determining a relationship between the polar angle $\eta$ and the azimuthal angle $\phi$ for the fully illuminated and un-illuminated zones where the penumbral zone lies between the two.

The polar angles of the fully illuminated and un-illuminated zones may be described as the spherical caps delimited by $ \eta_{zone} \leq \eta \leq \pi-\eta_{zone} $. The relationship of the polar and azimuthal angles is described by the base of the cone defined by the tangents between the star and exoplanet. The cone's base forms a circle of radius $ R = R_p\cos\eta_{zone} $ where the $y$-axis defines the axis of the cone and points from the center of the exoplanet to that of the host star. The Cartesian coordinates for a point on the exoplanet's surface obey  \cref{eq:planetOriginCoordinates}; therefore, a point along the base of the cone lies on the planet's surface such that its Cartesian coordinates obey $ x^2 +y^2 +z^2 =R_p^2 $ and  $ x^2 +z^2 =R^2 $. Combining these two relationships permits a determination of $\eta$ in terms of $\phi$ as follows:
\begin{equation}\label{eq:etaAndphi}
\begin{aligned}
	R_p^2 &= x^2 +y^2 +z^2\\
		  &= y^2+R^2\\
		  &= R_p^2\sin^2\eta\sin^2\phi + R_p^2\cos^2\eta_{zone}.
\end{aligned}
\end{equation}
\cref{eq:etaAndphi} may be solved for $\eta$ as follows:
\begin{equation} 
\begin{aligned} 
    1 &= \sin^2\eta\sin^2\phi + \cos^2\eta_{zone}\\
	1-\cos^2\eta_{zone} &= \sin^2\eta\sin^2\phi\\
	\sin^2\eta_{zone}&=\sin^2\eta\sin^2\phi\\
	\sin^2\eta &= \sin^2\eta_{zone}\csc^2\phi\\
	\sin\eta &= \pm\sin\eta_{zone}\csc\phi.
\end{aligned}    
\end{equation}
Finally, the relationship between $ \eta $ and $ \phi $ is
\begin{equation}\label{eq:etaLimit}
\eta = \invsin{\pm\sin\eta_{zone}\csc\phi},
\end{equation}
\mychange, The limits on $\eta $ for a given zone are then given by $ \pi - \sin^{-1}(\mySinEta_{i}\csc\phi) \leq \eta \leq \sin^{-1}(\mySinEta_i\csc\phi) $ where $ \mySinEta_1 = \sin\eta_1 $ in the fully illuminated zone and $-\mySinEta_2 = -\sin\eta_2$ in the un-illuminated zone. To maintain values of $ \eta $ between zero and $ \pi $ we see that the positive sign is taken in for the fully illuminated zone, for which $ \csc{\phi} $ is positive and the negative sign is used within the un-illuminated zone, for which $ \csc{\phi} $ is negative.

The limits on $\phi$ are more complicated. The location of the terminator is no longer $ \pi $ as in \cref{fig:planeparallelterminatorlabels}, but is now determined by the azimuthal angle of the un-illuminated zone at the intensity equator. In fact, if a zone lies between the two limbs of the exoplanet, then the entirety of the zone is visible to an observer. 

\myfillin, Let us now consider the azimuthal angle,  $ \phi_{zone} $, as measured from the $ +x- $direction along the intensity equator that delimits either the fully illuminated or un-illuminated zone. We may again use \cref{eq:etaAndphi}, but now we will consider $ y = R_p \sin{\phi_{zone}} $ where we have set $ \eta = \pi/2 $ so that we are considering points on the intensity equator. Therefore, we have
\begin{equation}
  \begin{aligned}
        &R_p^2 \sin^2 \phi_{zone}+R_p^2 \cos^2 \eta_{zone} =R_p^2\\
        &\sin^2 \phi_{zone} = \sin^2 \eta_{zone}\\
		&\sin{\phi_{zone}}=\pm \sin{\eta_{zone}}.
  \end{aligned}
\end{equation}

\begin{figure}[tbh]
\centering
\includegraphics[width=0.7\linewidth]{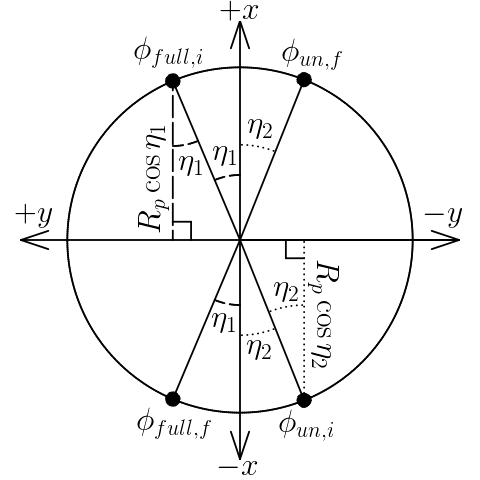}
\caption{Illustrated is a projection of the fully illuminated and un-illuminated zones onto the intensity equator of the exoplanet. Marked are the locations of the azimuthal angles delimiting said zones. Recall that the sub-stellar point is used to define the $ +y $ direction, see \cref{fig:planetcoordinateslabels}.}
\label{fig:philimitfiglabels}
\end{figure}

\mychange, The above equation does not provide enough information to determine the locations of each zone because there are two values of $\phi$ that satisfy the relationship within $ 0\le \phi \le 2\pi $. \cref{fig:philimitfiglabels}\ may be used as a geometrical tool to determine the azimuthal angles delimiting the fully illuminated and un-illuminated zones. The results are
\begin{equation}\label{eq:phizoneboth} 
    \begin{matrix}
		\phi_{full,i}=\eta_{1} &&\phi_{full,f} = \pi-\eta_{1} \\ 
		 \phi_{un,i}=\pi+\eta_{2} &&\phi_{un,f}=2\pi-\eta_{2}\\
	\end{matrix}
\end{equation}
which have ranges given by,
\begin{equation}\label{eq:phizoneranges}
	\begin{matrix}
		0\le \phi_{full,i} \le \piover2 &&\piover2\le  \phi_{full,f}\le \pi\\ 
		\pi\le \phi_{un,i}\le\frac{3\pi}{2}&& \frac{3\pi}{2}\le \phi_{un,f}\le 2\pi.
	\end{matrix}
\end{equation}
The foregoing limits are similar to those provided in Chapter IV.6 in \cite{Kopal1959}, see description of Equation (6-6) on pages 220 and 221, but here we have adapted them to our coordinate system in which the substellar point lies on the $ +y- $axis and have fully adapted to the fact that in the case of exoplanets $ R_p<R_s $ and the requirement that $ 0\le \eta\le \pi $. We have arranged our coordinate system such that the limits on the azimuthal angle will always proceed in order of smallest to largest and lie between zero and $ 2\pi $. 

In addition, it will be useful to define the azimuthal angles of the limbs of the exoplanet. From \cref{fig:planeparallelterminatorlabels}, we see that the first limb is still described by $\myPhase $. The second limb is located an angle of $\pi$ radians counterclockwise from the first limb at $ \pi+\myPhase $; therefore, we may write the azimuthal angles of the limbs of the exoplanet as
\begin{equation}\label{eq:philimbs}
	\begin{matrix}
		\phi_{limb,i}=\myPhase& &\phi_{limb,f}=\pi+\myPhase.
	\end{matrix}.
\end{equation}  

\cref{eq:phizoneranges}\ reveals that the azimuthal angles of the zones have the potential to include the entire exoplanet surface. In addition, \cref{eq:phizoneboth}\ indicates that as we approach the plane parallel ray case, in which both $ \eta_{1} $ and $ \eta_{2} $ approach zero, the fully illuminated  will extend from zero to $ \pi $ and the un-illuminated zone from $ \pi $ to $ 2\pi $. In such a situation the penumbral zone will not exist. %

\begin{figure}[hbt!]
	\centering
	\subfloat[][Case 1: secondary transit, $ 0\le \myPhase\le \eta_2 $. \label{fig:case1}]{\includegraphics[height = 0.24\textheight]{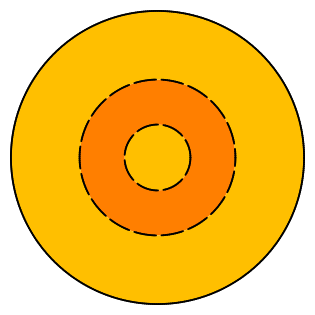}}
	\hfill
	\subfloat[][Case 2: secondary transit  ingress/egress, $ \eta_2\le \myPhase\le  \eta_{1}$. \label{fig:case2}]{\includegraphics[height = 0.24\textheight]{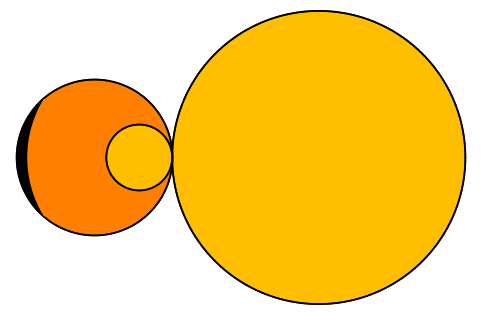}}

	\subfloat[][Case 3: $ \eta_{1}\le \myPhase \le \pi-\eta_{1}$. \label{fig:case3}]{\includegraphics[height = 0.23\textheight]{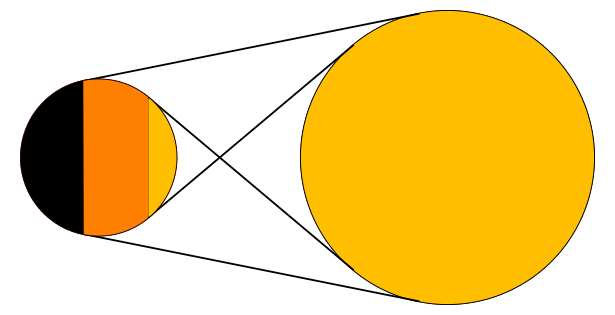}}

	\subfloat[][Case 4: primary transit ingress/egress, $ \pi-\eta_{1}\le \myPhase\le \pi-\eta_{2} $. \label{fig:case4}]{\includegraphics[height = 0.24\textheight]{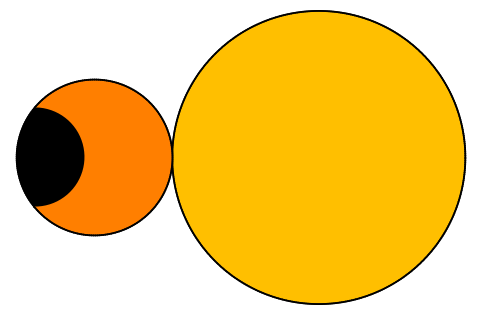}}
	\hfill
	\subfloat[][Case 5: primary transit, $ \pi-\eta_{2}\le \myPhase\le \pi $. \label{fig:case5}]{\includegraphics[height = 0.24\textheight]{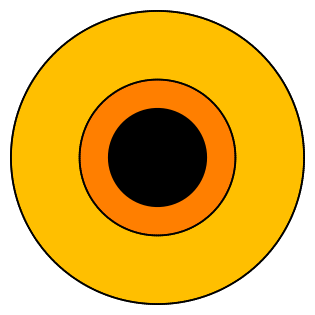}}
	\caption{Shown are illustrations of the zones visible to an observer for each of the five cases assuming there are only three three zones as in \cref{sec:incidentFluxPenKopalMethod}. In yellow is the fully illuminated zone, in orange the penumbral zone and in black the un-illuminated zone.}
	\label{fig:5cases}
\end{figure}

\myfillin, Let us now consider the conditions under which a given zone is visible to an observer. A zone is completely visible if $ \phi_{zone,i} $ and $ \phi_{zone,f} $ are both between $ \phi_{limb, i}$ and $ \phi_{limb, f} $; therefore, the fully illuminated zone is completely visible for phases between zero and $ \phi_{full,i}=\eta_{1} $ and the un-illuminated zone is completely visible for phases between $ \pi -\eta_{2} $ and $ \pi $. A zone will be partially visible if either its lower or upper limit is beyond one of the limbs of the exoplanet. From \cref{eq:fullarea,eq:nightarea}\ we see that the un-illuminated zone is always larger than the fully illuminated zone; therefore, there will be more cases in which the fully illuminated zone is completely visible than the un-illuminated zone as listed in \cref{fig:5cases}.

As described previously, the luminosity of the reflected light, $ \myLuminosity {(\myPhase)} $,  for a given phase may be determined by integration of the intensity of the reflected light at point $ P $ over the surface of the exoplanet visible along the line of sight. For convenience we will define the integral operator, $\myKintegral{\mySinEta}{\phi_i }{\phi_f }{\myIntensityDis_{zone}}$, as follows:
\begin{equation}\label{eq:Kx}
	\myKintegral{\mySinEta}{\phi_i }{\phi_f }{\myIntensityDis_{zone}} = R_p^2 \int_{\phi_i }^{\phi_f } \int_{\invsin{\mySinEta\csc\phi}}^{\pi - \invsin{\mySinEta\csc\phi}} \{\myIntensityDis_{zone}\}\sin^2\eta\sin(\phi-\myPhase)~d\eta~d\phi,
\end{equation}
where $\mySinEta$ may either be  $ \mySinEta_{1}=\sin\eta_{1} $,  $ -\mySinEta_{2} =-\sin\eta_{2} $ or zero. The upper and lower limits on $\phi $ will depend on the exoplanet's location within its orbit, as will be described by five unique cases for the phase angle $ \myPhase $ in \cref{tab:5cases}. It is also worth noting that \cref{eq:Kx}\ may be related to \cref{eq:Hphase}:
\begin{equation} 
    \myKintegral{0}{\myPhase}{\pi}{\mu}=R_p^2 H(\myPhase).
\end{equation}

\cref{fig:5cases}\ illustrates the five unique situations we must consider. The simplest case is shown in \cref{fig:case3}, in which case, the visible portion of fully illuminated zone lies between $ \phi_{limb,i} $ and $ \phi_{full,f} $, the penumbral zone lies between $  \phi_{full,f} $ and $  \phi_{un,i} $, and the un-illuminated zone lies between $  \phi_{un,i} $ and $   \phi_{limb,f} $. Recall that the reflected intensity distribution will depend on the zone in question as described by \cref{sec:incidentFluxFull,sec:incidentFluxPenKopalMethod}. The total reflected luminosity is determined by summation of  \cref{eq:Kx}\ for the fully illuminated and penumbral zones.

For the fully illuminated zone, the operator $ \myKintegral{\mySinEta_{1}}{\phi_{i,1}}{\phi_{f,1}}{\myIntensityDis_{refl, full}}$ will determine the reflected luminosity, \cite{Kopal1953,Kopal1959}. \mychange, The penumbral zone has a luminosity that is determined by first integrating over the visible portion of the exoplanet (limb to limb) and then subtracting the integration over the portions of the fully illuminated and un-illuminated zones that are visible for a given phase angle, i.e.
\begin{equation}\label{eq:Lpengeneral}
\begin{aligned}
	\myLuminosity_{refl,pen}(\myPhase)=&\myKintegral{0}{\phi_{limb,i}}{\phi_{limb,f}}{\myIntensityDis_{refl,pen}}-\\
&\myKintegral{\mySinEta_{1}}{\phi_{i,1}}{\phi_{f,1}}{\myIntensityDis_{refl, pen}}-\\
&\myKintegral{-\mySinEta_{2}}{\phi_{i,2}}{\phi_{f,2}}{\myIntensityDis_{refl, pen}},
\end{aligned}
\end{equation}
where the azimuthal limits in the second two lines will depend on the portion of the zone visible as described in \cref{tab:verbalcases,tab:5cases}. Here we are departing from the evaluation of the penumbral zone as described in \cite{Kopal1953,Kopal1959}\ because the expressions presented in either work only apply for cases in which the reflector, called the primary in Kopal's work, is larger than the emitter.

\begin{table}
\centering
\caption{\label{tab:verbalcases}Table of verbal descriptions of the five ranges of phase angles that must be evaluated to determine the total reflected luminosity of \myECIES. In the table the term ``Cap'' indicates that the entirety of a zone is visible as a spherical cap, the label ``Partial'' indicates that the zone is only partially visible, the label ``Not Visible'' is used to refer to situations in which the zone is not within either limb of the exoplanet, and finally the term ``Full Ring'' is used to describe situations in which the zone forms an unbroken ring about another zone.}
	\tabulinesep = 3mm
	\begin{tabu} to \textwidth {
			X[0.65,l]
			X[0.7,l]
			X[0.7,l]
			X[0.7,l]
		}
		\toprule
		\textbf{Range}&\textbf{Fully Illuminated}&\textbf{Un-illuminated} &\textbf{Penumbral} \\
		\midrule
		$ 0\le \myPhase\le \eta_{2} $&Cap&Not Visible&Full Ring\\ 
		$ \eta_{2}\le \myPhase\le \eta_{1} $&Cap&Partial & Partial\\
		$ \eta_{1}\le \myPhase\le \pi-\eta_{1} $&Partial&Partial& Partial\\
		$ \pi-\eta_{1}\le \myPhase\le \pi-\eta_{2} $& Not Visible&Partial&  Partial\\
		$ \pi- \eta_{2}\le \myPhase\le \pi$& Not Visible & Cap &Full Ring\\ 
		\bottomrule	
	\end{tabu}
\end{table}
\vspace{-3em}

\begin{figure}[hbt]
	\centering
	\includegraphics[width=0.9\linewidth]{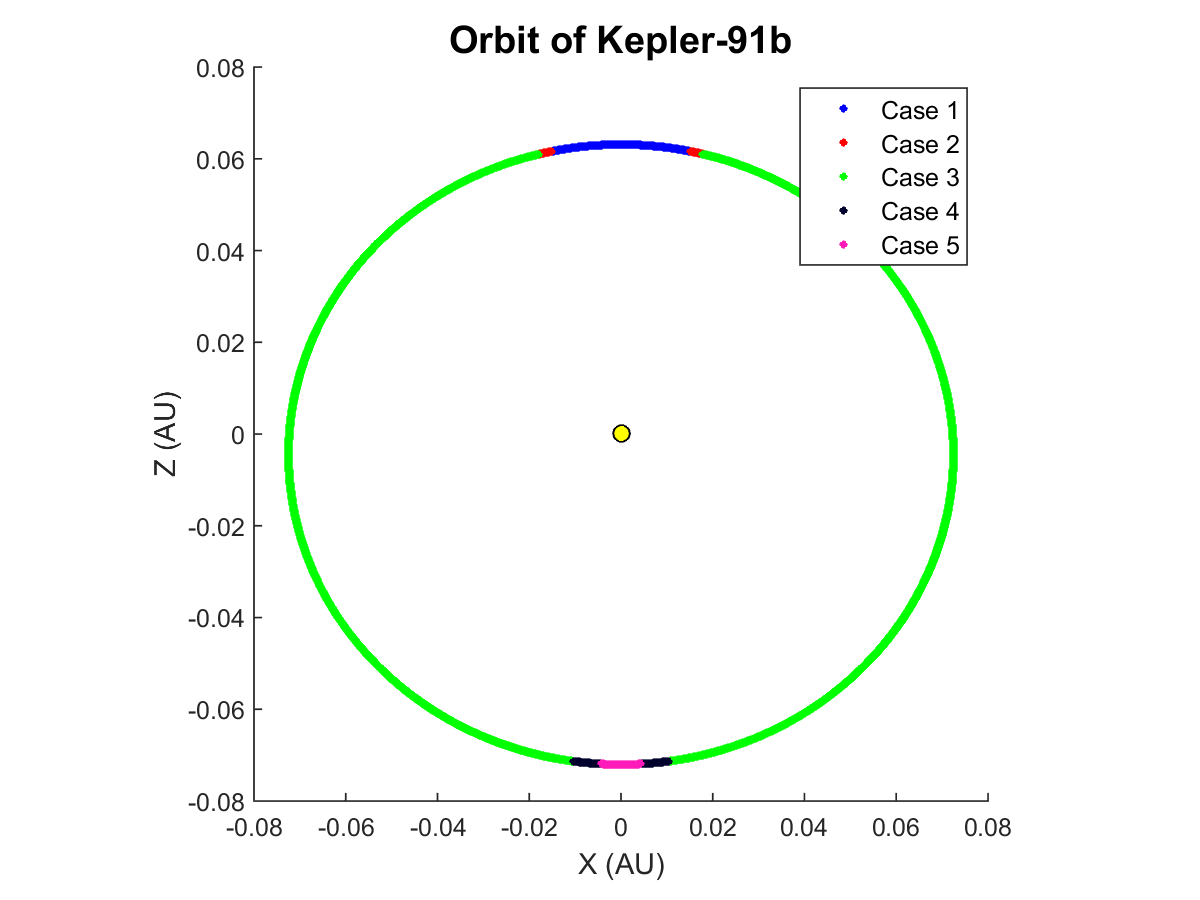}
	\caption{Shown is an illustration of the orbit of Kepler-91b using the parameters provided in \cref{tab:K91params}. The colors correspond to each of the five cases as described in the text. The observer is viewing the orbit from the bottom of the page and $ \myPhase = 0 $ occurs at  $ X $ = 0 and  $ Z >$ 0.}
	\label{fig:plotkepler-91bOrbit}
\end{figure}

\cref{fig:5cases}\ shows an illustration of the exoplanet and host star for each of the five cases for an observer with an edge-on line of sight. \cref{fig:plotkepler-91bOrbit}\ depicts each portion of the orbit of Kepler-91b for the five cases. In this work, we will consider a phase angle in the ranges of zero to $ \pi$, where a phase angle of zero corresponds to the full phase of the exoplanet as illustrated in \cref{fig:losfig3} and described by  \cref{eq:phaseangle}. This situation corresponds to $ X>0 $ in  \cref{fig:plotkepler-91bOrbit}. For  $ X>0 $ , \cref{eq:phaseangle}\ gives the same value of $ \myPhase $; therefore, we need only consider $ \myPhase \in [0,\pi] $. 

\myfillin, To determine the $ \myLuminosity_{refl,full}(\myPhase) $, we must substitute \cref{eq:fullLegendre}\ in \cref{eq:Kx}\ in place of $ \myIntensityDis_{zone} $. If we assume that the reflection coefficient $ \myRefCo $ is equal to the single scattering albedo, $ \myScatteringAlbedo $, then the integration that must be performed for the fully illuminated zone is
\begin{equation}\label{eq:Fullgeneral}
	\begin{aligned}
		\myLuminosity_{refl,full} (\myPhase)&= \myKintegral{\mySinEta_{1}}{\phi_{i,1}}{\phi_{f,1}}{\frac{L_s\myScatteringAlbedo}{\pi r^2}\sum_{\ell=1}^3 \ell \left( \frac{R_p}{r} \right)^{\ell-1}P_\ell(\mu)}\\
		&\approx \left( \frac{L_s \myScatteringAlbedo}{\pi r^2 }\right)  \myKintegral{\mySinEta_{1}}{\phi_{i,1}}{\phi_{f,1}}{P_1(\mu)+2\frac{R_p}{r}P_2(\mu)+ 3 \left( \frac{R_p}{r}\right)^2P_3(\mu)}\\
%
	\end{aligned}
\end{equation}
where the limits on $ \phi$ of $\myKintegral{\mySinEta_{1}}{\phi_{i,1}}{\phi_{f,1}}{\myIntensityDis_{refl, full}}$ are determined by the phase, $ \myPhase $, and are summarized in \cref{tab:5cases}. Maintaining up to the third term of $ R_p/r $ for \cref{eq:fullLegendre}\ is sufficient to maintain the order of accuracy of which we are concerned, which is  $ (R_{p,s}/r)^4 $, in the final expression for the exoplanet's reflected flux, see Equation (6-61) in \cite{Kopal1959}.

\begin{figure}[hbt]
\centering
\includegraphics[width=0.95\linewidth]{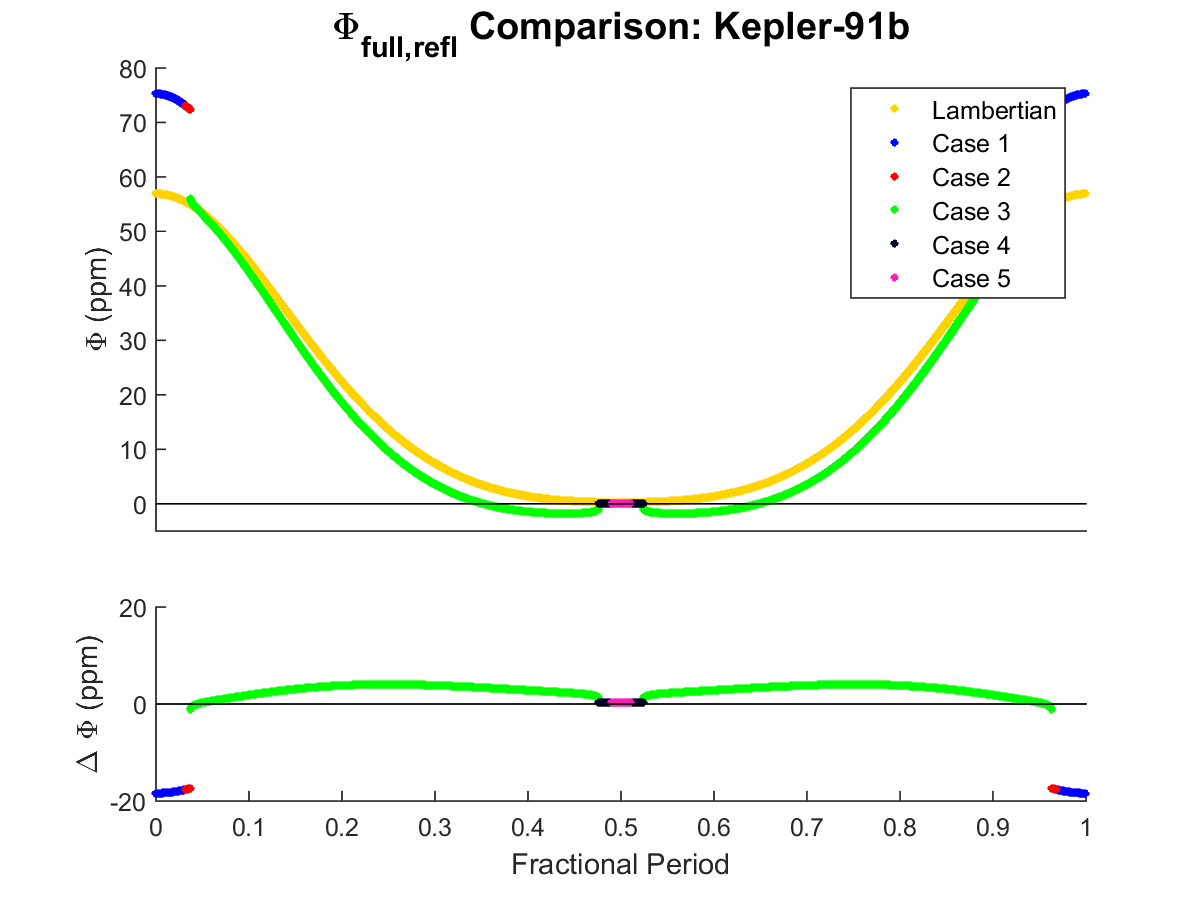}
\caption{The plot shows the results of using Equations (6-62)-(6-66) in \protect\cite{Kopal1959}\ to evaluate the reflected luminosity of Kepler-91b using the parameter is given in \cref{tab:K91params}. Note that for \mycase{3}\ the luminosity becomes negative near the start of \mycase{4}. In addition, the transition from \mycase{2}\ to \mycase{3}\ is not continuous.}
\label{fig:OGfullK91plot}
\end{figure}

\mynew, It was discovered that if one attempts to use the equations given in Equations (6-62)-(6-66) in \cite{Kopal1959}\ to evaluate the reflected luminosity of an exoplanet that negative luminosity in the fully illuminated zone can occur near $ \myPhase =\pi $ as shown in \cref{fig:OGfullK91plot}. Plotted is the fractional flux due to the fully illuminated zone
\begin{equation}\label{eq:phifull}
	\Phi_{full,refl}(\myPhase) =  \frac{F_{full,refl}(\myPhase)}{F_s} =  \frac{\myLuminosity_{full,refl}(\myPhase)}{L_s},
\end{equation}
versus the fractional period. We have switched to unitless axes to allow for easy comparison of exoplanets with different star-planet separations.
In \cref{sec:luminosityFiniteSizenew}, we will present a new evaluation of the integrals described by Equation (6-61) in \cite{Kopal1959}, i.e. \cref{eq:Fullgeneral}. Said analysis will produce the integrals given by \cref{sec:fulledgetoedge,sec:fulllimbtoedge}\ which exhibit smoother transitions and do not allow for negative luminosity.

\mynew, In general, the determination of the reflected luminosity as a function of phase within the penumbral zone, $ \myLuminosity_{refl,pen}(\myPhase) $, may be determined via the integration described in \cref{eq:Lpengeneral}, 
where $\myIntensityDis_{refl, pen}$ is given by  \cref{eq:intensityDisPen,eq:JUDSum,eq:CUs,eq:CDs}. We will again assume that $ \myRefCo = \myScatteringAlbedo$ so that we may pull out $ \myRefCo $ from the integration. Here we will evaluate the integrals described in \cref{tab:5cases}\ exactly as opposed to using the approximations given in \cite{Kopal1953,Kopal1959}, \cf\ Equation (86) in \cite{Kopal1953}. In addition, it will be shown that the approximations used to determine \cref{eq:intensityDisPen,eq:JUDSum,eq:CUs,eq:CDs}\ will produce negative luminosity within the penumbral zone; however, the fully illuminated zone is positive.

\begin{table}
	\centering
	\tabulinesep = 3mm
	\caption{\label{tab:5cases}\myfillin, Shown is a table describing each of the five cases and the required application of the operator $ \myKintegral{\mySinEta}{\phi_i}{\phi_f}{\myIntensityDis_{zone}}$, defined by \cref{eq:Kx}, to determine the reflected luminosity is a function of phase, $ \myLuminosity_{refl,zone}(\myPhase)$, for the fully illuminated and penumbral zones. The azimuthal angles shown in $ \myKintegral{\mySinEta}{\phi_i}{\phi_f}{\myIntensityDis_{zone}}$ are defined in \cref{eq:phizoneboth,eq:philimbs}
		. For each case the range of phase angles for which it applies is listed. The total reflected luminosity is the sum of the luminosity from the fully illuminated and penumbral zones, $ \myLuminosity_{refl}(\myPhase) = \myLuminosity_{full}(\myPhase)+\myLuminosity_{pen}(\myPhase) $. Here $ \myIntensityDis_{full} $ is given by \cref{eq:fullLegendre}\ and $ \myIntensityDis_{pen} $ by \cref{eq:intensityDisPen,eq:JUDSum,eq:CUs,eq:CDs}.}
\begin{tabu} to \textwidth {
			X[0.25,l]
			X[0.4,c]
			X[0.4,l]
		}
		\toprule
		\textbf{Range}&\textbf{Fully Illuminated}, $ \myLuminosity_{full} $ &\multicolumn{1}{c}{\textbf{Penumbral}, $ \myLuminosity_{pen} $ } \\
		\midrule
		$ 0\le \myPhase\le \eta_{2} $& $ \myKintegral{\mySinEta_{1}}{\phi_{full,i}}{\phi_{full,f}}{\myIntensityDis_{full}} $ &  $ \myKintegral{0}{\phi_{limb,i}}{\phi_{limb,f}}{\myIntensityDis_{pen}} -\myKintegral{\mySinEta_{1}}{\phi_{full,i}}{\phi_{full,f}}{\myIntensityDis_{pen}}$  \\ 
		$ \eta_{2}\le \myPhase\le \eta_{1} $&$ \myKintegral{\mySinEta_{1}}{\phi_{full,i}}{\phi_{full,f}}{\myIntensityDis_{full}} $ &$\myKintegral{0}{\phi_{limb,i}}{\phi_{limb,f}}{\myIntensityDis_{pen}} -\myKintegral{\mySinEta_{1}}{\phi_{full,i}}{\phi_{full,f}}{\myIntensityDis_{pen}}- \myKintegral{-\mySinEta_{2}}{\phi_{un,i}}{\phi_{limb,f}}{\myIntensityDis_{pen}} $\\
		$ \eta_{1}\le \myPhase\le \pi-\eta_{1} $&$ \myKintegral{\mySinEta_{1}}{\phi_{limb,i}}{\phi_{full,f}}{\myIntensityDis_{full}} $ &$\myKintegral{0}{\phi_{limb,i}}{\phi_{limb,f}}{\myIntensityDis_{pen}}-\myKintegral{\mySinEta_{1}}{\phi_{limb,i}}{\phi_{full,f}}{\myIntensityDis_{pen}} - \myKintegral{-\mySinEta_{2}}{\phi_{un,i}}{\phi_{limb,f}}{\myIntensityDis_{pen}} $\\
		$ \pi-\eta_{1}\le \myPhase\le \pi-\eta_{2} $& 0&$\myKintegral{0}{\phi_{limb,i}}{\phi_{limb,f}}{\myIntensityDis_{pen}} -\myKintegral{-\mySinEta_{2}}{\phi_{un,i}}{\phi_{limb,f}}{\myIntensityDis_{pen}}$\\
		$ \pi- \eta_{2}\le \myPhase\le \pi$& 0 &$ \myKintegral{0}{\phi_{limb,i}}{\phi_{limb,f}}{\myIntensityDis_{pen}} -\myKintegral{-\mySinEta_{2}}{\phi_{un,i}}{\phi_{un,f}}{\myIntensityDis_{pen}} $\\ 	
		\bottomrule	
	\end{tabu}
\vspace{-3em}
\end{table}

Inspection of \cref{eq:CUs,eq:CDs}\ reveals that our task is then to evaluate $ \myKintegral{\mySinEta}{\phi_i }{\phi_f }{\mu^n} $ for each of the cases listed \cref{tab:5cases}. Furthermore, the definition of $ P_\ell (\mu) $ reveals that evaluation of \cref{eq:Fullgeneral}\ requires similar evaluations of $ \myKintegral{\mySinEta_{1}}{\phi_{i,1} }{\phi_{f,1} }{\mu^n} $ up to  $ n=3 $. The evaluation of the penumbral zone will require evaluation up to $ n=4 $; therefore, we will now consider a strategy for the evaluation of integrals of the form of
\begin{equation}\label{eq:generalInt}
	\begin{aligned}
		\myKintegral{\mySinEta}{\phi_{i}}{\phi_{f}}{\mu^n}& =R_p^2 \int_{\phi_i}^{\phi_f} \int_{\invsin{\mySinEta\csc\phi}}^{\pi - \invsin{\mySinEta\csc\phi}}\lbrace\mu^n\rbrace\sin^2{\eta}\sin(\phi-\myPhase)~d\eta~d\phi\\
		& =R_p^2\int_{\phi_i}^{\phi_f} \int_{\invsin{\mySinEta\csc\phi}}^{\pi - \invsin{\mySinEta\csc\phi}}\sin^{n+2}\eta\sin^n{\phi}\sin(\phi-\myPhase)~d\eta~d\phi,
	\end{aligned}
\end{equation}
and evaluate these up to  $ n=4 $. To aid in the evaluation of the following integrals it will be convenient to create a table listing the ranges and relationships between various trigonometric functions for each of the five cases, see \cref{tab:ranges}\ in \cref{app:tables}.  We will now proceed with the evaluation of \cref{eq:generalInt}\ for  $ n=0 $  to four.

It can be shown that
\begin{equation} 
    \int \sin^m{\eta}~d\eta = {}_{2}{F}_{1}\left( \1over2,\frac{1-m}{2};\frac{3}{2};\cos^2 {\eta}\right) (-\cos{\eta}) \left( \sin^{1+m}{\eta}\right)\left( \sin^2{\eta}\right)^{-(1+m)/2},
\end{equation}
where $ {}_{2}{F}_{1}(a,b;c;x) $ is the hypergeometric function of $ x $.
If the sign of $ \sin\eta $ is known then we may use 
\begin{equation} 
    \int \sin^m{\eta}~d\eta = {}_{2}{F}_{1} \left( \1over2,\frac{1-m}{2};\frac{3}{2};\cos^2 {\eta}\right) (-\cos{\eta}) \textnormal{Sign}[\sin\eta]^{1+m}.  
\end{equation}
The coordinate system in this work was designed such that for all cases $ \sin{\eta}\ge 0 $; therefore, we may conclude that
\begin{equation} \label{eq:sinM}
    \int \sin^m{\eta}~d\eta =( -\cos{\eta})\, {}_{2}{F}_{1} \left( \1over2,\frac{1-m}{2};\frac{3}{2};\cos^2 {\eta}\right).  
\end{equation}
We will now consider integrations of this type first over the visible disk of the exoplanet, then over the visible portion of the fully illuminated zone, and finally over the visible portion of the un-illuminated zone.

To integrate over the visible portion of the exoplanet disk, one must first integrate over the polar angle from zero to $ \pi $ radians; therefore, we may set $ \mySinEta = 0 $ and  \cref{eq:sinM}\ simplifies to
\begin{equation} 
\begin{aligned}
	\int_{0}^{\pi}\sin^{m}\eta~d\eta=\frac{\sqrt{\pi}\Gamma\left( (1+ m)/2\right)}{\Gamma\left( 1+m/2\right)}
\end{aligned}
\end{equation}
revealing that
\begin{equation} \label{eq:sinMK0}
    \int_{0}^{\pi} \sin^{n+2}{\eta}~d\eta = 
		\begin{cases}
			\piover2&,\, n=0\\
			\frac{4}{3}&, \,n=1\\
			\frac{3\pi}{8}&, \,n=2\\
			\frac{16}{15}&, \,n=3\\
			\frac{5\pi}{16}&, \,n=4.
		\end{cases}
\end{equation}
Substitution of \cref{eq:sinMK0}\ into \cref{eq:generalInt}\ and using the definition of the $K$-operator given in \cref{eq:Kx}\ produces 
\begin{equation}\label{eq:K0results}
	\myKintegral{0}{\phi_{limb,i}}{\phi_{limb,f}}{\mu^n} = R_{p}^2 
  \begin{cases}
        \pi&,\, n=0\\
		\frac{2}{3} \pi  \cos \myPhase &, \,n=1\\
		\frac{1}{8} \pi  (\cos (2 \myPhase )+3)&, \,n=2\\
		\frac{2}{5} \pi  \cos\myPhase&, \,n=3\\
		\frac{1}{192} \pi  (20 \cos (2 \myPhase )-\cos (4 \myPhase )+45)&, \,n=4,
  \end{cases}
\end{equation}
which applies for each of the five cases listed in \cref{tab:verbalcases,tab:5cases}.

Within the fully illuminated zone we set $ \mySinEta =\mySinEta_{1} $ and we must now consider the definite integral
\begin{equation} \label{eq:sinMdef}
    \int_{\invsin{\mySinEta_{1}\csc\phi}}^{\pi - \invsin{\mySinEta_{1}\csc\phi}} \sin^m{\eta}~d\eta =2 \sqrt{1-\mySinEta_{1}^2 \csc ^2\phi} \, _2F_1\left(\frac{1}{2},\frac{1-m}{2};\frac{3}{2};1-\csc ^2\phi
   \mySinEta_{1}^2\right)
\end{equation}

Next, we may evaluate \cref{eq:sinMdef}\ for  $ m=n+2 $ and where  $ n=0,1,2,3 $ and 4 to determine that
\begin{equation}\label{eq:sinMsolns}
\begin{aligned}
	\int_{\invsin{\mySinEta_{1}\csc\phi}}^{\pi - \invsin{\mySinEta_{1}\csc\phi}}& \sin^{n+2}{\eta}~d\eta=\\
	&\begin{cases}
		\frac{\mySinEta_{1}}{\sin^2{\phi}}\sqrt{\sin ^2\phi-\mySinEta_{1}^2}+\invcos{\frac{\mySinEta_{1}}{\sin\phi}}&, \, n=0\\
		\frac{2}{3} \frac{\sqrt{\sin ^2\phi-\mySinEta_{1}^2} }{\sin{\phi}} \left(\frac{\mySinEta_{1}^2}{\sin^2\phi}+2\right) &, \, n=1\\
		\frac{1}{4} \left(3 \cos ^{-1}\left(\frac{\mySinEta_{1}}{\sin{\phi}}\right)+\mySinEta_{1} \frac{\sqrt{\sin ^2\phi-\mySinEta_{1}^2} }{\sin^2{\phi}}
   \left(2 \frac{\mySinEta_{1}^2}{\sin^2\phi}+3\right)\right) &, \, n=2\\
	\frac{2}{15} \csc \phi \sqrt{\sin ^2\phi-\mySinEta_{1}^2} \left(3 \mySinEta_{1}^4 \csc ^4\phi+4 \frac{\mySinEta_{1}^2}{\sin^2\phi}+8\right)&, \, n=3\\
	\frac{1}{24} \left(15 \cos ^{-1}\left(\frac{\mySinEta_{1}}{\sin{\phi}}\right)+\mySinEta_{1} \frac{\sqrt{\sin ^2\phi-\mySinEta_{1}^2} }{\sin^2{\phi}}
   \left(8 \frac{\mySinEta_{1}^4}{\sin^4\phi}+10 \frac{\mySinEta_{1}^2}{\sin^2\phi}+15\right)\right)&, \, n=4,
	\end{cases}
\end{aligned}
\end{equation}
where we have used $ \sqrt{1 -\mySinEta_{1}^2\csc\phi} =\csc\phi\sqrt{\sin^2\phi -\mySinEta_{1}^2} $.

Substituting \cref{eq:sinMsolns}\ into the integration given in \cref{eq:generalInt}, separating the integration into parts involving  $ \sin{\myPhase} $ and $ \cos{\myPhase} $ terms, and by using the identity $ \sin{(\phi-\myPhase)} =\sin{\phi}\cos{\myPhase}-\cos{\phi}\sin{\myPhase} $ it can be shown that we may rewrite \cref{eq:generalInt}\ as follows
\begin{equation}
  \begin{aligned}
       & \int_{\phi_i}^{\phi_f} \int_{\invsin{\mySinEta_{1}\csc\phi}}^{\pi - \invsin{\mySinEta_{1}\csc\phi}}\lbrace\mu^n\rbrace\sin^2{\eta}\sin(\phi-\myPhase)~d\eta~d\phi\\
		=&\int_{\phi_i}^{\phi_f} \int_{\invsin{\mySinEta_{1}\csc\phi}}^{\pi - \invsin{\mySinEta_{1}\csc\phi}}\sin^{n+2}\eta\sin^n{\phi}(\sin{\phi}\cos{\myPhase}-\cos{\phi}\sin{\myPhase} )~d\eta~d\phi\\
		= &\int_{\phi_i}^{\phi_f} \int_{\invsin{\mySinEta_{1}\csc\phi}}^{\pi - \invsin{\mySinEta_{1}\csc\phi}}\sin^{n+2}\eta(\sin^{n+1}{\phi}\cos{\myPhase}-\sin^n{\phi}\cos{\phi}\sin{\myPhase})~d\eta~d\phi.
  \end{aligned}
\end{equation}
Substituting our results from  \cref{eq:sinMsolns}\ and some algebra reveals that
\begin{equation}\label{eq:fullQRequiv}
  \begin{aligned}
        \int_{\phi_i}^{\phi_f} &\int_{\invsin{\mySinEta_{1}\csc\phi}}^{\pi - \invsin{\mySinEta_{1}\csc\phi}}\lbrace\mu^n\rbrace\sin^2{\eta}\sin(\phi-\myPhase)~d\eta~d\phi=\\
		&\begin{cases}
			\left( \myQnCosepsshort{1}+\mySinEta_{1} \myRnCosepsshort{-1} \right)\cos{\myPhase}-\left( \myQnSinepsshort{0}+\mySinEta_{1}\myRnSinepsshort{-2}\right)\sin{\myPhase}&, \, n= 0\\
			\left(\frac{4}{3}\myRnCosepsshort{1} +\frac{2}{3} \mySinEta_{1}^2  \myRnCosepsshort{-1}\right) \cos{\myPhase} - \left(\frac{4}{3}\myRnSinepsshort{0} + \frac{2}{3} \mySinEta_{1}^2 \myRnSinepsshort{-2}\right)\sin{\myPhase} &, \, n= 1\\
			 \left(\frac{3}{4} \myQnCosepsshort{3} + \frac{3}{4} \mySinEta_{1} \myRnCosepsshort{1} + \1over2\mySinEta_{1}^3 \myRnCosepsshort{-1}\right)\cos{\myPhase}- 
    	\left(\frac{3}{4} \myQnSinepsshort{2} + \frac{3}{4} \mySinEta_{1}\myRnSinepsshort{0}+ \1over2 \mySinEta_{1}^3 \myRnSinepsshort{-2}\right)\sin{\myPhase} &, \, n= 2\\
		\left( \frac{16}{15} \myRnCosepsshort{3}+ \frac{8}{15} \mySinEta_{1}^2 \myRnCosepsshort{1} + \frac{2}{5}\mySinEta_{1}^4 \myRnCosepsshort{-1}\right)\cos{\myPhase}-\left(\frac{16}{15} \myRnSinepsshort{2}+ \frac{8}{15} \mySinEta_{1}^2 \myRnSinepsshort{0} + \frac{2}{5}\mySinEta_{1}^4 \myRnSinepsshort{-2} \right)\sin{\myPhase} &, \, n= 3\\
		\begin{aligned}
			&\left( \frac{5}{8}\myQnCosepsshort{5}+\frac{5}{8} \mySinEta_{1} \myRnCosepsshort{3} + \frac{5}{12} \mySinEta_{1}^3\myRnCosepsshort{1} + \frac{1}{3}\mySinEta_{1}^5\myRnCosepsshort{-1}\right)\cos{\myPhase}-\\
			&\left( \frac{5}{8}\myQnSinepsshort{4}+\frac{5}{8} \mySinEta_{1} \myRnSinepsshort{2} + \frac{5}{12} \mySinEta_{1}^3\myRnSinepsshort{0} + \frac{1}{3}\mySinEta_{1}^5\myRnSinepsshort{-2} \right)\sin{\myPhase}
		\end{aligned}
		&, \, n=4.
		\end{cases}
  \end{aligned}
\end{equation}
where
\begin{equation}\label{eq:Qns}
  \begin{aligned}
        \myQnCoseps{\mySinEta}{\phi_{i}}{\phi_f}{n}&=\int_{\phi_i}^{\phi_f}\sin^n{\phi}\invcos{\frac{\mySinEta}{\sin{\phi}}}~d\phi\\
		\myQnSineps{\mySinEta}{\phi_{i}}{\phi_f}{n}&=\int_{\phi_i}^{\phi_f}\sin^n{\phi}\cos{\phi}\invcos{\frac{\mySinEta}{\sin{\phi}}}~d\phi
  \end{aligned}
\end{equation}
and 
\begin{equation}\label{eq:Rns}
  \begin{aligned}
        \myRnCoseps{\mySinEta}{\phi_{i}}{\phi_f}{n}&=\int_{\phi_i}^{\phi_f}\sin^n{\phi}\sqrt{\sin^2\phi-\mySinEta^2}~d\phi\\
		\myRnSineps{\mySinEta}{\phi_{i}}{\phi_f}{n}&=\int_{\phi_i}^{\phi_f}\sin^n{\phi}\cos{\phi}\sqrt{\sin^2\phi-\mySinEta^2}~d\phi.
  \end{aligned}
\end{equation}
In \cref{eq:fullQRequiv}\ It is understood that the integral operators share the azimuthal limits of the original integration and that one is to insert $ \mySinEta =\mySinEta_{1} $.

Inspection of the foregoing equations reveals that our task is to evaluate the following:
\begin{enumerate}
	\item  $ \myQnCoseps{\mySinEta_{1}}{\phi_{full,i}}{\phi_{full,f}}{n} $ and $ \myQnCoseps{\mySinEta_{1}}{\phi_{limb,i}}{\phi_{full,f}}{n} $ for  $ n=1,3,5 $ ,
	\item  $ \myQnSineps{\mySinEta_{1}}{\phi_{full,i}}{\phi_{full,f}}{n} $ and $ \myQnSineps{\mySinEta_{1}}{\phi_{limb,i}}{\phi_{full,f}}{n} $ for  $ n=0,2,4 $ ,
	\item  $ \myRnCoseps{\mySinEta_{1}}{\phi_{full,i}}{\phi_{full,f}}{n} $ and $ \myRnCoseps{\mySinEta_{1}}{\phi_{limb,i}}{\phi_{full,f}}{n} $ for  $ n=-1,1,3 $ , and
	\item  $ \myRnSineps{\mySinEta_{1}}{\phi_{full,i}}{\phi_{full,f}}{n} $ and $ \myRnSineps{\mySinEta_{1}}{\phi_{limb,i}}{\phi_{full,f}}{n} $ for  $ n=-2,0,2 $.
\end{enumerate}
%
See \cref{app:integrations}\ for the evaluation of each of the indefinite and definite integrals. The final determinations of the integrations over the spherical cap of the  fully illuminated zone are given in \cref{sec:fulledgetoedge,sec:fulllimbtoedge}. 


Following the strategy used in \cref{eq:fullQRequiv}\ and substituting $ \mySinEta = -\mySinEta_{2} $ we find that integration over the un-illuminated zone may be described as
\begin{equation}\label{eq:unQRequiv}
  \begin{aligned}
        \int_{\phi_i}^{\phi_f} &\int_{\invsin{-\mySinEta_{2}\csc\phi}}^{\pi - \invsin{-\mySinEta_{2}\csc\phi}}\lbrace\mu^n\rbrace\sin^2{\eta}\sin(\phi-\myPhase)~d\eta~d\phi=\\
		&\begin{cases}
			\left( \myQnCosepsshort{1}+\mySinEta_{2} \myRnCosepsshort{-1} \right)\cos{\myPhase}-\left( \myQnSinepsshort{0}+\mySinEta_{2}\myRnSinepsshort{-2}\right)\sin{\myPhase}&, \, n= 0\\
			-\left(\frac{4}{3}\myRnCosepsshort{1} +\frac{2}{3} \mySinEta_{2}^2  \myRnCosepsshort{-1}\right) \cos{\myPhase} + \left(\frac{4}{3}\myRnSinepsshort{0} + \frac{2}{3} \mySinEta_{2}^2 \myRnSinepsshort{-2}\right)\sin{\myPhase} &, \, n= 1\\
			 \left(\frac{3}{4} \myQnCosepsshort{3} + \frac{3}{4} \mySinEta_{2} \myRnCosepsshort{1} + \1over2\mySinEta_{2}^3 \myRnCosepsshort{-1}\right)\cos{\myPhase}- 
    	\left(\frac{3}{4} \myQnSinepsshort{2} + \frac{3}{4} \mySinEta_{2}\myRnSinepsshort{0}+ \1over2 \mySinEta_{2}^3 \myRnSinepsshort{-2}\right)\sin{\myPhase} &, \, n= 2\\
		-\left( \frac{16}{15} \myRnCosepsshort{3}+ \frac{8}{15} \mySinEta_{2}^2 \myRnCosepsshort{1} + \frac{2}{5}\mySinEta_{2}^4 \myRnCosepsshort{-1}\right)\cos{\myPhase}+\left(\frac{16}{15} \myRnSinepsshort{2}+ \frac{8}{15} \mySinEta_{2}^2 \myRnSinepsshort{0} + \frac{2}{5}\mySinEta_{2}^4 \myRnSinepsshort{-2} \right)\sin{\myPhase} &, \, n= 3\\
		\begin{aligned}
			&\left( \frac{5}{8}\myQnCosepsshort{5}+\frac{5}{8} \mySinEta_{2} \myRnCosepsshort{3} + \frac{5}{12} \mySinEta_{2}^3\myRnCosepsshort{1} + \frac{1}{3}\mySinEta_{2}^5\myRnCosepsshort{-1}\right)\cos{\myPhase}-\\
			&\left( \frac{5}{8}\myQnSinepsshort{4}+\frac{5}{8} \mySinEta_{2} \myRnSinepsshort{2} + \frac{5}{12} \mySinEta_{2}^3\myRnSinepsshort{0} + \frac{1}{3}\mySinEta_{2}^5\myRnSinepsshort{-2} \right)\sin{\myPhase}
		\end{aligned}
		&, \, n=4.
		\end{cases}
  \end{aligned}
\end{equation} 
While evaluating the integrals in \cref{eq:unQRequiv}\ we must keep in mind that $ -1\le \csc{\phi}\le 0 $, for example it is now the case that $ \sqrt{1 -\mySinEta_{1}^2\csc\phi} =-\csc\phi\sqrt{\sin^2\phi -\mySinEta_{1}^2} $. In \cref{eq:unQRequiv}\ it is understood that the integral operators share the azimuthal limits of the original integration and that one is to insert $ \mySinEta =-\mySinEta_{2} $.
See \cref{app:integrations}\ for the evaluation of each of the indefinite and definite integrals. The final determinations of the integrations over the spherical cap of the un-illuminated zone are given in \cref{sec:unedgetoedge,sec:unedgetolimb}.

\begin{figure}
\centering
\subfloat[][\label{fig:correct5full}Corrected $ \myLuminosity_{full}/L_s $ ]{\includegraphics[trim={40 0 40 0}, clip,width=0.5\linewidth]{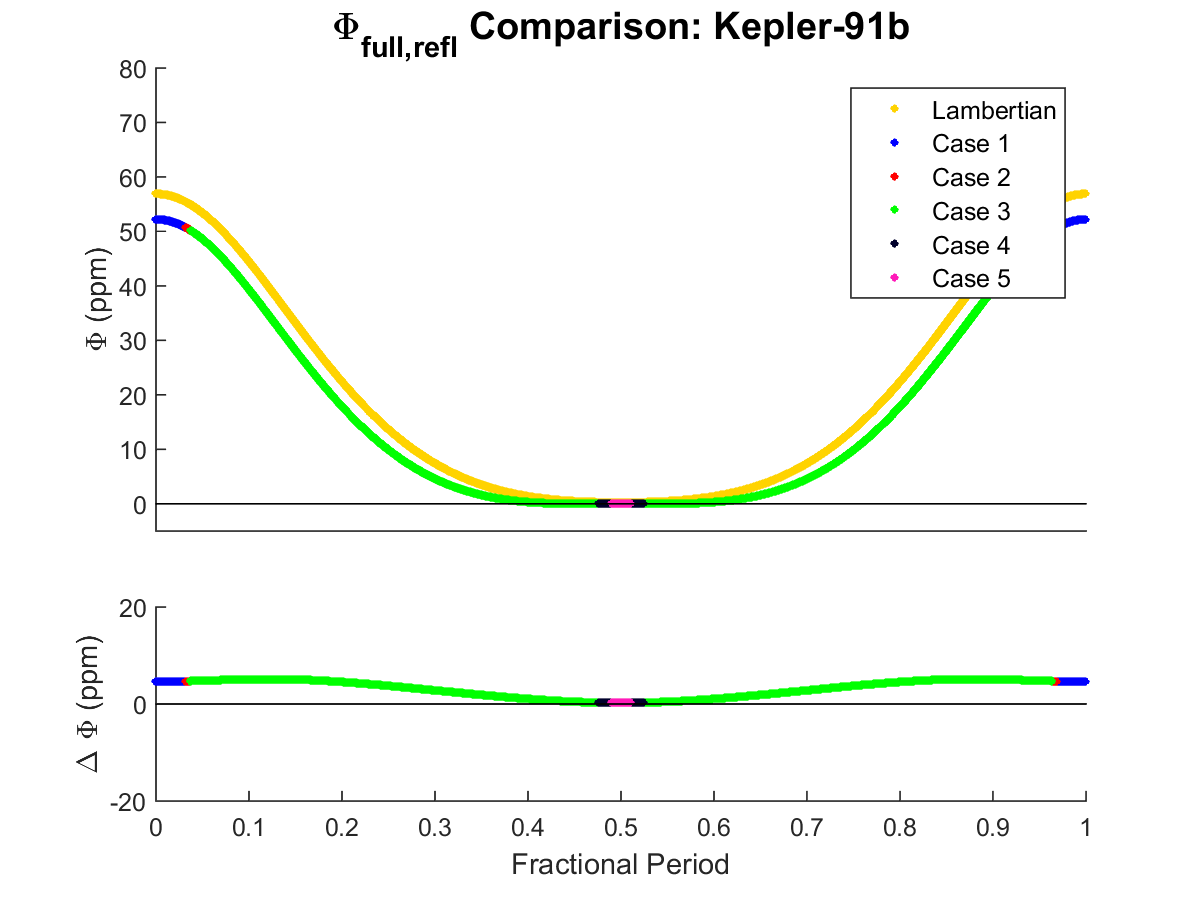}}
\hfill
\subfloat[][\label{fig:OGpen}$ \myLuminosity_{pen}/L_s $]{\includegraphics[trim={40 0 40 0}, clip,width=0.5\linewidth]{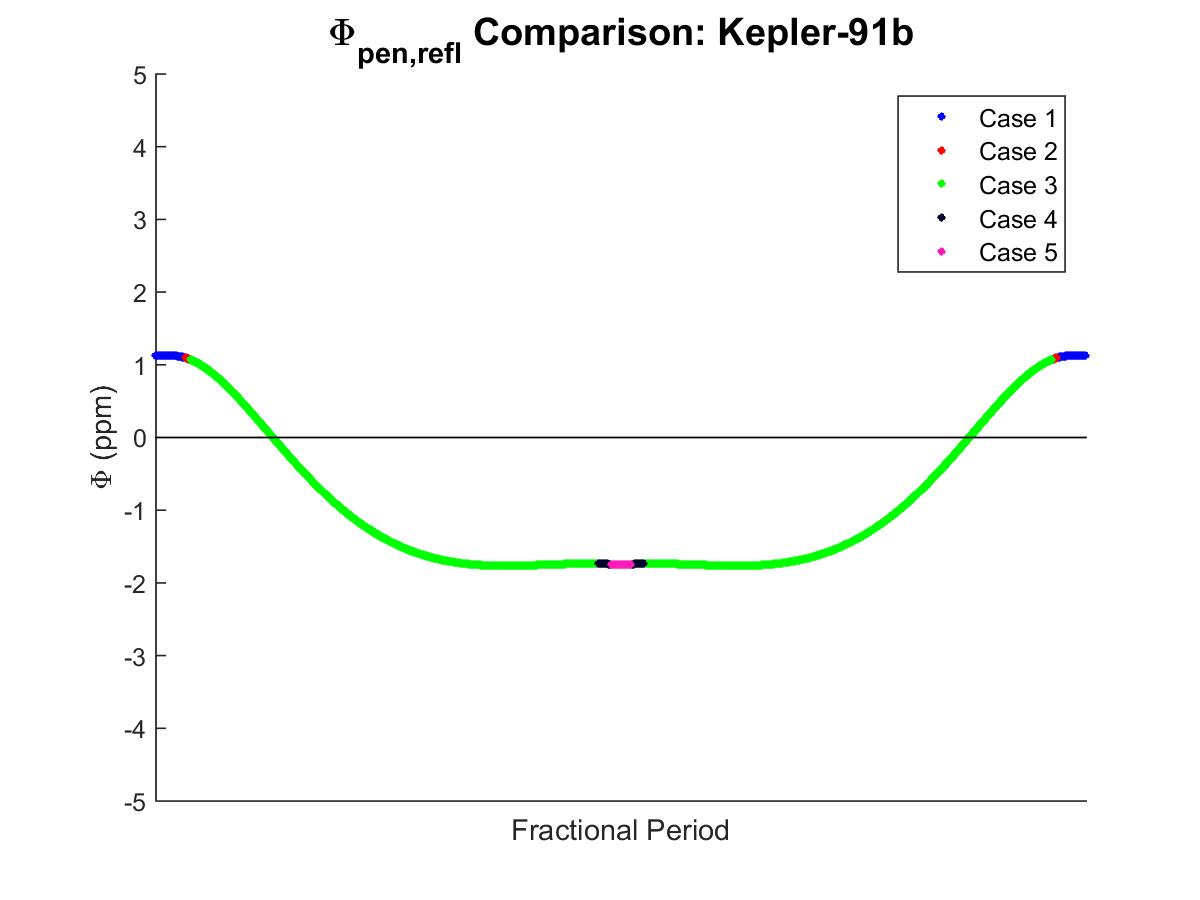}}
\caption{\label{fig:fractionalfluxfigs}Shown in the above plots is the fractional luminosity of the fully illuminated zone in \cref{fig:correct5full}\ and the penumbral zone in \cref{fig:OGpen}. Here we have used the reflected intensity distributions described in \cref{eq:fullLegendre}\ for the fully illuminated zone and \cref{eq:intensityDisPen,eq:JUDSum,eq:CUs,eq:CDs}\ for the penumbral zone. Notice that the use of the equations given in \cref{app:integrations}\ produces positive luminosities throughout the entire orbit for the fully illuminated zone, but not for the penumbral zone.}
\label{fig:full5casesnewreflcomparek91paramslimbdark}
\end{figure}

Here we present the results of using the integrations presented in \cref{app:integrations}\ for both the fully illuminated and penumbral zones if we consider the five cases as described in  \cref{tab:5cases} for the parameters given in \cref{tab:K91params}\ for which \cref{eq:intensityDisPen,eq:JUDSum,eq:CUs,eq:CDs}\ is used to describe the reflected intensity distribution in the penumbral zone. As shown in \cref{fig:full5casesnewreflcomparek91paramslimbdark}, the fully illuminated zone is now positive throughout the entire orbit, whereas there are large portions of the orbit for which the penumbral zone exhibits negative luminosity. Furthermore, we see that the luminosity of the fully illuminated zone never exceeds that of a lossless Lambertian sphere with the same orbital parameter values. This seems only appropriate because we are comparing the luminosity of a spherical cap that is smaller and exposed to less incident radiation than the equivalent Lambertian exoplanet. The difference between the luminosity of the lossless Lambertian sphere and that of the fully illuminated zone, $ \Delta \Phi $, ranges between a minimum of 0.258 ppm and a maximum of 5.15 ppm. In \cref{sec:luminosityFiniteSizenew}\ we will present a deeper look at the fully illuminated zone and present possible solutions to the determination of the luminosity within the penumbral zone using the intensity distribution described in \cref{sec:incidentFluxPennew}.

\section{New Derivation for Finite Angular Size}\label{sec:luminosityFiniteSizenew}

\mynew, The definitions of the limits on the fully illuminated and un-illuminated zones do not change and are given by \cref{eq:phizoneboth}, but now we must take into the account the fact that the penumbral zone is comprised of two distinct zones. If the fully illuminated zone were not present, then Penumbral Zone One would form a spherical cap with polar limits between $ \eta_{\mypen1limit}\le \eta\le \pi -\eta_{\mypen1limit} $ and azimuthal angles
\begin{equation}\label{eq:phizonepen1} 
    \begin{matrix}
		\phi_{pen1,i}=\eta_{\mypen1limit} &&\phi_{pen1,f} = \pi-\eta_{\mypen1limit},
	\end{matrix}
\end{equation}
where $ \eta_{\mypen1limit} $ is given by \cref{eq:pen12boundary}. To determine the reflected luminosity within this zone we will treat it as a spherical cap and then subtract from our results the integration of the reflected intensity distribution,  $ \myIntensityDis_{pen1} $, over the fully illuminated zone.
%
%
Finally, \cref{eq:phizoneboth,eq:phizonepen1}\ indicates that as we approach the plane parallel ray case where $ r\gg R_{p,s} $, in which $\eta_{\mypen1limit}, \eta_{1} $, and $ \eta_{2} $ approach zero, the fully illuminated zone will extend from zero to $ \pi $ and the un-illuminated zone from $ \pi $ to $ 2\pi $. In such a situation the penumbral zones will not exist.%

\begin{figure}[hbt!]
	\centering
	\subfloat[][Case 1: $ 0\le \myPhase\le \eta_{\mypen1limit} $. \label{fig:7case1}]{\includegraphics[height = 0.2\textheight]{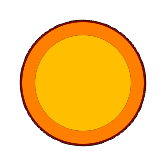}}
	\hfill
	\subfloat[][Case 2: $ \eta_{\mypen1limit}\le \myPhase\le  \eta_{2}$. \label{fig:7case2}]{\includegraphics[height = 0.2\textheight]{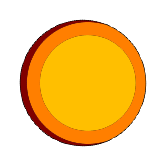}}
	\hfill
	\subfloat[][Case 3: $ \eta_{2}\le \myPhase \le \eta_{1}$. \label{fig:7case3}]{\includegraphics[height = 0.2\textheight]{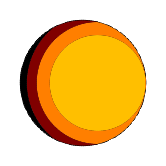}}

	\subfloat[][Case 4: $ \eta_{1}\le \myPhase\le \pi-\eta_{1} $. \label{fig:7case4}]{\includegraphics[height = 0.2\textheight]{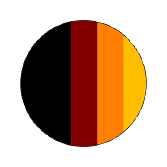}}
	\hfill
	\subfloat[][Case 5: $ \pi-\eta_{1}\le \myPhase\le \pi-\eta_{2} $. \label{fig:7case5}]{\includegraphics[height = 0.2\textheight]{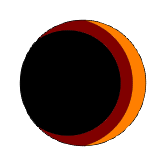}}
	\hfill
	\subfloat[][Case 6: $ \pi-\eta_{2}\le \myPhase\le \pi-\eta_{\mypen1limit} $. \label{fig:7case6}]{\includegraphics[height = 0.2\textheight]{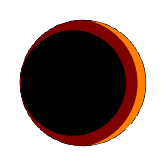}}

	\subfloat[][Case 7: $ \pi-\eta_{\mypen1limit}\le \myPhase\le \pi $. \label{fig:7case7}]{\includegraphics[height = 0.2\textheight]{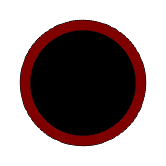}}
	\caption{Shown are example illustrations of the zones visible to an observer for each of the seven cases if one assumes that the exoplanet consists of four zones as described in \cref{sec:incidentFluxPennew}. In yellow is the fully illuminated zone, in orange is Penumbral Zone One, in dark-red is Penumbral Zone Two, and in black is the un-illuminated zone. See \cref{tab:verbalcases}\ for a verbal description of the cases and \cref{tab:7cases}\ for the integrals required to determine the luminosity of each zone. Note that Penumbral Zone One is visible throughout most of the orbit, but not during \mycase{7}. In contrast, Penumbral Zone Two is visible throughout the orbit because it extends from positive $ y $ to negative $ y $ in the planetocentric coordinate system.}
	\label{fig:7cases}
\end{figure}

\begin{figure}[hbt]
	\centering
	\includegraphics[width=0.9\linewidth]{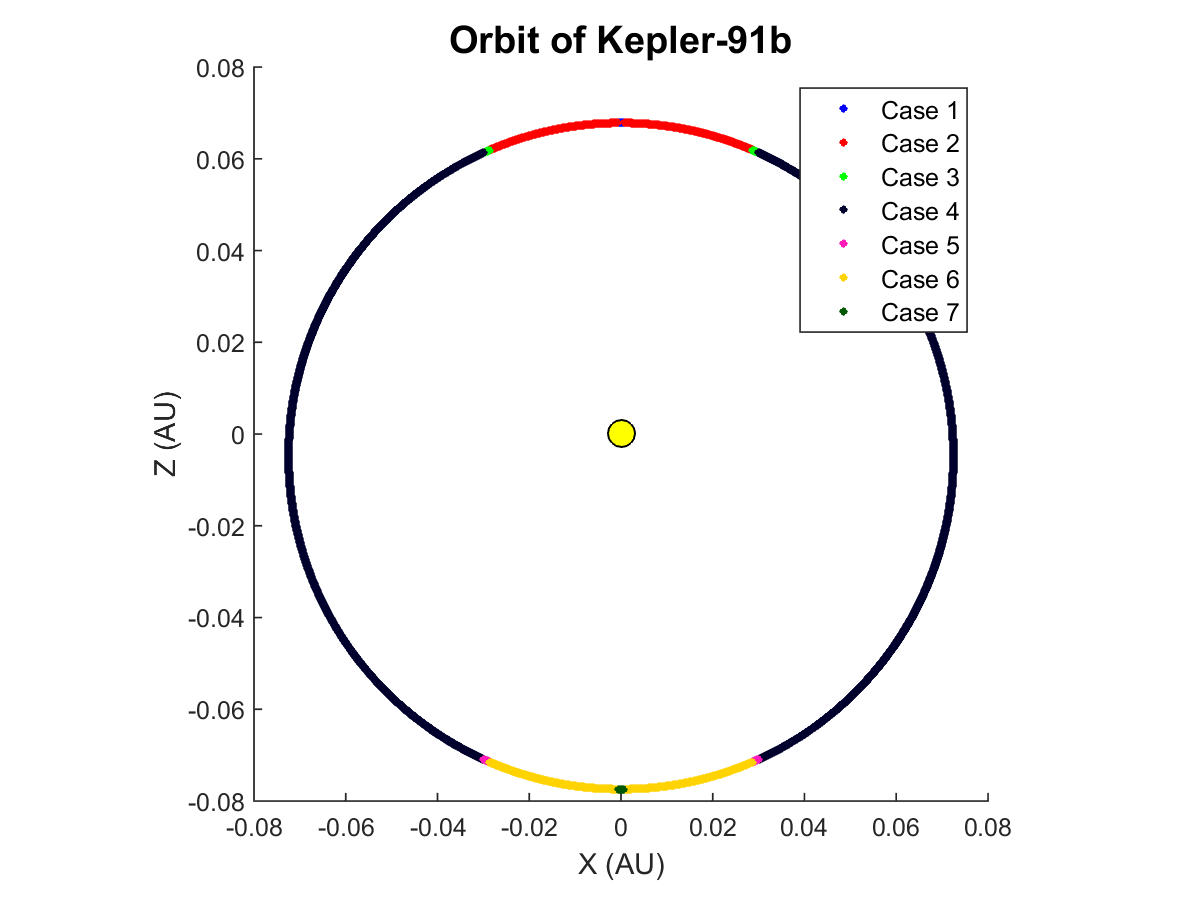}
	\caption{Shown is an illustration of the orbit of Kepler-91b using the parameters given in \cref{tab:K91params}\ with the adjustment that the inclination is now set to 90\degree. For some exoplanets, not all seven cases will occur because their orbits may be inclined such that the phase angle is never within a given case. For example, Kepler-91b does not actually have a phase angle within \mycase{1}\ or \mycase{7}; therefore, we have adjusted the inclination to 90\degree\ for illustrative purposes. The colors correspond to each of the seven cases as described in \cref{tab:7verbalcases}. The observer is viewing the orbit from the bottom of the page and $ \myPhase = 0 $ occurs at  $ X $ = 0 and  $ Z >$ 0.}
	\label{fig:7plotkepler-91bOrbit}
\end{figure}

The new geometry we must consider will result in seven, as opposed to five, unique cases whose descriptions are given in \cref{tab:7cases}. \cref{fig:7cases}\ illustrates the seven unique situations we must consider where the case shown in \cref{fig:7case4}\ is analogous to that shown in \cref{fig:case3}\ as is \cref{fig:7case7}\ to \cref{fig:case5}. \cref{fig:7plotkepler-91bOrbit}\ depicts each portion of the orbit of Kepler-91b for the seven cases. We note here that because $ \eta_{\mypen1limit} <\eta_{2} <\eta_{1} $ we will have an asymmetry in the amounts of the orbit for which the fully illuminated and un-illuminated zones are completely visible to an observer. Furthermore, the first penumbral zone will be visible throughout most of the orbit because $ \eta_{\mypen1limit} $ is nearly zero. As was done in \cref{sec:luminosityFiniteSizeog}, we will consider phase angles in the ranges of zero to $ \pi$ radians in our analysis.

\begin{table}[hbt]
	\centering
	\tabulinesep = 3mm
	\caption{\label{tab:7verbalcases}Table of verbal descriptions of the seven ranges of phase angles that must be evaluated to determine the total reflected luminosity of \myECIES\ if one assumes that the exoplanet may be described by four zones, see \cref{sec:incidentFluxPennew}. In the table the term ``Cap'' indicates that the entirety of a zone is visible as a spherical cap, the label ``Partial'' indicates that the zone is only partially visible, the label ``Not Visible'' is used to refer to situations in which the zone is not within either limb of the exoplanet, and finally the term ``Full Ring'' is used to describe situations in which the zone forms an unbroken ring about another zone.
		}
	\begin{tabu} to \textwidth {
			X[0.9,l]
			X[0.7,l]
			X[0.7,l]
			X[0.6,l]
			X[0.6,l]
		}
		
		\toprule
		\textbf{Range}&\textbf{Full}&\textbf{Un-ill} &\textbf{Pen 1} & \textbf{Pen 2} \\
		\midrule
		$ 0\le \myPhase\le \eta_{\mypen1limit} $&Cap&Not visible& Full ring&Full ring\\ 
		$ \eta_{\mypen1limit}\le \myPhase\le \eta_{2} $&Cap&Not Visible &Partial&Partial\\
		$ \eta_{2}\le \myPhase\le \eta_{1} $&Cap &Partial & Partial & Partial\\
		$\eta_{1}\le \myPhase\le \pi-\eta_{1} $&Partial &Partial &Partial &Partial\\
		$ \pi- \eta_{1}\le \myPhase\le \pi-\eta_{2}$&Not visible  & Partial & Partial & Partial\\ 
		$ \pi-\eta_{2}\le \myPhase\le \pi-\eta_{\mypen1limit} $ &Not visible &Cap &Partial &Partial \\
		 $ \pi-\eta_{\mypen1limit}\le \myPhase\le \pi $ & Not visible& Cap&Not Visible&Full ring\\
		\bottomrule
	\end{tabu}
\end{table}

\mynew, In general, the first penumbral zone's reflected luminosity is given by
\begin{equation}\label{eq:pen1general}
	\begin{aligned}
		\myLuminosity_{refl,pen1} (\myPhase)=&\myKintegral{\mySinEta_{\mypen1limit}}{\phi_{i,\mypen1limit}}{\phi_{f,\mypen1limit}}{\myIntensityDis_{refl, pen1}(\mu)}-\\
				&\myKintegral{\mySinEta_{1}}{\phi_{i,1}}{\phi_{f,1}}{\myIntensityDis_{refl, pen1}(\mu)}.
	\end{aligned}
\end{equation}
In addition, the determination of the reflected luminosity as a function of phase within the second penumbral zone, $ \myLuminosity_{refl,pen2}(\myPhase) $, may be determined via the integration
\begin{equation}\label{eq:pen2general}
	\begin{aligned}
		\myLuminosity_{refl,pen2} (\myPhase)=&\myKintegral{0}{\phi_{limb,i}}{\phi_{limb,f}}{\myIntensityDis_{refl,pen2}(\mu)}-\\
&\myKintegral{\mySinEta_{\mypen1limit}}{\phi_{i,\mypen1limit}}{\phi_{f,\mypen1limit}}{\myIntensityDis_{refl, pen2}(\mu)}-\\
		& \myKintegral{-\mySinEta_{2}}{\phi_{i,2}}{\phi_{f,2}}{\myIntensityDis_{refl, pen2}(\mu)}
	\end{aligned}
\end{equation}
where $\myIntensityDis_{refl, pen1,2}(\mu)  $ is given by \cref{eq:pen1intensitydistribution,eq:pen2intensitydistribution}. We will again assume that $ \myRefCo = \myScatteringAlbedo$ so that we may pull out $ \myRefCo $ from the integration. The integrations necessary to determine the luminosity of the three zones that exhibit reflection are presented in \cref{tab:7cases}. 

\clearpage
\begin{sidewaystable}
\centering
	\tabulinesep = 3mm
	\begin{longtabu} to \textwidth {
			X[0.25,l]
			X[0.35,c]
			X[0.4,l]
			X[0.4,l]
		}
		\caption{\label{tab:7cases}Shown is a table describing each of the seven cases and the required application of the operator $ \myKintegral{\mySinEta}{\phi_i}{\phi_f}{\myIntensityDis_{zone}}$, defined by \cref{eq:Kx}, to determine the reflected luminosity is a function of phase, $ \myLuminosity_{refl,zone}(\myPhase)$, for the fully illuminated and penumbral zones. The azimuthal angles shown in $ \myKintegral{\mySinEta}{\phi_i}{\phi_f}{\myIntensityDis_{zone}}$ are defined in \cref{eq:phizoneboth,eq:philimbs}
		. For each case the range of phase angles for which it applies is listed. The total reflected luminosity is the sum of the luminosity from the fully illuminated and penumbral zones, $ \myLuminosity_{refl}(\myPhase) = \myLuminosity_{full}(\myPhase)+\myLuminosity_{pen1}(\myPhase)+\myLuminosity_{pen2}(\myPhase) $.}\\
		\toprule
		\textbf{Range}&\textbf{Fully Illuminated}, $ \myLuminosity_{full} $ &\multicolumn{1}{c}{\textbf{Penumbral 1}, $ \myLuminosity_{pen1} $} &\multicolumn{1}{c}{\textbf{Penumbral 2}, $ \myLuminosity_{pen2} $ } \\
		\midrule
		\endfirsthead
		\multicolumn{4}{c}
		{\textbf{Integral Case Descriptions} -- \textit{Continued from previous page}} \\
		\midrule
		\textbf{Range}&\textbf{Fully Illuminated}, $ \myLuminosity_{full} $ &\multicolumn{1}{c}{\textbf{Penumbral}, $ \myLuminosity_{pen1} $ } &\multicolumn{1}{c}{\textbf{Penumbral 2}, $ \myLuminosity_{pen2} $ } \\
		\midrule
		\endhead
		\midrule
		\multicolumn{4}{c}{\textit{Continued on next page}} \\ 
		\endfoot
		\bottomrule
		\multicolumn{4}{c}{\textit{End of  $ \myIntensityDis_{zone} $ table}}
		\endlastfoot
		$ 0\le \myPhase\le \eta_{\mypen1limit} $& $ \myKintegral{\mySinEta_{1}}{\phi_{full,i}}{\phi_{full,f}}{\myIntensityDis_{full}} $&$ \myKintegral{\mySinEta_{\mypen1limit}}{\phi_{\mypen1limit,i}}{\phi_{\mypen1limit,f}}{\myIntensityDis_{pen1}} - \myKintegral{\mySinEta_{1}}{\phi_{full,i}}{\phi_{full,f}}{\myIntensityDis_{pen1}} $ &$\myKintegral{0}{\phi_{limb,i}}{\phi_{limb,f}}{\myIntensityDis_{pen2}}- \myKintegral{\mySinEta_{\mypen1limit}}{\phi_{\mypen1limit,i}}{\phi_{\mypen1limit,f}}{\myIntensityDis_{pen2}}$\\ 
		$ \eta_{\mypen1limit}\le \myPhase\le \eta_{2} $& $ \myKintegral{\mySinEta_{1}}{\phi_{full,i}}{\phi_{full,f}}{\myIntensityDis_{full}} $ &$ \myKintegral{\mySinEta_{\mypen1limit}}{\phi_{limb,i}}{\phi_{\mypen1limit,f}}{\myIntensityDis_{pen1}} - \myKintegral{\mySinEta_{1}}{\phi_{full,i}}{\phi_{full,f}}{\myIntensityDis_{pen1}} $& $\myKintegral{0}{\phi_{limb,i}}{\phi_{limb,f}}{\myIntensityDis_{pen2}} -\myKintegral{\mySinEta_{\mypen1limit}}{\phi_{limb,i}}{\phi_{\mypen1limit,f}}{\myIntensityDis_{pen1}}$\\ 
		$ \eta_{2}\le \myPhase\le \eta_{1} $& $ \myKintegral{\mySinEta_{1}}{\phi_{full,i}}{\phi_{full,f}}{\myIntensityDis_{full}} $  &  $ \myKintegral{\mySinEta_{\mypen1limit}}{\phi_{limb,i}}{\phi_{\mypen1limit,f}}{\myIntensityDis_{pen1}} - \myKintegral{\mySinEta_{1}}{\phi_{full,i}}{\phi_{full,f}}{\myIntensityDis_{pen1}}$  & $\myKintegral{0}{\phi_{limb,i}}{\phi_{limb,f}}{\myIntensityDis_{pen2}} -\myKintegral{\mySinEta_{\mypen1limit}}{\phi_{limb,i}}{\phi_{\mypen1limit,f}}{\myIntensityDis_{pen2}}-\myKintegral{-\mySinEta_{2}}{\phi_{un,i}}{\phi_{limb,f}}{\myIntensityDis_{pen2}} $\\ 
		$\eta_{1}\le \myPhase\le \pi-\eta_{1} $&$ \myKintegral{\mySinEta_{1}}{\phi_{limb,i}}{\phi_{full,f}}{\myIntensityDis_{full}} $ &$ \myKintegral{\mySinEta_{\mypen1limit}}{\phi_{limb,i}}{\phi_{\mypen1limit,f}}{\myIntensityDis_{pen1}} - \myKintegral{\mySinEta_{1}}{\phi_{limb,i}}{\phi_{full,f}}{\myIntensityDis_{pen1}}$ &$\myKintegral{0}{\phi_{limb,i}}{\phi_{limb,f}}{\myIntensityDis_{pen2}}  -\myKintegral{\mySinEta_{\mypen1limit}}{\phi_{limb,i}}{\phi_{\mypen1limit,f}}{\myIntensityDis_{pen2}}-\myKintegral{-\mySinEta_{2}}{\phi_{un,i}}{\phi_{limb,f}}{\myIntensityDis_{pen2}} $\\ 
		$ \pi- \eta_{1}\le \myPhase\le \pi-\eta_{2}$&0  &  $  \myKintegral{\mySinEta_{\mypen1limit}}{\phi_{limb,i}}{\phi_{\mypen1limit,f}}{\myIntensityDis_{pen1}} $  & $\myKintegral{0}{\phi_{limb,i}}{\phi_{limb,f}}{\myIntensityDis_{pen2}}  -\myKintegral{\mySinEta_{\mypen1limit}}{\phi_{limb,i}}{\phi_{\mypen1limit,f}}{\myIntensityDis_{pen2}}-\myKintegral{-\mySinEta_{2}}{\phi_{un,i}}{\phi_{limb,f}}{\myIntensityDis_{pen2}}$\\ 
		$ \pi-\eta_{2}\le \myPhase\le \pi-\eta_{\mypen1limit} $ &0 & $  \myKintegral{\mySinEta_{\mypen1limit}}{\phi_{limb,i}}{\phi_{\mypen1limit,f}}{\myIntensityDis_{pen1}} $  &$\myKintegral{0}{\phi_{limb,i}}{\phi_{limb,f}}{\myIntensityDis_{pen2}}  -\myKintegral{\mySinEta_{\mypen1limit}}{\phi_{limb,i}}{\phi_{\mypen1limit,f}}{\myIntensityDis_{pen2}}-\myKintegral{-\mySinEta_{2}}{\phi_{un,i}}{\phi_{un,f}}{\myIntensityDis_{pen2}}$ \\ 
		 $ \pi-\eta_{\mypen1limit}\le \myPhase\le \pi $ &0&\multicolumn{1}{c}{0}&$\myKintegral{0}{\phi_{limb,i}}{\phi_{limb,f}}{\myIntensityDis_{pen2}}-\myKintegral{-\mySinEta_{2}}{\phi_{un,i}}{\phi_{un,f}}{\myIntensityDis_{pen2}} $ 		
	\end{longtabu}
\end{sidewaystable}


The determination of the luminosity from the two penumbral zones will be left for future work after the final equations for the intensity distributions of said zones has been determined. If it is possible to determine the reflected intensity distribution in terms of $ \mu^n $, then we may use the integrals provided in \cref{app:integrations}. Otherwise, it may be necessary to reevaluate the integration of the reflected luminosity of the penumbral zones in terms of $ \myexpansionz $ or perhaps numerically.

\begin{figure}[htb]
\centering
\includegraphics[width=0.95\linewidth]{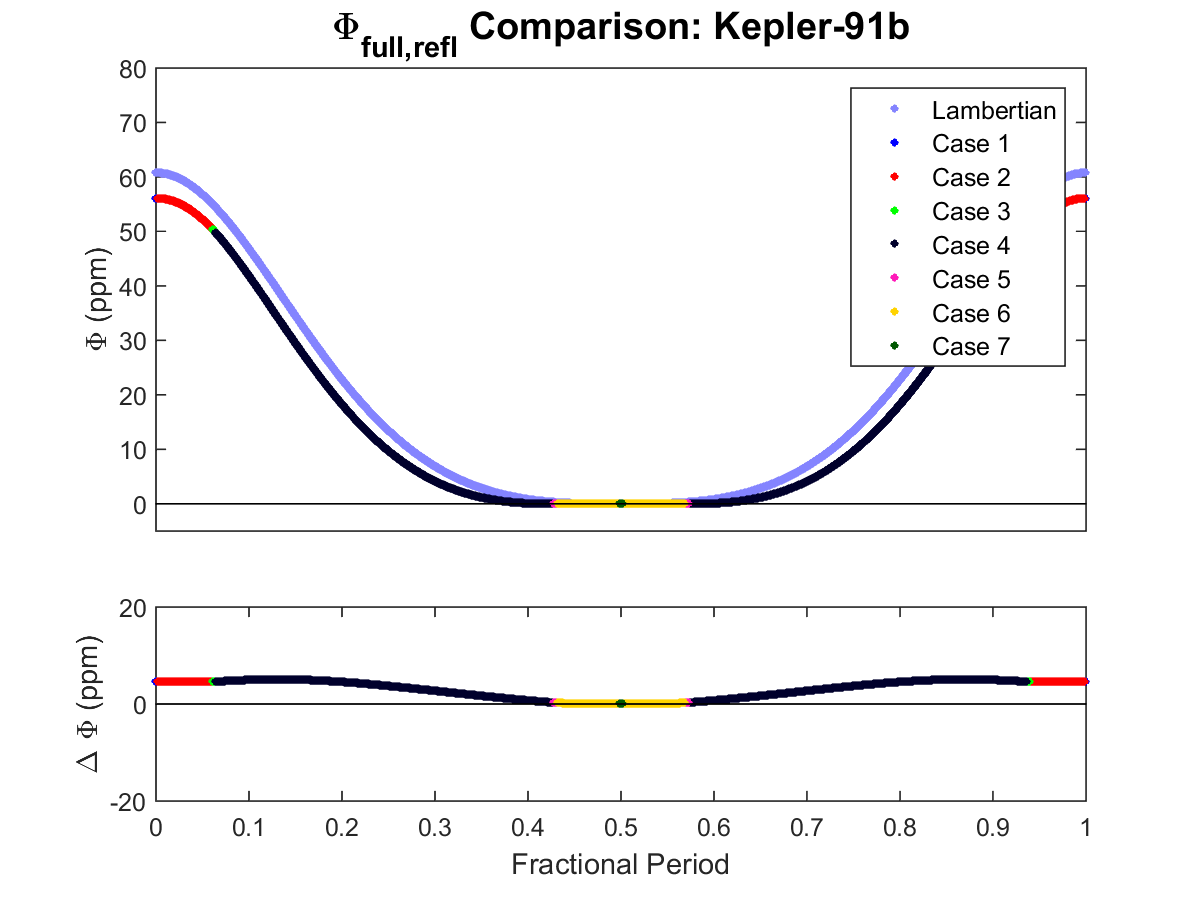}
\caption{Shown is a plot comparing the fractional flux of a Kepler-91b analog using the parameters given in \cref{tab:K91params}\ with the exception that we now have set the inclination to 90\degree. Also shown is the fractional flux of a lossless Lambertian sphere with the same parameters. The bottom panel displays the difference between these two fractional fluxes which ranges between  $ 3.87\times 10^{-11} $ ppm and 5.09 ppm.}
\label{fig:fullreflcomparek91paramslimbdarkedgeon}
\end{figure}

Here we present the evaluation of the luminosity of the fully illuminated zone using the seven cases described in the section. We begin by considering an analog to Kepler-91b, \cref{fig:fullreflcomparek91paramslimbdarkedgeon}\ is a plot of the fractional luminosity from the fully illuminated zone using the parameters given in  \cref{tab:K91params}\ where we have changed the inclination to 90\degree\ for illustrative purposes. The true inclination of Kepler-91b is such that no portion of its phase is within \textbf{Cases 1} or \textbf{7}. The plot is much the same as that of \cref{fig:correct5full} because very little has changed between the two plots. For the fully illuminated zone we can consider \textbf{Cases 1-3}\ in \cref{fig:fullreflcomparek91paramslimbdarkedgeon}\ to be equivalent to \textbf{Cases 1}\ and \textbf{2}\ of  \cref{fig:correct5full}. Furthermore, \mycase{4}\ corresponds to \mycase{3}\ of \cref{sec:luminosityFiniteSizeog}. Finally, \textbf{Cases 5-6}\ are equivalent to \textbf{Cases 4} and \textbf{5} of the previous analysis.

\begin{figure}[htb]
\centering
\includegraphics[width=0.95\linewidth]{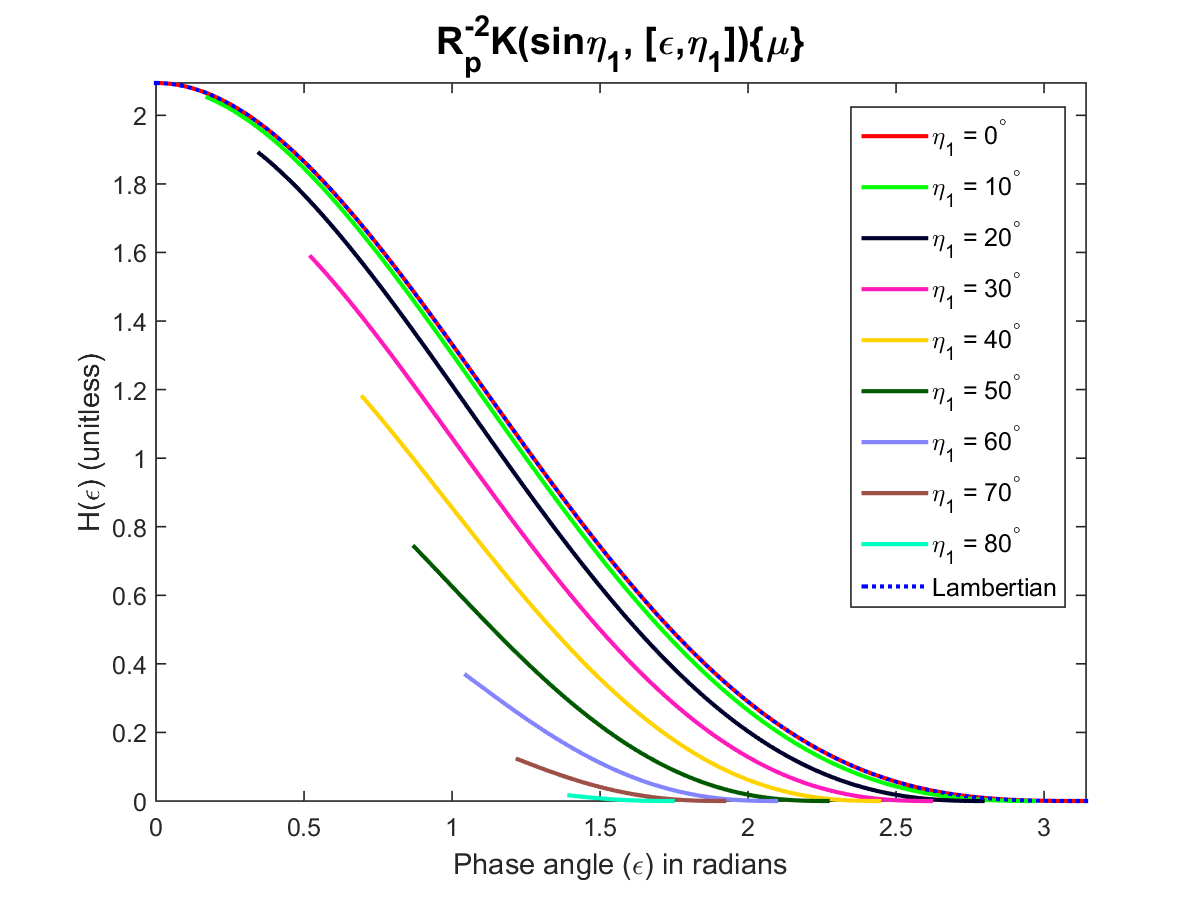}
\caption{In the figure we have plotted the unitless quantity $ \myKintegral{\sin{\eta_{1}}}{\myPhase}{\eta_{1}}{\mu}/R_p^2  $ as a function of the phase angle $ \myPhase $ for multiple values of $ \eta_{1} $ for \mycase{4}. Here we see that the extent of this case decreases as $ \eta_{1} $ increases. In addition, we find that for $ \eta_{1} = 0 $ the function is equal to $ H_{Lambert}(\myPhase) $ for a lossless Lambertian sphere as expected. Finally, we note that as $\eta_{1}$ increases the size of the fully illuminated zone decreases such that the overall luminosity decreases in magnitude. It is worth noting that the exoplanets in \cref{tab:targets}\ do not contain values of $ \eta_{1} $ greater than 41.0\degree.}
\label{fig:kfullcase4plot}
\end{figure}

Let us now consider the extent of \mycase{4}\ for multiple values of $ \eta_{1} $ as in \cref{fig:kfullcase4plot}\ where we have plotted the first term of \cref{eq:Fullgeneral}\ multiplied by $ \pi r^2/(L_s \myScatteringAlbedo R_p^2 ) $ to provide a unitless quantity. As expected for $ \eta_{1} = 0 $  $ \myKintegral{\sin{\eta_{1}}}{\myPhase}{\eta_{1}}{\mu}/R_p^2  $ produces the Lambertian phase function given in \cref{eq:HphaseLambert}\ for $ \myScatteringAlbedo=1 $ . Furthermore, we see that as the value of $ \eta_{1} $ increases the extent of \mycase{4}\ decreases. In addition, the magnitude of the reflected luminosity decreases because the size of the fully illuminated zone decreases more quickly than the increase in incident radiation from the host star due to the decrease in the star-planet separation. Finally, we note here that of the confirmed exoplanets listed in \cref{tab:targets}, none exhibit a value of $\eta_{1}$ greater than 41.0\degree. Curves in \cref{fig:kfullcase4plot}\ for which $ \eta_{1} $ is greater than 40\degree\ are presented for illustrative purposes and to check for the existence of nonphysical results of which there are none.

\begin{figure}[htb]
\centering
\includegraphics[width=0.95\linewidth]{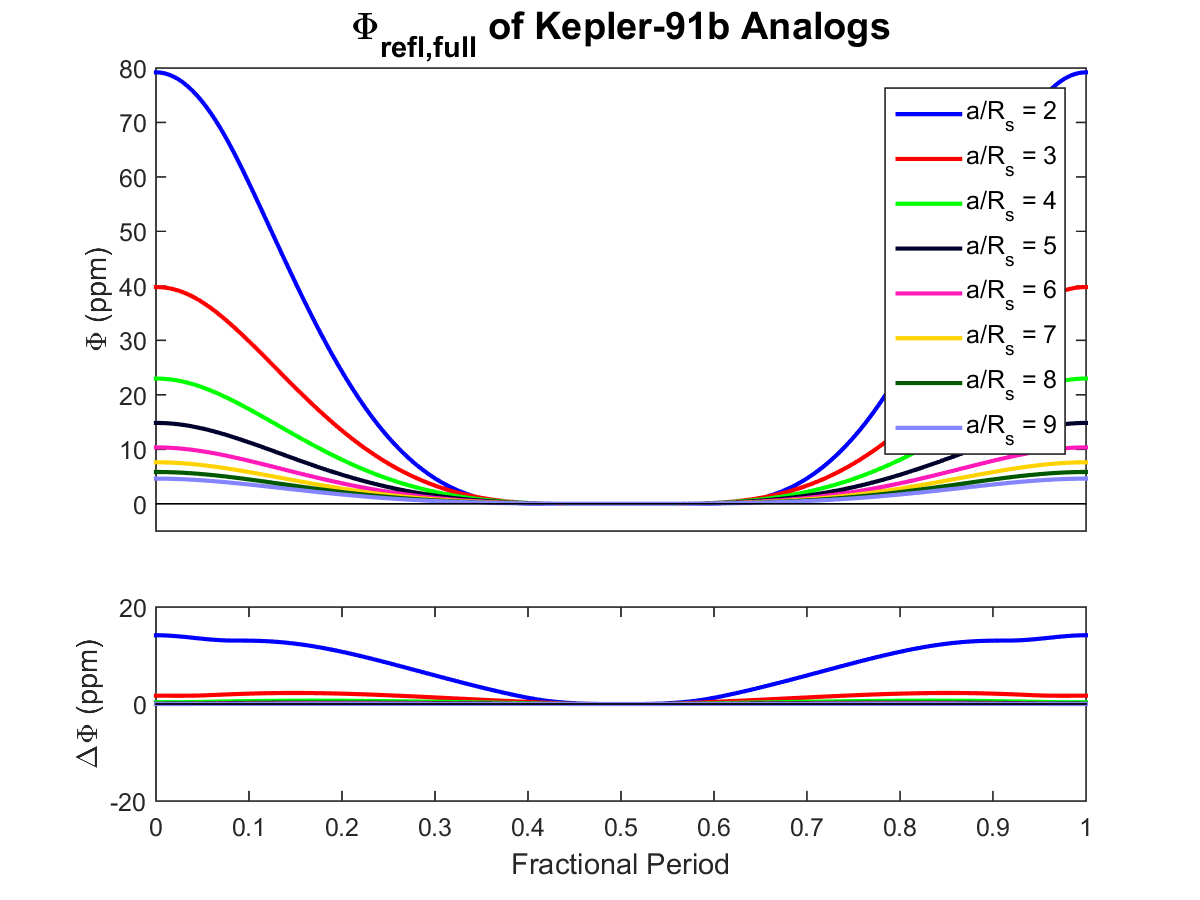}
\caption{Plotted in the figure are the fractional fluxes of the fully illuminated zone of Kepler-91b analogs for an edge-on orbit using the parameters given in \cref{tab:K91params}\ where we have also varied the normalized semi-major axis $ a/R_s $. The bottom panel plots the difference between the fractional flux of a lossless Lambertian sphere and the fractional flux of the fully illuminated zone.
}
\label{fig:multiplotfullreflk91paramslimbdarkedgeon}
\end{figure}

We see a similar pattern when we consider multiple Kepler-91b analogs as in \cref{fig:multiplotfullreflk91paramslimbdarkedgeon}\ where we are considering an edge-on orbit and have varied the normalized semi-major axis. In addition, we see that the difference between the fractional flux of a lossless Lambertian sphere with equivalent parameters and that of the fully illuminated zone decreases as the star-planet separation increases. It is likely that for other exoplanets we will find that the luminosity of the fully illuminated zone quickly approaches that given by the plane parallel ray approximation for star-planet separations greater than about 3 stellar radii. 

We note that for Kepler- 91b the difference between the incident stellar radiation modeled as plane parallel rays as described in \cref{sec:planeparallel}\ and that modeled using the finite angular size of the host star as described in \cref{sec:finitesize}\ is not detectable using current technology. For example, \myKepler\ exoplanets are characterized to a maximum precision of 29 ppm,  \cite{KeplerPrecision}. Should future missions be capable of detecting changes in flux on the order of a few parts per million one would be able to detect the influence of the finite angular size of the host star on the luminosity of the exoplanet.

Finally, let us consider the possible contributions of the penumbral zones to the overall reflected luminosity of an exoplanet by considering the difference between the luminosity of a lossless Lambertian sphere and the fully illuminated zone. Because the penumbral zones receive less incident stellar radiation than the fully illuminated zone one should expect that the inclusion of the zones would not increase the total luminosity of the exoplanet beyond that of a lossless Lambertian sphere for \textbf{Cases 1-3}. For such cases the maximum difference is 14.2 ppm, or about 17.9\%\ of the luminosity of the fully illuminated zone during the secondary eclipse, for the Kepler91-b analog plotted in \cref{fig:fullreflcomparek91paramslimbdarkedgeon}. Thus, the increase in the luminosity one may expect from the inclusion of the penumbral zones for cases near the new phase of the exoplanet, i.e. \textbf{Cases 5-7}\ for which the exoplanet is near the primary transit, will likely be of a similar magnitude. Again, changes in luminosity of this order are not yet detectable with current instruments, but maybe so in the future. For example, PLATO is a planned ESA mission intended to provide up to 2\%\ precision in the determination of planet radii, \cite{Plato2Mission}.

\section{Geometric Albedo for Finite Angular Size}\label{sec:geometricalbedo}
\mynew, Thus far we have been considering the reflected luminosity of an exoplanet in terms of its single scattering albedo, $ \myScatteringAlbedo $. For a Lambertian sphere it was shown in \cref{sec:Lforplaneparallel}\ that the geometric albedo, defined by  \cref{eq:geometricAlbedo}, was equal to
\begin{equation} \label{eq:Aglambert}
    A_{g, Lambert} =\frac{2 }{3}\myScatteringAlbedo.
\end{equation}
We will now consider the meaning of the geometric albedo for the case in which one models the incident stellar radiation by considering the finite angular size of the host star.

From the definition of the geometric albedo we see that we should be considering the luminosity of the exoplanet for phase angles within \mycase{1}; therefore, the total luminosity of the exoplanet according to line one of \cref{tab:7cases}\ is given by
\begin{equation}
  \begin{aligned}
        \myLuminosity_{refl}(0) = &\left( \myKintegral{\mySinEta_{1}}{\eta_{1}}{\pi-\eta_{1}}{\myIntensityDis_{full}}\right)+\\
&\left(\myKintegral{\mySinEta_{\mypen1limit}}{\eta_{\mypen1limit}}{\pi-\eta_{\mypen1limit}}{\myIntensityDis_{pen1}} - \myKintegral{\mySinEta_{1}}{\eta_{1}}{\pi-\eta_{1}}{\myIntensityDis_{pen1}}\right)+\\
&\left(\myKintegral{0}{0}{\pi}{\myIntensityDis_{pen2}}- \myKintegral{\mySinEta_{\mypen1limit}}{\eta_{\mypen1limit}}{\pi-\eta_{\mypen1limit}}{\myIntensityDis_{pen2}}\right)
  \end{aligned}
\end{equation}
evaluated at $ \myPhase=0 $. 

For now, let us consider the expression for the geometric albedo of an exoplanet if we only include the luminosity of the fully illuminated zone:
\begin{equation} \label{eq:Agfull}
    A_{g,full}=\frac{\myLuminosity_{full}(0)}{L_s(R_p/r)^2}.
\end{equation}
For \mycase{1}\ the luminosity of the fully illuminated zone is given by
\begin{equation}\label{eq:fullcase1}
  \begin{aligned}
        \myLuminosity_{full,case1}(\myPhase)=\left( \frac{\myScatteringAlbedo L_s R_p^2}{6r^4}\right) \Big(&3 r \left(-3 \mySinEta_{1}^4+2 \mySinEta_{1}^2+1\right) R_p-\\
&18 \left(\mySinEta_{1}^2-1\right) \mySinEta_{1}^3 R_p^2-4 r^2 \left(\mySinEta_{1}^3-1\right)\Big)\cos{\epsilon},
  \end{aligned}
\end{equation}
from \cref{eq:Fullgeneral,eq:fulledgetoedge}.
Inserting \cref{eq:fullcase1}\ into \cref{eq:Agfull}\ and setting $ \myPhase $ to zero we find that the geometric albedo is given by
\begin{equation}
  \begin{aligned}
        A_{g,full}=\frac{\myScatteringAlbedo}{6r^2}\Big(&\mySinEta_{1}^3 \left(18 R_p^2-4 r^2\right)-9 r \mySinEta_{1}^4 R_p+6 r \mySinEta_{1}^2 R_p+3 r R_p-18 \mySinEta_{1}^5 R_p^2+4 r^2\Big).
  \end{aligned}
\end{equation}
To get a better understanding of this quantity it is helpful to use \cref{eq:sineta1}\ to set $ \mySinEta_{1}=(R_s+R_p)/r $, revealing that
\begin{equation}\label{eq:Agfullfinal}
	\begin{aligned}
		A_{g,full}=\myScatteringAlbedo\bigg(&\frac{2}{3}+\frac{1}{r}\left(\frac{R_p}{2} \right)+\frac{1}{r^3}\left(-R_p R_s^2+\frac{R_p^3}{3}-\frac{2 R_s^3}{3} \right)+\\
&\frac{1}{r^5}\left( 3 R_p^4 R_s-3 R_p^2 R_s^3-\frac{3}{2} R_p R_s^4+\frac{3 R_p^5}{2}\right)+\\
&\frac{1}{r^7}\left(-15 R_p^6 R_s-30 R_p^5 R_s^2-30 R_p^4 R_s^3-15 R_p^3 R_s^4-3 R_p^2 R_s^5-3 R_p^7 \right)\bigg).
	\end{aligned}
\end{equation}

\cref{eq:Agfullfinal}\ reveals that the geometric albedo of an exoplanet for which we take into account the finite angular size of the host star not only depends on the reflective properties of the exoplanet as described by the single scattering albedo, but also on the geometry of the system. The geometry of the system adds corrective terms to the expected geometric albedo of $ 2\myScatteringAlbedo/3 $ for a Lambertian sphere of the form $ (R_{p,s}/r)^n  $ where $ n $ is odd. Finally, we see that as the star-planet separation,  $ r $, increases the geometric albedo quickly approaches that of a Lambertian sphere.

In \cref{sec:Lforplaneparallel}\ we saw that both the geometric and spherical albedos played an important role in characterizing the reflective properties of exoplanets. For the plane parallel ray modeling of incident stellar radiation the two variables depended only on the directional scattering properties of the exoplanets surface or atmosphere. We see here that when the finite angular size of the host star is considered, as is necessary for \myECIES, that the albedo of an exoplanet cannot be separated from geometrical properties of the star-planet system. This fact was hinted at in \cref{sec:Lforplaneparallel}\ when we stated that the parameter $ \sqrt{A_g}(R_p) $, or the minimum radius, was the only one that could be determined from the reflected luminosity, \cref{eq:phirPhaseFunc}, alone. The parameter of interest now has shifted from the geometric albedo to the single scattering albedo $ \myScatteringAlbedo $.

Within the previous chapter we have outlined the relationship between the reflected intensity distribution and the reflected luminosity of an exoplanet,  \cite{seager}. In addition, we have reviewed the luminosity as a function of phase in the case of plane parallel ray illumination, \cf\ \cite{seager,sobolev}. We then turned the luminosity for the fully illuminated zone of an exoplanet in which we have taken the finite angular size of the host star into account when describing the illumination. We saw that the equations given in previous work, \cite{Kopal1953,Kopal1959}, produced nonphysical luminosity in that it exhibited negative luminosity. New to this work is a correction to that error as described in \cref{sec:luminosityFiniteSizenew}. We also compared the reflected luminosity using this new model and the plane parallel ray model and found that the two differ by no more than 5.09 ppm for the case of Kepler-91b, which is not detectable within the \myKepler\ data set. We concluded with a discussion of the effect of this model on the determination of the geometric albedo of exoplanets. We will now shift our attention to the determination of the luminosity due to thermal radiation of an exoplanet.

\chapter{Photometric Planetary Emissions: Thermal Light}\label{ch:thermalradiation}
The luminosity of the thermal light emitted by an exoplanet may be determined by calculating the contribution from each temperature zone as a function of phase. For the case of plane parallel rays illuminating the exoplanet, it can be assumed that there are two temperature zones for the exoplanet: the day side and night side temperatures, \cf\ Placek, Knuth and Angerhausen \cite{Placek2014}. If we instead take a new approach in which we consider the finite angular size of the host star the third penumbral zone temperature(s) must be considered. In this chapter we will first review the method to determining the luminosity due to thermal radiation for a two zone model from previous literature, and then we will consider the new three zone model from \cref{sec:incidentFluxPenKopalMethod}, and finally describe the four zone model presented in \cref{sec:incidentFluxPennew}.

\section{Day and Night Side}\label{sec:thermDaynight}
\myfillin, The luminosity of thermal light emitted by an area $dA$ within a given zone of an exoplanet is given by
\begin{equation}
	\begin{aligned}\label{eq:dLtherm}
		d\mathscr{L}_\textnormal{Th,zone} & = F(T_\textnormal{zone})\hat{n}\cdot\hat{r}~dA\\
			& = F(T_\textnormal{zone})\cos\gamma~dA,
	\end{aligned}
\end{equation}
where  $ F(T_\textnormal{zone}) $ is the flux produced by a surface with effective temperature $ T_\textnormal{zone} $, \cf\ Chapter 3 of Seager's textbook \cite{seager}. 
To determine $ F(T_\textnormal{}{zone})$, we will assume that both the star and the exoplanet are behaving like black bodies with radiance given by
\begin{equation} 
	B (T,\lambda) =\left( \frac{2hc^2 }{\lambda^5}\right)\left( \exp\left( \frac{hc}{\lambda k_{B}T}\right)-1\right)^{-1},	
\end{equation}
where $ h $ is Planck's constant, $ \lambda  $ is the wavelength, $ k_{B} $ is Boltzmann's constant, and $ T $ is the temperature. The temperature dependent flux is then given by
\begin{equation} \label{eq:fluxKeplerResponse}
    F (T) =\int B (T,\lambda) K (\lambda)~d\lambda,
\end{equation}
where $ K (\lambda) $ is the \myKepler\ response function as described in the Kepler Instrument Handbook by Cleve and Caldwell, \cite{KeplerHandbook}. Note that other response functions could be inserted into \cref{eq:fluxKeplerResponse}\ to calculate the flux measured by other instruments. 

To determine the total flux of the day side of the exoplanet, one must first integrate \cref{eq:dLtherm}\ over the day side of the exoplanet that is visible along the LOS in a similar fashion to the method given in \cref{sec:planeparallel}: 
\begin{equation}\label{eq:luminositythermalday} 
\begin{aligned} 
    \myLuminosity_\textnormal{Th,day}(\myPhase) &= F_p(T_\textnormal{day})\int_{\myPhase}^{\pi}\int_{0}^{\pi}\sin\eta\sin(\phi-\myPhase)~dA_p\\
		& =F_p(T_\textnormal{day})R_p^2\int_{\myPhase}^{\pi} \int_{0}^{\pi}\sin^2\eta\sin(\phi-\myPhase)~d\eta~d\phi\\
		&=F_p(T_\textnormal{day})\pi R_{p}^2\left(\frac{1+\cos\myPhase}{2}\right).
\end{aligned}
\end{equation}
As described in \cref{eq:fluxAtEarth,eq:phirPhaseFunc}, we may write the flux received at Earth from the exoplanet as 
\begin{equation}
\begin{aligned} 
    F_{Th, day}(\myPhase)&=\frac{\myLuminosity_\textnormal{Th,day}(\myPhase)}{\pi d^2}\\
			   &=F_p(T_{day})\left( \frac{R_{p}}{d}\right)^2\left(\frac{1+\cos\myPhase}{2}\right);
\end{aligned}
\end{equation}
therefore, the normalized flux received at Earth is then
\begin{equation} \label{eq:normalizedfluxthermalday}
\begin{aligned} 
    \Phi_{Th,day} (\myPhase) &= \frac{F_{Th, day} (\myPhase)}{F_{s}}\\
						   &= \left( \frac{F_{p} (T_{day})}{F_s (T_{eff})}\right)\left( \frac{R_{p} }{R_{s}}\right)^2\left( \frac{1+\cos{\myPhase}}{2}\right),
\end{aligned}
\end{equation}
where $ F (T) $ is given by  \cref{eq:fluxKeplerResponse}, and $ F_{s}( T_{eff}) $ is the constant flux of the host star if it is treated like a blackbody with effective temperature $ T_{eff} $.

The thermal luminosity from the night side of the exoplanet is determined in a similar fashion to that of the day side thermal luminosity. From \cref{fig:planeparallelterminatorlabels}\ we see that the limits on the azimuthal angle of the night side range from the terminator to the limb of the exoplanet farthest from the sub-stellar point, namely $ \pi $ to $ \pi +\myPhase $; therefore,
\begin{equation}\label{eq:luminositythermalnight}
  \begin{aligned}
        \myLuminosity_\textnormal{Th,night}&= F_p(T_\textnormal{night})\int_{\pi}^{\pi+\myPhase}\int_{0}^{\pi}\sin\eta\sin(\phi-\myPhase)~dA_p\\
		& =F_p(T_\textnormal{night})R_p^2\int_{\pi}^{\pi+\myPhase} \int_{0}^{\pi}\sin^2\eta\sin(\phi-\myPhase)~d\eta~d\phi\\
		& =F_p(T_\textnormal{night}) R_{p}^2\pi\sin^2\left( \frac{\myPhase}{2}\right)\\
		&=F_p(T_\textnormal{night})\pi R_{p}^2\left(\frac{1+\cos(\myPhase-\pi)}{2}\right).
  \end{aligned}
\end{equation} 
Inspection of \cref{eq:normalizedfluxthermalday,eq:luminositythermalnight} reveals that the normalized flux of the night side of the exoplanet is exactly out of phase of that of the day side.:
\begin{equation}\label{eq:thermnight}
	\Phi_{Th,night} (\myPhase) =\left( \frac{F_{p} (T_{night})}{F_s (T_{eff})}\right)\left( \frac{R_{p} }{R_{s}}\right)^2\left( \frac{1+\cos{(\myPhase-\pi)}}{2}\right),
\end{equation}
\cf\ Knuth et al. Equations (18)-(21) in \cite{2017Entropy}\ or Charbonneau et al Equations (18)-(21) in\cite{charbonneau05}.

The total fractional flux due to thermal radiation is then
\begin{equation} \label{eq:thermaltotal}
    \Phi_{therm}(\myPhase) =\Phi_{Th,day} (\myPhase) +\Phi_{Th,night} (\myPhase).
\end{equation}

\section{Finite Angular Size}
Inspection of \cref{eq:luminositythermalday,eq:luminositythermalnight}\ and comparison to \cref{eq:Kx}\ reveals that the equations are equivalent to
\begin{equation}
\begin{aligned}
	\myLuminosity_\textnormal{Th,day}&=\myKintegral{0}{\myPhase}{\pi}{F_p(T_\textnormal{day})}\\
			&=F_p(T_\textnormal{day})\myKintegral{0}{\myPhase}{\pi}{1}
\end{aligned}
\end{equation}
and 
\begin{equation}
\begin{aligned}
	\myLuminosity_\textnormal{Th,night}&=\myKintegral{0}{\myPhase}{\pi}{F_p(T_\textnormal{night})}\\
			&=F_p(T_\textnormal{night})\myKintegral{0}{\pi}{\pi+\myPhase}{1}
\end{aligned}
\end{equation}
respectively. With this knowledge we see that we will be able to use the integration strategies presented in \cref{sec:luminosityFiniteSizeog,sec:luminosityFiniteSizenew}\ to determine the thermal luminosity originating from the fully illuminated, penumbral, and un-illuminated zones for the case of \myECIES. Specifically, we may use the integrations of $ \mu^n $ for the case of $ n=0 $ in our evaluation, where $ \mu $ is given by \cref{eq:mudef}. First, we will explore the thermal emissions as if the exoplanet is modeled as the three zones described in \cref{sec:luminosityFiniteSizeog}\ and then consider the four zone model of \cref{sec:luminosityFiniteSizenew}.

\subsection{Three Zones}\label{sec:thermal3zones}
We will now present the thermal luminosity as a function of phase of each of the three zones and explore the results for each of the five cases described in \cref{tab:5cases}. It should be noted that in each of the following cases the total luminosity due to thermal radiation is given by
\begin{equation}
  \begin{aligned}
        \myLuminosity_{Th,total}(\myPhase)= \myLuminosity_{Th,full}(\myPhase)+ \myLuminosity_{Th,pen}(\myPhase) + \myLuminosity_{Th,un}(\myPhase).
  \end{aligned}
\end{equation}
In addition, we will be making use of the equations defined in \cref{eq:myFi,eq:thephis}.

\begin{table}
	\centering
	\tabulinesep = 3mm
	\caption{\label{tab:thermal5cases} Table of integrals required to determine the thermal luminosity of \myECIES. The variables $ \mySinEta_{1} $ and $ \mySinEta_{2} $ represent $ \sin{\eta_{1}} $ and $ \sin{\eta_{2}} $ defined in \cref{eq:sineta1,eq:sineta2}\ respectively. Each of the azimuthal coordinates are defined in \cref{eq:phizoneboth,eq:philimbs}.}
	\begin{tabu} to \textwidth {
			X[0.7,c]
			X[0.7,c]
			X[0.8,l]
		}
		
		\toprule
		\textbf{Fully Illuminated}&\textbf{Un-illuminated} &\multicolumn{1}{c}{\textbf{Penumbral}} \\
		\midrule
		$\myKintegral{\mySinEta_{1}}{\phi_{full,i}}{\phi_{full,f}}{F_{full}} $ &0&$ \myKintegral{0}{\phi_{limb,i}}{\phi_{limb,f}}{F_{pen}} -\myKintegral{\mySinEta_{1}}{\phi_{full,i}}{\phi_{full,f}}{F_{pen}}$\\ 
		$ \myKintegral{\mySinEta_{1}}{\phi_{full,i}}{\phi_{full,f}}{F_{full}} $&$\myKintegral{-\mySinEta_{2}}{\phi_{un,i}}{\phi_{limb,f}}{F_{un}} $ &$\myKintegral{0}{\phi_{limb,i}}{\phi_{limb,f}}{F_{pen}} -\myKintegral{\mySinEta_{1}}{\phi_{full,i}}{\phi_{full,f}}{F_{pen}}- \myKintegral{-\mySinEta_{2}}{\phi_{un,i}}{\phi_{limb,f}}{F_{pen}} $\\
		$ \myKintegral{\mySinEta_{1}}{\phi_{limb,i}}{\phi_{full,f}}{F_{full}} $&$\myKintegral{-\mySinEta_{2}}{\phi_{un,i}}{\phi_{limb,f}}{F_{un}} $& $\myKintegral{0}{\phi_{limb,i}}{\phi_{limb,f}}{F_{pen}}-\myKintegral{\mySinEta_{1}}{\phi_{limb,i}}{\phi_{full,f}}{F_{pen}} - \myKintegral{-\mySinEta_{2}}{\phi_{un,i}}{\phi_{limb,f}}{F_{pen}} $\\
		 0&$\myKintegral{-\mySinEta_{2}}{\phi_{un,i}}{\phi_{limb,f}}{F_{un}} $&$\myKintegral{0}{\phi_{limb,i}}{\phi_{limb,f}}{F_{pen}} -\myKintegral{-\mySinEta_{2}}{\phi_{un,i}}{\phi_{limb,f}}{F_{pen}}$\\
		 0 & $\myKintegral{-\mySinEta_{2}}{\phi_{un,i}}{\phi_{un,f}}{F_{un}} $ &$ \myKintegral{0}{\phi_{limb,i}}{\phi_{limb,f}}{F_{pen}} -\myKintegral{-\mySinEta_{2}}{\phi_{un,i}}{\phi_{un,f}}{F_{pen}} $\\ 
		\bottomrule
	\end{tabu}
\end{table}

\subsubsection{Case 1}
First, let us consider the thermal luminosity of the fully illuminated zone
\begin{equation}\label{eq:thermfull1}
  \begin{aligned}
        \myLuminosity_\textnormal{Th,full}(\myPhase) &=\myKintegral{\mySinEta_{1}}{\phi_{full,i}}{\phi_{full,f}}{F_{full}}\\
        & = F(T_\textnormal{full})\pi R_p^2(1-\mySinEta_{1}^2 )\cos{\myPhase}.
  \end{aligned}
\end{equation}
The un-illuminated zone is not visible for this particular case; therefore, we need only concern ourselves with the penumbral zone whose thermal luminosity is given by
\begin{equation}\label{eq:thermpen1}
  \begin{aligned}
        \myLuminosity_\textnormal{Th,pen}(\myPhase)& =\myKintegral{0}{\phi_{limb,i}}{\phi_{limb,f}}{F_{pen}} -\myKintegral{\mySinEta_{1}}{\phi_{full,i}}{\phi_{full,f}}{F_{pen}}\\
		& =F(T_\textnormal{pen})\pi R_p^2\left[1-(1-\mySinEta_{1}^2 )\cos{\myPhase}\right].
  \end{aligned}
\end{equation}

As the star-planet separation increases, both $ \mySinEta_{1} $ and $ \mySinEta_{2} $ approach zero. In this limit, we expect the luminosity of the fully illuminated zone to approach \cref{eq:luminositythermalday}, that of the un-illuminated zone to approach \cref{eq:luminositythermalnight}, and that of the penumbral zone to approach zero. Furthermore, in this limit the portion of the orbit governed by each of the five cases will change such that the entirety of the orbit resides within \mycase{3}.

We may assume that as the star-planet separation increases $ \eta_{1} $ and $ \eta_{2} $  approach zero. In addition, the limits on the phase angle, that is $0\le \myPhase\le \eta_{1}$, for which \mycase{1}\ applies shrink such that we may assume that $ \myPhase $ is approximately zero for \mycase{1}\ in this limit. If we now consider the limiting cases for each zone, we find that the fully illuminated zone is approximately given by
\begin{equation}\label{eq:thermfull1limit}
  \begin{aligned}
        \lim_{r\to \infty}\myLuminosity_\textnormal{Th,full}& =F(T_\textnormal{full})\pi R_p^2\cos{\myPhase}\\
			& \approx F(T_\textnormal{full})\pi R_p^2.
  \end{aligned}
\end{equation}
In other words, the fully illuminated zone occupies the entire half hemisphere of the exoplanet visible during full phase, for which $ \myPhase\approx 0 $. In addition, we find that the penumbral zone obeys
\begin{equation}\label{eq:thermpen1limit}
  \begin{aligned}
        \lim_{r\to\infty}\myLuminosity_\textnormal{Th,pen} &=F(T_\textnormal{pen})\pi R_p^2\left[1-\cos{\myPhase}\right]\\
			& \approx 0
  \end{aligned}
\end{equation}
as expected.

\subsubsection{Case 2}
The luminosity due to thermal radiation of the fully illuminated zone for this case is given by \cref{eq:thermfull1}, but the un-illuminated zone is now partially visible such that its luminosity is given by
\begin{equation}\label{eq:thermun2}
  \begin{aligned}
        \myLuminosity_\textnormal{Th,un}(\myPhase)& =\myKintegral{-\mySinEta_{2}}{\phi_{un,i}}{\phi_{limb,f}}{F_{un}} \\
			&\begin{aligned}
				=F(T_\textnormal{un})R_p^2 \bigg[&\myKopalPhi{1}{2}{\myPhase}-\mySinEta_{2}\myF{2}{\myPhase}+\\				
				&\begin{aligned}
					\cos{\myPhase} \bigg(&\myKopalPhi{2}{2}{\myPhase}-\piover2+ \left(\myKopalPhi{4}{2}{\myPhase}-\myKopalPhi{3}{2}{\myPhase}\right)\mySinEta_{2}+ \\
					& \left( \piover2-\myKopalPhi{2}{2}{\myPhase} \right)\mySinEta_{2}^2\bigg) \bigg],
				\end{aligned}
			\end{aligned}
  \end{aligned}
\end{equation}
where $\myF{i}{\vartheta} $ and $ \myKopalPhi{j}{i}{\vartheta} $ are defined in \cref{eq:myFi,eq:thephis}\ respectively. In addition, the luminosity of the penumbral zone is
\begin{equation}\label{eq:thermpen2}
  \begin{aligned}
        \myLuminosity_\textnormal{Th,pen}(\myPhase)& =\myKintegral{0}{\phi_{limb,i}}{\phi_{limb,f}}{F_{pen}} -\\
			&\hspace{0.2in}\myKintegral{\mySinEta_{1}}{\phi_{full,i}}{\phi_{full,f}}{F_{pen}}-\\
			&\hspace{0.2in} \myKintegral{-\mySinEta_{2}}{\phi_{un,i}}{\phi_{limb,f}}{F_{pen}}\\
			&\begin{aligned}
				=F(T_\textnormal{pen})R_p^2&\bigg[\myF{2}{\myPhase}\mySinEta_{2}+ \myKopalPhi{4}{2}{\myPhase}+\piover2+\\
				&\begin{aligned}
				\cos\myPhase \bigg(&\pi(2\mySinEta_{1}^2-1)-\myKopalPhi{2}{2}{\myPhase}+\left(\myKopalPhi{3}{2}{\myPhase}-\myKopalPhi{4}{2}{\myPhase} \right)\mySinEta_{2} + \\
					&\left(\myKopalPhi{2}{2}{\myPhase}-\piover2\right)\mySinEta_{2}^2  \bigg)\bigg].
				\end{aligned}
			\end{aligned}
  \end{aligned}
\end{equation}

Let us now explore the limiting behavior of the un-illuminated and penumbral zones as the star-planet separation becomes large. For large star-planet separations, i.e. for $ \mySinEta_2\approx 0 $, \cref{eq:thermun2}\ does approach  \cref{eq:luminositythermalnight}\ so that the un-illuminated zone behaves like the night side described in \cref{sec:thermDaynight}. Likewise, the luminosity of the penumbral zone becomes
\begin{equation} \label{eq:thermpen2limit}
    \lim_{r\to\infty}\myLuminosity_\textnormal{Th,pen} = F(T_\textnormal{pen})\pi R_p^2\left( \frac{1+\cos{(\myPhase-\pi)}}{2}\right);
\end{equation}
however, in the limit that both $ \mySinEta_{1} $ and $ \mySinEta_{2} $ approach zero the phase angle for which \mycase{2} applies approaches zero. In this case, \cref{eq:thermpen2limit} is approximately zero.

\subsubsection{Case 3}
\mycase{3} includes the majority of the exoplanet's orbit. The luminosity of the fully illuminated zone is similar to that given by \cref{eq:thermun2}:
\begin{equation}\label{eq:thermfull3}
  \begin{aligned}
        \myLuminosity_\textnormal{Th,full}& =\myKintegral{\mySinEta_{1}}{\phi_{limb,i}}{\phi_{full,f}}{F_{full}} \\
			&\begin{aligned}
				=F(T_\textnormal{Full})R_p^2 &\bigg[\myKopalPhi{1}{1}{\myPhase}-\mySinEta_{1}\myF{1}{\myPhase}+ \\
&\begin{aligned}
	\cos{\myPhase} \bigg(&\piover2+ \myKopalPhi{2}{1}{\myPhase}+ \left( \myKopalPhi{4}{1}{\myPhase}-\myKopalPhi{3}{1}{\myPhase}\right)\mySinEta_{1}-\\
		&\left( \piover2+\myKopalPhi{2}{1}{\myPhase}\right)\mySinEta_{1}^2\bigg) \bigg].
\end{aligned}	 
			\end{aligned}
  \end{aligned}
\end{equation}
The luminosity due the un-illuminated zone is still given by \cref{eq:thermun2}. Finally, the luminosity of the thermal radiation from the penumbral zone is given by
\begin{equation}\label{eq:thermpen3}
  \begin{aligned}
        \myLuminosity_\textnormal{Th,pen}(\myPhase)& =\myKintegral{0}{\phi_{limb,i}}{\phi_{limb,f}}{F_{pen}} -\\
			&\hspace{0.2in}\myKintegral{\mySinEta_{1}}{\phi_{limb,i}}{\phi_{full,f}}{F_{pen}}-\\
			&\hspace{0.2in} \myKintegral{-\mySinEta_{2}}{\phi_{un,i}}{\phi_{limb,f}}{F_{pen}}\\
			&\begin{aligned}
				=F(T_\textnormal{pen})R_p^2&\bigg[\myF{1}{\myPhase} \mySinEta_{1}+\myF{2}{\myPhase} \mySinEta_{2}+\myKopalPhi{5}{1}{\myPhase}+\myKopalPhi{5}{2}{\myPhase}+\\
				&\begin{aligned}
					\cos\myPhase \bigg( &\left(\myKopalPhi{2}{1}{\myPhase}+\frac{\pi }{2}\right) \mySinEta_{1}^2-\myKopalPhi{2}{1}{\myPhase}+\left(\myKopalPhi{2}{2}{\myPhase}-\frac{\pi }{2}\right)\mySinEta_{2}^2-\\
					&\myKopalPhi{2}{2}{\myPhase}+\mySinEta_{1} (\myKopalPhi{3}{1}{\myPhase}-\myKopalPhi{4}{1}{\myPhase})+\\
					&\mySinEta_{2} (\myKopalPhi{3}{2}{\myPhase}-\myKopalPhi{4}{2}{\myPhase}) \bigg)\bigg].
				\end{aligned}
			\end{aligned}
  \end{aligned}
\end{equation}

As expected, as the star-planet separation becomes very large compared to the radius of the host star or exoplanet the thermal luminosity of the un-illuminated zone described by \cref{eq:thermun2}\ approaches \cref{eq:luminositythermalnight}, and that of the fully illuminated zone described by \cref{eq:thermfull3}\ approaches \cref{eq:luminositythermalday}. In addition, \cref{eq:thermpen3}\ cancels out exactly. This is in contrast to the other four cases in which the penumbral zone does not cancel exactly, but rather approaches zero because the phase angle is close to zero or $ \pi $. In \mycase{3}\ the penumbral zone must cancel exactly because we may no longer approximate the phase angle as being close to zero or $ \pi $, but rather \mycase{3}\ occupies the entire orbit of the exoplanet. 

\subsubsection{Case 4}
In the preceding cases, at least some portion of the fully illuminated zone was visible to the observer. It is now the case that the fully illuminated zone is no longer visible, but the un-illuminated is still visible to the observer as a consequence of its being larger than the fully illuminated zone. The thermal luminosity from the un-illuminated zone is given by \cref{eq:thermun2}\ and the penumbral zone obeys
\begin{equation}\label{eq:thermpen4}
  \begin{aligned}
        \myLuminosity_\textnormal{Th,pen}(\myPhase)& =\myKintegral{0}{\phi_{limb,i}}{\phi_{limb,f}}{F_{pen}} - \myKintegral{-\mySinEta_{2}}{\phi_{un,i}}{\phi_{limb,f}}{F_{pen}}\\
			&\begin{aligned}
				=F(T_\textnormal{pen})R_p^2&\bigg[\myF{2}{\myPhase}\mySinEta_{2}-\myKopalPhi{1}{2}{\myPhase}+\pi +\\
				&\begin{aligned}
					\cos\myPhase \bigg(&\left(\myKopalPhi{2}{2}{\myPhase}-\frac{\pi }{2}\right)\mySinEta_{2}^2-\\
					&\myKopalPhi{2}{2}{\myPhase}+\mySinEta_{2}(\myKopalPhi{3}{2}{\myPhase}-\myKopalPhi{4}{2}{\myPhase})+\pi\bigg)\bigg].
				\end{aligned}				
			\end{aligned}
  \end{aligned}
\end{equation}

For this case, the limiting behavior of the three zones should be such that the exoplanet is described only by the un-illuminated zone which occupies the half hemisphere visible to the observer. During this portion of the orbit the fully illuminated zone is not visible, which is consistent with the desired limiting behavior. Given that \cref{eq:thermun2}\ from \mycase{2}\ describes the un-illuminated zone we see that its limiting behavior does in fact match that of the night side of the exoplanet given in  \cref{eq:thermnight}. The limiting behavior of the penumbral zone is as follows
\begin{equation}\label{eq:thermpen4limit}
  \begin{aligned}
        \lim_{r\to\infty}\myLuminosity_\textnormal{Th,pen} = F(T_\textnormal{pen})\pi R_p^2\left( \frac{1+\cos{(\myPhase)}}{2}\right),
  \end{aligned}
\end{equation}
but it is clear from the limits on \mycase{4}\ listed in \cref{tab:5cases}\ that as the star-planet separation increases the value of the phase angle for \mycase{4}\ is approximately $ \pi $.  \cref{eq:thermpen4limit}\ is zero for $\myPhase =\pi  $; therefore, the penumbral zone does disappear for large star-planet separations as expected.

\subsubsection{Case 5}
We will conclude our discussion of the derivation of the equations for the luminosity due to the thermal radiation of the exoplanet by considering phase angles between $ \pi-\eta_{2} $ and $ \pi $ radians. During such phases of the exoplanet's orbit the fully illuminated zone is not visible. However, it is now the case that the un-illuminated zone is now completely visible to an observer and its luminosity is given by
\begin{equation}\label{eq:thermun5}
  \begin{aligned}
        \myLuminosity_\textnormal{Th,un}(\myPhase)& =\myKintegral{-\mySinEta_{2}}{\phi_{un,i}}{\phi_{un,f}}{F_{un}} \\
				&=F(T_\textnormal{un})\pi R_p^2(1-\mySinEta_{2}^2)\cos{(\myPhase-\pi)}.
  \end{aligned}
\end{equation}
Finally, the luminosity of the penumbral zone due to thermal radiation is
\begin{equation}\label{eq:thermpen5}
  \begin{aligned}
         \myLuminosity_\textnormal{Th,pen}(\myPhase)& =\myKintegral{0}{\phi_{limb,i}}{\phi_{limb,f}}{F_{pen}} - \myKintegral{-\mySinEta_{2}}{\phi_{un,i}}{\phi_{un,f}}{F_{pen}}\\
			&=F(T_\textnormal{pen})\pi R_p^2\left[ 1+\cos{\myPhase}\left( 1-\mySinEta_{2}^2 \right) \right].
  \end{aligned}
\end{equation}

As in \mycase{4}, we may assume that as the star-planet separation increases the phase angle for which \mycase{5}\ applies approaches  $ \pi $. In such a situation, \cref{eq:thermun5}\ approaches $ \pi F(T_{un}) R_p^2 $, i.e. it behaves like a half hemisphere with an effective temperature equal to the un-illuminated zone's temperature. In addition, \cref{eq:thermpen5}\ approaches zero as required.

\subsubsection{Fractional Flux Comparison}
\begin{figure}[tbh!]
\centering
\vspace{-1em}
\subfloat[][\label{fig:K91totaltherm5cases}Total fractional flux due to thermal radiation.]{\includegraphics[trim={15 0 15 0},clip,width=0.5\linewidth]{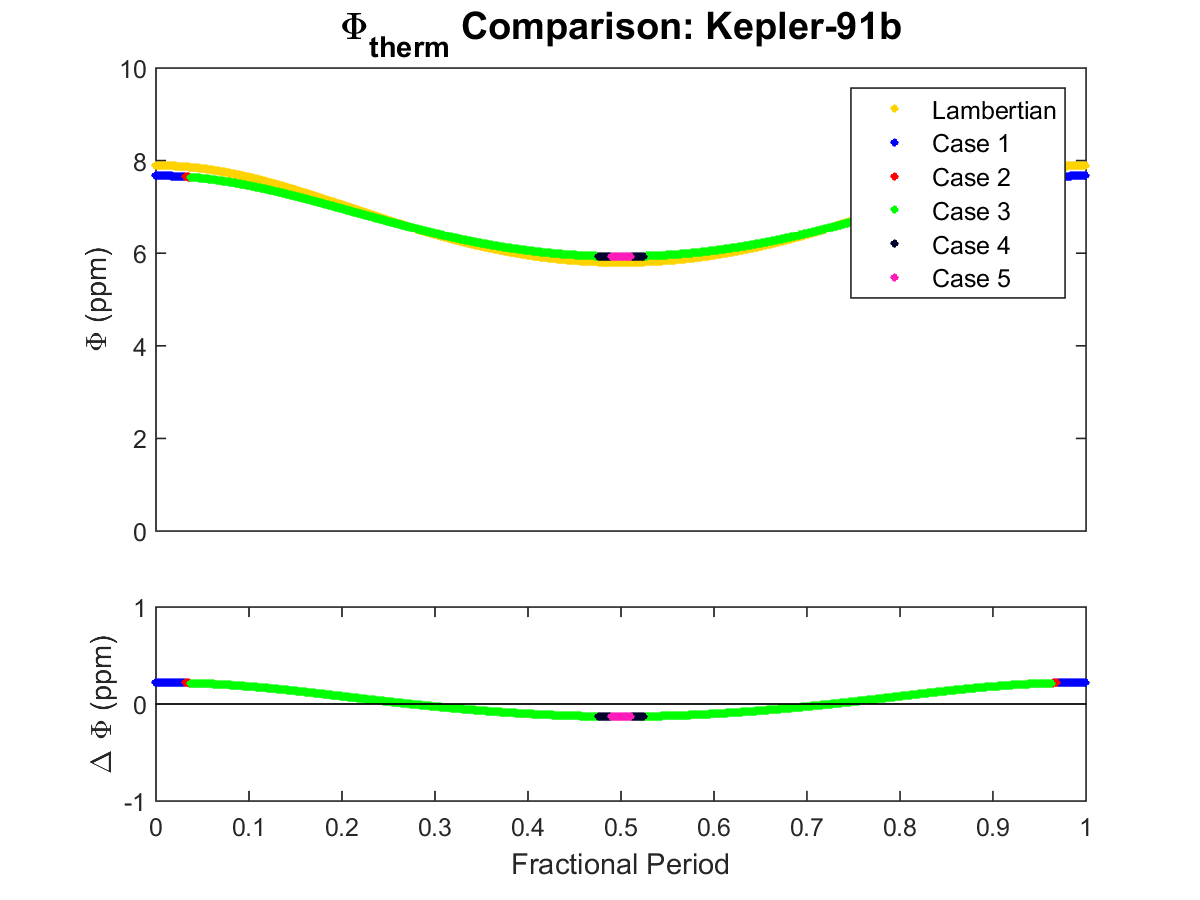}} 
\hfill
\subfloat[][\label{fig:K91fulltherm5cases}Fractional flux due to the fully illuminated zone.]{\includegraphics[trim={15 0 15 0},clip,width=0.5\linewidth]{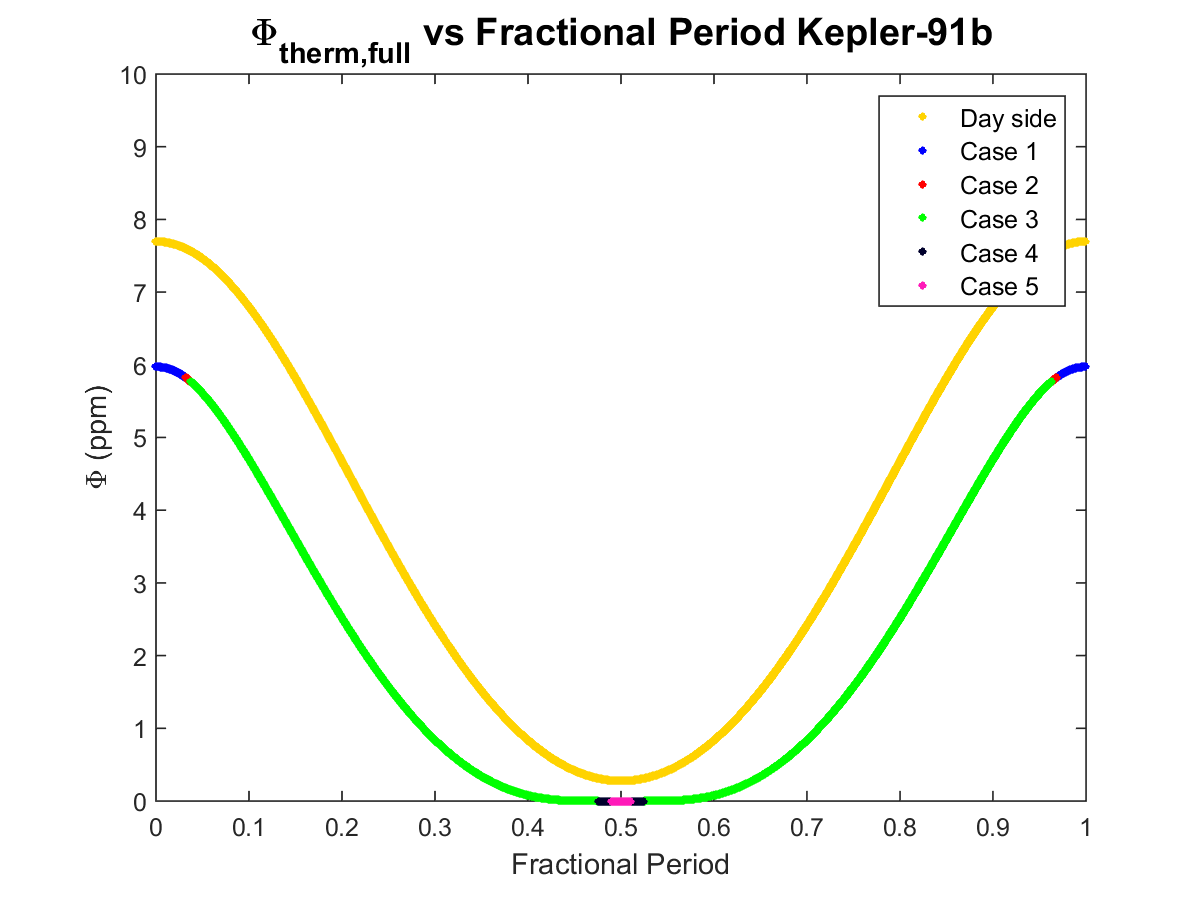}}

\vspace{-1em}
\subfloat[][\label{fig:K91untherm5cases}Fractional flux due to the un-illuminated zone.]{\includegraphics[trim={15 0 15 0},clip,width=0.5\linewidth]{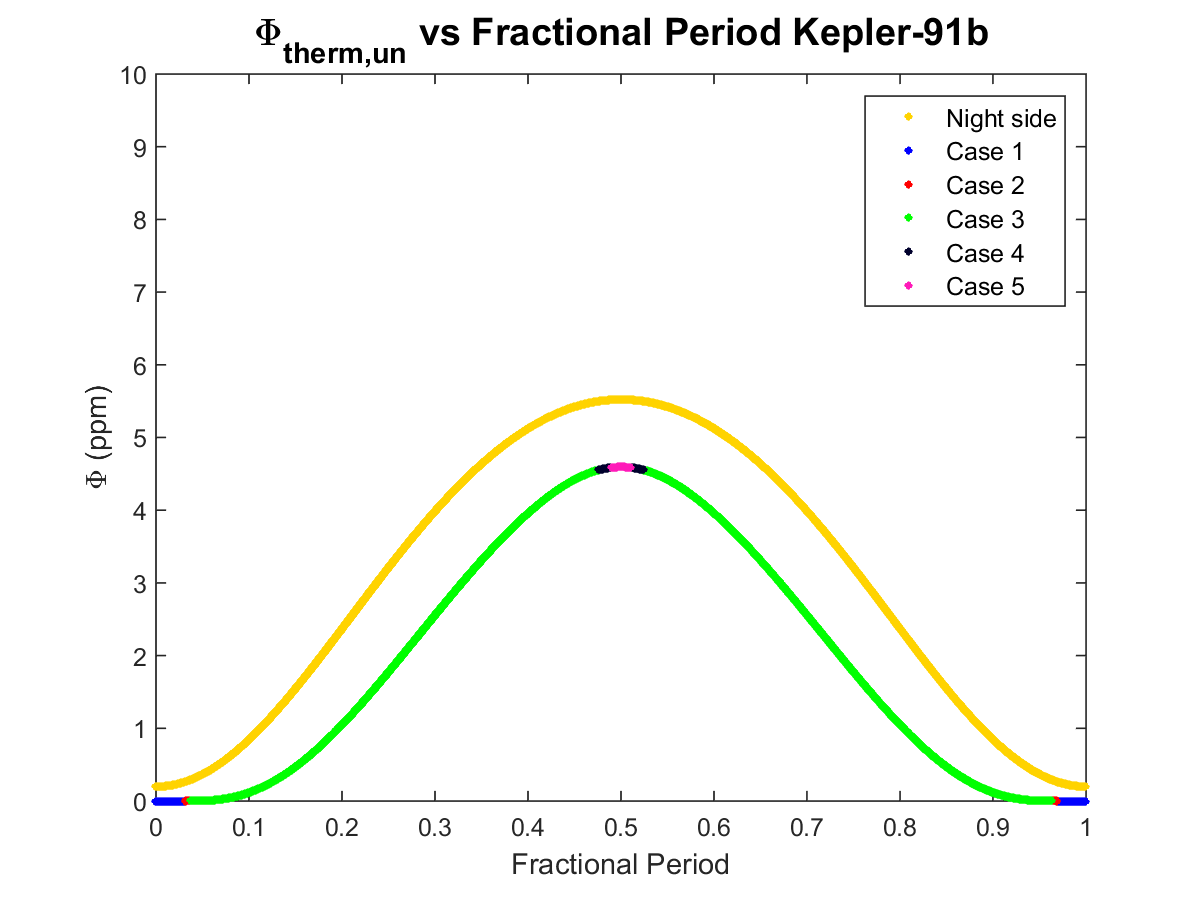}}
\hfill
\subfloat[][\label{fig:K91pentherm5cases}Fractional flux due to the penumbral zone.]{\includegraphics[trim={15 0 15 0},clip,width=0.5\linewidth]{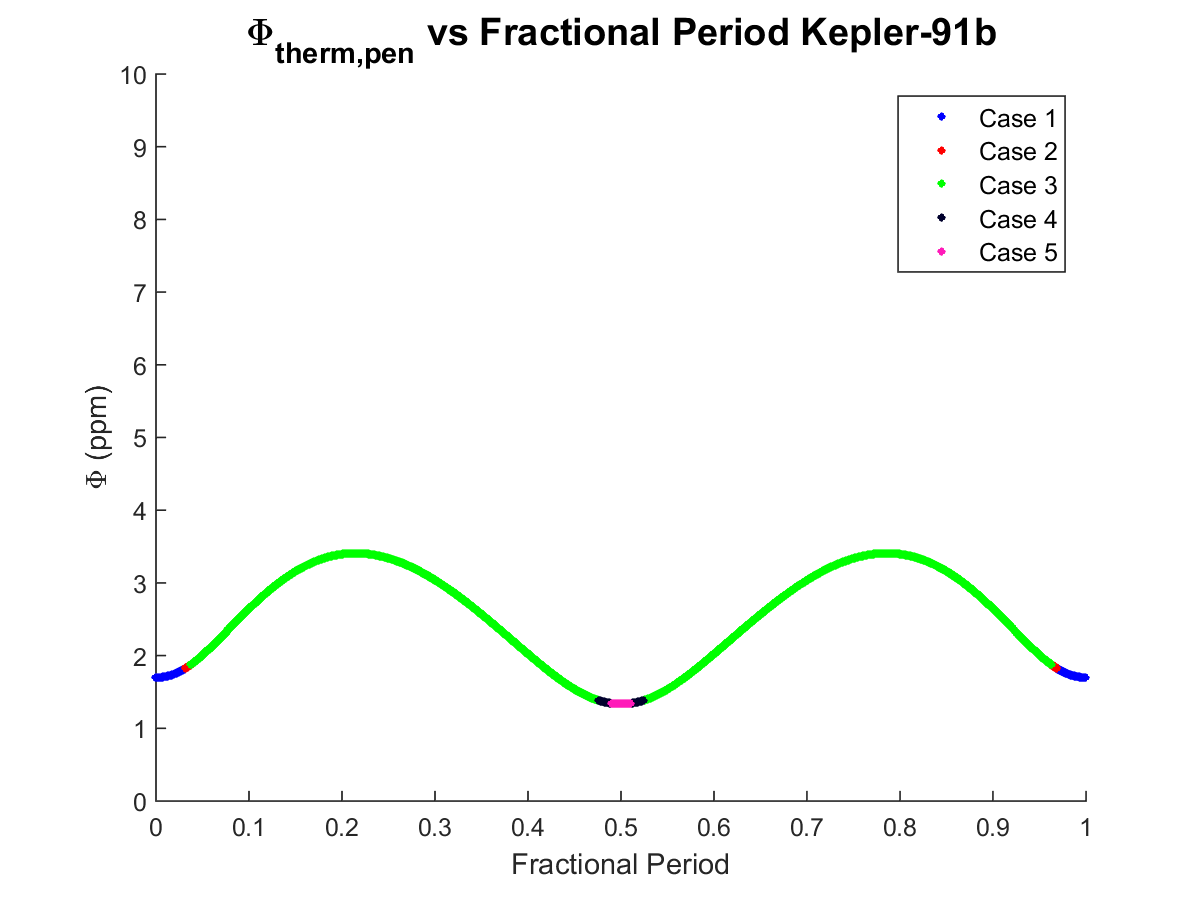}}
\vspace{-0.25em}
\caption{\label{fig:K91thermal5cases}In the above figures we have plotted the fractional flux due to thermal radiation of an exoplanet. \cref{fig:K91totaltherm5cases}\ shows a comparison between the total fractional flux due to thermal radiation from all three zones as a function of phase. Also plotted is the fractional flux due to thermal radiation assuming Lambertian plane parallel ray illumination such that the thermal radiation may be described by \cref{eq:thermaltotal}. The bottom panel of \cref{fig:K91totaltherm5cases}\ shows the difference between the fractional flux of the plane parallel ray model and that of the three zone model. Next, in \cref{fig:K91fulltherm5cases}\ we have plotted a comparison between the fractional flux due to the fully illuminated zone and that of the dayside of the exoplanet if we were to assume a plane parallel ray model described by \cref{eq:normalizedfluxthermalday}. Similarly, in \cref{fig:K91untherm5cases}\ we have a comparison between the fractional flux due to the un-illuminated zone and the night side of an exoplanet described by \cref{eq:thermnight}. Finally, in \cref{fig:K91pentherm5cases}\ is a plot of the fractional flux due to the penumbral zone. To produce the above plots we used the parameters given in \cref{tab:K91params}\ with the exception of exoplanet temperature. The day side temperature of the fully illuminated zone temperature are both 2441.7 K, and the un-illuminated zone and night side temperatures are both 2348.2 K, Table 3 of \protect\cite{Placek2015}. Finally, the penumbral zone temperature is the average of the day side and night side temperatures reported in Table 3 of \protect\cite{Placek2015}\ at 2395.0 K.}
\end{figure}

For each of the foregoing cases one may determine the total normalized flux due to the thermal radiation of the exoplanet via,
\begin{equation}
	\Phi_{therm}(\myPhase) =\frac{\myLuminosity_{Th,full}(\myPhase)+ \myLuminosity_{Th,pen}(\myPhase) + \myLuminosity_{Th,un}(\myPhase)}{F_s(T_{eff})\pi R_s^2 }
\end{equation} 
as was done to create \cref{fig:K91totaltherm5cases}. Also shown in \cref{fig:K91thermal5cases}\ are comparisons to the fractional flux produced by the fully illuminated zone, un-illuminated zone and the penumbral zone for Kepler-91b using the parameters given in \cref{tab:K91params}. From these plots we see that with current technology one would not be able to distinguish the two models for the thermal radiation of the exoplanet. For exoplanets with greater temperature differences it may be possible to distinguish the two models. In future work, we will explore the possibility further.

Let us now take a deeper look at the plot shown in \cref{fig:K91thermal5cases}. First, note that the fractional luminosity of the fully illuminated zone is always less than that produced by the day side of the exoplanet in  \cref{fig:K91fulltherm5cases}\ because the fractional area of the fully illuminated zone is always less than that of a half hemisphere. The same is true of the fractional luminosity of the un-illuminated zone as compared to the night side of the exoplanet in \cref{fig:K91untherm5cases}. We note here that neither the fractional flux due to the day side or the night side of the exoplanet are zero at any point in the exoplanet's orbit because the inclination of Kepler-91b is such that the half hemisphere of either side is never outside of the line of sight. This is not the case for either the fully illuminated or un-illuminated zones. 

In \cref{fig:K91pentherm5cases}, we see that the fractional flux due to the thermal radiation of the penumbral zone is never zero because some portion of the penumbral zone is always visible. In addition, we see that it's maximum occurs twice per orbit during phases within \mycase{3}. We note here that the fractional flux due to the reflected light of the penumbral zone will likely follow a similarly shaped curve.

\begin{figure}[hbt!]
\centering
\subfloat[][\label{fig:K91totalthermMulti5cases}Total fractional flux due to thermal radiation.]{\includegraphics[trim={40 0 40 0},clip,width=0.5\linewidth]{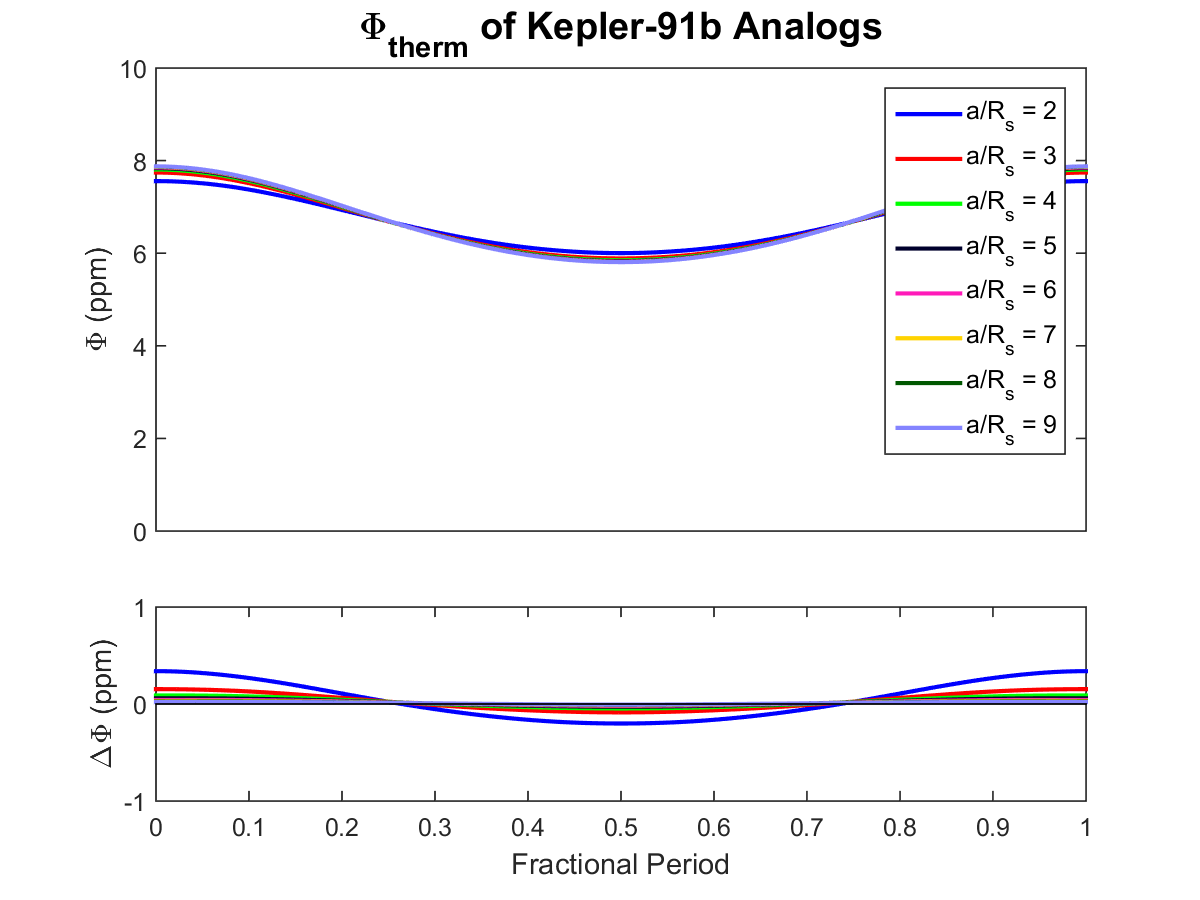}}
\hfill
\subfloat[][\label{fig:K91fullthermMulti5cases}Fractional flux due to the fully illuminated zone.]{\includegraphics[trim={40 0 40 0},clip,width=0.5\linewidth]{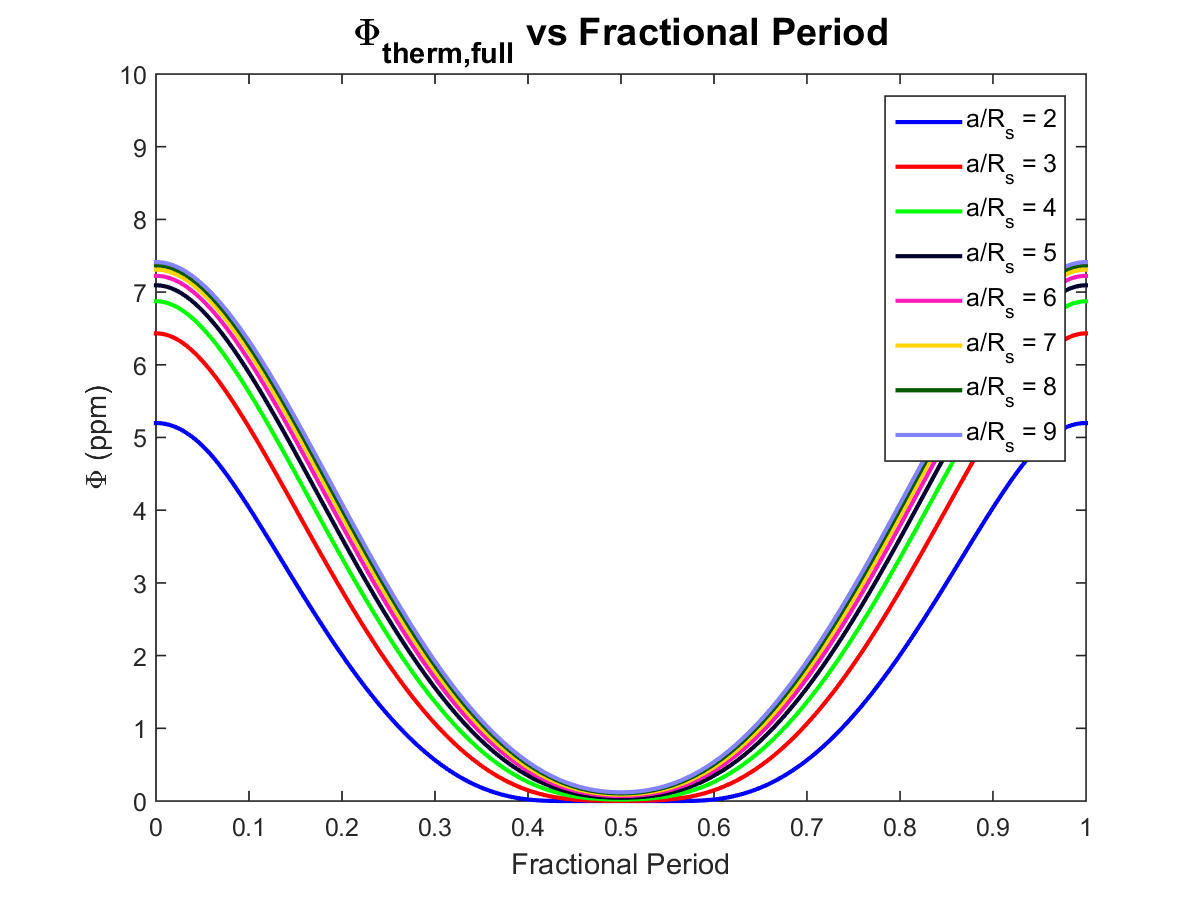}}

\subfloat[][\label{fig:K91unthermMulti5cases}Fractional flux due to the un-illuminated zone.]{\includegraphics[trim={40 0 40 0},clip,width=0.5\linewidth]{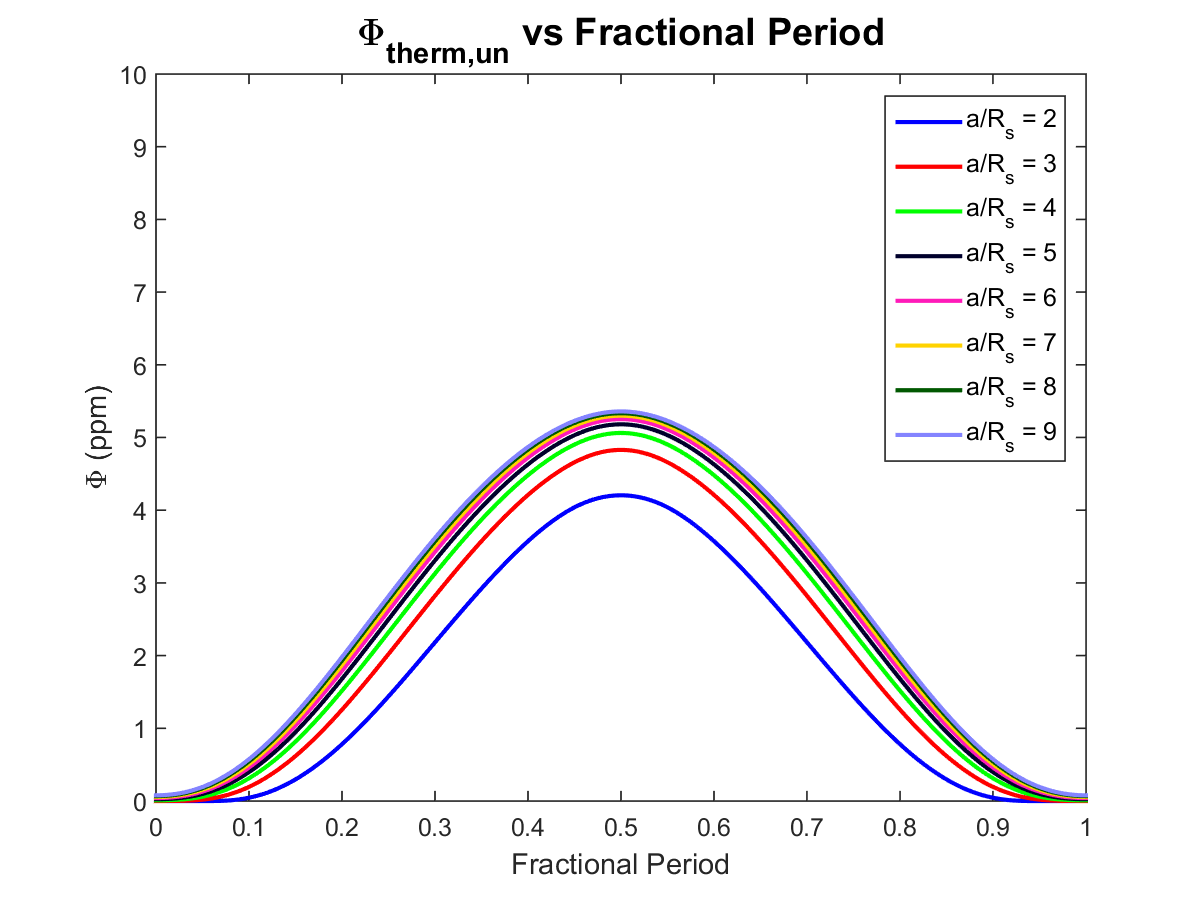}}
\hfill
\subfloat[][\label{fig:K91penthermMulti5cases}Fractional flux due to the penumbral zone.]{\includegraphics[trim={40 0 40 0},clip,width=0.5\linewidth]{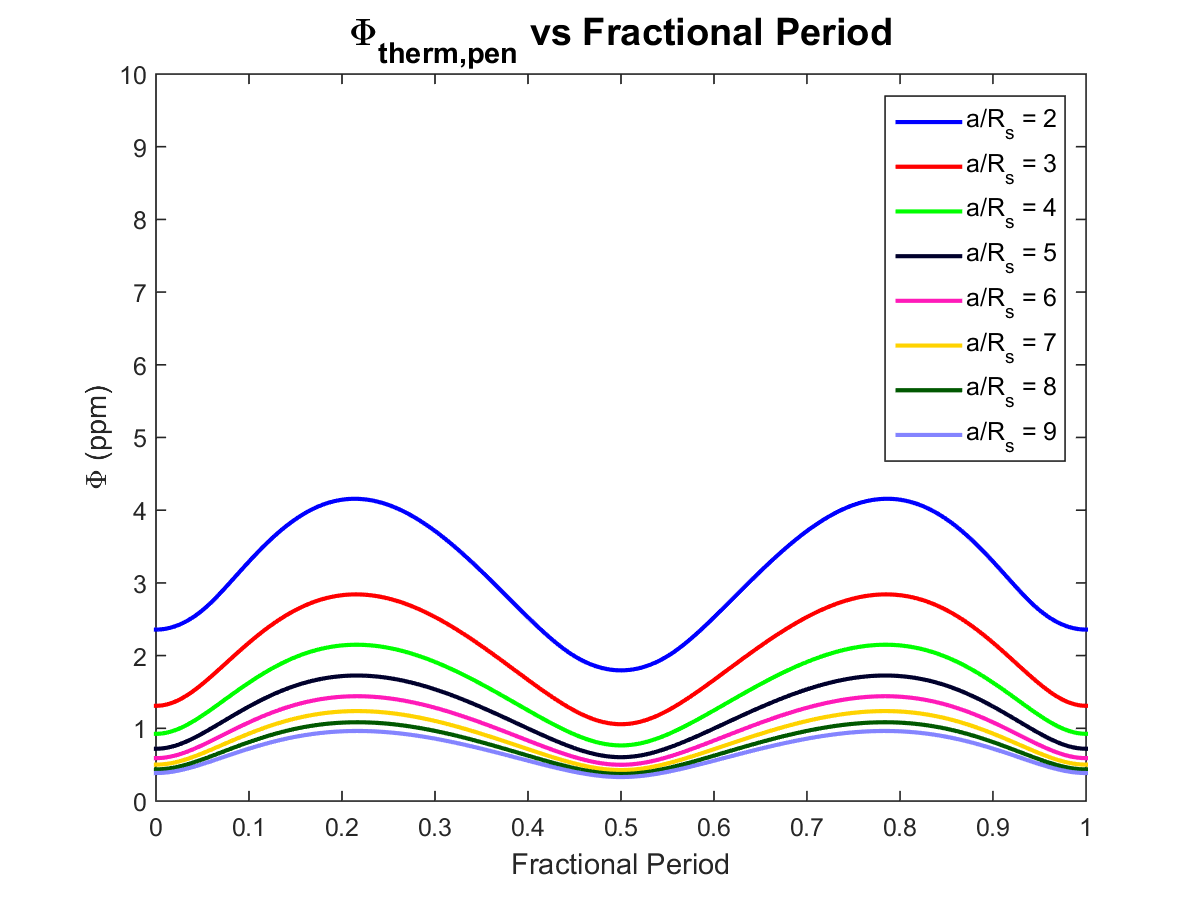}}
\caption{\label{fig:K91thermalMulti5cases} Each of the above plots the fractional flux due to thermal radiation with parameter values described in \cref{tab:K91params}\ and for the same temperatures as those used in \cref{fig:K91thermal5cases}\ where we have varied the normalized semi-major axis, $ a/R_s $. Here we see that the fractional flux emitted by the penumbral zone decreases as the star-planet separation increases because the size of the zone is inversely proportional to the star-planet separation. Furthermore, note that the opposite is true for both the fully illuminated and un-illuminated zones.}
\end{figure}
\cref{fig:K91thermalMulti5cases}\ contains plots of the fractional flux due to the total thermal radiation and due to each of the three zones for Kepler-91b analogs. Here we see that the difference between modeling the thermal radiation with three zones and modeling it as a dayside and night side decreases quickly as the normalized semi-major axis increases. In addition, we see that both the fractional flux due to the fully illuminated and un-illuminated zones increases with increasing star-planet separation. This is because the fractional area of each zone increases and approaches one half with increasing star-planet separation. Likewise, we see that the fractional flux due to thermal radiation within the penumbral zone decreases because the size of the penumbral zone decreases with increasing star-planet separation.

\subsection{Four Zones}\label{sec:thermal4zones}
The principles discussed in \cref{sec:thermal3zones}\ still apply now that we are considering four zones, for example, each of the two penumbral zones still vanish as the star-planet separation increases. To determine the total luminosity of the exoplanet due to thermal emissions we may use the integrations described in \cref{tab:7casesthermal}\ where the total luminosity is given by the sum
\begin{equation}
	\myLuminosity_{therm}(\myPhase) = \myLuminosity_{therm,full}(\myPhase)+\myLuminosity_{therm,un}(\myPhase)+\myLuminosity_{therm,pen1}(\myPhase)+\myLuminosity_{therm,pen2}(\myPhase).
\end{equation}
In addition, the fractional flux of a given zone is given by
\begin{equation}
	\Phi_{therm,zone}(\myPhase) = \frac{\myLuminosity_{therm,zone}(\myPhase)}{F_s(T_{eff})\pi R_s^2 }
\end{equation}
so that the total fractional flux due to thermal radiation is given by
\begin{equation} \label{eq:phithermfinitetotal}
    \Phi_{therm}(\myPhase)=\Phi_{therm,full}(\myPhase)+\Phi_{therm,un}(\myPhase)+
\Phi_{therm,pen1}(\myPhase)+
\Phi_{therm,pen2}(\myPhase).
\end{equation}
\begin{sidewaystable}
\centering\small
	\tabulinesep = 3mm
	\begin{longtabu} to \textwidth {
			X[0.37,l]
			X[0.35,l]
			X[0.45,l]
			X[0.45,l]
		}
		\caption{\label{tab:7casesthermal}Shown is a table describing each of the seven cases and the required application of the operator $ \myKintegral{\mySinEta}{\phi_i}{\phi_f}{F_{zone}}$. The total thermal luminosity is the sum of the luminosity from the fully illuminated and penumbral zones, $ \myLuminosity_{therm}(\myPhase) = \myLuminosity_{therm,full}(\myPhase)+\myLuminosity_{therm,un}(\myPhase)+\myLuminosity_{therm,pen1}(\myPhase)+\myLuminosity_{therm,pen2}(\myPhase) $.}\\
		\toprule
		\multicolumn{1}{c}{\textbf{Fully Illuminated}, $ \myLuminosity_{therm,full}$ } &\multicolumn{1}{c}{\textbf{Un-illuminated}, $ \myLuminosity_{therm,un} $ }&\multicolumn{1}{c}{\textbf{Penumbral 1}, $ \myLuminosity_{therm,pen1}$ } &\multicolumn{1}{c}{\textbf{Penumbral 2}, $ \myLuminosity_{therm,pen2}$ } \\
		\midrule
		\endfirsthead
		\multicolumn{4}{c}
		{\textbf{Integral Case Descriptions} -- \textit{Continued from previous page}} \\
		\midrule
		\multicolumn{1}{c}{\textbf{Fully Illuminated}, $ \myLuminosity_{therm,full}$ } &\multicolumn{1}{c}{\textbf{Un-illuminated}, $ \myLuminosity_{therm,un}$ }&\multicolumn{1}{c}{\textbf{Penumbral 1}, $ \myLuminosity_{therm,pen1} $ } &\multicolumn{1}{c}{\textbf{Penumbral 2}, $ \myLuminosity_{therm,pen2}$ } \\
		\midrule
		\endhead
		\midrule
		\multicolumn{4}{c}{\textit{Continued on next page}} \\ 
		\endfoot
		\bottomrule
		\multicolumn{4}{c}{\textit{End of  $ \myIntensityDis_{therm,zone} $ table}}
		\endlastfoot
		$ \myKintegral{\mySinEta_{1}}{\phi_{full,i}}{\phi_{full,f}}{F_{full}} $& \multicolumn{1}{c}{0}&$ \myKintegral{\mySinEta_{\mypen1limit}}{\phi_{\mypen1limit,i}}{\phi_{\mypen1limit,f}}{F_{pen1}} - \myKintegral{\mySinEta_{1}}{\phi_{full,i}}{\phi_{full,f}}{F_{pen1}} $ &$\myKintegral{0}{\phi_{limb,i}}{\phi_{limb,f}}{F_{pen2}}- \myKintegral{\mySinEta_{\mypen1limit}}{\phi_{\mypen1limit,i}}{\phi_{\mypen1limit,f}}{F_{pen2}}$\\ 
		$ \myKintegral{\mySinEta_{1}}{\phi_{full,i}}{\phi_{full,f}}{F_{full}} $ & \multicolumn{1}{c}{0}&$ \myKintegral{\mySinEta_{\mypen1limit}}{\phi_{limb,i}}{\phi_{\mypen1limit,f}}{F_{pen1}} - \myKintegral{\mySinEta_{1}}{\phi_{full,i}}{\phi_{full,f}}{F_{pen1}} $& $\myKintegral{0}{\phi_{limb,i}}{\phi_{limb,f}}{F_{pen2}} -\myKintegral{\mySinEta_{\mypen1limit}}{\phi_{limb,i}}{\phi_{\mypen1limit,f}}{F_{pen1}}$\\ 
		$ \myKintegral{\mySinEta_{1}}{\phi_{full,i}}{\phi_{full,f}}{F_{full}} $  &   $ \myKintegral{-\mySinEta_{2}}{\phi_{un,i}}{\phi_{limb,f}}{F_{un}} $&$ \myKintegral{\mySinEta_{\mypen1limit}}{\phi_{limb,i}}{\phi_{\mypen1limit,f}}{F_{pen1}} - \myKintegral{\mySinEta_{1}}{\phi_{full,i}}{\phi_{full,f}}{F_{pen1}}$  & $\myKintegral{0}{\phi_{limb,i}}{\phi_{limb,f}}{F_{pen2}} -\myKintegral{\mySinEta_{\mypen1limit}}{\phi_{limb,i}}{\phi_{\mypen1limit,f}}{F_{pen2}}-\myKintegral{-\mySinEta_{2}}{\phi_{un,i}}{\phi_{limb,f}}{F_{pen2}} $\\ 
		$ \myKintegral{\mySinEta_{1}}{\phi_{limb,i}}{\phi_{full,f}}{F_{full}} $ & $ \myKintegral{-\mySinEta_{2}}{\phi_{un,i}}{\phi_{limb,f}}{F_{un}} $&$ \myKintegral{\mySinEta_{\mypen1limit}}{\phi_{limb,i}}{\phi_{\mypen1limit,f}}{F_{pen1}} - \myKintegral{\mySinEta_{1}}{\phi_{limb,i}}{\phi_{full,f}}{F_{pen1}}$ &$\myKintegral{0}{\phi_{limb,i}}{\phi_{limb,f}}{F_{pen2}}  -\myKintegral{\mySinEta_{\mypen1limit}}{\phi_{limb,i}}{\phi_{\mypen1limit,f}}{F_{pen2}}-\myKintegral{-\mySinEta_{2}}{\phi_{un,i}}{\phi_{limb,f}}{F_{pen2}} $\\ 
		\multicolumn{1}{c}{0}  &  $ \myKintegral{-\mySinEta_{2}}{\phi_{un,i}}{\phi_{limb,f}}{F_{un}} $& $  \myKintegral{\mySinEta_{\mypen1limit}}{\phi_{limb,i}}{\phi_{\mypen1limit,f}}{F_{pen1}} $  & $\myKintegral{0}{\phi_{limb,i}}{\phi_{limb,f}}{F_{pen2}}  -\myKintegral{\mySinEta_{\mypen1limit}}{\phi_{limb,i}}{\phi_{\mypen1limit,f}}{F_{pen2}}-\myKintegral{-\mySinEta_{2}}{\phi_{un,i}}{\phi_{limb,f}}{F_{pen2}}$\\ 
		\multicolumn{1}{c}{0} & $ \myKintegral{-\mySinEta_{2}}{\phi_{un,i}}{\phi_{un,f}}{F_{un}} $& $  \myKintegral{\mySinEta_{\mypen1limit}}{\phi_{limb,i}}{\phi_{\mypen1limit,f}}{F_{pen1}} $  &$\myKintegral{0}{\phi_{limb,i}}{\phi_{limb,f}}{F_{pen2}}  -\myKintegral{\mySinEta_{\mypen1limit}}{\phi_{limb,i}}{\phi_{\mypen1limit,f}}{F_{pen2}}-\myKintegral{-\mySinEta_{2}}{\phi_{un,i}}{\phi_{un,f}}{F_{pen2}}$ \\ 
		 \multicolumn{1}{c}{0}& $ \myKintegral{-\mySinEta_{2}}{\phi_{un,i}}{\phi_{un,f}}{F_{un}} $ &\multicolumn{1}{c}{0}&$\myKintegral{0}{\phi_{limb,i}}{\phi_{limb,f}}{F_{pen2}}-\myKintegral{-\mySinEta_{2}}{\phi_{un,i}}{\phi_{un,f}}{F_{pen2}} $ 		
	\end{longtabu}
\vspace{-3em}
\end{sidewaystable}

\begin{figure}[tbh!]
\centering
\subfloat[][\label{fig:K91totaltherm7cases}Kepler-91b analog  $i$ = 90\degree.]{\includegraphics[trim={40 0 40 0}, clip, width=0.5\linewidth]{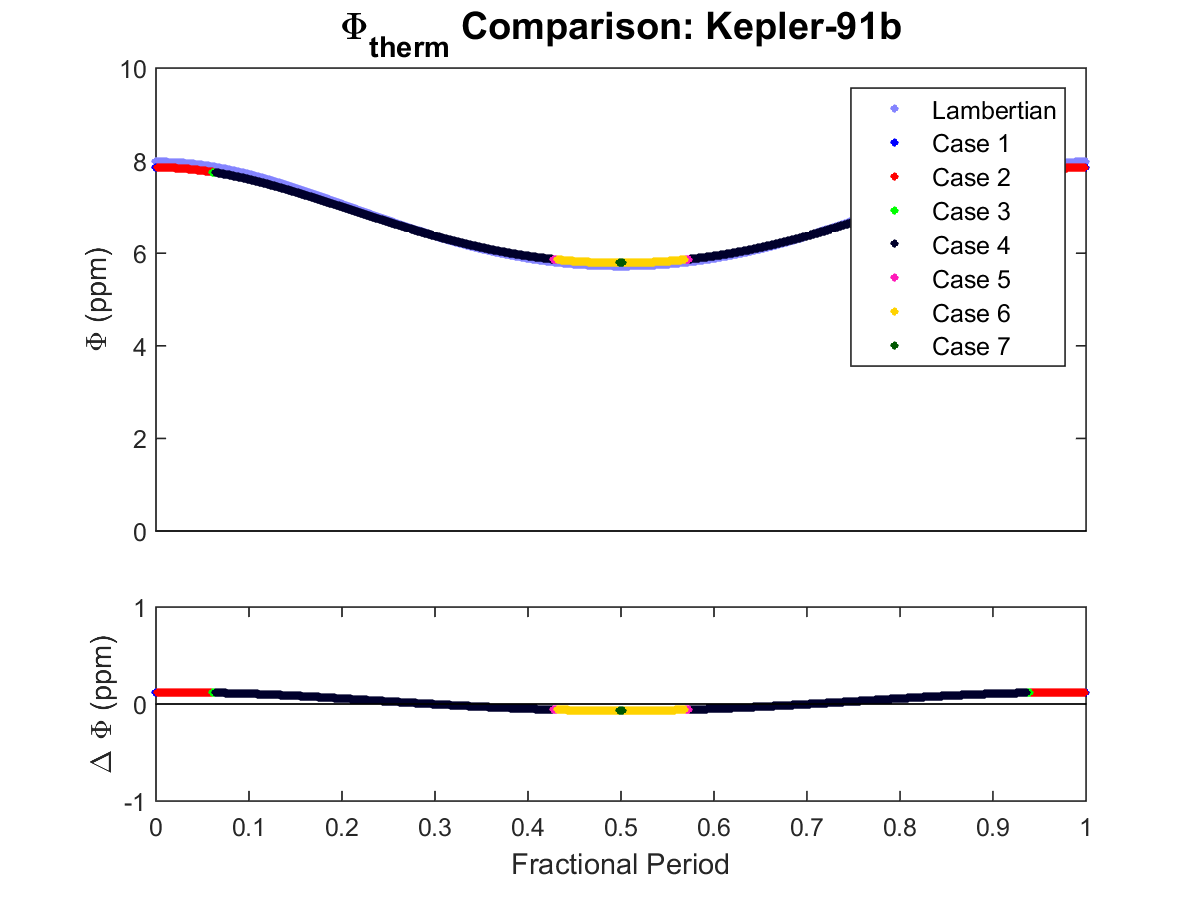}}
\hfill
\subfloat[][\label{fig:K91totalthermMulti7cases}Kepler-91b analogs with $i$ = 90\degree\ and varying star-planet separation.]{\includegraphics[trim={40 0 40 0}, clip, width=0.5\linewidth]{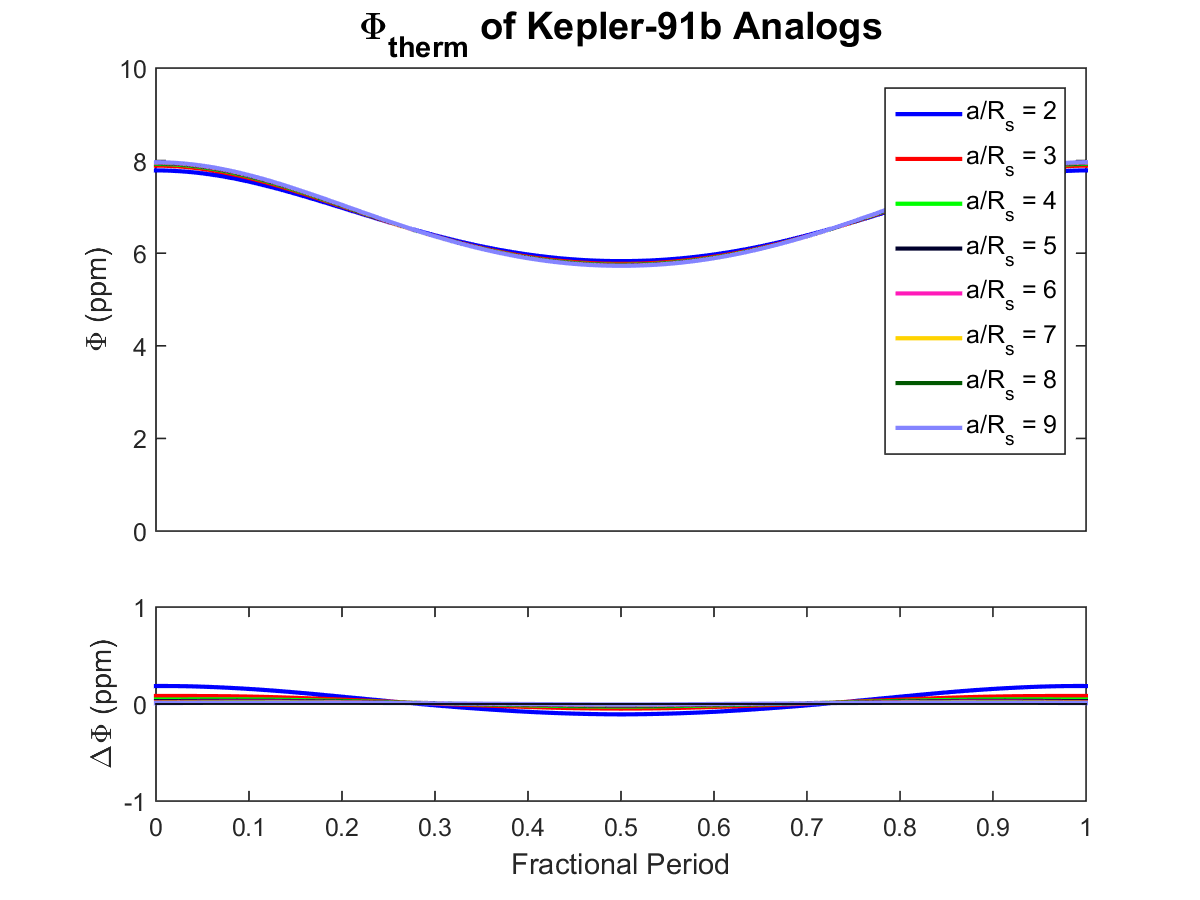}}
\caption{\label{fig:K91thermaltotal7cases}In each of the above figures we have plotted the total fractional flux due to the thermal radiation of Kepler-91b, see \cref{eq:phithermfinitetotal}. Plotted in \cref{fig:K91totaltherm7cases}\ is the fractional flux given the parameters in \cref{tab:K91params}\ with the exception that the inclination is now 90\degree. For the Lambertian curve the two zone model described in \cref{sec:thermDaynight} was used with a day side temperature of 2441.7 K and night size temperature of 2348.2 K, Table 3 of \cite{Placek2015}. In both plots the fully illuminated zone temperature was set to 2441.K and the un-illuminated zone to 2348.2 K. To approximate a gradual temperature gradient the temperature of Penumbral Zone One was set to 2418.3 K and that of Penumbral Zone Two was set to 2371.6 K. In addition,  \cref{fig:K91totalthermMulti7cases}\ considers edge-on Kepler-91b analogs where we have also varied the normalized semi-major axis, $ a/R_s$.}
\end{figure}

\cref{fig:K91thermaltotal7cases}\ shows the effect of modeling the thermal radiation using four zones for Kepler-91b for an edge-on orbit. We are considering an edge-on orbit for illustrative purposes because the inclination of Kepler-91b is such that phase angles within \mycase{1}\ and \mycase{7}\ are not present. The first plot compares all seven cases to that of the fractional luminosity produced by an exoplanet experiencing Lambertian plane parallel incident radiation as in \cref{eq:thermaltotal}. In the second plot we have also varied the star-planet separation. Here we have not considered how the star-planet separation may influence the exoplanet's temperature; therefore, we see very little difference between the curves shown in \cref{fig:K91thermaltotal7cases}. The primary reason behind the differences between each curve is the influence the star-planet separation has on the size of each of the four zones. As the star-planet separation increases the size of the fully illuminated and un-illuminated zones increase until the two penumbral zones no longer exist and we can model the exoplanet using only a day side and night side temperature.

\begin{figure}[tbh!]
\centering
\subfloat[][\label{fig:K91fulltherm7cases}Fractional flux due to the fully illuminated zone.]{\includegraphics[trim={40 0 40 0}, clip, width=0.5\linewidth]{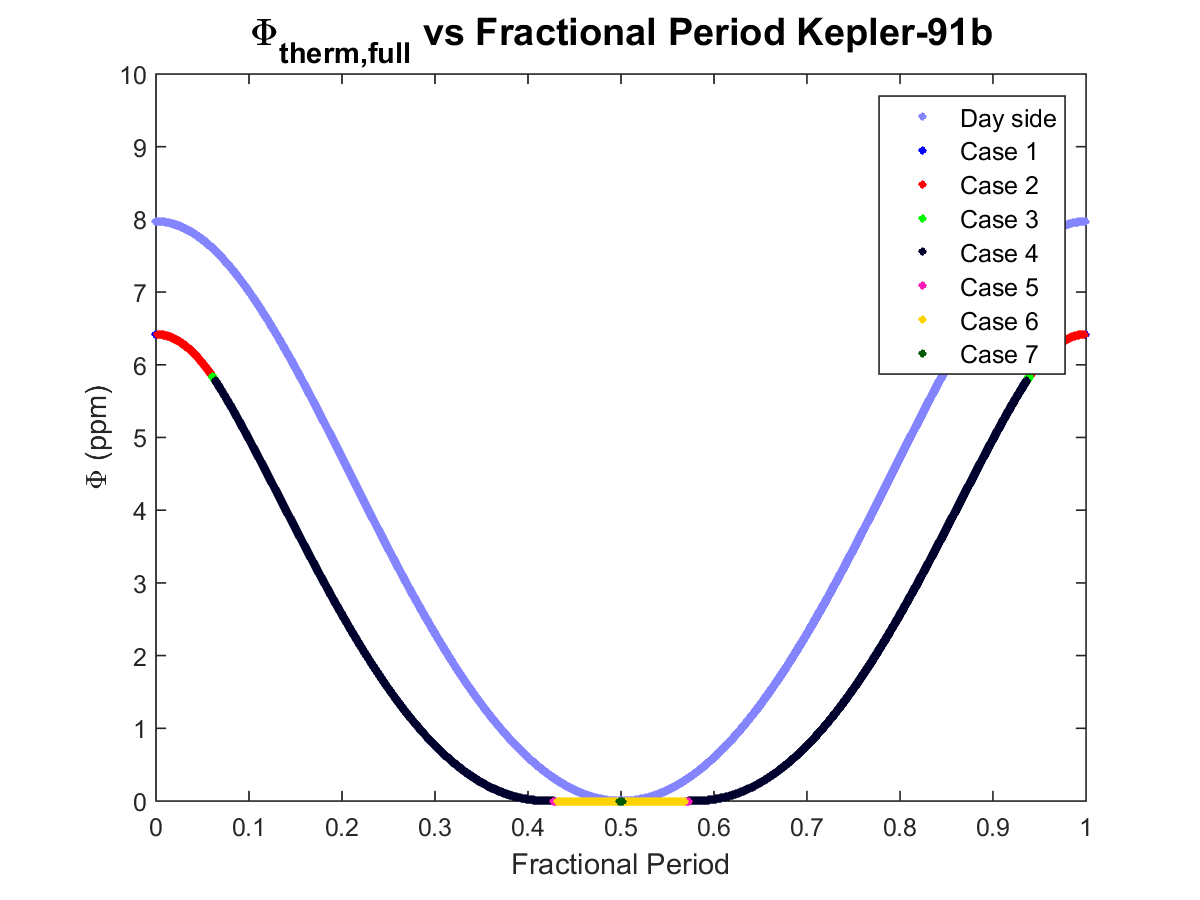}}
\hfill
\subfloat[][\label{fig:K91untherm7cases}Fractional flux due to the un-illuminated zone.]{\includegraphics[trim={40 0 40 0}, clip,width=0.5\linewidth]{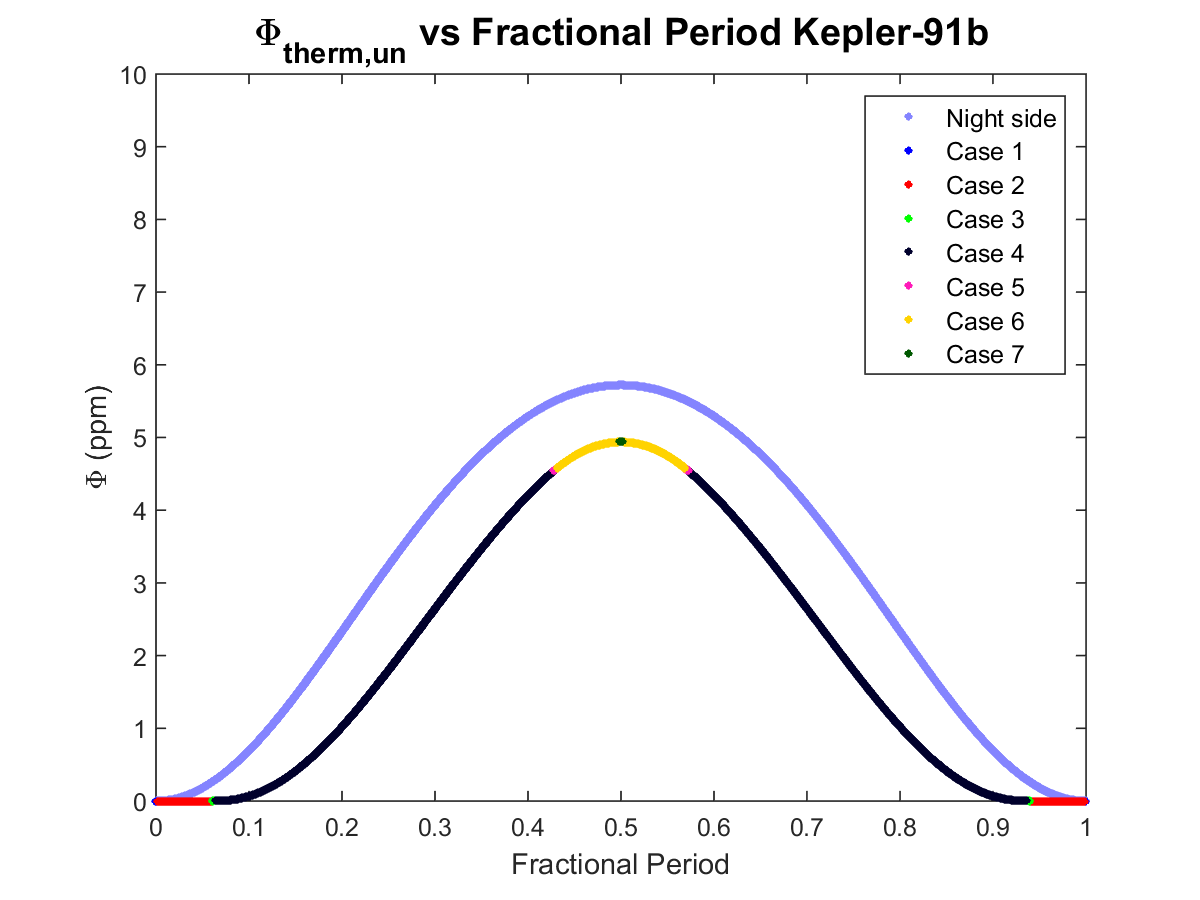}}

\subfloat[][\label{fig:K91pen1therm7cases}Fractional flux due to the Penumbral Zone One.]{\includegraphics[trim={40 0 40 0}, clip,width=0.5\linewidth]{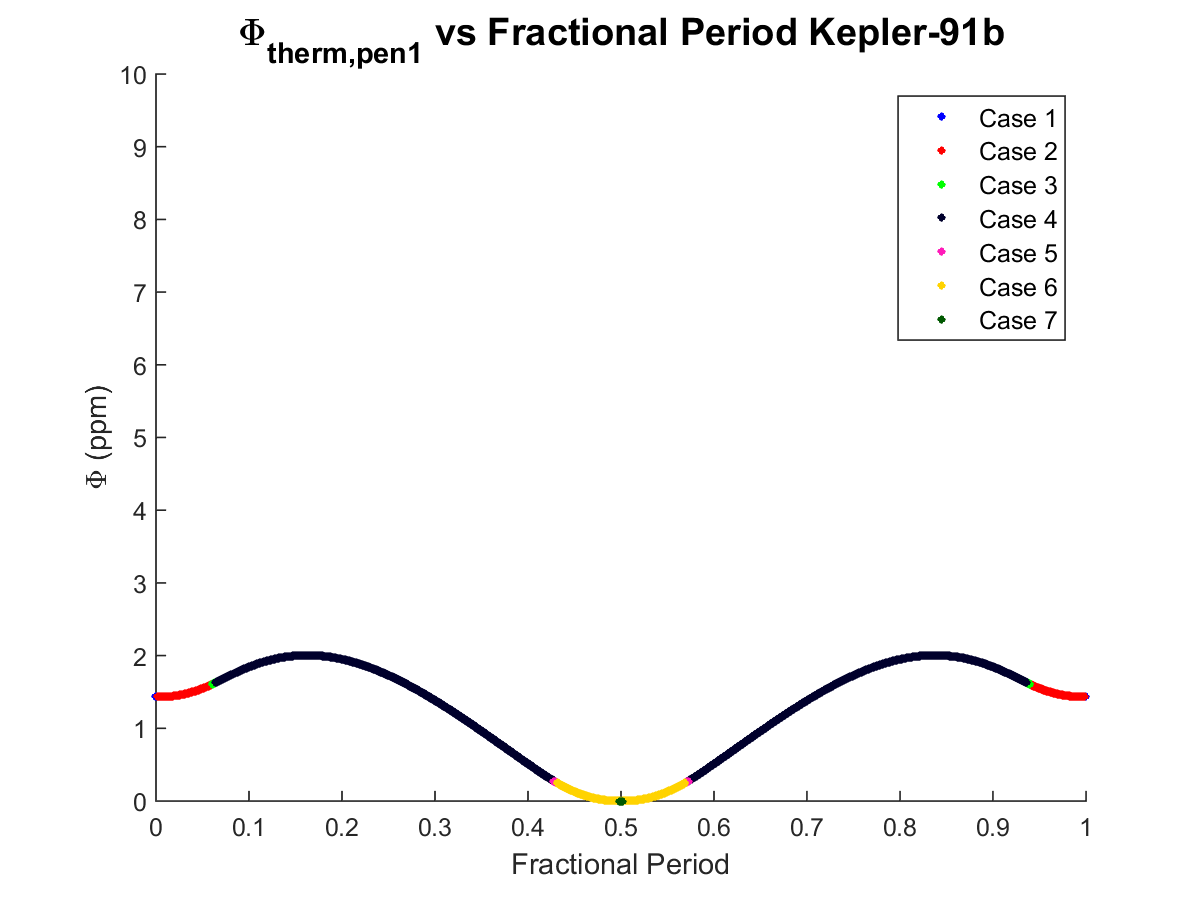}}
\hfill
\subfloat[][\label{fig:K91pen2therm7cases}Fractional flux due to the Penumbral Zone Two.]{\includegraphics[trim={40 0 40 0}, clip,width=0.5\linewidth]{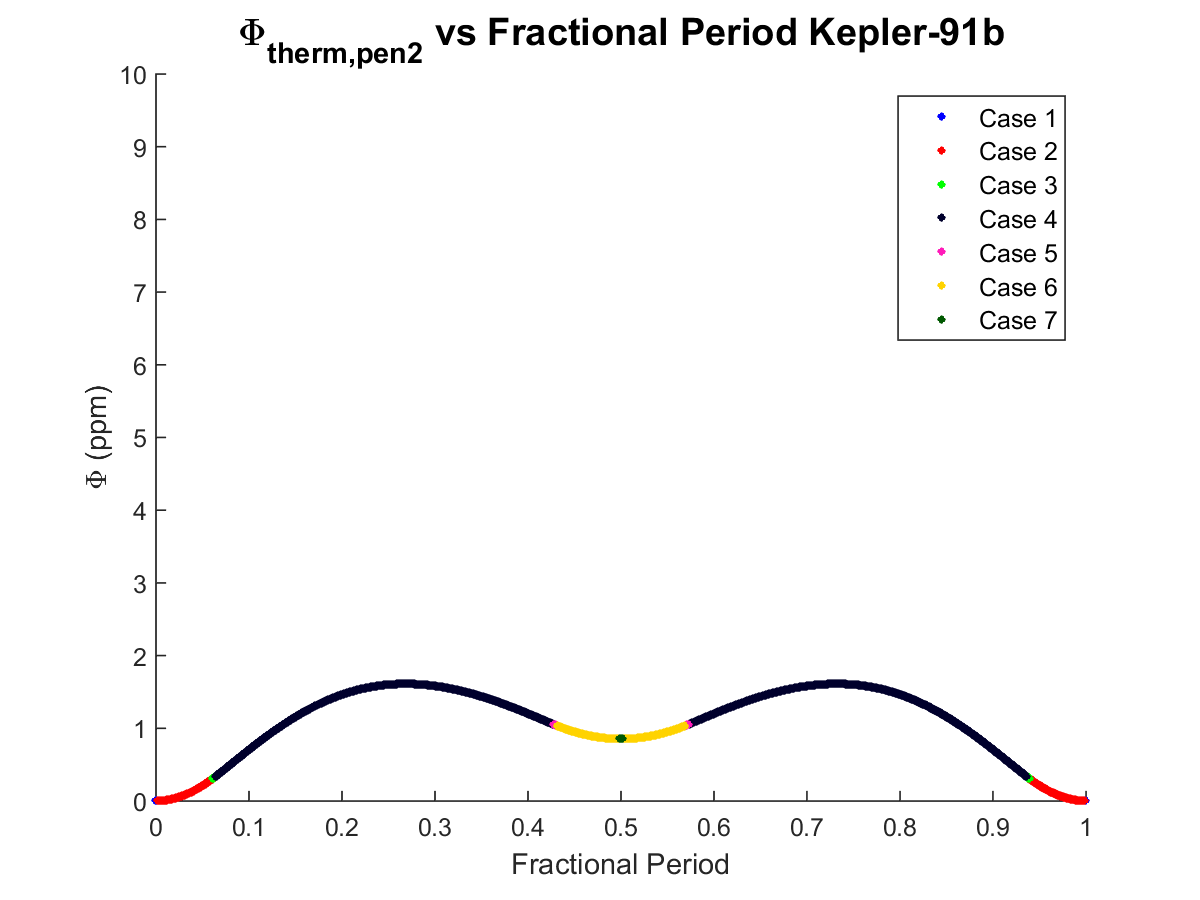}}
\caption{\label{fig:K91thermal7cases}Each of the above plots shows the fractional flux due to thermal radiation from one of the four zones described in \cref{tab:7casesthermal}\ using the parameters given in \cref{tab:K91params}\ with the exceptions noted in \cref{fig:K91thermaltotal7cases}. For comparative purposes the fractional flux due to the dayside temperature is shown in \cref{fig:K91fullthermmulti7cases}\ and that of the night side in \cref{fig:K91unthermmulti7cases}. Note that the fractional flux emitted by the fully illuminated and un-illuminated zones always remain less than that emitted by the day side or nightside of an exoplanet with equivalents parameters, but neglecting the finite angular size of the host star. In addition, we see that the maximum amount of flux from either penumbral zone occurs during \mycase{4}\ of the orbit.}
\end{figure}

In \cref{fig:K91thermal7cases}\ we have plotted the fractional flux due to thermal radiation of each of the four zones for Kepler-91b assuming an edge-on orbit and using the same temperatures as those used in \cref{fig:K91thermaltotal7cases}. In \cref{fig:K91fulltherm7cases,fig:K91untherm7cases}\ we have included the fractional thermal flux of the day side, as given in \cref{eq:normalizedfluxthermalday}, and night side, \cref{eq:thermnight}, of the exoplanet respectively. As before neither the fully illuminated or un-illuminated zones' flux ever exceeds that of the day or night side of the exoplanet because their fractional areas are less than that of a half hemisphere. \cref{fig:K91pen1therm7cases}\ shows that the luminosity of the first penumbral zone is at a minimum when it is not visible during \mycase{7}\ and at a maximum when its maximum fractional area is visible during part of \mycase{4}. In contrast, Penumbral Zone Two is at a minimum and approaches zero during \mycase{1}\ rather than \mycase{7}. Like Penumbral Zone One, Penumbral Zone Two is at a maximum sometime during \mycase{3}.

\begin{figure}[tbh!]
\centering
\subfloat[][\label{fig:K91fullthermmulti7cases}Fractional flux due to the fully illuminated zone.]{\includegraphics[trim={40 0 40 0}, clip,width=0.5\linewidth]{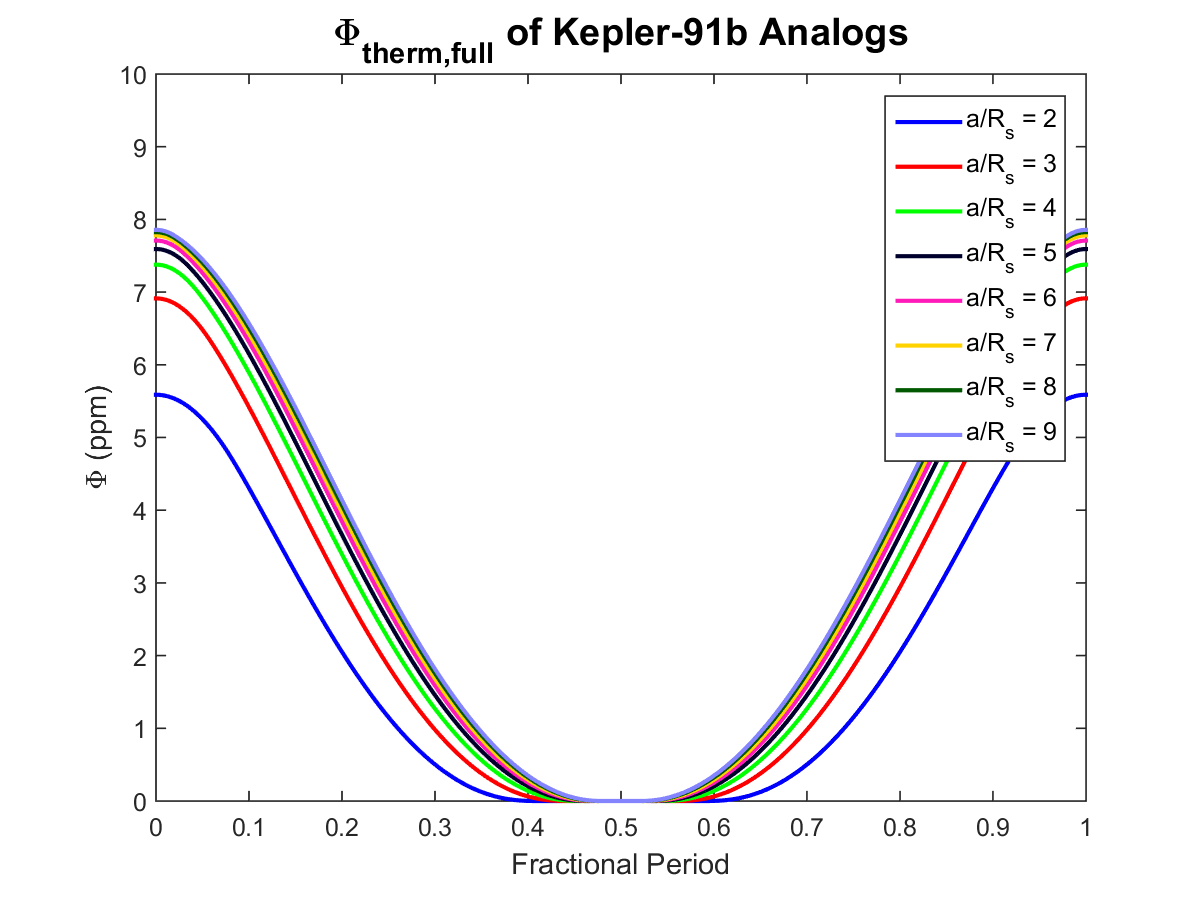}}
\hfill
\subfloat[][\label{fig:K91unthermmulti7cases}Fractional flux due to the un-illuminated zone.]{\includegraphics[trim={40 0 40 0}, clip,width=0.5\linewidth]{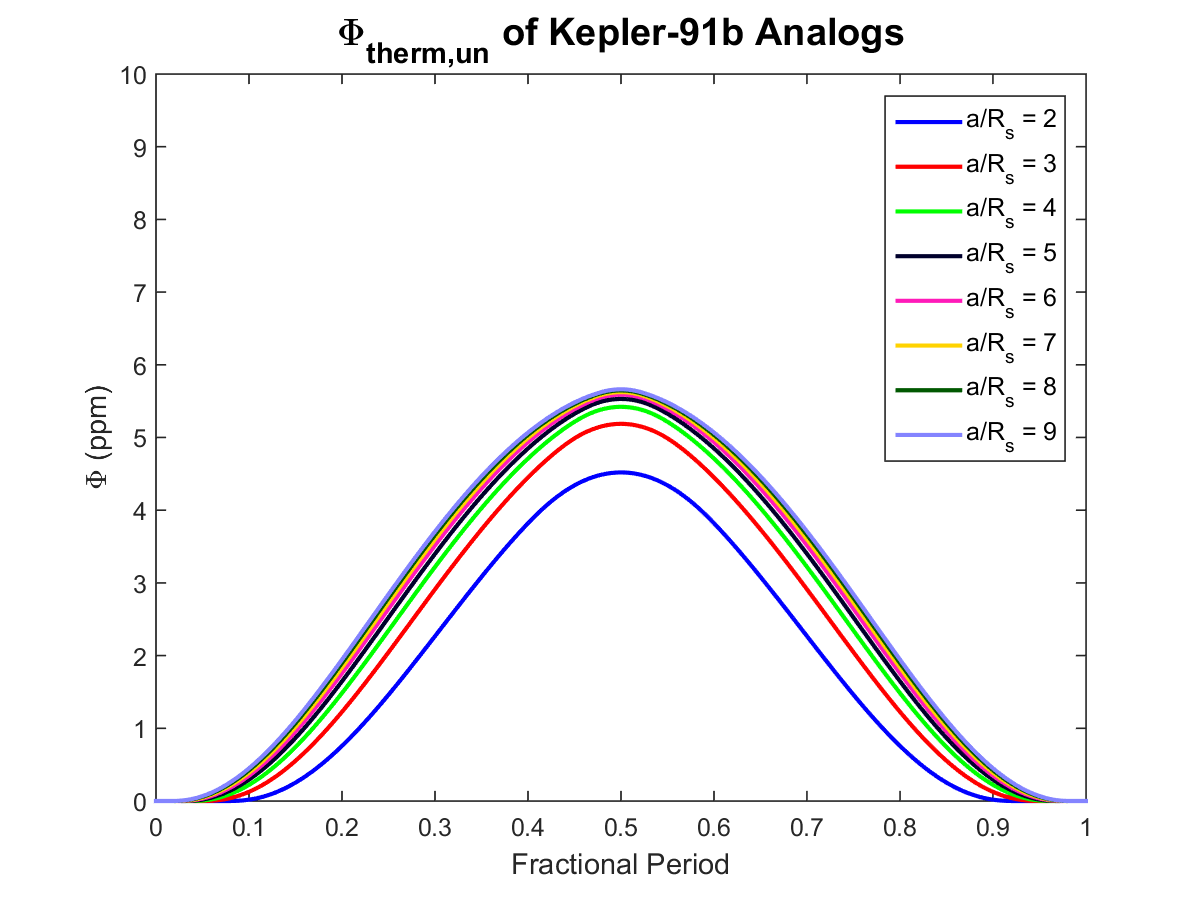}}

\subfloat[][\label{fig:K91pen1thermmulti7cases}Fractional flux due to the Penumbral Zone One.]{\includegraphics[trim={40 0 40 0}, clip,width=0.5\linewidth]{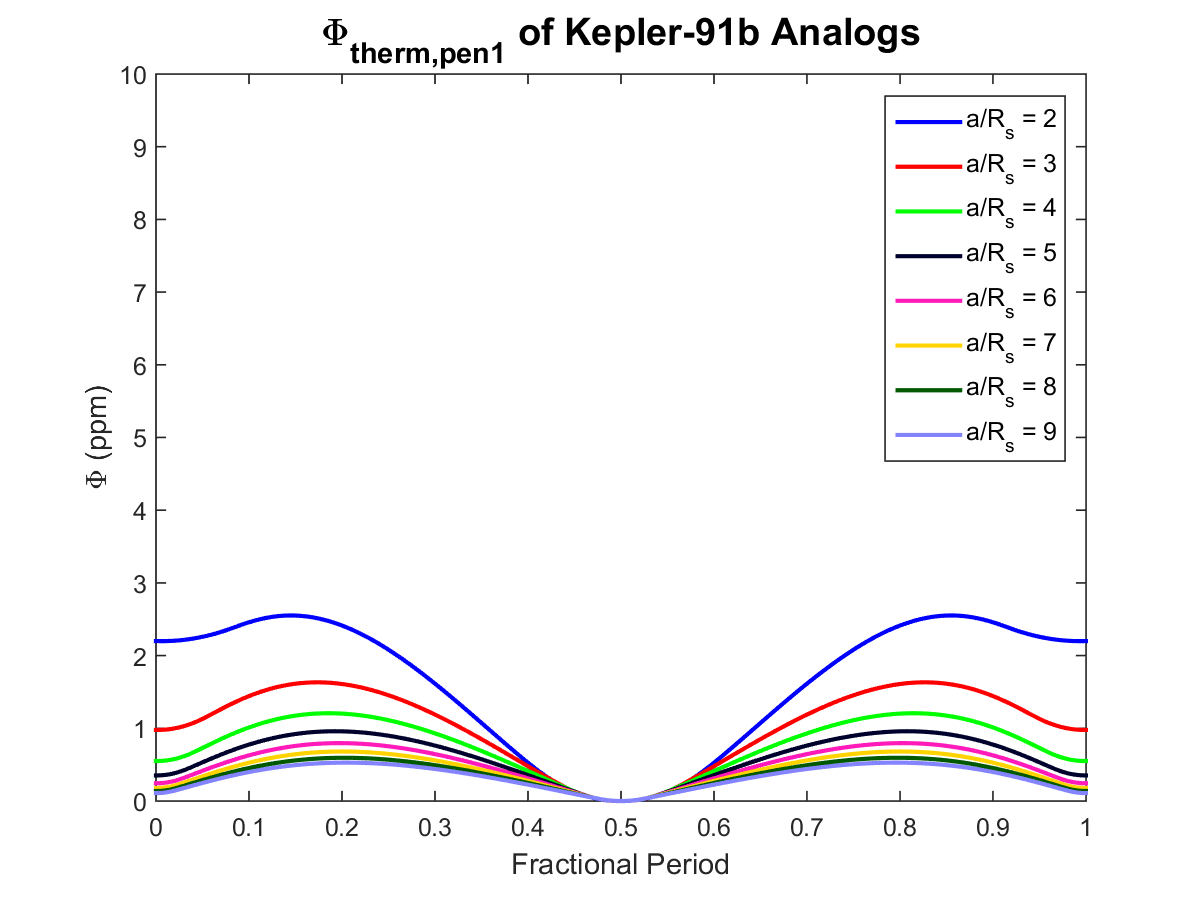}}
\hfill
\subfloat[][\label{fig:K91pen2thermmulti7cases}Fractional flux due to the Penumbral Zone Two.]{\includegraphics[trim={40 0 40 0}, clip,width=0.5\linewidth]{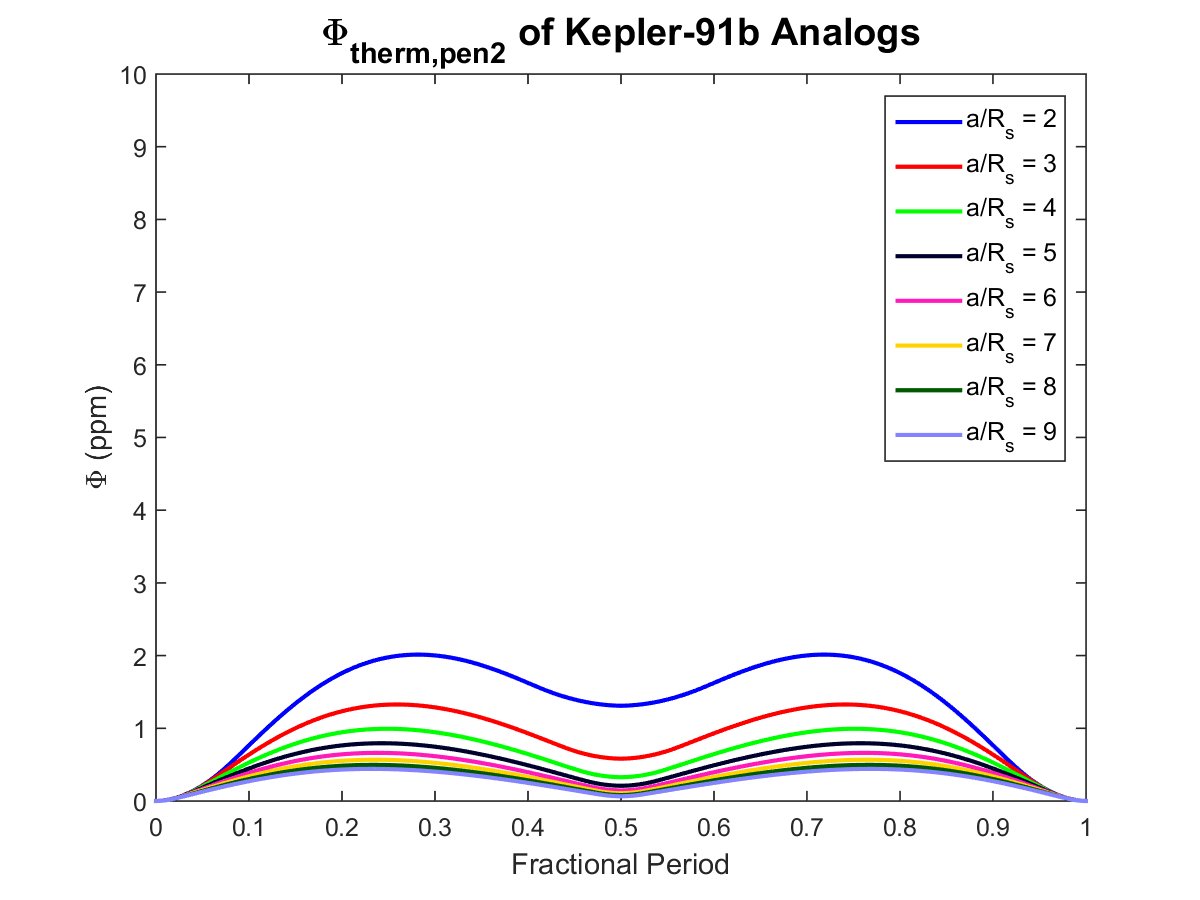}}
\caption{\label{fig:K91thermalmulti7cases}Shown are plots of the fractional flux due to the thermal radiation of each of the four zones described in \cref{tab:7casesthermal}\ where we have varied the star-planet separation by varying the normalized semi-major axis, $ a/R_s $. Here we see patterns similar to those observed in \cref{fig:K91thermalMulti5cases}. The amount of thermal radiation received from both penumbral zones decreases as the star planet separation increases, but the amount from both the fully illuminated and un-illuminated zones increases.}
\end{figure}

Finally, let us consider the effect of star-planet separation on the fractional flux due to thermal radiation from the four zones as shown in \cref{fig:K91thermalmulti7cases}. As was the case in \cref{sec:thermal3zones}, the fractional flux from the fully illuminated and un-illuminated zones increases with star-planet separation because the size of each zone increases with increasing star-planet separation and we have not taken into account the influence of star-planet separation on the temperature of each zone. The opposite is true of the two penumbral zones and we find that as star-planet separation increases the thermal radiation from these two zones decreases.

As was the case for the three zone model described in \cref{sec:thermal3zones}\ the four zone model described here is not distinguishable from the day/night side model described in \cref{sec:thermDaynight}\ for Kepler-91b with current technology. For exoplanets with greater temperature gradients it may be possible to distinguish between the two models and to characterize the change in temperature over multiple zones.

\section{Future Work}
Efforts are now under way to map features of exoplanets, \cf\ Cowan and Fujii's work \cite{Cowan2017}, using a variety of methods including mapping the temperature of an exoplanet's atmosphere as a function of longitude and latitude. 

\begin{figure}[bth]
\centering
\includegraphics[width=0.7\linewidth]{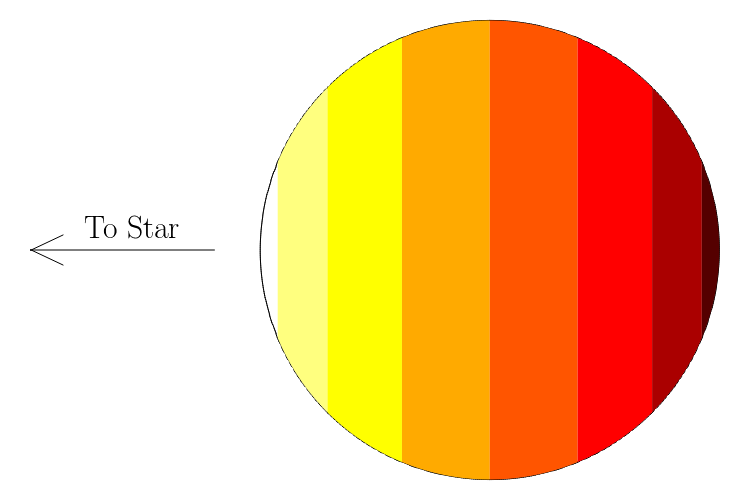}
\caption{Example spacing for modeling the temperature gradient of an exoplanet using equally spaced rings of equal temperature. Shown here is the case for eight zones where each zone is separated by 22.5\degree\ along the polar angle where white represents the highest temperature and dark red the lowest temperature.}
\label{fig:thermal8zones}
\end{figure}

In a future work, we will seek to model an exoplanet's temperature variations by treating it as a series of rings centered along the line connecting the center of the exoplanet to the sub-stellar point. Each of these rings would be treated as a blackbody with constant temperature. The temperature would be greatest at the sub-stellar point at $\phi=\pi/2$ and decrease to a minimum on the other side of the exoplanet at $ \phi =3\pi/2 $. The temperature would be constant within each ring which will be evenly spaced according to the polar angle delimiting each zone. An example is shown in \cref{fig:thermal8zones}\ for the case of eight zones. 

The spacing, i.e. the number of concentric rings $N $, required to describe the temperature gradient of the exoplanet could be determined via model testing, \cf\ Sivia and Skllings work\cite{siviaSkilling}. As the number of rings required to accurately describe the temperature gradient or observational resolution increases there would be an increase in computation time. Fortunately, the integrals required have already been solved and the increase in computational time arises due to the increase in the number of times the functions must be evaluated. 

As was the situation for the models described in the foregoing sections, the luminosity due to thermal light will have to be evaluated differently based on the number of rings used and the phase angle of the exoplanet. Unlike the previous descriptions, the total number of rings in this model will always be even and the rings are always evening spaced. The symmetric nature of this model will result in fewer cases that will need to be evaluated. For example, for $ N = 4 $ only five cases are required to describe the exoplanet's orbit as opposed to the seven cases required to describe the asymmetric situation in \cref{sec:thermal4zones}.

One advantage of modeling the temperature gradient of an exoplanet in this manner is the fact that it offers an analytical approach to the problem. For the purposes of characterizing exoplanets using nested sampling techniques, see \cite{siviaSkilling}, such as those used in \exonest\footnote{\exonest\ is a software package designed to work with \myKepler\ or other photometric exoplanet data to detect and characterize exoplanets, \cite{Placek2014,Placek2015,2017Entropy}.}, analytical solutions are preferred because they are often faster than numerical approaches. When using nested sampling techniques a model exoplanet orbit may be run a large number of times and any increase in efficiency is desirable.

Within this chapter, we compared the thermal luminosity of an exoplanet using three different to purchase. The first approach considered the dayside and nightside of the exoplanet to be at constant temperature, \cf\, \cite{Placek2014}. In the second to repurchase we treated the exoplanet as having either three or four temperature zones in which each zone was delimited according to how much of the host star is visible within the zone. A comparison between the two models reveals that they differ by less than 1 ppm for Kepler-91b, which is not detectable within the \myKepler\ dataset; however, the difference may be detectable for exoplanets that exhibit a greater temperature gradient. We concluded this chapter with an outline of future work in which an exoplanet's temperature gradient may be modeled as having $ N $ equally spaced temperature zones. We will now move on to consider the overall light curve of an exoplanet which includes both the primary and secondary transits and the four photometric variations discussed in this thesis.
\chapter{The Light Curve}\label{ch:lightcurvechapter}
Within this chapter we will discuss the properties of observed light curves that include the transits described in  \cref{ch:transitchapter}\ and the photometric variations discussed in \cref{ch:hoststar,ch:reflectedlightluminosity,ch:thermalradiation}. We will begin by considering how the photometric variations are added to produce the final light curve as described in previous literature and then describe the features of said light curves. We will conclude with a brief discussion of the effect of modeling a finite angular size of the host star on the light curve as well as in researchers' ability to model the stellar and planetary radii.
\section{Adding Up The Variations}\label{sec:addingVariations}
The total flux observed from the planetary system is the sum of the fluxes due to the primary transit of the host star by the exoplanet, the secondary transit of the exoplanet by the host star and the sum of each of the photometric variations. The photometric variations include those induced by the presence of an exoplanet, which includes the boosted light and ellipsoidal variations, and the emissions of the exoplanet itself, which includes reflected light and thermal emissions. The predicted flux is then given by
\begin{equation} 
    F_{pred}(t) = F_s\left(\frac{F_{transits}(t)}{F_s}+\frac{F_{photo}(t)}{F_s}\right)
\end{equation}
where
\begin{equation} 
    F_{transits}(t)= \left( F^s(t)+ F^p(t)\right)F_s
\end{equation}
where $ F^s(t) $ is given by \cref{eq:Fquadratic}\ and $ F^{p}(t) $ by \cref{eq:Fp}. In addition the term $ F_{photo}(t) $ is the sum of the photometric of variations and is given by
\begin{equation} 
    F_{photo}(t) = F_{boost}(t)+ F_{ellip}(t)+ F_{refl}(t) + F_{therm}(t)
\end{equation}
where $ F_{boost}(t) $ is given by \cref{eq:phiBoost},  $ F_{ellip}(t) $ by \cref{eq:KGmodified}, $ F_{refl}(t) $ is the reflected light described in \cref{ch:reflectedlightluminosity}, and $ F_{therm}(t) $ is the thermal radiation described in \cref{ch:thermalradiation}.

It is common practice for researchers to normalize the photometric timeseries by dividing the observed flux, $ F_{obs}(t) $, by the mean flux, $ \langle F_{obs}\rangle $, and then subtracting the mean flux so that the normalized flux is described by, \cite{Placek2014}
\begin{equation}\label{eq:Fnorm}
    F_{norm}(t) =\frac{F_{obs}(t)}{\langle F_{obs}\rangle}-\langle F_{obs}\rangle.
\end{equation}
The use of \cref{eq:Fnorm}\ requires that one models the mean flux unless it is possible to say that the mean of all of the photometric variations, including the primary and secondary transits, are zero, in which case the mean flux corresponds to the stellar flux. Such a situation is never the case, yet, it is common practice to approximate the mean flux, $ \langle F_{obs}\rangle $, as the stellar flux, $F_s$, and we shall do so here to write the predicted normalized flux as
\begin{equation}\label{eq:normlightcurve}
  \begin{aligned}
        F_{normpred}(t)&\approx \frac{F_{transits}(t)}{F_s}+\frac{F_{photo}(t)}{F_s}\\
			&=F^s(t)+F^p(t)+\frac{F_{boost}(t)}{F_s} +\frac{F_{ellip}(t)}{F_s}+\frac{F_{p}(t)}{F_s}\\
            &= F^s(t)+F^p(t)+\Phi_{boost}(t)+\Phi_{ellip}(t)+\Phi_{p}(t)
  \end{aligned}
\end{equation}
where 
\begin{equation}\label{eq:fracphplanet} 
\begin{aligned}
	\Phi_{p}(t)&=\frac{F_p(t)}{F_s}\\
		& =\frac{F_{refl}(t)}{F_s}+\frac{F_{therm}(t)}{F_s}\\
		&= \Phi_{refl}(t)+\Phi_{therm}(t)
\end{aligned}
\end{equation}
is the fractional flux due to the reflected and thermal emissions of the exoplanet. We will see one of the issues that can arise from the approximation that $\langle F_{obs} \rangle \approx F_s$ in \cref{sec:completelightcurve}.

We note here that in \cref{eq:Fp}, describing $ F^p(t) $, we are only considering the first-order effect of the star eclipsing the exoplanet by implicitly treating the exoplanet as a uniform emitter. For the case in which reflected light is modeled using plane parallel rays and the thermal radiation is modeled using only a day side and a nightside this approximation is well justified, but for \myECIES\ this is not the case. If the face of the exoplanet along the line of sight is non-uniform one must consider more exactly which parts of the exoplanet are being obscured. In \cite{planetPlanetOccultation}, Luger, Lustig-Yaeger and Agol describe methods to model planet-planet occultations and such methods could be applied in future work to properly account for the portion of each zone of the exoplanet being obscured by the host star.

In addition to normalizing the light curve, it is common practice for exoplanet researchers to phase-fold the timeseries because many timeseries data sets are years in length and include many orbital periods worth of data. To phase-fold a data set one folds the data set about an excepted orbital period which may be determined via periodogram. The advantage of phase folding is that it lowers computation time because only one orbit must be evaluated as opposed to years worth of orbits; however, for systems with multiple exoplanets or other many-body systems phase-folding is not appropriate because the light curves will not be periodic.

For a known orbital period, $ T $, the phase-folded data set can be obtained by iterating
\begin{equation}
  \begin{aligned}
        \theta_i =&\frac{t_i -t_0}{T}\\
        \theta_i=&\theta_i-\textnormal{floor}(\theta_i)
  \end{aligned}
\end{equation}
where $ \theta_i $ is the $i^{th}  $ fractional period,  $ t_i $ is the  $i^{th}  $ observed time and $ t_0 $ is the first observed time. Here the value of $ \theta_i $ ranges from zero to one.

Let us now consider the light curve for Kepler-91b using the parameters in \cref{tab:K91params}. In addition, the geometric albedo from Table 3 of \cite{Placek2015}\ was used to determine the single scattering albedo. Assuming Lambertian reflection allows us to calculate the single scattering albedo, see \cref{eq:Aglambert}, as
\begin{equation} 
    \myScatteringAlbedo =\frac{3}{2} A_g
\end{equation}
and using a geometric albedo of 0.39 allows us to estimate the single scattering albedo to be 0.585.

\section{Photometric Variation Light Curve}
\begin{figure}
\centering
\subfloat[][\label{fig:ogphotocircular}Circular orbit.]{\includegraphics[width=0.5\linewidth]{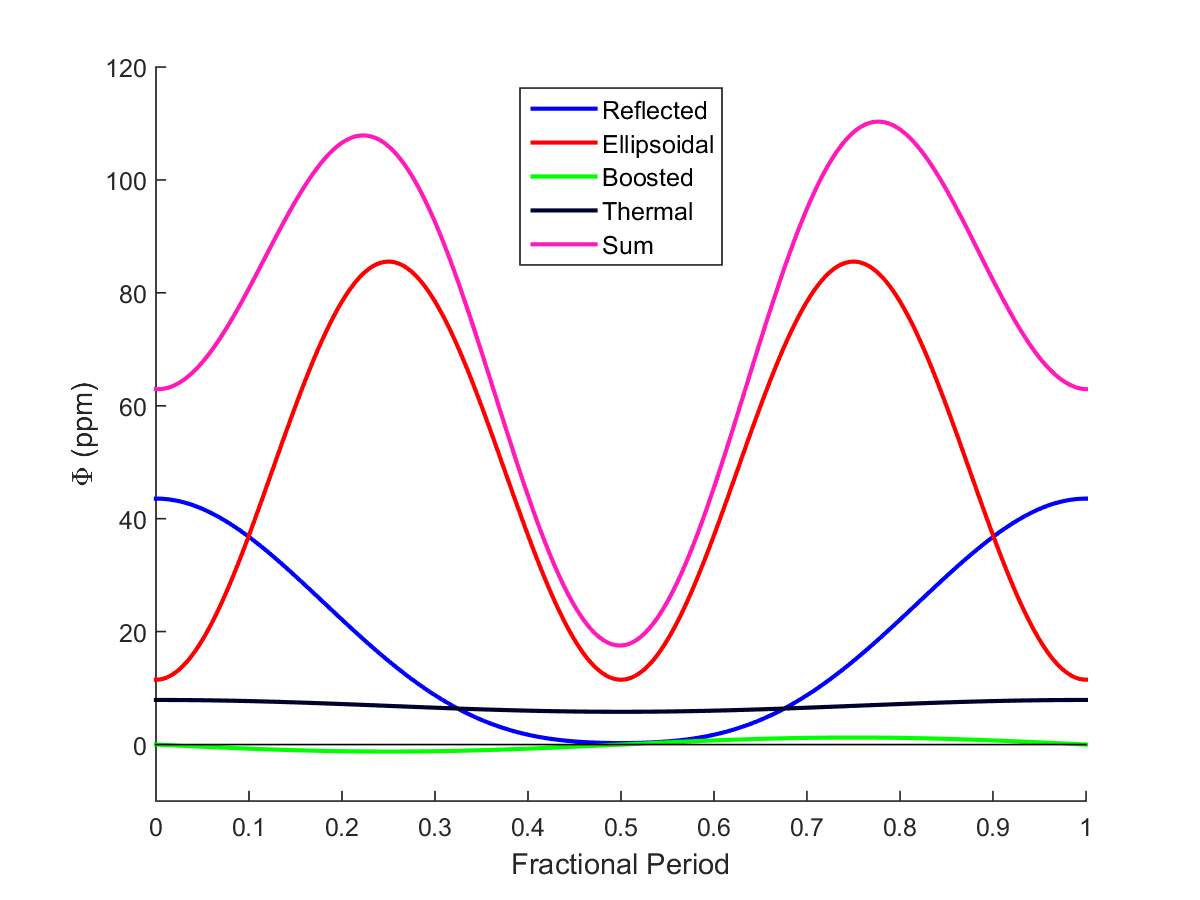}}
\hfill
\subfloat[][\label{fig:ogphotoecc}Eccentric orbit.]{\includegraphics[width=0.5\linewidth]{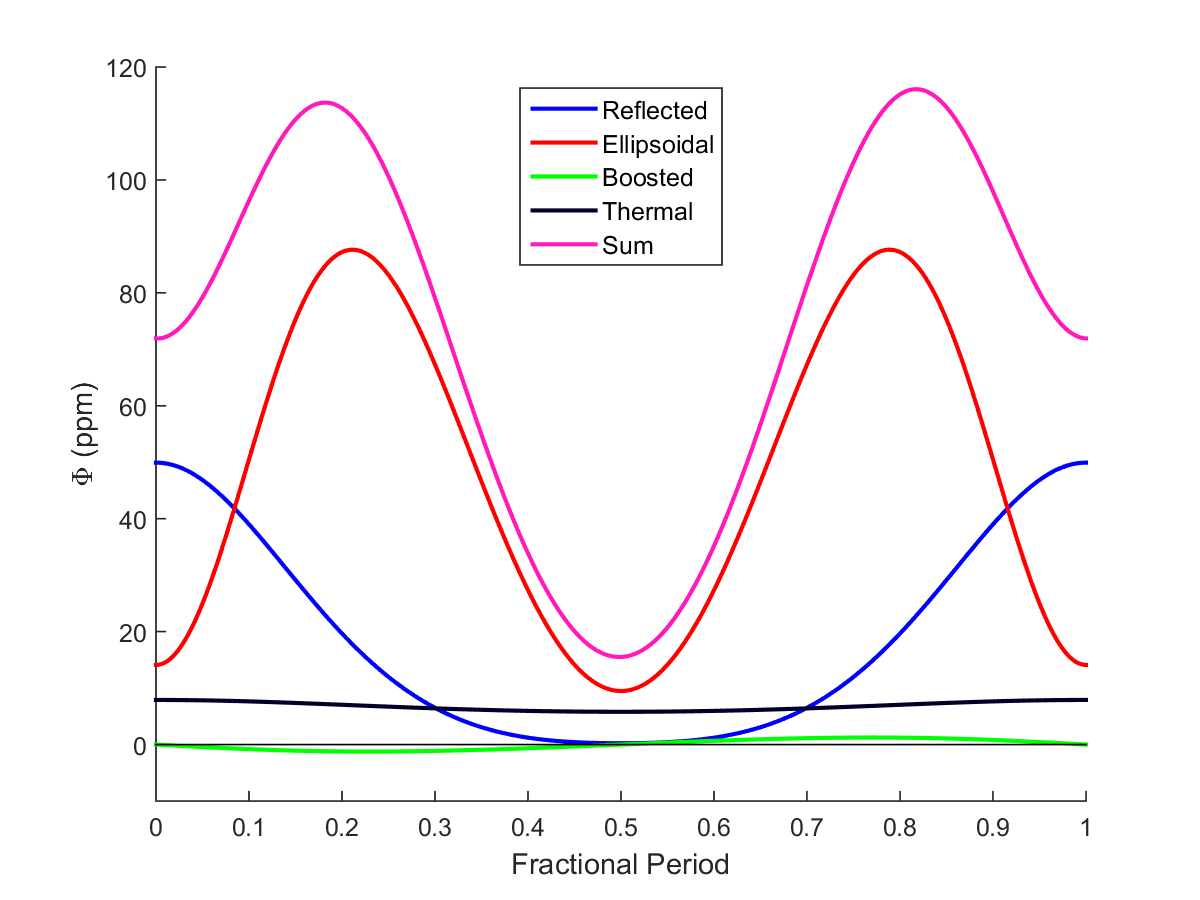}}
\caption{\label{fig:originalphotometric}Each of the plots in the figures were produced using the parameters given in \cref{tab:K91params}, with the exception that the eccentricity was set to zero for the circular orbit in \cref{fig:ogphotocircular}. The single scattering albedo is 0.585 for both plots. To produce the reflected light and thermal radiation a plane parallel ray model was assumed so that the reflected light is given by \cref{eq:phirLambertPlaneParallel}\ and the thermal radiation by \cref{eq:thermaltotal}.}
\end{figure}

The light curve due to photometric variations alone is given by
\begin{equation}\label{eq:fracphoto}
	\Phi_{photo}(t) =\Phi_{boost}(t)+\Phi_{ellip}(t)+\Phi_{refl}(t)+\Phi_{therm}(t).
\end{equation}
\cref{fig:originalphotometric}\ compares the photometric emissions for Kepler-91b assuming either a circular or eccentric orbit and a single scattering albedo of 0.585. Assuming a precision of about 29 ppm we see that both the reflected light and the ellipsoidal variations are detectable from Kepler light curves. Notice that the reflected light, boosted light, and thermal radiation have a period equal to the orbital period of the exoplanet, but the ellipsoidal variation has a period half that of the orbital period. As a result, the sum of the photometric variations has two peaks which are equal in height for circular orbits, but unequal for eccentric orbits for certain values of the argument of periastron. The difference in height for eccentric orbits is due to the misalignment of the ellipsoidal variation to that of the other three photometric variations for eccentric orbits.

\begin{figure}
\centering
\subfloat[][\label{fig:newphotocircular}Circular orbit.]{\includegraphics[width=0.5\linewidth]{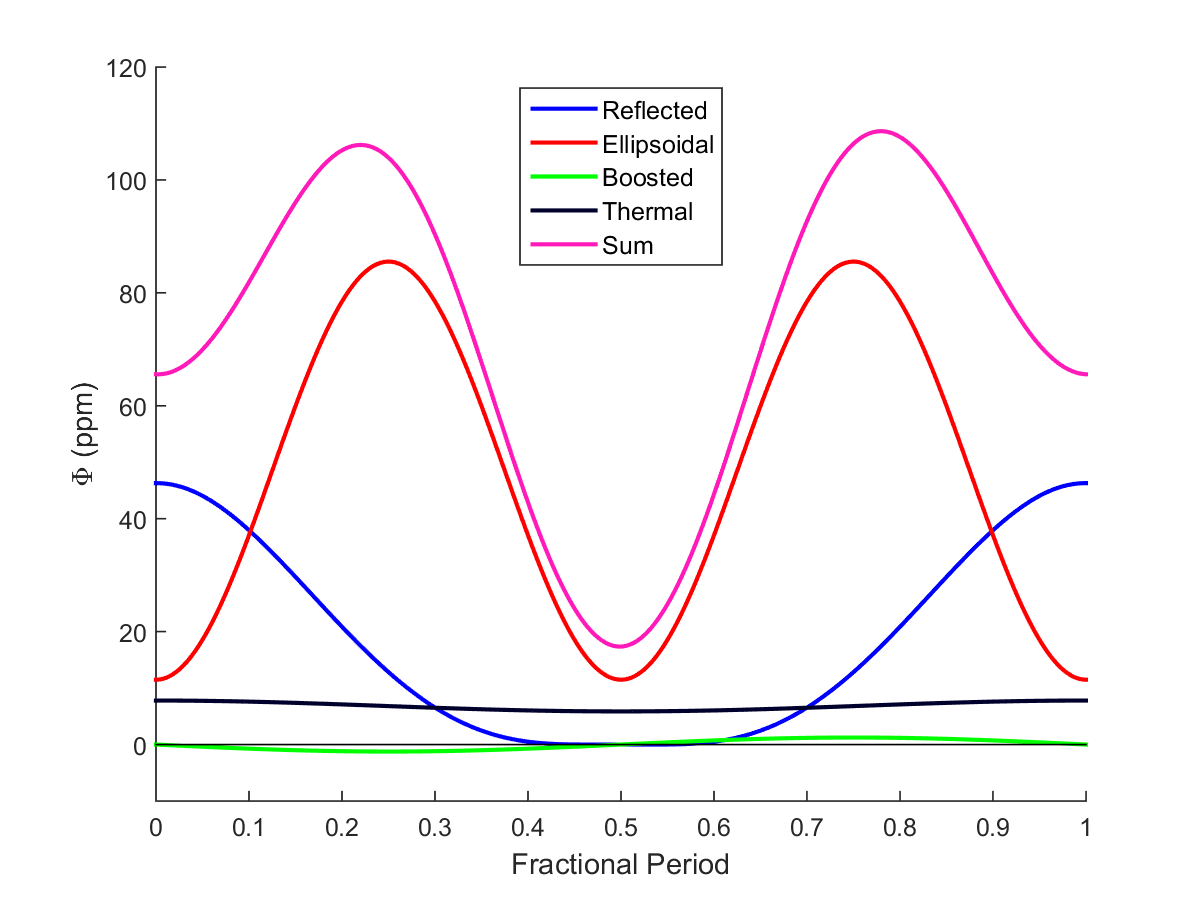}}
\hfill
\subfloat[][\label{fig:newphotoecc}Eccentric orbit.]{\includegraphics[width=0.5\linewidth]{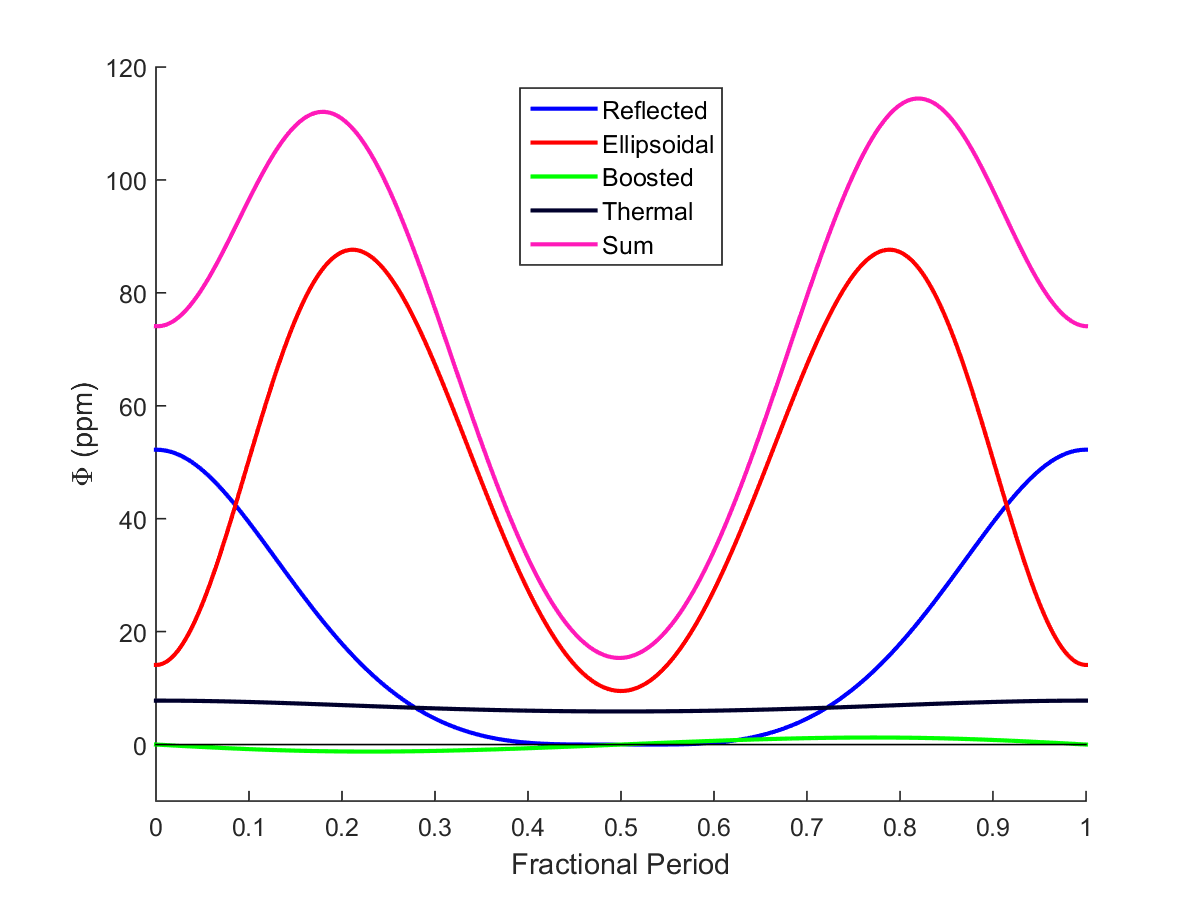}}
\caption{\label{fig:newphotometric}The parameters to produce the above plots are given in \cref{tab:K91params}\ where the single scattering albedo is equal to 0.585. For comparative purposes the eccentricity has been set to zero in \cref{fig:newphotocircular}. In this case the fractional flux due to reflected light is that only of the fully illuminated zone given by \cref{eq:phifull}. The thermal radiation was modeled using \cref{eq:phithermfinitetotal}\ to account for the finite angular size of the host star.}
\end{figure}

Illustrated in \cref{fig:newphotometric}\ are the photometric variations assuming that the reflected light is produced using the model described in \cref{sec:luminosityFiniteSizenew}\ where we have only plotted the fractional flux of the fully illuminated zone and leave a fuller description that includes the penumbral zones for a later work. In addition, the thermal radiation was determined from \cref{eq:phithermfinitetotal}. The general features of the two plots do not differ significantly from \cref{fig:originalphotometric}, but there is a slight difference noticeable near a fractional period of 0.6 in which the reflected light is slightly less than the boosted light whereas in \cref{fig:originalphotometric}\ at the same location was slightly greater than the boosted light. This is not surprising because it is expected that the reflected light from the fully illuminated zone alone will always be less than that of an exoplanet modeled using the plane parallel ray of illumination as described in \cref{sec:luminosityFiniteSizenew}\ in our exploration of \cref{fig:fullreflcomparek91paramslimbdarkedgeon}.

\section{Complete Light Curve}\label{sec:completelightcurve}
\begin{figure}
\centering
\subfloat[][\label{fig:circular}Circular orbit.]{\includegraphics[width=0.5\linewidth]{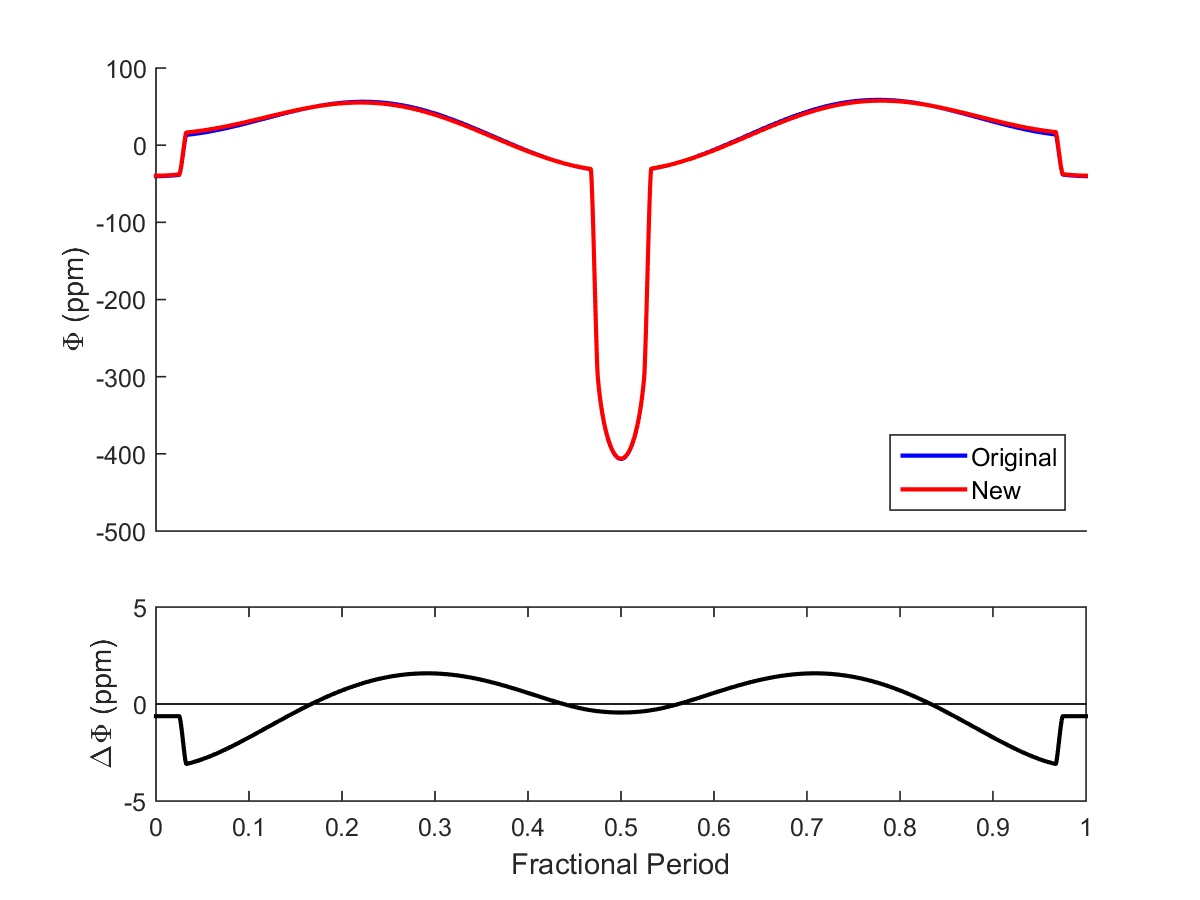}}
\hfill
\subfloat[][\label{fig:ecc}Eccentric orbit.]{\includegraphics[width=0.5\linewidth]{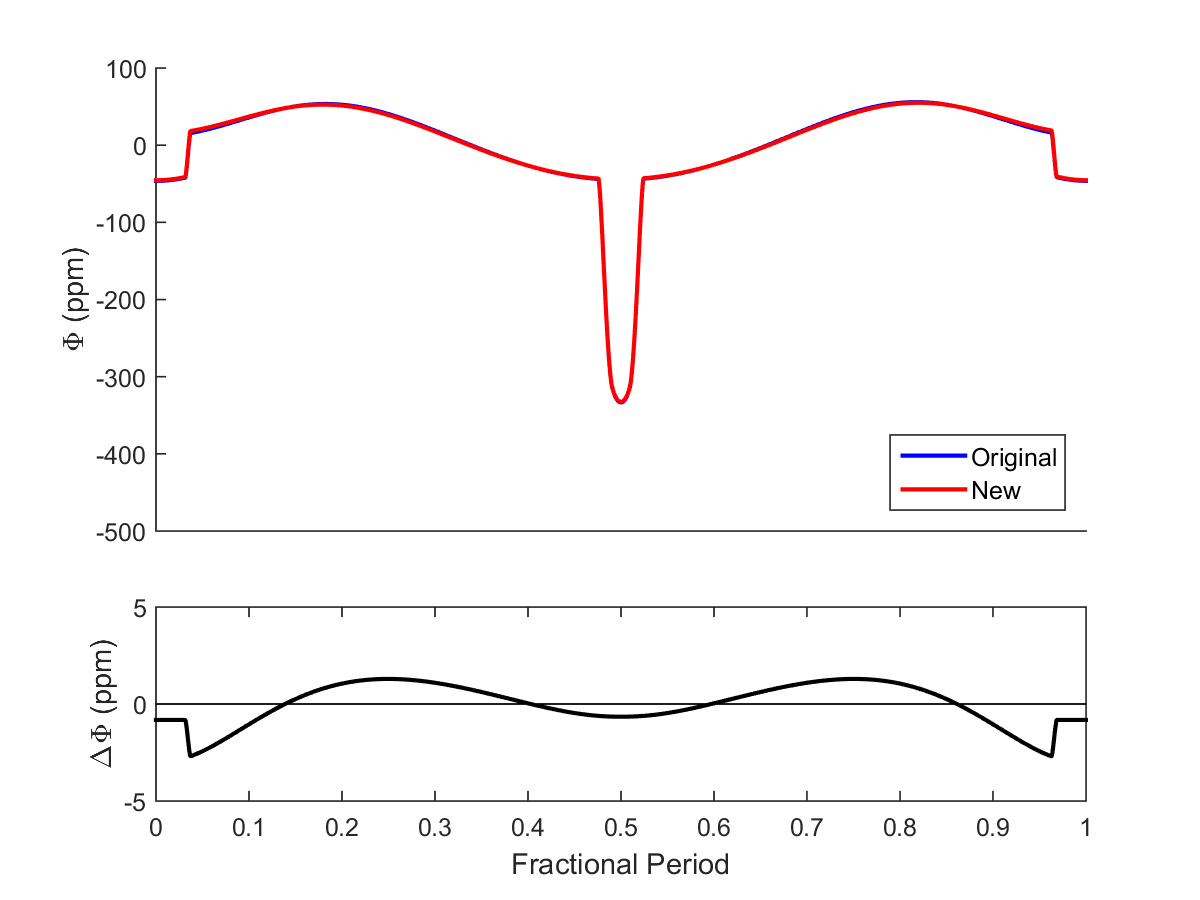}}
\caption{\label{fig:transitcompare}The above plots were produced using parameters given in \cref{tab:K91params}\ where the single scattering albedo is equal to 0.585. For comparative purposes the eccentricity has been set to zero in \cref{fig:circular}. The curve labeled as ``Original'' assumes that the planetary photometric emissions were produced by an exoplanet that is illuminated by plane parallel rays so that the reflected light is given by \cref{eq:phirLambertPlaneParallel}\ and the thermal radiation by \cref{eq:thermaltotal}. The curve labeled as ``New'' was produced by accounting for the finite angular size of the host star. Only the reflected light of the fully illuminated zone, \cref{eq:phifull}, is included in these plots, but all four zones are included within the thermal radiation, \cref{eq:phithermfinitetotal}. The bottom panel shows the difference between the fractional flux using the original model and the new model, $ \Delta\Phi $. Note that the maximum of $ \Delta\Phi $ in this case is not detectable from \myKepler\ data.}
\end{figure}

We will now explore the total light curve described by \cref{eq:normlightcurve}\ as shown in \cref{fig:transitcompare}. First, we see that the transit depth appears to differ between the circular orbit and eccentric orbit. This is due to the normalization process described in \cref{sec:addingVariations}. The mean flux, $\langle F_{obs}\rangle $, differs between the two models because it is not actually the case that $ \langle F_{obs}\rangle=F_s $ and this alters the final transit depth. Properly accounting for the mean flux of the system will be left for future work. We also see that the transit time is longer for the circular orbit than the eccentric orbit because the argument of periastron is such that the primary transit occurs near periastron. 

The bottom panel of each plot shows the difference between the fractional flux using the plane parallel ray model versus one that accounts for the finite angular size of the host star,  $ \Delta\Phi $. The general shape is the same between the circular orbit and eccentric orbit. Note that $ \Delta\Phi $ is constant during the secondary transit because the primary difference between the two models during the secondary eclipse is the total planetary emissions during the full phase of the exoplanet which is being blocked by the host star. Furthermore, we see that $ \Delta\Phi $ is on the order of only a few ppm and is therefore not detectable using current technology. Exoplanets with greater temperature gradients and larger thermal emissions may produce light curves such that the two models could be distinguished and new exoplanets will be tested in later work. Finally, a proper analysis of the penumbral zones may produce reflected light that would be detectable with current technology.

\section{Determination of Stellar Radius}
The determination of the planetary radius from transit depth requires that the stellar radius be well known, see \cref{eq:transitdepth}. Mathur et. al. in \cite{2017revisedStar}\ produced a revised stellar catalog of \myKepler\ candidates for which the typical uncertainty in stellar radius was 27\%, in part due to the lack of spectroscopic data for some host stars. Further efforts were made by Berger et. al. in \cite{gaiaStarRadii}\ to improve stellar radius estimations of \myKepler\ stars using \textit{Gaia} photometry and parallax measurements. These efforts produced a median uncertainty in stellar radius of 8\%, which corresponds to an improvement of about a factor of 4-5 in the determination of stellar radii. It may also be possible to use photometric variations to estimate stellar radius by including it in the set of model parameters used in \exonest\ and modeling for it directly.

In the absence of spectroscopic data one may be able to disentangle the planetary and stellar radii by considering the transit duration as described in \cref{sec:tranSepDuration}\ and the  photometric variations described in \cref{ch:hoststar,ch:reflectedlightluminosity,ch:thermalradiation} because there are unique dependences of the planetary photometric emissions on the two radii. To begin, the transit depth and the thermal variation depend on the planetary and stellar radius according to $R_p/R_s$ as described by \cref{eq:transitdepth,eq:normalizedfluxthermalday,eq:luminositythermalnight}. The transit separation, as described by \cref{eq:tranDuration}, depends on the stellar radius and the semi-major axis in the form of $R_s/a$, as do the ellipsoidal variations via \cref{eq:amplitude}. In addition, the plane parallel ray model for reflected light includes a factor of $R_p/r$, see \cref{eq:phirLambertPlaneParallel}. If some of these photometric variations are detectable along with the transit depth it is possible to disentangle the stellar radius from the planetary radius by determining the star-planet separation from the period of the orbit if the mass of the star is well known. The stellar mass may be determined via a combination of the boosted light and ellipsoidal variations if they are detectable. 

If the planetary photometric emissions are modeled by taking into account the finite angular size of the host star we find a new dependence on stellar radius, star-planet separation, and planetary radius. According to the equations described in \cref{sec:koperators}, the terms of $ R_p/R_s $ and $ R_s/r $ now appear as part of inverse trigonometric function. Finally, from \cref{eq:Agfullfinal}\ we see that the reflected light now has unique dependencies on combinations of $R_{s,p}/r$. Thus, the inclusion of the finite size of the host star in light curve models provides new relationships between stellar, planetary and orbital radius will grant researchers a new tool to determining stellar and exoplanetary characteristics for \myECIES.

To summarize the foregoing chapter, we have described the relationship between total light curves and the photometric variations of an exoplanetary system. First, the predicted light curve is a summation of the primary and secondary transit, and the photometric variations. Next, the measures data is normalized to the observed mean flux and it is assumed that this is approximately equal to the stellar flux. Typically, the data is phase folded over the period of the orbit. We concluded the chapter with a comparison between the predicted light curves using the plane parallel ray approximation of incident stellar radiation and those predicted by taking into account the finite angular size of the host star. We saw that the two differ the most just before and after the secondary transit and that the difference is only a few parts per million for Kepler-91b. This difference will be greater for exoplanets with larger penumbral zones or temperature gradients.
%
%

\chapter{Conclusion}\label{ch:conclusion}
In the foregoing chapters we have explored the nature of photometric light curves produced by exoplanetary systems. We have connected the orbital and physical parameters of exoplanets to their signatures on the transit and photometric light variations. Most notably we have addressed the problem of analyzing planetary photometric emissions for extremely close-in exoplanets, \myECIES, for which the plane parallel ray model of incident stellar radiation breaks down. For such exoplanets the finite angular size of the host star must be taken into account so that the relationship between the radius of the exoplanet and its temperature and reflectivity is properly model.

The work of Kopal \cite{Kopal1953,Kopal1959}, which was concerned with modeling the reflection effect for binary stars, was used as a template to determine the reflected luminosity as a function of phase for \myECIES. It was discovered that his analysis of the fully illuminated zone, for example \cite[Eq's (62)-(69)]{Kopal1953}, actually produces negative luminosity near the full phase of an exoplanet. A reevaluation of the integrals described in \cref{tab:5cases}\ was presented and shown to produce luminosity that is never negative and always less than that of the Lambertian phase function given in \cref{eq:phirPhaseFunc}.

In addition, a more complete analysis of the reflected luminosity of the penumbral zone was presented as an improvement to the approximations used to produce \cite[Eq's (86) and (90)]{Kopal1953}; however, it was found that such an analysis produces negative luminosity within the penumbral zone. Possible reasons for this phenomenon include the poor approximations described in \cref{eq:approximation}, but more likely is the fact that part of the integration used to produce the reflected intensity distribution within the penumbral zone is imaginary. Efforts to correct these errors led to the discovery that the exoplanet may be more accurately described using two penumbral zones as opposed to the single penumbral zone presented in \cite{Kopal1953,Kopal1959}. 

Finally, we have presented an analysis of the thermal emissions of an exoplanet if one models it as four zones each with their own effective temperature as opposed to the two zone model used if one considers the exoplanet as having only a day side and nightside. One can consider the four zone model as being one that roughly accounts for the temperature gradient of an exoplanet produced by accounting for the finite angular size of the host star. We have also shown that the difference between modeling the exoplanet using the plane parallel ray approximation and properly accounting for the finite angular size of the host star is only on the order of a few parts per million and is not detectable using current technology. Future instruments, such as PLATO, may be capable of distinguishing the two models.

In future work we will complete the analysis of the luminosity of the two penumbral zones and extend the analysis to linear limb darkening. Once completed, we will add the model to the \exonest software package's suite of plug-and-play models of photometric effects to determine if any exoplanets have light curves that support modeling their reflected light and thermal emissions using the four zones described in this work.

Also of interest is the improvement of the analysis of the secondary eclipse. Within this work we have used the method presented in \cite{Mandel2002}\ and treated the occultation of the exoplanet by the host star like that of a planet transiting a uniform star. The luminosity of an exoplanet is not uniform even for the plane parallel ray approximation, but rather falls with the cosine of the angle from the sub-planetary point, $ \cos{\gamma} $. For \myECIES, the luminosity of the exoplanet near the secondary transit will also vary between the fully illuminated and penumbral zones. Properly accounting for the portion of each of the zones visible during the ingress and egress of the secondary transit may require the techniques presented in \cite{planetPlanetOccultation}.

Finally, in future work we will develop a method of modeling the temperature gradient of an exoplanet as a set of concentric rings centered about the line connecting the center of the exoplanet to the sub-stellar point. Such a model will add an analytical approach to efforts to produce heat maps of exoplanets such as those described in \cite{Cowan2017}.

\begin{appendices}
	\chapter{Tables}\label[app]{app:tables}

\begin{center}
	\tabulinesep = 2.5mm
	\begin{longtabu} to \textwidth {
			X[0.3,c]
			X[0.4,c]
			X[0.4,c]
		}
		\caption{\label{tab:ranges}
		Shown is a table describing the values of various angles and trigonometric functions of interest for each of the cases and their relationships to each other. This table describes the relationships between variables for the fully illuminated and un-illuminated zones, for which there are only five cases to consider. The extension to seven cases requires only that one make the appropriate replacements of $ \mySinEta_{1} $ with $ \mySinEta_{\mypen1limit} $.}\\
		\toprule
		\textbf{Angles}& $ \mathbf{\sin^nA} $& $ \mathbf{\cos^nA} $   \\
		\midrule
		\endfirsthead
		\multicolumn{3}{c}
		{\textbf{Ranges} -- \textit{Continued from previous page}} \\
		\midrule
		\textbf{Angles}& $ \mathbf{\sin^nA} $& $ \mathbf{\cos^nA} $   \\
		\midrule
		\endhead
		\midrule
		\multicolumn{3}{c}{\textit{Continued on next page}} \\ 
		\endfoot
		\bottomrule
		\multicolumn{3}{c}{\textit{End table}}
		\endlastfoot
		\multicolumn{3}{c}{ \textbf{All Cases}: $ \forall\myPhase $ }\\
		$ 0\le \eta_{\mypen1limit}\le \piover2 $ & $ 0\le \mySinEta_{\mypen1limit}\le 1 $ & $ 0\le \cos{\eta_{\mypen1limit}}\le 1 $ \\ 
		$ 0\le \eta_{1}\le \piover2 $ & $ 0\le \mySinEta_{1}\le 1 $ & $ 0\le \cos{\eta_{1}}\le 1 $ \\ 
		$ 0\le \eta_{2}\le \piover2 $ & $ 0\le \mySinEta_{2}\le 1 $& $ 0\le \cos\eta_{2}\le 1 $  \\ 
		$ \eta_{\mypen1limit}<\eta_{2}\le \eta_{1} $ & $\mySinEta_{\mypen1limit}< \mySinEta_{2}\le \mySinEta_{1} $ & $ \cos\eta_{1}\le \cos\eta_{2}<\cos{\eta_{\mypen1limit}} $\\
		$ 0\le \phi_{\lambda,i}\le \eta_{\mypen1limit} $&  $ 0\le \sin\phi_{\lambda,i}\le 1 $ & $ 0\le \cos\phi_{\lambda,i}\le 1 $ \\
		$ \piover2\le \phi_{\lambda,f}\le \pi $ & $ 0\le \sin\phi_{\lambda,f}\le 1 $ & $ -1\le \cos\phi_{\lambda,f}\le0  $ \\
		$ 0\le \phi_{full,i}\le \eta_{1} $&  $ 0\le \sin\phi_{full,i}\le 1 $ & $ 0\le \cos\phi_{full,i}\le 1 $ \\
		$ \piover2\le \phi_{full,f}\le \pi $ & $ 0\le \sin\phi_{full,f}\le 1 $ & $ -1\le \cos\phi_{full,f}\le0  $ \\
		$ \pi\le \phi_{un,i}\le \frac{3\pi}{2}$ & $ -1\le \sin\phi_{un,i}\le 0 $ & $ -1\le \cos\phi_{un,i}\le 0 $ \\
		$  \frac{3\pi}{2}\le \phi_{un,f}\le 2\pi $ & $ -1\le \sin\phi_{un,f}\le 0 $ & $ 0\le \cos\phi_{un,f}\le 1 $ \\
		$ \phi_{full,f}=\pi-\phi_{full,i} $& $ \sin{\phi_{full,f}}= \sin{\phi_{full,i}} $ & $ \cos{\phi_{full,f}}=-\cos{\phi_{full,i}} $ \\
		$ \phi_{un,f}= 3\pi-\phi_{un,i} $ & $ \sin{\phi_{un,f}}= \sin{\phi_{un,i}} $ &$ \cos{\phi_{un,f}}=-\cos{\phi_{un,i}} $ \\
		\midrule \midrule
		\multicolumn{3}{c}{\mycase{1}: $ 0\le \myPhase\le \eta_{2} $}\\
			&$ 0\le \sin\myPhase\le \mySinEta_{2} $&  $ \cos\eta_{2}\le \cos\myPhase\le 1 $\\ 
			& $ 0\le \sin^2\myPhase\le \mySinEta_{2}^2 $& $ \cos^2\eta_{2}\le \cos^2\myPhase\le 1 $\\ 
			\midrule 
			$0\le \phi_{limb,i}\le \piover2$& $ 0\le \sin\phi_{limb,i}\le 1 $ & $ 0\le \cos\phi_{limb,i}\le 1 $ \\ 
			 $\pi\le \phi_{limb,f}\le \frac{3\pi}{2}$& $ -1\le \sin\phi_{limb,f}\le 0 $ & $ -1\le \cos\phi_{limb,f}\le 0 $ \\ 
			\midrule
			 \multicolumn{3}{c}{For $ \myKintegral{0}{\phi_{limb,i}}{\phi_{limb,f}}{\myIntensityDis_{pen}}$ }\\ 
			  $ 0\le \phi\le \frac{3\pi}{2} $& $ -1\le \sin\phi\le 1 $ &$ -1\le \cos\phi\le 1$ \\
			\midrule
			 \multicolumn{3}{c}{For $ \myKintegral{\mySinEta_{1}}{\phi_{full,i}}{\phi_{full,f}}{\myIntensityDis_{pen}} $ }\\ 
			  $ 0\le \phi\le \pi $ & $ 0\le \sin\phi\le 1 $  &$ -1\le \cos\phi\le 1$\\
		\midrule\midrule
		\multicolumn{3}{c}{\mycase{2}: $ \eta_{2}\le \myPhase\le \eta_{1} $}\\
		& $\mySinEta_{2}\le \sin\myPhase\le \mySinEta_{1}$ & $ \cos\eta_{1}\le \cos\myPhase\le \cos\eta_{2} $ \\
		&  $ \mySinEta_{2}^2\le \sin^2\myPhase\le \mySinEta_{1}^2 $ & $ \cos^2\eta_{1}\le \cos^2\myPhase\le \cos^2\eta_{2} $ \\
		\midrule
		 $0\le \phi_{limb,i}\le \piover2$& $ 0\le \sin\phi_{limb,i}\le 1 $ & $ 0\le \cos\phi_{limb,i}\le 1 $ \\
		 $\pi\le \phi_{limb,f}\le \frac{3\pi}{2}$& $ -1\le \sin\phi_{limb,f}\le 0 $ & $ -1\le \cos\phi_{limb,f}\le 0 $ \\ 
		\midrule
		\multicolumn{3}{c}{For $ \myKintegral{0}{\phi_{limb,i}}{\phi_{limb,f}}{\myIntensityDis_{pen}}$ }\\ 
        $ 0\le \phi\le \frac{3\pi}{2} $&$ -1\le \sin\phi\le 1 $  & $ -1\le \cos\phi\le 1$\\
		\midrule 
		\multicolumn{3}{c}{For $ \myKintegral{\mySinEta_{1}}{\phi_{full,i}}{\phi_{full,f}}{\myIntensityDis_{pen}} $ }\\
		 $0\le \phi\le \piover2 $ & $ 0\le \sin\phi\le 1$  &$ 0\le \cos\phi\le 1 $\\
		\midrule 
		\multicolumn{3}{c}{For $ \myKintegral{-\mySinEta_{2}}{\phi_{un,i}}{\phi_{limb,f}}{\myIntensityDis_{pen}} $ }\\
         $\pi\le \phi\le \frac{3\pi}{2}  $ & $ -1\le \sin\phi\le 0 $ &$ -1\le \cos{\phi}\le 0 $\\ 
		\midrule\midrule
		\multicolumn{3}{c}{\mycase{3}: $ \eta_{1}\le \myPhase\le \pi-\eta_{1} $}\\
		& $ \mySinEta_{1}\le \sin\myPhase\le 1 $ & $ -\cos{\eta_{1}}\le \cos{\myPhase}\le \cos{\eta_{1}} $ \\
		&  $ \mySinEta_{1}^2\le \sin^2\myPhase\le 1 $ & $ 0\le \cos^2\myPhase\le \cos^2\eta_{1} $ \\
		\midrule
		 $0\le \phi_{limb,i}\le \pi$& $ 0\le \sin\phi_{limb,i}\le 1 $ & $ -1\le \cos\phi_{limb,i}\le 1 $ \\
		 $\pi\le \phi_{limb,f}\le 2\pi$& $ -1\le \sin\phi_{limb,f}\le 0 $ & $ -1\le \cos\phi_{limb,f}\le 1 $ \\ 
		\midrule 
		\multicolumn{3}{c}{For $ \myKintegral{0}{\phi_{limb,i}}{\phi_{limb,f}}{\myIntensityDis_{pen}}$ }\\
		 $ 0\le \phi \le  2\pi  $&  $ -1\le \sin{\phi}\le 1 $ & $ -1\le \cos{\phi}\le 1 $ \\ 
		\midrule 
		\multicolumn{3}{c}{For$ \myKintegral{\mySinEta_{1}}{\phi_{limb,i}}{\phi_{full,f}}{\myIntensityDis_{pen}} $ }\\
		 $ 0\le \phi\le \pi $ & $ 0\le \sin\phi\le 1 $  & $-1\le \cos\phi\le 1 $\\
		\midrule 
		\multicolumn{3}{c}{For $ \myKintegral{-\mySinEta_{2}}{\phi_{un,i}}{\phi_{limb,f}}{\myIntensityDis_{pen}} $ }\\
		 $ \pi\le \phi\le 2\pi $ & $ -1\le \sin\phi\le 0 $  & $-1\le \cos{\phi}\le 1$\\ 
		\midrule\midrule
		\multicolumn{3}{c}{\mycase{4}: $ \pi-\eta_{1}\le \myPhase\le \pi-\eta_{2} $}\\
		& $ \mySinEta_{2}\le \sin\myPhase\le \mySinEta_{1} $ & $ -\cos\eta_{2}\le \cos\myPhase\le -\cos{\eta_{1}} $ \\
		& $ \mySinEta_{2}^2\le \sin^2\myPhase\le \mySinEta_{1}^2 $ & $ \cos^2\eta_{1}\le \cos^2\myPhase\le \cos^2\eta_{2} $\\
		\midrule 
		 $ \piover2\le \phi_{limb,i}\le \pi $ & $ 0\le\sin\phi_{limb,i}\le 1 $& $ -1\le \cos\phi_{limb,i}\le 0 $ \\
		 $ \frac{3\pi}{2}\le \phi_{limb,f}\le 2\pi $ & $ 0\le \sin\phi_{limb,f}\le 1 $& $0 \le \cos\phi_{limb,f}\le 1$\\
		\midrule 
		\multicolumn{3}{c}{For $ \myKintegral{0}{\phi_{limb,i}}{\phi_{limb,f}}{\myIntensityDis_{pen}}$ }\\
		 $ \piover2\le \phi 2\pi  $&  $ -1\le \sin{\phi}\le 1 $ & $ -1\le \cos{\phi}\le 1 $ \\ 
		\midrule 
		\multicolumn{3}{c}{For $ \myKintegral{-\mySinEta_{2}}{\phi_{un,i}}{\phi_{limb,f}}{\myIntensityDis_{pen}} $ }\\
		 $ \pi\le \phi\le 2\pi $ & $ -1\le \sin\phi\le 0 $ & $ -1\le \cos\phi\le 1  $ \\ 
		\midrule\midrule
		\multicolumn{3}{c}{\mycase{5}: $ \pi-\eta_{2}\le \myPhase\le \pi $}\\
		&  $ 0\le \sin{\myPhase}\le \mySinEta_{2} $ &  $ -1\le \cos{\myPhase}\le -\cos{\eta_{2}} $ \\
		& $ 0\le \sin^2\myPhase\le \mySinEta_{2}^2 $ & $ \cos^2\eta_{2}\le \cos^2\myPhase\le 1 $ \\
		\midrule 
		 $\piover2\le \phi_{limb,i}\le \pi $ & $ 0\le \sin\phi_{limb,i}\le 1 $& $ 0\le \cos\phi_{limb,i}\le 1 $\\
		 $  \frac{3\pi}{2}\le \phi_{limb,f}\le 2\pi $ & $ -1\le \sin{\phi_{limb,f}}\le 0 $ &$ 0\le \cos{\phi_{limb,f}}\le 1 $\\
		\midrule 
		\multicolumn{3}{c}{For $ \myKintegral{0}{\phi_{limb,i}}{\phi_{limb,f}}{\myIntensityDis_{pen}} $ }\\
		 $ \piover2 \le \phi\le  2\pi$ & $ -1\le \sin\phi\le 1 $& $ -1\le \cos\phi\le 1 $ \\		
		\midrule 
		\multicolumn{3}{c}{For $ \myKintegral{-\mySinEta_{2}}{\phi_{un,i}}{\phi_{un,f}}{\myIntensityDis_{pen}} $ }\\
		 $ \pi\le \phi\le 2\pi $& $ -1\le \sin{\phi}\le 0 $& $ -1\le \cos{\phi}\le 1 $\\
	\end{longtabu}
\end{center}
\vspace{-3em}

	\chapter{Integrations}\label[app]{app:integrations}
Here we present the evaluation of the integrals described in \cref{ch:reflectedlightluminosity}. The strategy will be to evaluate each $ \myQnCosepsshort{n}, \myQnSinepsshort{n},\myRnCosepsshort{n} $, and $ \myRnSinepsshort{n} $ and then to combine the indefinite integrals according to \cref{eq:fullQRequiv,eq:unQRequiv}. Finally, we will evaluate each equation describing $ \mu^n $ for the appropriate limits as described in \cref{tab:5cases,tab:thermal5cases}. 

For the evaluation of the integrals it will be convenient to define the following
\begin{equation}\label{eq:myFi}
	\myF{i}{\vartheta}=\sqrt{\cos^2{\eta_{i}}-\cos^2{\vartheta}}=\sqrt{\sin^2{\vartheta}-\mySinEta_{i}^2},
\end{equation}
and
\begin{equation}\label{eq:thephis}
	\begin{aligned}
		\myKopalPhi{1}{i}{\vartheta}&=\invcos{\frac{\mySinEta_i}{\sin{\vartheta}}}\\
		\myKopalPhi{2}{i}{\vartheta}& =\invtan{\frac{\mySinEta_i\cos{\vartheta}}{\myF{i}{\vartheta}}}\\
		\myKopalPhi{3}{i}{\vartheta}& =\invtan{\frac{\cos{\vartheta}}{\myF{i}{\vartheta}}}\\
		\myKopalPhi{4}{i}{\vartheta}& =\invsin{\frac{\cos{\vartheta}}{\sqrt{1 -\mySinEta_i^2 }}}=\invsin{\frac{\cos{\vartheta}}{\cos{\eta_i}}}\\
		\myKopalPhi{5}{i}{\vartheta}&=\invsin{\frac{\mySinEta_i}{\sin{\vartheta}}}=\piover2-\myKopalPhi{1}{i}{\vartheta}\\
		\myKopalPhi{6}{i}{\vartheta}&=\invcos{\frac{\vartheta}{\sqrt{1 -\mySinEta_i^2 }}}=\piover2-\myKopalPhi{4}{i}{\vartheta}.
	\end{aligned}
\end{equation}

\section{Integration of Fully Illuminated Zone}\label{sec:QRfull}
Recall that $ 0\le \phi\le \pi $ for the following integrations; therefore, both $ \sin{\phi} $ and $ \csc{\phi} $ are positive.
\subsection{Evaluation of $  \myQnCoseps{\mySinEta_{1}}{\phi_{i}}{\phi_{f}}{n} $ }
%
%
We begin with the evaluation of $ \myQnCoseps{\mySinEta_{1}}{\phi_{i}}{\phi_{f}}{1} $ 
\begin{equation}
	\begin{aligned}
		\myQnCoseps{\mySinEta_{1}}{\phi_{i}}{\phi_{f}}{1} &=\int_{\phi_i}^{\phi_f}\invcos{\frac{\mySinEta_{1}}{\sin\phi}}\sin{\phi}~d\phi\\
			& =\left.\mySinEta_{1}\myKopalPhi{3}{1}{\phi}-\myKopalPhi{2}{1}{\phi}-\myKopalPhi{1}{1}{\phi}\cos\phi\right\vert_{\phi_i}^{\phi_f}.
	\end{aligned}
\end{equation}
%
In the case of $ n=3 $ we have
\begin{equation}
	\begin{aligned}
		\myQnCoseps{\mySinEta_{1}}{\phi_{i}}{\phi_{f}}{3} & =\int_{\phi_i}^{\phi_f}\invcos{\frac{\mySinEta_{1}}{\sin\phi}}\sin^3{\phi}~d\phi\\
			& = \frac{1}{12}\left[-8\myKopalPhi{2}{1}{\phi}+(\cos{(3\phi)}-9\cos{\phi})\myKopalPhi{1}{1}{\phi}+\right.\\
			&\hspace{0.5in}\left.\left.2\mySinEta_{1}\left( \myF{1}{\phi}\cos{\phi}+(3+\mySinEta_{1}^2 )\myKopalPhi{3}{1}{\phi} \right) \right] \right\vert_{\phi_i}^{\phi_f}.\\
	\end{aligned}
\end{equation}
%
%
Finally, we may evaluate for the case in which $ n =5 $ as follows:
\begin{equation}
  \begin{aligned}
        	&\myQnCoseps{\mySinEta_{1}}{\phi_{i}}{\phi_{f}}{5} =\int_{\phi_i}^{\phi_f}\invcos{\frac{\mySinEta_{1}}{\sin\phi}}\sin^5{\phi}~d\phi\\
			& \qquad = \frac{1}{240} \left[(25 \cos (3 \phi )-3 \cos (5 \phi )-150 \cos \phi)\myKopalPhi{1}{1}{\phi}-128 \myKopalPhi{2}{1}{\phi}+\right.\\
			&\hspace{0.8in}  \mySinEta_{1} \left(\myF{1}{\phi} \left(18 \mySinEta_{1}^2+29\right) \cos\phi+2\left(9\mySinEta_{1}^4+10\mySinEta_{1}^2+45\right)\myKopalPhi{3}{1}{\phi}-\right.\\
			&\hspace{0.8in} \left. \left.3 \myF{1}{\phi} \cos (3 \phi )\right)\right]\big\vert_{\phi_i}^{\phi_f}\\
  \end{aligned}
\end{equation}

\subsection{Evaluation of $  \myQnSineps{\mySinEta_{1}}{\phi_{limb,i}}{\phi_{full,f}}{n} $ }
Let us continue our evaluation with integrals of the form of $ \myQnSineps{\mySinEta_{1}}{\phi_{full,i}}{\phi_{full,f}}{n}$. In the case of $ n=0 $ we have
\begin{equation}
	\begin{aligned}
			\myQnSineps{\mySinEta_{1}}{\phi_{i}}{\phi_{f}}{0}&= \int_{\phi_i}^{\phi_f}\cos{\phi}\invcos{\frac{\mySinEta_{1}}{\sin\phi}}~d\phi\\
			& = \Eval{\sin\phi\invcos{\frac{\mySinEta_{1}}{\cos{\phi}}} -\mySinEta_{1}\coth^{-1}\left(\frac{\myF{1}{\phi}}{\sin{\phi}}\right)}{\phi_i}{\phi_f}.
	\end{aligned}
\end{equation}
%
Following is an evaluation for $ n = 2 $ 
\begin{equation}
	\begin{aligned}
			\myQnSineps{\mySinEta_{1}}{\phi_{i}}{\phi_{f}}{2}&= \int_{\phi_i}^{\phi_f}\sin^2\phi\cos{\phi}\invcos{\frac{\mySinEta_{1}}{\sin\phi}}~d\phi\\
			& = \Eval{\frac{1}{6} \left(2\sin^3\phi\myKopalPhi{1}{1}{\phi}-\mySinEta_{1} \myF{1}{\phi}\sin\phi-\mySinEta_{1}^3  \coth^{-1}\left(\frac{\myF{1}{\phi}}{\sin{\phi}}\right)\right)}{\phi_i}{\phi_f}.
	\end{aligned}
\end{equation}
Finally, the following is an evaluation for $ n =4 $ 
\begin{equation}
	\begin{aligned}
			\myQnSineps{\mySinEta_{1}}{\phi_{i}}{\phi_{f}}{4}&= \int_{\phi_f}^{\phi_i}\sin^4\phi\cos{\phi}\invcos{\frac{\mySinEta_{1}}{\sin\phi}}~d\phi\\
			& =\frac{1}{40} \left(8 \sin ^5\phi \cos^{-1}\left(\frac{\mySinEta_{1}}{\sin{\phi}}\right)-\right.\\
			&\hspace{0.5in}	\left.\left. 2 \mySinEta_{1} \myF{1}{\phi} \sin ^3\phi-3\mySinEta_{1}^3\myF{1}{\phi} \sin \phi  - 3 \mySinEta_{1}^5 \coth ^{-1}\left(\frac{\myF{1}{\phi}}{\sin{\phi}}\right)\right) \right\vert_{\phi_i}^{\phi_f}
	\end{aligned}
\end{equation}

\subsection{Evaluation of $  \myRnCoseps{\mySinEta_{1}}{\phi_{i}}{\phi_{f}}{n} $ }
We begin with the evaluation of $ \myRnCoseps{\mySinEta_{1}}{\phi_{i}}{\phi_{f}}{-1} $ 
\begin{equation}
	\begin{aligned}
		\myRnCoseps{\mySinEta_{1}}{\phi_{i}}{\phi_{f}}{-1} &=\int_{\phi_i}^{\phi_f}\csc{\phi}\sqrt{\sin^2\phi-\mySinEta_{1}^2 }
~d\phi\\
			&=\Eval{\mySinEta_{1} \myKopalPhi{2}{1}{\phi}+\frac{1}{2} i \log \left(2-2 \mySinEta_{1}^2\right)-\myKopalPhi{4}{1}{\phi}}{\phi_i}{\phi_f}
	\end{aligned}
\end{equation}
%
In the case of $ n=1 $ we have
\begin{equation}
	\begin{aligned}
		\myRnCoseps{\mySinEta_{1}}{\phi_{i}}{\phi_{f}}{1} & =\int_{\phi_i}^{\phi_f} \sin{\phi}\sqrt{\sin^2\phi-\mySinEta_{1}^2 }
~d\phi\\
		& =\Eval{\frac{1}{4} \left(-2\myF{1}{\phi}\cos (\phi )-\left(1-\mySinEta_{1}^2\right) \left(2 \myKopalPhi{4}{1}{\phi}-i \log \left(2-2 \mySinEta_{1}^2\right)\right)\right)}{\phi_i}{\phi_f}
	\end{aligned}
\end{equation}
%
%
Finally, we may evaluate for the case in which $ n =3 $ as follows:
\begin{equation}
  \begin{aligned}
        	\myRnCoseps{\mySinEta_{1}}{\phi_{i}}{\phi_{f}}{3}& =\int_{\phi_i}^{\phi_f} \sin^3\phi\sqrt{\sin^2\phi-\mySinEta_{1}^2 }
~d\phi\\
		& =\frac{1}{16} \Bigg[\myF{1}{\phi}\cos (\phi ) \left(\cos (2 \phi )-2 \myF{1}{\phi}^2-7\right)+ \\
		& \left. \hspace{0.5in}\left(\mySinEta_{1}^4+2 \mySinEta_{1}^2-3\right) \left(2 \myKopalPhi{4}{1}{\phi}-i \log \left(2-2 \mySinEta_{1}^2\right)\right)\Bigg] \right\vert_{\phi_i}^{\phi_f}.
  \end{aligned}
\end{equation}

\subsection{Evaluation of $  \myRnSineps{\mySinEta_{1}}{\phi_{i}}{\phi_{f}}{n} $ }
We begin with the evaluation of $ \myRnSineps{\mySinEta_{1}}{\phi_{i}}{\phi_{f}}{-2} $ 
\begin{equation}
	\begin{aligned}
		\myRnSineps{\mySinEta_{1}}{\phi_{i}}{\phi_{f}}{-2} &=\int_{\phi_i}^{\phi_f}\left( \frac{\cos{\phi}}{\sin^2\phi}\right)\sqrt{\sin^2 {\phi}-\mySinEta_{1}^2 }
~d\phi\\
		& =\Eval{\invcoth{\frac{\myF{1}{\phi}}{\sin{\phi}}}-\frac{\myF{1}{\phi}}{\sin{\phi}}}{\phi_i}{\phi_f}
	\end{aligned}
\end{equation}
%
In the case of $ n=0 $ we have
\begin{equation}
	\begin{aligned}
		\myRnSineps{\mySinEta_{1}}{\phi_{i}}{\phi_{f}}{0} & =\int_{\phi_i}^{\phi_f}\cos{\phi}\sqrt{\sin^2 {\phi}-\mySinEta_{1}}
~d\phi\\
		& =\Eval{\frac{1}{2} \left[\myF{1}{\phi}\sin (\phi )-\mySinEta_{1}^2 \left(\log \left(\mySinEta_{1}\right)+\invcosh{\frac{\sin{\phi}}{\mySinEta_{1}}}\right)\right]}{\phi_i}{\phi_f}
	\end{aligned}
\end{equation}
where we have used the definition for the hyperbolic inverse cosine and \cref{eq:myFi}\ to show that 
\begin{equation*}
	 \log\left( \sin{\phi}+\myF{1}{\phi}\right)= \log\left( \mySinEta_{1}\right)+ \invcosh{\frac{\sin{\phi}}{\mySinEta_{1}}}.
\end{equation*} 
%
Finally, we may evaluate for the case in which $ n =2 $ as follows:
\begin{equation}
  \begin{aligned}
        	\myRnSineps{\mySinEta_{1}}{\phi_{i}}{\phi_{f}}{2}& =\int_{\phi_i}^{\phi_f} \sin^2\phi\cos{\phi}\sqrt{\sin^2 {\phi}-\mySinEta_{1}}
~d\phi\\
		& =\frac{1}{8} \bigg[2 \myF{1}{\phi}\sin ^3(\phi )-\\
		& \hspace{0.4in}\mySinEta_{1}^2 \left(\mySinEta_{1}^2 \invcoth{\frac{\myF{1}{\phi}}{\sin{\phi}}}+\myF{1}{\phi}\sin (\phi )+\log \left(i\mySinEta_{1}\right)\right)\bigg] \bigg\vert_{\phi_i}^{\phi_f}.
  \end{aligned}
\end{equation}

\subsubsection{Summation of $ \cos{\myPhase} $ Terms}\label{sec:x1cospart}
Here we list the terms of the form $ \mu^ n\cos{\myPhase} $ as described in \cref{eq:fullQRequiv}.

For  $ n=0 $ we have 
\begin{equation}
	\begin{aligned}
		\left( \myQnCosepsshort{1}+\mySinEta_{1} \myRnCosepsshort{-1} \right)\cos{\myPhase}= \bigg[&\mySinEta_{1}\left(\myKopalPhi{3}{1}{\phi}-\myKopalPhi{4}{1}{\phi}+\frac{1}{2} i \log \left(2-2 \mySinEta_{1}^2\right)\right)-\\
		&\myKopalPhi{1}{1}{\phi}\cos\phi-(1-\mySinEta_{1}^2)\myKopalPhi{2}{1}{\phi}\bigg]\cos{\myPhase}
	\end{aligned}
\end{equation}
and for $ n=1 $ 
\begin{equation}
  \begin{aligned}
        \left(\frac{4}{3}\myRnCosepsshort{1} +\frac{2}{3} \mySinEta_{1}^2  \myRnCosepsshort{-1}\right) \cos{\myPhase}=\frac{1}{3} \bigg[&-2 (\myF{1}{\phi}\cos (\phi )+\myKopalPhi{4}{1}{\phi})+\\
		&2 \myKopalPhi{2}{1}{\phi} \mySinEta_{1}^3+i \log \left(2-2 \mySinEta_{1}^2\right)\bigg]\cos{\myPhase}
  \end{aligned}
\end{equation}
Next, the $ n=2 $ terms simplify to
\begin{equation}
	\begin{aligned}
		&\left(\frac{3}{4} \myQnCosepsshort{3} + \frac{3}{4} \mySinEta_{1} \myRnCosepsshort{1} + \1over2\mySinEta_{1}^3 \myRnCosepsshort{-1}\right)\cos{\myPhase}=\frac{1}{16} \bigg[\myKopalPhi{1}{1}{\phi} (\cos (3 \phi )-9 \cos \phi)-8 \myKopalPhi{2}{1}{\phi}+\\
		&\hspace{0.5in}\mySinEta_{1} \left(6 \myKopalPhi{3}{1}{\phi}-6  \myKopalPhi{4}{1}{\phi}-4\myF{1}{\phi} \cos \phi+3 i \log \left(2-2 \mySinEta_{1}^2\right)+\right.\\
		&\hspace{0.5in}\left.\mySinEta_{1}^2 \left(8 \myKopalPhi{2}{1}{\phi} \mySinEta_{1}+2 \myKopalPhi{3}{1}{\phi}-2\myKopalPhi{4}{1}{\phi}+i \log \left(2-2 \mySinEta_{1}^2\right)\right)\right)\bigg]\cos{\myPhase}
	\end{aligned}
\end{equation}
and $ n=3 $ to
\begin{equation}
	\begin{aligned}
		&\left( \frac{16}{15} \myRnCosepsshort{3}+ \frac{8}{15} \mySinEta_{1}^2 \myRnCosepsshort{1} + \frac{2}{5}\mySinEta_{1}^4 \myRnCosepsshort{-1}\right)\cos{\myPhase}=\frac{1}{15} \bigg[\myF{1}{\phi} \cos \phi \left(-2 \myF{1}{\phi}^2+\cos (2 \phi )-7\right)-\\
		&\hspace{0.5in}4 \myF{1}{\phi} \mySinEta_{1}^2 \cos \phi+6 \myKopalPhi{2}{1}{\phi} \mySinEta_{1}^5-6 \myKopalPhi{4}{1}{\phi}+3 i \log \left(2-2 \mySinEta_{1}^2\right)\bigg]\cos{\myPhase}
	\end{aligned}
\end{equation}
Finally, $ n = 4 $ is given by 
\begin{equation}
  \begin{aligned}
        &\left( \frac{5}{8}\myQnCosepsshort{5}+\frac{5}{8} \mySinEta_{1} \myRnCosepsshort{3} + \frac{5}{12} \mySinEta_{1}^3\myRnCosepsshort{1} + \frac{1}{3}\mySinEta_{1}^5\myRnCosepsshort{-1}\right)\cos{\myPhase} =\\
		&\hspace{0.5in}\frac{1}{384} \bigg[\mySinEta_{1} \Big(\myF{1}{\phi} \cos\phi \left(-30 \myF{1}{\phi}^2+9 \cos (2 \phi )-73\right)+90 \myKopalPhi{3}{1}{\phi}-\\
		&\hspace{1.2in}90 \myKopalPhi{4}{1}{\phi}+45 i \log \left(2-2 \mySinEta_{1}^2\right)\Big)+\\
		&\hspace{0.8in}2 \mySinEta_{1}^3\left(-31 \myF{1}{\phi} \cos\phi+10 \myKopalPhi{3}{1}{\phi}-10 \myKopalPhi{4}{1}{\phi}+5 i \log \left(2-2 \mySinEta_{1}^2\right)\right)+\\
		&\hspace{0.8in}\myKopalPhi{1}{1}{\phi} (-150 \cos\phi+25 \cos (3 \phi )-3 \cos (5 \phi ))-\\
		&\hspace{0.8in}128\myKopalPhi{2}{1}{\phi}(1-\mySinEta_{1}^6)+9 \mySinEta_{1}^5 \left(2 \myKopalPhi{3}{1}{\phi}-2 \myKopalPhi{4}{1}{\phi}+i \log \left(2-2 \mySinEta_{1}^2\right)\right)\bigg]\cos{\myPhase}
  \end{aligned}
\end{equation}

\subsubsection{Summation of $ \sin{\myPhase} $ Terms}\label{sec:x1sinpart}
For  $ n=0 $ we have 
\begin{equation}
	\begin{aligned}
		\left( \myQnSinepsshort{0}+\mySinEta_{1}\myRnSinepsshort{-2}\right)\sin{\myPhase}&= \bigg[\myKopalPhi{1}{1}{\phi} \sin\phi-\myF{1}{\phi} \chi _1 \csc\phi\bigg]\sin{\myPhase}
	\end{aligned}
\end{equation}
and for $ n=1 $ 
\begin{equation}
  \begin{aligned}
        \left(\frac{4}{3}\myRnSinepsshort{0} + \frac{2}{3} \mySinEta_{1}^2 \myRnSinepsshort{-2}\right) \sin{\myPhase}&=\frac{1}{3} \bigg[2 \myF{1}{\phi} \sin\phi-2 \mySinEta_{1}^2 \left(\myF{1}{\phi} \csc\phi+\log \left(\mySinEta_{1}\right)+\frac{i \pi }{2}\right)\bigg]\sin{\myPhase}
  \end{aligned}
\end{equation}
Next, the $ n=2 $ terms simplify to
\begin{equation}
	\begin{aligned}
	\left(\frac{3}{4} \myQnSinepsshort{2} + \frac{3}{4} \mySinEta_{1}\myRnSinepsshort{0}+ \1over2 \mySinEta_{1}^3 \myRnSinepsshort{-2}\right)	\sin{\myPhase}&=\frac{1}{16}\bigg[4 \myF{1}{\phi} \mySinEta_{1} \sin\phi+\\
		&\hspace{0.5in}\mySinEta_{1}^3 \left(-8 \myF{1}{\phi} \csc\phi-6 \log \left(\mySinEta_{1}\right)-3 i \pi \right)+\\
		&\hspace{0.5in}4 \myKopalPhi{1}{1}{\phi} \sin ^3(\phi )\bigg]\sin{\myPhase}
	\end{aligned}
\end{equation}
and $ n=3 $ to
\begin{equation}
	\begin{aligned}
	\left(\frac{16}{15} \myRnSinepsshort{2}+ \frac{8}{15} \mySinEta_{1}^2 \myRnSinepsshort{0} + \frac{2}{5}\mySinEta_{1}^4 \myRnSinepsshort{-2} \right)\sin{\myPhase}& =\frac{1}{15}\bigg[-2 \mySinEta_{1}^2 \left(-\myF{1}{\phi} \sin\phi+\log \left(i \mySinEta_{1}\right)\right)-\\
		&\hspace{0.5in}2 \mySinEta_{1}^4 \left(3 \myF{1}{\phi} \csc\phi+2 \log \left(\mySinEta_{1}\right)+i \pi \right)+\\
		&\hspace{0.5in}4 \myF{1}{\phi} \sin ^3(\phi )\bigg]\sin{\myPhase}
	\end{aligned}
\end{equation}
Finally, $ n = 4 $ is given by
\begin{equation}
  \begin{aligned}
        &\left( \frac{5}{8}\myQnSinepsshort{4}+\frac{5}{8} \mySinEta_{1} \myRnSinepsshort{2} + \frac{5}{12} \mySinEta_{1}^3\myRnSinepsshort{0} + \frac{1}{3}\mySinEta_{1}^5\myRnSinepsshort{-2} \right)\sin{\myPhase} =\\
		&\hspace{0.5in}\frac{1}{192}\bigg[24 \myF{1}{\phi} \mySinEta_{1} \sin ^3\phi+\mySinEta_{1}^3 \left(16 \myF{1}{\phi} \sin\phi-15 \log \left(i \mySinEta_{1}\right)\right)-\\
		&\hspace{0.9in}8 \mySinEta_{1}^5 \left(8 \myF{1}{\phi} \csc\phi+5 \log \left(\mySinEta_{1}\right)+\frac{5 i \pi
   }{2}\right)+24 \myKopalPhi{1}{1}{\phi} \sin ^5\phi\bigg]\sin{\myPhase}
  \end{aligned}
\end{equation}

\section{Integration of the Un-illuminated Zone}
Recall that $ \pi\le \phi\le 2\pi $ for the following integrations; therefore, both $ \sin{\phi} $ and $ \csc{\phi} $ are negative.

\subsection{Evaluation of $  \myQnCoseps{-\mySinEta_{2}}{\phi_{i}}{\phi_{f}}{n} $ }
We begin with the evaluation of $ \myQnCoseps{-\mySinEta_{2}}{\phi_{i}}{\phi_{f}}{1} $ 
\begin{equation}
	\begin{aligned}
		\myQnCoseps{-\mySinEta_{2}}{\phi_{i}}{\phi_{f}}{1}&=\int_{\phi_i}^{\phi_f}\invcos{\frac{-\mySinEta_{2}}{\sin{\phi}}}\sin{\phi}~d\phi\\
			& =\left.\mySinEta_{2}\myKopalPhi{3}{2}{\phi}-\myKopalPhi{2}{2}{\phi}-\myKopalPhi{1}{2}{-\phi}\cos\phi\right\vert_{\phi_i}^{\phi_f}.
	\end{aligned}
\end{equation}
where we had chosen the principal values for all branch cuts.
%
In the case of $ n=3 $ we have
\begin{equation}
	\begin{aligned}
		\myQnCoseps{-\mySinEta_{2}}{\phi_{i}}{\phi_{f}}{3} &=\int_{\phi_i}^{\phi_f}\invcos{\frac{-\mySinEta_{2}}{\sin{\phi}}}\sin^3{\phi}~d\phi\\
			& = \frac{1}{12}\left[-8\myKopalPhi{2}{2}{\phi}+(\cos{(3\phi)}-9\cos{\phi})\myKopalPhi{1}{2}{-\phi}+\right.\\
			&\hspace{0.5in}\left.\left.2\mySinEta_{2}\left( \myF{2}{\phi}\cos{\phi}+(3+\mySinEta_{2}^2 )\myKopalPhi{3}{2}{\phi} \right) \right] \right\vert_{\phi_i}^{\phi_f}.\\
	\end{aligned}
\end{equation}
%
%
Finally, we may evaluate for the case in which $ n =5 $ as follows:
\begin{equation}
  \begin{aligned}
        	&\myQnCoseps{-\mySinEta_{2}}{\phi_{i}}{\phi_{f}}{5}=\int_{\phi_i}^{\phi_f}\invcos{\frac{-\mySinEta_{2}}{\sin{\phi}}}\sin^5{\phi}~d\phi\\
			& \qquad = \frac{1}{240} \left[\textbf{}(25 \cos (3 \phi )-3 \cos (5 \phi )-150 \cos \phi)\myKopalPhi{1}{2}{-\phi}-128 \myKopalPhi{2}{2}{\phi}+\right.\\
			&\hspace{0.5in}\mySinEta_{2} \left(\myF{2}{\phi} \left(18 \mySinEta_{2}^2+29\right) \cos\phi+2\left(9\mySinEta_{2}^4+10\mySinEta_{2}^2+45\right)\myKopalPhi{3}{2}{\phi}-\right.\\
			&\hspace{0.5in}\left.\left.\left. 3 \myF{2}{\phi} \cos (3 \phi )\right)\right]\right\vert_{\phi_i}^{\phi_f}\\
  \end{aligned}
\end{equation}

\subsection{Evaluation of $  \myQnSineps{-\mySinEta_{2}}{\phi_{i}}{\phi_{f}}{n} $ }
First, let us consider the integration for $ n = 0 $, 
\begin{equation}
	\begin{aligned}
			\myQnSineps{-\mySinEta_{2}}{\phi_{i}}{\phi_{f}}{0}&=\int_{\phi_i}^{\phi_f}\cos{\phi}\invcos{\frac{-\mySinEta_{2}}{\sin{\phi}}}~d\phi\\
			& = \Eval{\sin\phi\myKopalPhi{1}{2}{-\phi} -\mySinEta_{2}\coth^{-1}\left(\frac{\myF{2}{\phi}}{\sin{\phi}}\right)}{\phi_i}{\phi_f}.
	\end{aligned}
\end{equation}
%
Next,  is an evaluation for $ n = 2 $ 
\begin{equation}
	\begin{aligned}
			\myQnSineps{-\mySinEta_{2}}{\phi_{i}}{\phi_{f}}{2}&=\int_{\phi_i}^{\phi_f}\sin^2\phi\cos{\phi}\invcos{\frac{-\mySinEta_{2}}{\sin{\phi}}}~d\phi\\
			& = \Eval{\frac{1}{6} \left(2\sin^3\phi\myKopalPhi{1}{2}{-\phi}-\mySinEta_{2} \myF{2}{\phi}\sin\phi-\mySinEta_{2}^3  \coth^{-1}\left(\frac{\myF{2}{\phi}}{\sin{\phi}}\right)\right)}{\phi_i}{\phi_f}.
	\end{aligned}
\end{equation}
%
%
Finally, we present the evaluation for $ n =4 $ 
\begin{equation}
	\begin{aligned}
			\myQnSineps{-\mySinEta_{2}}{\phi_{i}}{\phi_{f}}{4}&= \int_{\phi_f}^{\phi_i}\sin^4\phi\cos{\phi}\invcos{\frac{-\mySinEta_{2}}{\sin{\phi}}}~d\phi\\
			& =\frac{1}{40} \bigg(8 \sin ^5\phi \myKopalPhi{1}{2}{-\phi}-\\
			&\hspace{0.5in} 2 \mySinEta_{2} \myF{2}{\phi} \sin ^3\phi-3\mySinEta_{2}^3\myF{2}{\phi} \sin \phi  -\\
			&\hspace{0.5in}\left. 3 \mySinEta_{2}^5 \coth ^{-1}\left(\frac{\myF{2}{\phi}}{\sin{\phi}}\right)\bigg) \right\vert_{\phi_i}^{\phi_f}.
	\end{aligned}
\end{equation}

\subsection{Evaluation of $  \myRnCoseps{-\mySinEta_{2}}{\phi_{i}}{\phi_{f}}{n} $ }
We begin with the evaluation of $ \myRnCoseps{-\mySinEta_{2}}{\phi_{i}}{\phi_{f}}{-1} $ 
\begin{equation}
	\begin{aligned}
		\myRnCoseps{-\mySinEta_{2}}{\phi_{i}}{\phi_{f}}{-1} &\int_{\phi_i}^{\phi_f}\csc{\phi}\invcos{\frac{-\mySinEta_{2}}{\sin{\phi}}}~
d\phi\\
			&=\Eval{\mySinEta_{2} \myKopalPhi{2}{2}{\phi}+\frac{1}{2} i \log \left(2-2 \mySinEta_{2}^2\right)-\myKopalPhi{4}{2}{\phi}}{\phi_i}{\phi_f}
	\end{aligned}
\end{equation}
%
In the case of $ n=1 $ we have
\begin{equation}
	\begin{aligned}
		\myRnCoseps{-\mySinEta_{2}}{\phi_{i}}{\phi_{f}}{1} & =\int_{\phi_i}^{\phi_f} \sin{\phi}\invcos{\frac{-\mySinEta_{2}}{\sin{\phi}}}~
d\phi\\
		& =\Eval{\frac{1}{4} \left(-2\myF{2}{\phi}\cos \phi-\left(1-\mySinEta_{2}^2\right) \left(2 \myKopalPhi{4}{2}{\phi}-i \log \left(2-2 \mySinEta_{2}^2\right)\right)\right)}{\phi_i}{\phi_f}.
	\end{aligned}
\end{equation}
%
%
Finally, we may consider the case in which $ n =3 $ as follows:
\begin{equation}
  \begin{aligned}
        	\myRnCoseps{\mySinEta_{1}}{\phi_{i}}{\phi_{f}}{3}& =\int_{\phi_i}^{\phi_f} \sin^3\phi\invcos{\frac{-\mySinEta_{2}}{\sin{\phi}}}~
d\phi\\
		& =\frac{1}{16} \Bigg[\myF{2}{\phi}\cos \phi \left(\cos (2 \phi )-2 \myF{2}{\phi}^2-7\right)+ \\
		& \left. \hspace{0.5in}\left(\mySinEta_{2}^4+2 \mySinEta_{2}^2-3\right) \left(2 \myKopalPhi{4}{2}{\phi}-i \log \left(2-2 \mySinEta_{2}^2\right)\right)\Bigg] \right\vert_{\phi_i}^{\phi_f}.
  \end{aligned}
\end{equation}

\subsection{Evaluation of $  \myRnSineps{-\mySinEta_{2}}{\phi_{i}}{\phi_{f}}{n} $ }
We begin with the evaluation of $ \myRnSineps{-\mySinEta_{2}}{\phi_{i}}{\phi_{f}}{-2} $ 
\begin{equation}
	\begin{aligned}
		\myRnSineps{-\mySinEta_{2}}{\phi_{i}}{\phi_{f}}{-2} &= \int_{\phi_i}^{\phi_f}\left( \frac{\cos{\phi}}{\sin^2\phi}\right)\invcos{\frac{-\mySinEta_{2}}{\sin{\phi}}}~
d\phi\\
		& =\Eval{\invcoth{\frac{\myF{2}{\phi}}{\sin{\phi}}}-\frac{\myF{2}{\phi}}{\sin{\phi}}}{\phi_i}{\phi_f}.
	\end{aligned}
\end{equation}
%
In the case of $ n=0 $ we have
\begin{equation}
	\begin{aligned}
		\myRnSineps{-\mySinEta_{2}}{\phi_{i}}{\phi_{f}}{0} & =\int_{\phi_i}^{\phi_f}\cos{\phi}\invcos{\frac{-\mySinEta_{2}}{\sin{\phi}}}~
d\phi\\
		& =\Eval{\frac{1}{2} \left[\myF{1}{\phi}\sin \phi-\mySinEta_{1}^2\log\left(\myF{2}{\phi}+\sin{\phi}\right)\right]}{\phi_i}{\phi_f}.
	\end{aligned}
\end{equation}
Note, that  $ \sin{\phi}/\mySinEta_{2} $ is a negative quantity; therefore, we cannot rewrite $ \log\left( \sin{\phi}+\myF{2}{\phi}\right) $ as  $ \log\left( \mySinEta_{2}\right)+ \invcosh{\sin{\phi}/\mySinEta_{2}}$ as was done in \cref{sec:QRfull}.
%
Finally, we may evaluate for the case in which $ n =2 $ as follows:
\begin{equation}
  \begin{aligned}
        	\myRnSineps{-\mySinEta_{2}}{\phi_{i}}{\phi_{f}}{2}& =\int_{\phi_i}^{\phi_f} \sin^2\phi\cos{\phi}\invcos{\frac{-\mySinEta_{2}}{\sin{\phi}}}~
d\phi\\
		& =\frac{1}{8} \bigg[2 \myF{2}{\phi}\sin ^3\phi-\\
		&\hspace{0.4in}\mySinEta_{2}^2 \left(\mySinEta_{2}^2 \invcoth{\frac{\myF{2}{\phi}}{\sin{\phi}}}+\myF{2}{\phi}\sin \phi+\log \left(i\mySinEta_{2}\right)\right)\bigg] \bigg\vert_{\phi_i}^{\phi_f}.
  \end{aligned}
\end{equation}

\subsubsection{Summation of $ \cos{\myPhase} $ Terms}\label{sec:x2cospart}
For  $ n=0 $ we have 
\begin{equation}
	\begin{aligned}
		\left( \myQnCosepsshort{1}+\mySinEta_{2} \myRnCosepsshort{-1} \right)\cos{\myPhase}= \bigg[&\mySinEta_{2}\left(\myKopalPhi{3}{2}{\phi}-\myKopalPhi{4}{2}{\phi}+\frac{1}{2} i \log \left(2-2 \mySinEta_{2}^2\right)\right)-\\
		&\myKopalPhi{1}{2}{-\phi}\cos\phi-(1-\mySinEta_{2}^2)\myKopalPhi{2}{2}{\phi}\bigg]\cos{\myPhase}
	\end{aligned}
\end{equation}
and for $ n=1 $ 
\begin{equation}
  \begin{aligned}
        \left(\frac{4}{3}\myRnCosepsshort{1} +\frac{2}{3} \mySinEta_{2}^2  \myRnCosepsshort{-1}\right) \cos{\myPhase}=\frac{1}{3} \bigg[&-2 (\myF{2}{\phi}\cos \phi+\myKopalPhi{4}{2}{\phi})+\\
		&2 \myKopalPhi{2}{2}{\phi} \mySinEta_{2}^3+i \log \left(2-2 \mySinEta_{2}^2\right)\bigg]\cos{\myPhase}
	\end{aligned}
\end{equation}
Next, the $ n=2 $ terms simplify to
\begin{equation}
	\begin{aligned}
		&\left(\frac{3}{4} \myQnCosepsshort{3} + \frac{3}{4} \mySinEta_{2} \myRnCosepsshort{1} + \1over2\mySinEta_{2}^3 \myRnCosepsshort{-1}\right)\cos{\myPhase}=\frac{1}{16} \bigg[\myKopalPhi{1}{2}{-\phi} (\cos (3 \phi )-9 \cos \phi)-8 \myKopalPhi{2}{2}{\phi}+\\
		&\hspace{0.5in}\mySinEta_{2} \left(6 \myKopalPhi{3}{2}{\phi}-6  \myKopalPhi{4}{2}{\phi}-4\myF{1}{\phi} \cos \phi+3 i \log \left(2-2 \mySinEta_{2}^2\right)+\right.\\
		&\hspace{0.5in}\left.\mySinEta_{2}^2 \left(8 \myKopalPhi{2}{2}{\phi} \mySinEta_{2}+2 \myKopalPhi{3}{2}{\phi}-2\myKopalPhi{4}{2}{\phi}+i \log \left(2-2 \mySinEta_{2}^2\right)\right)\right)\bigg]\cos{\myPhase}
	\end{aligned}
\end{equation}
and $ n=3 $ to
\begin{equation}
	\begin{aligned}
		&\left( \frac{16}{15} \myRnCosepsshort{3}+ \frac{8}{15} \mySinEta_{2}^2 \myRnCosepsshort{1} + \frac{2}{5}\mySinEta_{2}^4 \myRnCosepsshort{-1}\right)\cos{\myPhase}=\frac{1}{15} \bigg[\myF{2}{\phi} \cos \phi \left(-2 \myF{2}{\phi}^2+\cos (2 \phi )-7\right)-\\
		&\hspace{0.5in}4 \myF{2}{\phi} \mySinEta_{2}^2 \cos \phi+6 \myKopalPhi{2}{2}{\phi} \mySinEta_{2}^5-6 \myKopalPhi{4}{2}{\phi}+3 i \log \left(2-2 \mySinEta_{2}^2\right)\bigg]\cos{\myPhase}
	\end{aligned}
\end{equation}
Finally, $ n = 4 $ is given by 
\begin{equation}
  \begin{aligned}
        &\left( \frac{5}{8}\myQnCosepsshort{5}+\frac{5}{8} \mySinEta_{2} \myRnCosepsshort{3} + \frac{5}{12} \mySinEta_{2}^3\myRnCosepsshort{1} + \frac{1}{3}\mySinEta_{2}^5\myRnCosepsshort{-1}\right)\cos{\myPhase} =\\
		&\hspace{0.5in}\frac{1}{384} \bigg[\mySinEta_{2} \Big(\myF{2}{\phi} \cos\phi \left(-30 \myF{2}{\phi}^2+9 \cos (2 \phi )-73\right)+\\
		&\hspace{1.2in}90 \myKopalPhi{3}{2}{\phi}-90 \myKopalPhi+45 i \log \left(2-2 \mySinEta_{2}^2\right)\Big)+\\
		&\hspace{0.8in}9 \mySinEta_{2}^5 \left(2 \myKopalPhi{3}{2}{\phi}-2 \myKopalPhi{4}{2}{\phi}+i \log \left(2-2 \mySinEta_{2}^2\right)\right)+\\
		&\hspace{0.8in}2 \mySinEta_{2}^3 \Big(-31 \myF{2}{\phi} \cos\phi+10 \myKopalPhi{3}{2}{\phi}-10 \myKopalPhi{4}{2}{\phi}+5 i \log \left(2-2 \mySinEta_{2}^2\right)\Big)-\\
		&\hspace{0.8in}128 \myKopalPhi{2}{2}{\phi}(1-\mySinEta_{2}^6)+\myKopalPhi{1}{2}{-\phi} (-150 \cos\phi+25 \cos (3 \phi )-3 \cos (5 \phi ))\bigg]\cos\myPhase
  \end{aligned}
\end{equation}

\subsubsection{Summation of $ \sin{\myPhase} $ Terms}\label{sec:x2sinpart}
For  $ n=0 $ we have 
\begin{equation}
	\begin{aligned}
		\left( \myQnSinepsshort{0}+\mySinEta_{2}\myRnSinepsshort{-2}\right)\sin{\myPhase}&= \bigg[\myKopalPhi{1}{2}{-\phi} \sin\phi-\myF{2}{\phi} \mySinEta_{2} \csc\phi\bigg]\sin{\myPhase}
	\end{aligned}
\end{equation}
and for $ n=1 $ 
\begin{equation}
  \begin{aligned}
        \left(\frac{4}{3}\myRnSinepsshort{0} + \frac{2}{3} \mySinEta_{2}^2 \myRnSinepsshort{-2}\right) \sin{\myPhase}&=\frac{1}{3}\bigg[2 \myF{2}{\phi} \sin\phi-\mySinEta_{2}^2 \left(2 \myF{2}{\phi} \csc\phi+\log \left(-\mySinEta_{2}^2\right)\right)\bigg]\sin{\myPhase}
  \end{aligned}
\end{equation}
Next, the $ n=2 $ terms simplify to
\begin{equation}
	\begin{aligned}
	\left(\frac{3}{4} \myQnSinepsshort{2} + \frac{3}{4} \mySinEta_{2}\myRnSinepsshort{0}+ \1over2 \mySinEta_{2}^3 \myRnSinepsshort{-2}\right)	\sin{\myPhase}&=\frac{1}{16}\bigg[4 \myF{2}{\phi} \mySinEta_{2} \sin\phi-\\
	&\hspace{0.5in}\mySinEta_{2}^3 \left(8 \myF{2}{\phi} \csc\phi+3 \log \left(-\mySinEta_{2}^2\right)\right)+\\
		&\hspace{0.5in}4 \myKopalPhi{1}{2}{-\phi} \sin ^3\phi\bigg]\sin{\myPhase}
	\end{aligned}
\end{equation}
and $ n=3 $ to
\begin{equation}
	\begin{aligned}
	\left(\frac{16}{15} \myRnSinepsshort{2}+ \frac{8}{15} \mySinEta_{2}^2 \myRnSinepsshort{0} + \frac{2}{5}\mySinEta_{2}^4 \myRnSinepsshort{-2} \right)\sin{\myPhase}& =\frac{1}{15}\bigg[-2 \mySinEta_{2}^2 \left(-\myF{1}{\phi} \sin\phi+\log \left(i \mySinEta_{2}\right)\right)-\\
		&\hspace{0.5in}2 \mySinEta_{2}^4 \left(3 \myF{1}{\phi} \csc\phi+\log \left(-\mySinEta_{2}^2\right)\right)+\\
		&\hspace{0.5in}4 \myF{1}{\phi} \sin ^3\phi\bigg]\sin{\myPhase}
	\end{aligned}
\end{equation}
Finally, $ n = 4 $ is given by
\begin{equation}
  \begin{aligned}
        &\left( \frac{5}{8}\myQnSinepsshort{4}+\frac{5}{8} \mySinEta_{2} \myRnSinepsshort{2} + \frac{5}{12} \mySinEta_{2}^3\myRnSinepsshort{0} + \frac{1}{3}\mySinEta_{2}^5\myRnSinepsshort{-2} \right)\sin{\myPhase} =\\
		&\hspace{0.5in}\frac{1}{192}\bigg[24 \myF{2}{\phi} \mySinEta_{2} \sin ^3\phi+\mySinEta_{2}^3 \left(16 \myF{2}{\phi} \sin\phi-15 \log \left(i \mySinEta_{2}\right)\right)-\\
		&\hspace{0.9in}4 \mySinEta_{2}^5 \left(16 \myF{2}{\phi} \csc\phi+5 \log \left(-\mySinEta_{2}^2\right)\right)+24
   \myKopalPhi{1}{2}{-\phi} \sin ^5\phi\bigg]\sin{\myPhase}
  \end{aligned}
\end{equation}

\section{Evaluation of the $K-$operators}\label{sec:koperators}
Here, we present the evaluation of the $ K $-operator for the fully illuminated and un-illuminated zones for the five values of $ \mu^n $. It should be noted that the results for the fully illuminated zone will apply to any spherical cap for which the azimuthal angles lie between zero and $ \pi $. Likewise, the results in the un-illuminated zone apply for any spherical cap for which the azimuthal angles lie between $ \pi $ and $ 2\pi $. Therefore, to evaluate the luminosity of Penumbral Zone One described in \cref{sec:luminosityFiniteSizenew}, we may simply use the equations given by $ \myKintegral{\mySinEta_{1}}{\phi_{i}}{\phi_{f}}{\mu^n} $ and replace $ \mySinEta_{1} $ with $ \mySinEta_{\lambda} $ if it is possible to determine the reflected intensity distribution as the sum of powers of $\mu$. 

The following equations were determined by combining the cosine and sine parts given in \cref{sec:x1cospart,sec:x1sinpart,sec:x2cospart,sec:x2sinpart}\ as described by \cref{eq:fullQRequiv,eq:unQRequiv}\ and then inserting the appropriate limits of $ \phi $ as given by the arguments described by the $ K $-operator. 

\subsection{Evaluation of $ \myKintegral{\mySinEta_{1}}{\phi_{full,i}}{\phi_{full,f}}{\mu^n} $ }\label{sec:fulledgetoedge}

\begin{equation}\label{eq:fulledgetoedge}
	 \myKintegral{\mySinEta_{1}}{\phi_{full,i}}{\phi_{full,f}}{\mu^n} =R_p^2\begin{cases}
		\pi(1-\mySinEta_{1}^2)\cos{\myPhase} &, \, n=0\\
		\frac{2}{3}\pi(1-\mySinEta_{1}^3)\cos{\myPhase}&,\, n=1\\
		\1over2\pi(1-\mySinEta_{1}^4)\cos{\myPhase}&,\, n=2\\
		\frac{2}{5}\pi(1-\mySinEta_{1}^5)\cos{\myPhase}&,\, n=3\\
		\frac{1}{3}\pi(1-\mySinEta_{1}^6)\cos{\myPhase}&,\, n=4.
	\end{cases}
\end{equation}

\subsection{Evaluation of $ \myKintegral{\mySinEta_{1}}{\phi_{limb,i}}{\phi_{full,f}}{\mu^n} $ }\label{sec:fulllimbtoedge}
We will address each integral in order, beginning with $\mu^0$:
\begin{equation}\label{eq:fulllimbtoedgemu0}
\begin{aligned}
	\myKintegral{\mySinEta_{1}}{\phi_{limb,i}}{\phi_{full,f}}{\mu^0}=R_p^2\bigg(&\myKopalPhi{1}{1}{\myPhase} -\mySinEta_{1} \myF{1}{\myPhase}+\\
		&\cos\myPhase\left[\myKopalPhi{2}{1}{\myPhase}+\frac{\pi }{2}+\right.\\
		&\hspace{0.45in}\mySinEta_{1} \left(\myKopalPhi{4}{1}{\myPhase}-\myKopalPhi{3}{1}{\myPhase}\right)-\\
		&\hspace{0.45in}\left.\mySinEta_{1}^2 \left(\myKopalPhi{2}{1}{\myPhase}+\frac{\pi }{2}\right)\right]\bigg).
\end{aligned}	 
\end{equation}

Next, we have the case for $ \mu^1 $ in the fully illuminated zone, 
\begin{equation}\label{eq:fulllimbtoedgemu1}
\begin{aligned}
	\myKintegral{\mySinEta_{1}}{\phi_{limb,i}}{\phi_{full,f}}{\mu^1}=R_p^2\left( \frac{2}{3}\right)\bigg(& (1-\mySinEta_{1}^2)\myF{1}{\myPhase} -\\
		&\left[\left( \myKopalPhi{2}{1}{\myPhase}+\piover2\right)  \mySinEta_{1}^3 + \left( \myKopalPhi{4}{1}{\myPhase}+\piover2\right) \right]\cos\myPhase\bigg).
\end{aligned}	 
\end{equation}

For the case for which $ \mu^2 $ in the fully illuminated zone,
\begin{equation}\label{eq:fulllimbtoedgemu2}
\begin{aligned}
	\myKintegral{\mySinEta_{1}}{\phi_{limb,i}}{\phi_{full,f}}{\mu^2}=\frac{R_p^2}{8}\bigg(&2\myKopalPhi{1}{1}{\myPhase}\left(1+\cos^2{\myPhase} \right)+2 \myF{1}{\myPhase} \mySinEta_{1}-4 \myF{1}{\myPhase} \mySinEta_{1}^3+\\
&\cos\myPhase \bigg[4 \left( \myKopalPhi{2}{1}{\myPhase}+\piover2\right) +3 \mySinEta_{1} (\myKopalPhi{4}{1}{\myPhase}-\myKopalPhi{3}{1}{\myPhase})\\
&\hspace{0.45in}\mySinEta_{1}^3 (\myKopalPhi{4}{1}{\myPhase}-\myKopalPhi{3}{1}{\myPhase})-4  \left( \myKopalPhi{2}{1}{\myPhase}+\piover2 \right) \mySinEta_{1}^4\bigg]\bigg).
\end{aligned}	 
\end{equation}

In addition, for $ \mu^3 $ in the fully illuminated zone,
\begin{equation}\label{eq:fulllimbtoedgemu3}
\begin{aligned}
	\myKintegral{\mySinEta_{1}}{\phi_{limb,i}}{\phi_{full,f}}{\mu^3}=R_p^2\left( \frac{2}{15}\right)\bigg(&\myF{1}{\myPhase}(2+\cos^2 {\myPhase})+ \myF{1}{\myPhase} \mySinEta_{1}^2-3 \myF{1}{\myPhase} \mySinEta_{1}^4+\\
&3\cos\myPhase\left[\left( \myKopalPhi{4}{1}{\myPhase}+\piover2\right)- \left( \myKopalPhi{2}{1}{\myPhase}+\piover2 \right) \mySinEta_{1}^5 \right]\bigg).
\end{aligned}	 
\end{equation}

Finally for $ \mu^4 $ in the fully illuminated zone, 
\begin{equation}\label{eq:fulllimbtoedgemu4}
\begin{aligned}
	\myKintegral{\mySinEta_{1}}{\phi_{limb,i}}{\phi_{full,f}}{\mu^4}=\frac{R_p^2}{192}\bigg(&8 \myF{1}{\myPhase} \mySinEta_{1} (3+\cos^2 {\myPhase})+16 \myF{1}{\myPhase} \mySinEta_{1}^3-\\
		&64 \myF{1}{\myPhase} \mySinEta_{1}^5+\myKopalPhi{1}{1}{\myPhase} (20 \cos (2 \myPhase )-\cos (4 \myPhase )+45)+\\
		&\cos\myPhase \bigg[64 \left(  \myKopalPhi{2}{1}{\myPhase}+\piover2\right)+\\
		&\hspace{0.45in}45\mySinEta_{1} ( \myKopalPhi{4}{1}{\myPhase}-\myKopalPhi{3}{1}{\myPhase})+\\
		&\hspace{0.45in}10\mySinEta_{1}^3 ( \myKopalPhi{4}{1}{\myPhase}- \myKopalPhi{3}{1}{\myPhase})+\\
		&\hspace{0.45in}9\mySinEta_{1}^5 ( \myKopalPhi{4}{1}{\myPhase}- \myKopalPhi{3}{1}{\myPhase})-\\
		&\hspace{0.45in}64 \left( \myKopalPhi{2}{1}{\myPhase}+\piover2\right)  \mySinEta_{1}^6\bigg]\bigg).
\end{aligned}	 
\end{equation}

\subsection{Evaluation of $ \myKintegral{\mySinEta_{2}}{\phi_{un,i}}{\phi_{un,f}}{\mu^n} $}\label{sec:unedgetoedge}
\begin{equation}\label{eq:unedgetoedge}
	 \myKintegral{\mySinEta_{1}}{\phi_{full,i}}{\phi_{full,f}}{\mu^n} =R_p^2\begin{cases}
		 -\pi(1-\mySinEta_{2}^2)\cos\myPhase &, \, n=0\\
		\frac{2}{3}\pi(1-\mySinEta_{2}^3)\cos{\myPhase}&,\, n=1\\
		-\1over2\pi(1-\mySinEta_{2}^4)\cos{\myPhase}&,\, n=2\\
		\frac{2}{5}\pi(1-\mySinEta_{2}^5)\cos{\myPhase}&,\, n=3\\
		 -\frac{1}{3}\pi(1-\mySinEta_{2}^6)\cos{\myPhase}&,\, n=4.
	\end{cases}
\end{equation}

\subsection{Evaluation of $ \myKintegral{\mySinEta_{2}}{\phi_{un,i}}{\phi_{limb,f}}{\mu^n} $ }\label{sec:unedgetolimb}

Let us begin with the results for integration over $\mu^0$ in the un-illuminated zone,
\begin{equation}\label{eq:unlimbtoedgemu0}
	\begin{aligned}
		\myKintegral{\mySinEta_{2}}{\phi_{un,i}}{\phi_{limb,f}}{\mu^0}= R_p^2\bigg(&\myKopalPhi{1}{2}{\myPhase}-\mySinEta_{2} \myF{2}{\myPhase}+\\
	&\cos\myPhase\left[\myKopalPhi{2}{2}{\myPhase}-\frac{\pi }{2}+\right.\\
&\hspace{0.45in}\mySinEta_{2} \left(\myKopalPhi{4}{2}{\myPhase}-\myKopalPhi{3}{2}{\myPhase}\right)+\\
&\hspace{0.45in}\left.\mySinEta_{2}^2\left(\frac{\pi}{2}-\myKopalPhi{2}{2}{\myPhase}\right).
\right]\bigg)
	\end{aligned}
\end{equation}

Next, we have the case for $ \mu^1 $ in the un-illuminated zone, 
\begin{equation}\label{eq:unlimbtoedgemu1}
\begin{aligned}
	\myKintegral{-\mySinEta_{2}}{\phi_{un,i}}{\phi_{limb,f}}{\mu^1}=R_p^2\left(\frac{2}{3}\right)\bigg(&2(1-\mySinEta_{2}^2)\myF{2}{\myPhase} -\\
		&\left[\left( \piover2 - \myKopalPhi{2}{2}{\myPhase}\right) \mySinEta_{2}^3 + \myKopalPhi{6}{2}{\myPhase}\right] \cos\myPhase\bigg).
\end{aligned}	 
\end{equation}

For the case for which $ \mu^2 $ in the un-illuminated zone, 
\begin{equation}\label{eq:unlimbtoedgemu2}
\begin{aligned}
	\myKintegral{-\mySinEta_{2}}{\phi_{un,i}}{\phi_{limb,f}}{\mu^2}=\frac{R_p^2}{8}\bigg(&2 \myKopalPhi{1}{2}{\myPhase}\left(\cos ^2\myPhase+1\right)+2 \myF{2}{\myPhase} \mySinEta_{2}-4 \myF{2}{\myPhase} \mySinEta_{2}^3+\\
	&\cos \myPhase \bigg[4\left(\myKopalPhi{2}{2}{\myPhase}-\piover2\right)+ 3 (\myKopalPhi{4}{2}{\myPhase}-\myKopalPhi{3}{2}{\myPhase})\mySinEta_{2}+ \\
	&\hspace{0.40in}(\myKopalPhi{4}{2}{\myPhase}-\myKopalPhi{3}{2}{\myPhase})\mySinEta_{2}^3+4 \left( \piover2 - \myKopalPhi{2}{2}{\myPhase}\right)\mySinEta_{2}^4\bigg]\bigg).
\end{aligned}	 
\end{equation}

In addition, for $ \mu^3 $ in the un-illuminated zone, 
\begin{equation}\label{eq:unlimbtoedgemu3}
\begin{aligned}
	\myKintegral{-\mySinEta_{2}}{\phi_{un,i}}{\phi_{limb,f}}{\mu^3}=R_p^2\left( \frac{2}{15}\right)\bigg(&-\myF{2}{\myPhase}\left(2+\cos^2\myPhase\right)- \\
	&\myF{2}{\myPhase} \mySinEta_{2}^2  + 3 \myF{2}{\myPhase} \mySinEta_{2}^4  +\\
&3\cos\myPhase\left[\myKopalPhi{6}{2}{\myPhase}-\left( \piover2 - \myKopalPhi{2}{2}{\myPhase}\right) \mySinEta_{2}^5\right]\bigg).
\end{aligned}	 
\end{equation}

Finally for $ \mu^4 $ in the un-illuminated zone, 
\begin{equation}\label{eq:unlimbtoedgemu4}
\begin{aligned}
	\myKintegral{-\mySinEta_{2}}{\phi_{un,i}}{\phi_{limb,f}}{\mu^4}=\frac{R_p^2}{192}\bigg(&8\myF{2}{\myPhase}\mySinEta_{2}\left(3+\cos^2\myPhase\right)+16 \myF{2}{\myPhase} \mySinEta_{2}^3-\\
&64 \myF{2}{\myPhase} \mySinEta_{2}^5 +\myKopalPhi{1}{2}{\myPhase}(20 \cos (2 \myPhase)-\cos (4 \myPhase)+45)+\\
&\cos \myPhase \bigg[64\left( \myKopalPhi{2}{2}{\myPhase}-\piover2 \right)+\\
&\hspace{0.45in}45 \mySinEta_{2}(\myKopalPhi{4}{2}{\myPhase}-\myKopalPhi{3}{2}{\myPhase})+\\
&\hspace{0.45in}10\mySinEta_{2}^3(\myKopalPhi{4}{2}{\myPhase}-\myKopalPhi{3}{2}{\myPhase}) +\\
&\hspace{0.45in}9\mySinEta_{2}^5(\myKopalPhi{4}{2}{\myPhase}-\myKopalPhi{3}{2}{\myPhase}) +\\
&\hspace{0.45in}64\left(\piover2-\myKopalPhi{2}{2}{\myPhase}\right) \mySinEta_{2}^6   \bigg]\bigg).
\end{aligned}	 
\end{equation}

\end{appendices}

\backmatter
%
%
%
\bibliographystyle{hunsrt} 
\bibliography{exoplanetReferences1}

\begin{thebibliography}{10}

\bibitem{exoEncyclopedia}
The extrasolar planet encyclopedia.

\bibitem{exohandbook}
Michael Perryman.
\newblock {\em The Exoplanet Handbook}.
\newblock Cambridge University Press, 2011.

\bibitem{KeplerPrecision}
Ronald~L. Gilliland, William~J. Chaplin, Edward~W. Dunham, Vic~S. Argabright,
  William~J. Borucki, Gibor Basri, Stephen~T. Bryson, Derek~L. Buzasi,
  Douglas~A. Caldwell, Yvonne~P. Elsworth, Jon~M. Jenkins, David~G. Koch,
  Jeffrey Kolodziejczak, Andrea Miglio, Jeffrey van Cleve, Lucianne~M.
  Walkowicz, and William~F. Welsh.
\newblock Kepler mission stellar and instrument noise properties.
\newblock {\em The Astrophysical Journal Supplement Series}, 197(1):6, 2011.

\bibitem{Placek2014}
Benjamin Placek, Kevin~H. Knuth, and Daniel Angerhausen.
\newblock Exonest: Bayesian model selection applied to the detection and
  characterization of exoplanets via photometric variations.
\newblock {\em The Astrophysical Journal}, 795:112, 2014.

\bibitem{seager}
Sara Seager.
\newblock {\em Exoplanet Atmospheres: Physical Processes}.
\newblock Princeton University Press, 2010.

\bibitem{Kopal1953}
Zden{\v e}k {\'K}opal.
\newblock Photometric effects of reflection in close binary systems.
\newblock {\em Monthly Notices of the Royal Astronomical Society},
  114(1):101--117, 1954.

\bibitem{Kopal1959}
Zden{\v e}k {\'K}opal.
\newblock {\em Close Binary Systems}.
\newblock New York John Wiley \& Sons Inc., 1959.

\bibitem{Mandel2002}
K.~Mandel and E.~Agol.
\newblock Analytic lightcurves for planetary transit searches.
\newblock {\em The Astrophysical Journal}, 580:L171--L175, 2002.

\bibitem{PlacekThesis}
Ben Placek.
\newblock {\em Bayesian Detection and Characterization of Extra-solar Planets
  Via Photometric Variations}.
\newblock {Ph.D.} thesis, University at Albany (SUNY), 2014.

\bibitem{landauMech}
L.D. Landau and E.M. Lifshitz.
\newblock {\em Mechanics, Third Edition}.
\newblock Elsevier Butterworth-Heinemann, 1976.

\bibitem{SeagerMallen2003}
S.~Seager and G.~Mall{\v e}n-Ornelas.
\newblock A unique solution of planet and star parameters from an extrasolar
  planet transit light curve.
\newblock {\em The Astrophysical Journal}, 585 (2):1038, 2003.

\bibitem{2017revisedStar}
S.~{Mathur}, D.~{Huber}, N.~M. {Batalha}, D.~R. {Ciardi}, F.~A. {Bastien},
  A.~{Bieryla}, L.~A. {Buchhave}, W.~D. {Cochran}, M.~{Endl}, G.~A. {Esquerdo},
  E.~{Furlan}, A.~{Howard}, S.~B. {Howell}, H.~{Isaacson}, D.~W. {Latham},
  P.~J. {MacQueen}, and D.~R. {Silva}.
\newblock {Revised Stellar Properties of Kepler Targets for the Q1-17 (DR25)
  Transit Detection Run}.
\newblock {\em Astrophysical Journal}, 229:30, April 2017, 1609.04128.

\bibitem{sudarsky}
{David Sudarsky, Adam Burrows, Ivan Hubeny and Aigen Li}.
\newblock Phase functions and light curves of wide separation extrasolar giant
  planets.
\newblock {\em The Astrophysical Journal}, 627:520--533, 2005.

\bibitem{Claret2000}
A.~Claret.
\newblock A new non-linear limb-darkening law for lte stellar atmosphere
  models.
\newblock {\em Astronomy \& Astrophysics}, 363:1081--1190, 2000.

\bibitem{Borucki2009}
{W. J. Borucki, D. Koch, J. Jenkins, D. Sasselov, R. Gilliland, N. Batalha,
  D.W. Latham, D. Caldwell, G. Basri, T. Brown, and et al.}
\newblock \textit{Kepler}’s optical phase curve of the exoplanet hat-p-7b.
\newblock {\em Science}, 325:709--709, 2009.

\bibitem{KallrathMilone99}
J.~Kallrath and E.~F. Milone.
\newblock {\em Eclipsing Binary Stars: Modeling and Analysis}.
\newblock New York: Springer, 1999.

\bibitem{charbonneau05}
David Charbonneau, Lori~E. Allen, S.~Thomas Megeath, Guillermo Torres, Roi
  Alonso, Timothy~M. Brown, Ronald~L. Gilliland, David~W. Latham, Georgi
  Mandushev, Francis~T. O’Donovan, and Alessandro Sozzetti.
\newblock Detection of thermal emission from an extrasolar planet.
\newblock {\em The Astrophysical Journal}, 626(1):523, 2005.

\bibitem{ford08}
Eric~B. Ford, Samuel~N. Quinn, and Dimitri Veras.
\newblock Characterizing the orbital eccentricities of transiting extrasolar
  planets with photometric observations.
\newblock {\em The Astrophysical Journal}, 678:1407, 2008a.

\bibitem{TingleySackett}
B.~Tingley and P.D. Sackett.
\newblock A photometric diagnostic to aid in the identification of transiting
  extrasolar planets.
\newblock {\em The Astrophysical Journal}, 627:1011, 2005.

\bibitem{Ford06MCMC}
Eric~B. Ford.
\newblock Improving the efficiency of markov chain monte carlo for analyzing
  the orbits of extrasolar planets.
\newblock {\em The Astrophysical Journal}, 642(1):505, 2006.

\bibitem{BoruckiSummers1984}
W.~J. {Borucki} and A.~L. {Summers}.
\newblock {The photometric method of detecting other planetary systems}.
\newblock {\em Icarus}, 58:121--134, 1984.

\bibitem{Barnes2007}
Jason~W. Barnes.
\newblock Effects of orbital eccentricity on extrasolar planet transit
  detectability and light curves.
\newblock {\em Publications of the Astronomical Society of the Pacific},
  119(859):986, 2007.

\bibitem{2017Entropy}
Kevin~H. Knuth, Ben Placek, Daniel Angerhausen, Jennifer~L. Carter, Bryan
  D’Angelo, Anthony~D. Gai, and Bertrand Carado.
\newblock Exonest: The bayesian exoplanetary explorer.
\newblock {\em Entropy}, 19(10), 2017.

\bibitem{kipping2014}
David~M. Kipping.
\newblock Bayesian priors for the eccentricity of transiting planets.
\newblock {\em Monthly Notices of the Royal Astronomical Society}, 444(3):2263,
  2014.

\bibitem{rybickiandlightman}
George~B. Rybicki and Alan~P. Lightman.
\newblock {\em Radiative Processes in Astrophysics}.
\newblock John Wiley \& Sons, Inc., 2008.

\bibitem{Loeb2003}
A.~Loeb and S.~B. Gaudi.
\newblock Period flux variability of stars due to the reflex doppler effect
  induced by planetary companions.
\newblock {\em The Astrophysical Journal}, 588(2):L117--L120, 2003.

\bibitem{Morris1985}
S.~L. Morris.
\newblock The ellipsoidal variable stars.
\newblock {\em The Astrophysical Journal}, 295:143--152, 1985.

\bibitem{ClaretBloemen2011}
A.~Claret and S.~Bloemen.
\newblock Gravity and limb-darkening coefficients for the \textit{Kepler},
  corot, \textit{Spitzer}, \textit{uvby}, \textit{UBVRIJHK}, and sloan
  photometric systems.
\newblock {\em Astronomy \& Astrophysics}, 529:A75, 2011.

\bibitem{Faigler2010}
S.~Faigler and T.~Mazeh.
\newblock Detection of the ellipsoidal and the relativistic beaming effects in
  the corot-3 lightcurve.
\newblock {\em Astronomy \& Astrophysics}, 15550, 2010.

\bibitem{KaneGelino2012}
S.~R. Kane and D.~M. Gelino.
\newblock Distinguishing between stellar and planetary companions with phase
  monitoring.
\newblock {\em Monthly Notices of the Royal Astronomical Society},
  424(1):779–788, 2012.

\bibitem{Jackson2012}
B.~K. Jackson, N.K. Lewis, J.W. Barnes, L.D. Deming, A.P. Showman, and J.J.
  Fortney.
\newblock The evil-mc model for ellipsoidal variations of planet-hosting stars
  and applications to the hat-p-7 system.
\newblock {\em The Astrophysical Journal}, 751:112 (13pp), 2012.

\bibitem{gaiandknuth2018}
Anthony~D. Gai and Kevin~H. Knuth.
\newblock Bayesian model testing of ellipsoidal variations on stars due to hot
  jupiters.
\newblock {\em The Astrophysical Journal}, 853(1):49, 2018.

\bibitem{gaimasters}
Anthony~D. Gai.
\newblock Bayesian model testing of models for ellipsoidal variation on stars
  due to hot jupiters.
\newblock Master's thesis, University at Albany, Albany, NY, 2016.

\bibitem{sobolev}
V.~V. Sobolev.
\newblock {\em Light Scattering in Planetary Atmospheres}.
\newblock Oxford: Pergamon, 1975.

\bibitem{fairbairn}
M.B. Fairbairn and J.B.~Tatum (editor).
\newblock Max {F}airbairn's planetary photometry, 2005.

\bibitem{lilloBox2013}
J.~{Lillo-Box}, D.~{Barrado}, A.~{Moya}, B.~{Montesinos}, J.~{Montalb{\'a}n},
  A.~{Bayo}, M.~{Barbieri}, C.~{R{\'e}gulo}, L.~{Mancini}, H.~{Bouy}, and
  T.~{Henning}.
\newblock {Kepler-91b: a planet at the end of its life. Planet and giant host
  star properties via light-curve variations}.
\newblock {\em Astronomy \& Astrophysics}, 562:A109, 2014.

\bibitem{guillot2010}
{Guillot, T.}
\newblock On the radiative equilibrium of irradiated planetary atmospheres.
\newblock {\em Astronomy \& Astrophysics}, 520:A27, 2010.

\bibitem{2017robinson}
Tyler~D. Robinson.
\newblock A theory of exoplanet transits with light scattering.
\newblock {\em The Astrophysical Journal}, 836(2):236, 2017.

\bibitem{K91Trojan}
B.~{Placek}, K.~H. {Knuth}, D.~{Angerhausen}, and J.~M. {Jenkins}.
\newblock {Characterization of {K}epler-91b and the Investigation of a
  Potential Trojan Companion Using EXONEST}.
\newblock {\em The Astrophysical Journal}, 814:147, December 2015, 1511.01068.

\bibitem{lilloBox2014}
{Lillo-Box, J.}, {Barrado, D.}, {Henning, Th.}, {Mancini, L.}, {Ciceri, S.},
  {Figueira, P.}, {Santos, N. C.}, {Aceituno, J.}, and {Sánchez, S.}
\newblock Radial velocity confirmation of {K}epler-91 b - additional evidence
  of its planetary nature using the calar alto/cafe instrument.
\newblock {\em Astronomy \& Astrophysics}, 568:L1, 2014.

\bibitem{boas}
Mary~L. Boas.
\newblock {\em Mathematical Methods is the Physical Sciences}.
\newblock Wiley, 3 edition, 2006.

\bibitem{mathTable}
I.S. Gradshteyn, I.M. Ryzhik, and Alan~Jeffrey (editor).
\newblock {\em Table of Integrals, Series and Products, 5th ed.}
\newblock Academic Press, 1994.

\bibitem{leigh2003a}
C.~{Leigh}, A.~{Collier Cameron}, S.~{Udry}, J.-F. {Donati}, K.~{Horne},
  D.~{James}, and A.~{Penny}.
\newblock {A search for starlight reflected from HD 75289b}.
\newblock {\em Monthly Notices of the Royal Astronomical Society},
  346:L16--L20, 2003a.

\bibitem{Plato2Mission}
H.~{Rauer}, C.~{Catala}, C.~{Aerts}, T.~{Appourchaux}, W.~{Benz},
  A.~{Brandeker}, J.~{Christensen-Dalsgaard}, M.~{Deleuil}, L.~{Gizon}, M.-J.
  {Goupil}, M.~{G{\"u}del}, E.~{Janot-Pacheco}, M.~{Mas-Hesse}, I.~{Pagano},
  G.~{Piotto}, D.~{Pollacco}, {\.C}.~{Santos}, A.~{Smith}, J.-C. {Su{\'a}rez},
  R.~{Szab{\'o}}, S.~{Udry}, V.~{Adibekyan}, Y.~{Alibert}, J.-M. {Almenara},
  P.~{Amaro-Seoane}, M.~A.-v. {Eiff}, M.~{Asplund}, E.~{Antonello},
  S.~{Barnes}, F.~{Baudin}, K.~{Belkacem}, M.~{Bergemann}, G.~{Bihain}, A.~C.
  {Birch}, X.~{Bonfils}, I.~{Boisse}, A.~S. {Bonomo}, F.~{Borsa}, I.~M.
  {Brand{\~a}o}, E.~{Brocato}, S.~{Brun}, M.~{Burleigh}, R.~{Burston},
  J.~{Cabrera}, S.~{Cassisi}, W.~{Chaplin}, S.~{Charpinet}, C.~{Chiappini},
  R.~P. {Church}, S.~{Csizmadia}, M.~{Cunha}, M.~{Damasso}, M.~B. {Davies},
  H.~J. {Deeg}, R.~F. {D{\'{\i}}az}, S.~{Dreizler}, C.~{Dreyer},
  P.~{Eggenberger}, D.~{Ehrenreich}, P.~{Eigm{\"u}ller}, A.~{Erikson},
  R.~{Farmer}, S.~{Feltzing}, F.~{de Oliveira Fialho}, P.~{Figueira},
  T.~{Forveille}, M.~{Fridlund}, R.~A. {Garc{\'{\i}}a}, P.~{Giommi},
  G.~{Giuffrida}, M.~{Godolt}, J.~{Gomes da Silva}, T.~{Granzer}, J.~L.
  {Grenfell}, A.~{Grotsch-Noels}, E.~{G{\"u}nther}, C.~A. {Haswell}, A.~P.
  {Hatzes}, G.~{H{\'e}brard}, S.~{Hekker}, R.~{Helled}, K.~{Heng}, J.~M.
  {Jenkins}, A.~{Johansen}, M.~L. {Khodachenko}, K.~G. {Kislyakova}, W.~{Kley},
  U.~{Kolb}, N.~{Krivova}, F.~{Kupka}, H.~{Lammer}, A.~F. {Lanza},
  Y.~{Lebreton}, D.~{Magrin}, P.~{Marcos-Arenal}, P.~M. {Marrese}, J.~P.
  {Marques}, J.~{Martins}, S.~{Mathis}, S.~{Mathur}, S.~{Messina}, A.~{Miglio},
  J.~{Montalban}, M.~{Montalto}, M.~J.~P.~F.~G. {Monteiro}, H.~{Moradi},
  E.~{Moravveji}, C.~{Mordasini}, T.~{Morel}, A.~{Mortier}, V.~{Nascimbeni},
  R.~P. {Nelson}, M.~B. {Nielsen}, L.~{Noack}, A.~J. {Norton}, A.~{Ofir},
  M.~{Oshagh}, R.-M. {Ouazzani}, P.~{P{\'a}pics}, V.~C. {Parro}, P.~{Petit},
  B.~{Plez}, E.~{Poretti}, A.~{Quirrenbach}, R.~{Ragazzoni}, G.~{Raimondo},
  M.~{Rainer}, D.~R. {Reese}, R.~{Redmer}, S.~{Reffert}, B.~{Rojas-Ayala},
  I.~W. {Roxburgh}, S.~{Salmon}, A.~{Santerne}, J.~{Schneider}, J.~{Schou},
  S.~{Schuh}, H.~{Schunker}, A.~{Silva-Valio}, R.~{Silvotti}, I.~{Skillen},
  I.~{Snellen}, F.~{Sohl}, S.~G. {Sousa}, A.~{Sozzetti}, D.~{Stello}, K.~G.
  {Strassmeier}, M.~{{\v S}vanda}, G.~M. {Szab{\'o}}, A.~{Tkachenko},
  D.~{Valencia}, V.~{Van Grootel}, S.~D. {Vauclair}, P.~{Ventura}, F.~W.
  {Wagner}, N.~A. {Walton}, J.~{Weingrill}, S.~C. {Werner}, P.~J. {Wheatley},
  and K.~{Zwintz}.
\newblock {The PLATO 2.0 mission}.
\newblock {\em Experimental Astronomy}, 38:249--330, November 2014, 1310.0696.

\bibitem{KeplerHandbook}
J.~E.~Van Cleve and D.~A. Caldwell.
\newblock Kepler: A search for terrestrial planets.
\newblock {\em Kepler Instrument Handbook}, NASA/Ames Research Center, Moffett
  Field, California:KSCI--19033, 2009.

\bibitem{Placek2015}
Ben Placek, Kevin~H. Knuth, Daniel Angerhausen, and Jon~M. Jenkins.
\newblock Characterization of {K}epler-91b and the investigation of a potential
  trojan companion using exonest.
\newblock {\em The Astrophysical Journal}, 814(2):147, 2015.

\bibitem{Cowan2017}
Nicolas~B. Cowan and Yuka Fujii.
\newblock Mapping exoplanets.
\newblock In Hans~J. Deeg and Juan~Antonio Belmonte, editors, {\em Handbook of
  Exoplanet}, pages 1--16. Springer International Publishing, 2017.

\bibitem{siviaSkilling}
D.S. Sivia and J.~Skilling.
\newblock {\em Data Analysis: A Bayesian Tutorial, ed. 2}.
\newblock Oxford University Press, 2006.

\bibitem{planetPlanetOccultation}
Rodrigo Luger, Jacob Lustig-Yaeger, and Eric Agol.
\newblock Planet–planet occultations in trappist-1 and other exoplanet
  systems.
\newblock {\em The Astrophysical Journal}, 851(2):94, 2017.

\bibitem{gaiaStarRadii}
T.~A. {Berger}, D.~{Huber}, E.~{Gaidos}, and J.~L. {van Saders}.
\newblock {Revised Radii of Kepler Stars and Planets Using Gaia Data Release
  2}.
\newblock {\em ArXiv e-prints}, May 2018, 1805.00231.

\end{thebibliography}
\addcontentsline{toc}{section}{{\bfseries Bibliography}}

\end{document}